\documentclass[twocolumn]{aa}
\pdfoutput=1                       
\usepackage[multidot]{grffile}     
\usepackage{graphicx}
\usepackage{txfonts}
\usepackage{natbib}

\bibpunct{(}{)}{;}{a}{}{,}

\begin{document}

\title
{
Uncertainties and biases of source masses derived\\
from fits of integrated fluxes or image intensities
}


\author
{
A.~Men'shchikov 
}


\institute
{
Laboratoire AIM Paris--Saclay, CEA/DSM--CNRS--Universit{\'e} Paris Diderot, IRFU, Service d'Astrophysique, Centre d'Etudes de 
Saclay, Orme des Merisiers, 91191 Gif-sur-Yvette, France\\
\email{alexander.menshchikov@cea.fr}
}

\date{Received 14 January 2016 / Accepted 12 June 2016}

\offprints{Alexander Men'shchikov}
\titlerunning{Uncertainties and biases of derived masses}
\authorrunning{Men'shchikov}


\abstract{ 

Fitting spectral distributions of total fluxes or image intensities are two standard methods for estimating the masses of starless
cores and protostellar envelopes. These mass estimates, which are the main source and basis of our knowledge of the origin and
evolution of self-gravitating cores and protostars, are uncertain. It is important to clearly understand sources of statistical and
systematic errors stemming from the methods and minimize the errors. In this model-based study, a grid of radiative transfer models
of starless cores and protostellar envelopes was computed and their total fluxes and image intensities were fitted to derive the
model masses. To investigate intrinsic effects related to the physical objects, all observational complications were explicitly
ignored. Known true values of the numerical models allow assessment of the qualities of the methods and fitting models, as well as
the effects of nonuniform temperatures, far-infrared opacity slope, selected subsets of wavelengths, background subtraction, and
angular resolutions. The method of fitting intensities gives more accurate masses for more resolved objects than the method of
fitting fluxes. With the latter, a fitting model that assumes optically thin emission gives much better results than the one
allowing substantial optical depths. Temperature excesses within the objects above the mass-averaged values skew their spectral
shapes towards shorter wavelengths, leading to masses underestimated typically by factors $2\,{-}\,5$. With a fixed opacity slope
deviating from the true value by a factor of $1.2$, masses are inaccurate within a factor of $2$. The most accurate masses are
estimated by fitting just two or three of the longest wavelength measurements. Conventional algorithm of background subtraction is
a likely source of large systematic errors. The absolute values of masses of the unresolved or poorly resolved objects in
star-forming regions are uncertain to within at least a factor of $2\,{-}\,3$.

}


\maketitle


\section{Introduction}
\label{introduction}

Significant technological advances in the astronomical instrumentation during the last four decades enabled measurements of the
far-infrared thermal dust emission (usually optically thin in that wavelength range) and hence estimates of the masses of dusty
objects. Fitting the far-infrared and submillimeter flux or intensity distributions of optically thin sources can give their
average temperatures and masses \citep{Hildebrand1983}. This simple method has become standard in studies of Galactic star
formation and a major source of our knowledge of the physical properties and evolution of self-gravitating cores and protostars.
Although there are more sophisticated approaches \citep[e.g.,][]{Kelly_etal2012}, a simple fitting of the observed spectral shapes
remains the most widely used method in the observational studies of star formation \citep[e.g.,][]{{Ko"nyves_etal2015}}. Its
inaccuracies, biases, and limitations need to be carefully investigated before reliable conclusions can be made on the physical
properties and evolution of the observed objects.

Mass derivation from fitting total fluxes or pixel intensities involves a strong assumption of a constant temperature within an
object. In addition to such poorly known parameters as the distance, the far-infrared opacity and its power-law slope, and the
dust-to-gas mass ratio, the most problematic assumption is that a single color temperature obtained from the fitting is a good
approximation of the mass-averaged physical dust temperatures. This may be true for only the simplest case of the lowest density
starless cores, transparent in the visible wavelength range and thus practically isothermal, but it is clearly invalid for the
protostellar envelopes that are centrally heated by accretion luminosity. Very sensitive dependence of the emission of dust grains
on their temperature warrants careful investigation of the effects of nonuniform temperatures. There are papers that have
investigated some aspects of the problem, notably the correlation between the estimated temperatures and power-law opacity slopes
\citep[e.g.,][and references therein]{Shetty_etal2009a,Shetty_etal2009b,JuvelaYsard2012a} and the inaccuracies of mass derivation
and their effect on the resulting core mass function \citep[][]{Malinen_etal2011}.

The present purely model-based study simplifies the problem by removing the ``observational layer'' between the physical reality
and observers. To investigate the intrinsic effects related to the physical objects, all observational intricacies (the complex
filamentary backgrounds, instrumental noise, calibration errors, different angular resolutions across wavebands, etc.) were
explicitly ignored. Measurement errors in intensities and fluxes are assumed to be nonexistent and the radiation emitted by the
model objects is known to a high precision, limited only by their numerical accuracy. A grid of radiative transfer models of
starless cores and protostellar envelopes was computed and their total fluxes and image intensities were fitted to derive the model
masses. Known true values of the numerical models allow us to assess the qualities of the methods and fitting models, as well as
the effects of nonuniform temperatures, far-infrared opacity slope, selected subsets of wavelengths, background subtraction, and
angular resolutions. The main goal was to quantify how much the mass derivation methods are affected, what the realistic
uncertainties of the temperatures and masses are, and what one could possibly do to improve the estimates. Although this study is
completely independent of the instruments and wavebands used in actual observations, it employs six \emph{Herschel} wavebands
\citep[70, 100, 160, 250, 350, and 500\,{${\mu}$m};][]{Pilbratt_etal2010}, for which a wealth of recent results in star formation
has been obtained \citep[e.g.,][and references therein]{Ko"nyves_etal2015}.

\begin{figure}  
\centering
\centerline{\resizebox{0.695\hsize}{!}{\includegraphics{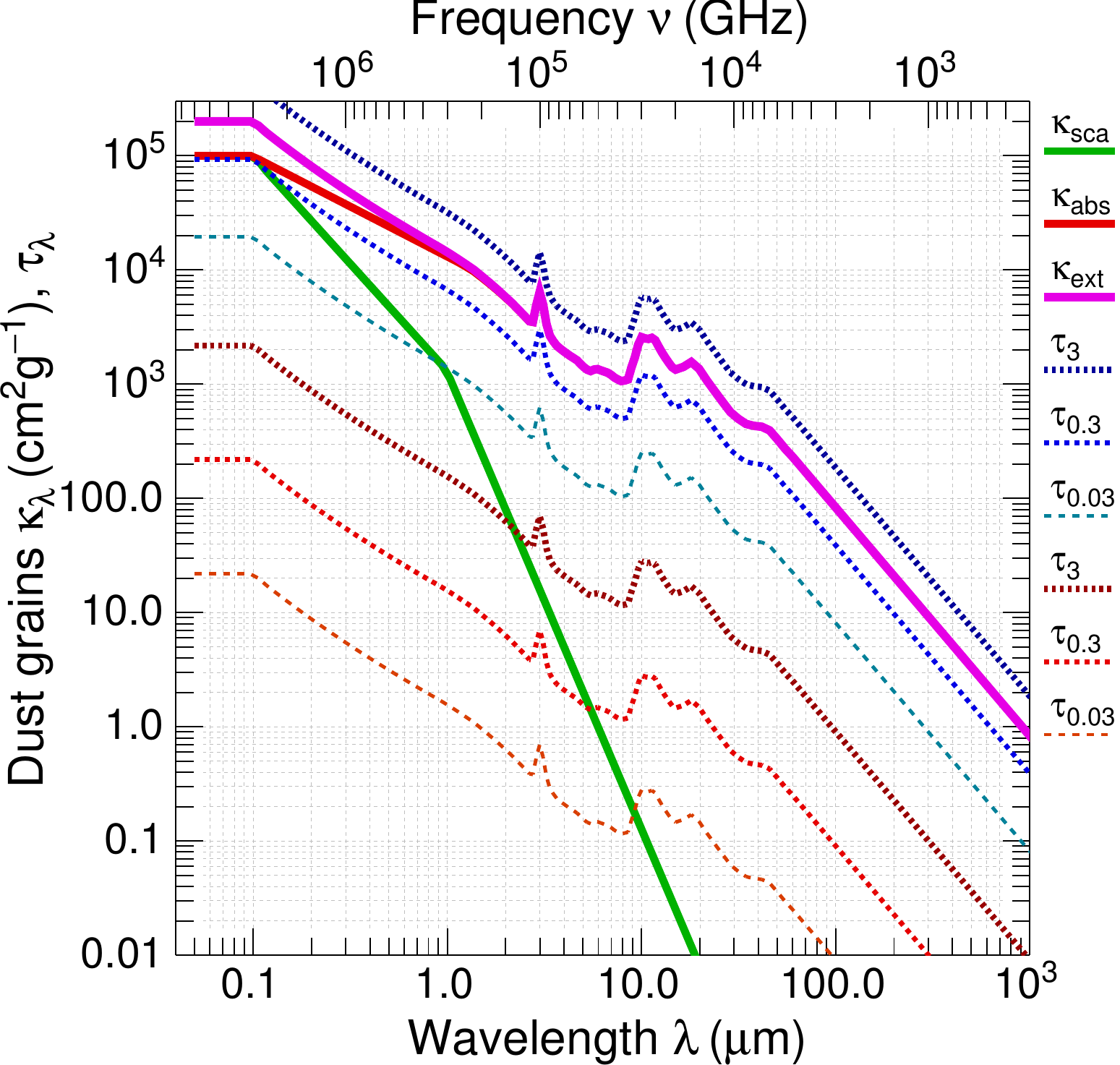}}}
\caption{
Opacities of grains (per gram of dust) and model radial optical depths. Subscripts on the curve labels ($n_{30}$ to $n_{0.03}$) 
indicate the model mass $M$ (in $M_{\sun}$). The wavelength dependence of $\kappa_{\rm sca}$, $\kappa_{\rm abs}$, and $\kappa_{\rm 
ext}$ is shown by thick solid lines (see Sect.~\ref{dust.properties} for details). The extinction optical depths $\tau_{\rm ext}$ 
of the protostellar envelopes ($L_{\star}\,{=}\,0.3\,L_{\sun}$) and starless cores are indicated respectively by the three sets of 
dashed blue and red lines. 
} 
\label{loo}
\end{figure}

The radiative transfer models of starless cores and protostellar envelopes are presented in Sect.~\ref{rtmodels}, the methods of
mass derivation from fitting far-infrared and submillimeter observations are introduced in Sect.~\ref{fitting.methods}, the results
of this work are presented in Sect.~\ref{results} and discussed in Sect.~\ref{discussion}, the conclusions are outlined in
Sect.~\ref{conclusions}, and further details are found in Appendices \ref{AppendixA}\,{--}\,\ref{AppendixE}.


\section{Radiative transfer models}
\label{rtmodels}

The models were computed with the 3D Monte Carlo radiative transfer code \textsl{RADMC-3D} by C.\,Dullemond\footnote{
\url{http://www.ita.uni-heidelberg.de/~dullemond/software/radmc-3d}}. Spherical model geometry was chosen to simplify the problem
by reducing the number of free parameters involved in the study: asymmetries in model density distribution would introduce
dependence on viewing angle \citep[e.g.,][]{Men'shchikovHenning1997,Men'shchikov_etal1999,Stamatellos_etal2004} and hence increase
the uncertainties of derived parameters. Isotropic scattering by dust grains was considered.

Grids of models for starless cores and protostellar envelopes were constructed, covering the ranges of masses $M$ ($0.03\,{-}\,30$
$M_{\sun}$) and luminosities $L$ ($0.03\,{-}\,30$\,$L_{\sun}$) relevant for both low- and intermediate-mass star formation. The
masses and luminosities were sampled at the values of $0.0316, 0.1, 0.316, 1, 3.16, 10, 31.6$ (separated by a factor of
$\sqrt{10}$); for simplicity, they will be referred to as $0.03, 0.1, 0.3, 1, 3, 10, 30$ ($M_{\sun}$, $L_{\sun}$). Although the
luminosity of an accreting protostar depends on its mass, the goal is to separate the effects of masses and luminosities.

In addition to isolated models, their embedded variants were constructed by implanting the isolated models into the centers of
larger spherical background shells of uniform densities, in order to simulate the fact that stars form within their dense parental
clouds that shield the embedded objects from the interstellar radiation field. All models were put at a distance $D{\,=\,}140$ pc
of the nearest star-forming regions.

\begin{figure}
\centering
\centerline{\resizebox{0.695\hsize}{!}{\includegraphics{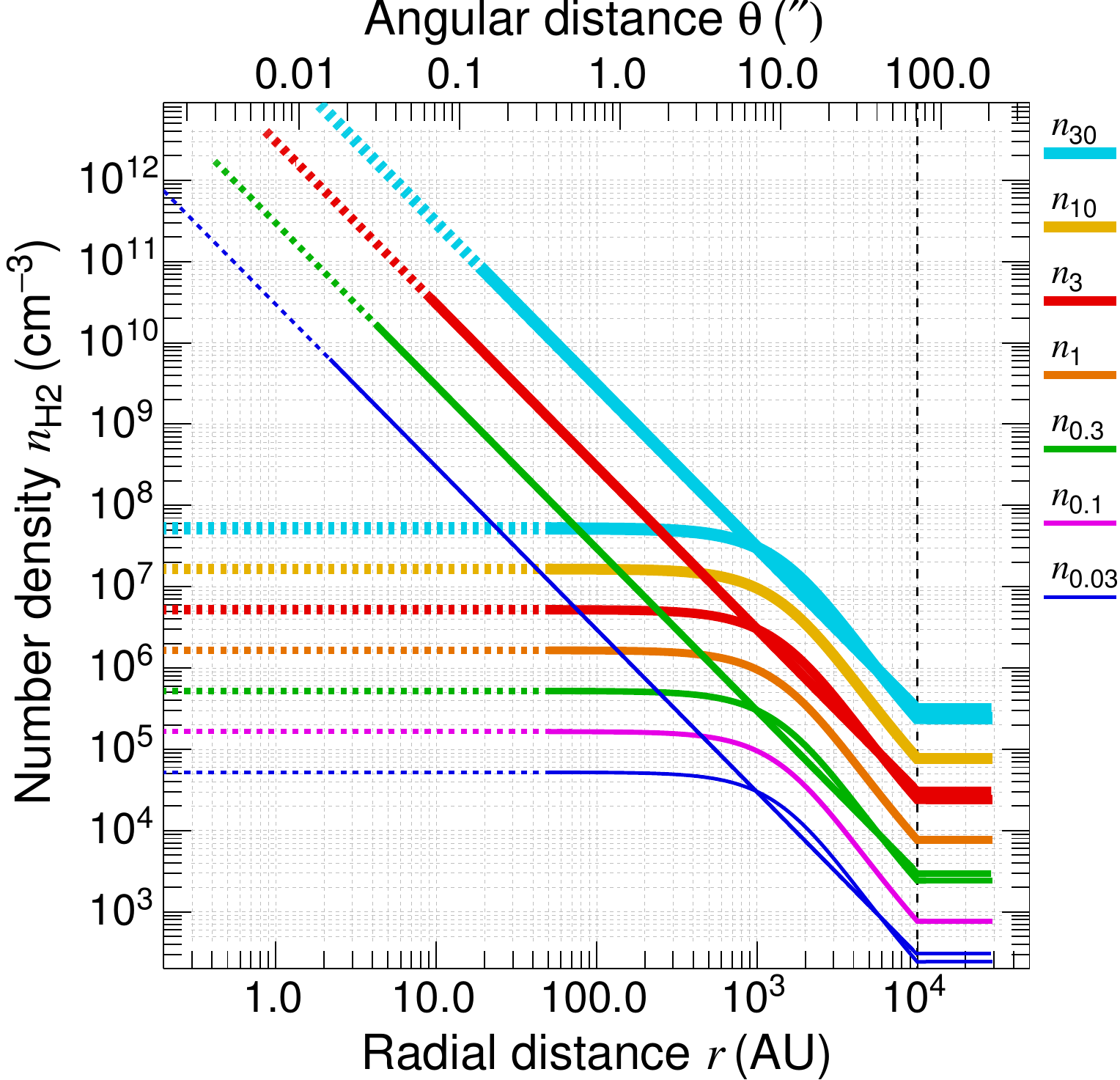}}}
\caption{
Density structure of the model starless cores and protostellar envelopes. Subscripts on the curve labels ($n_{30}$ to $n_{0.03}$)
indicate the model mass $M$ (in $M_{\sun}$). The dashed vertical line shows the outer boundary radius $R\,{=}\,10^{4}$\,AU for all
models. Embedded models are implanted in larger uniform-density clouds with an outer boundary at $3\,{\times}\,10^{4}$\,AU. The
dashed horizontal lines continue the densities of starless cores within the innermost radial zone. The dashed diagonal lines 
continue the densities of protostellar envelopes towards the radius of the inner dust-free cavity $R_{0}$, whose size depends on 
the model temperature profile $T_{\rm d}(r)$ (cf. Fig.~\ref{trp.bes.pro}) and adopted dust sublimation temperature ($T_{\rm 
S}\,{=}$ $\,10^{3}$\,K). In other words, the dashed diagonal lines visualize the range of densities and radial distances over which
the inner boundary $R_{0}$ is located for $L_{\star}$ spanning the entire range $0.03\,{-}\,30$\,$L_{\sun}$ (see 
Eq.~(\ref{inner.boundary})).
} 
\label{drp.bes.pro}
\end{figure}

\subsection{Dust properties}
\label{dust.properties}

Properties of the real astrophysical dust grains are poorly known and they are unlikely to be universal in the different
star-forming regions observed. The standard mass derivation methods ignore many complications related to the cosmic dust grains,
assuming just a simple power-law opacity across all bands being fitted. For example, the presence of very small, stochastically
heated grains is neglected \citep[e.g.,][]{Desert_etal1990}; the contribution of these grains to the emission of starless cores and
protostellar envelopes can become significant at $\lambda\,{\la}\,100$\,{${\mu}$m}
\citep[e.g.,][]{Bernard_etal1992,Siebenmorgen_etal1992}. For consistency with the mass derivation methods and previous studies of
star formation, this model study adopts tabulated absorption opacities $\kappa_{\rm abs}$ for grains with thin ice mantles
\citep{OssenkopfHenning1994}, corresponding to coagulation time $t\,{=}\,10^5$\,yr and number density $n_{\rm
H}\,{=}\,10^6$\,cm$^{-3}$ (Fig.~\ref{drp.bes.pro}).

The opacity values at long wavelengths $\lambda\,{>}\,70$\,{${\mu}$m} were replaced with a power law
$\kappa_{\lambda}\,{\propto}\,\lambda^{-2}$; the modification aimed at testing the widely used assumption on the power-law
far-infrared opacities $\kappa_{\lambda}\,{=}\,\kappa_{0}\left(\lambda_0/\lambda\right)^{\,\beta}$. At short wavelengths
(0.1$\,{<}\,\lambda\,{<}\,1$\,{${\mu}$m}), the opacities were extrapolated with a power law
$\kappa_{\lambda}\,{\propto}\,\lambda^{-0.87}$ based on the last tabulated values. Although dust scattering is unimportant in the
far-infrared, scattering opacities were constructed to resemble the values and wavelength dependence $\kappa_{\rm
sca}\,{\propto}\,\lambda^{-4}$ of typical dust grains. The resulting dust opacity at $\lambda\,{>}\,70$\,{${\mu}$m} was
parameterized by $\kappa_0\,{=}\,$9.31\,cm$^{2}$g$^{-1}$ (per gram of dust), $\lambda_0\,{=}\,$300\,{${\mu}$m}, and
$\beta\,{=}\,2$, with the maximum opacities limited by $10^5$\,cm$^{2}$g$^{-1}$ (Fig.~\ref{loo}).

\begin{figure*}
\centering
\centerline{\resizebox{0.3386\hsize}{!}{\includegraphics{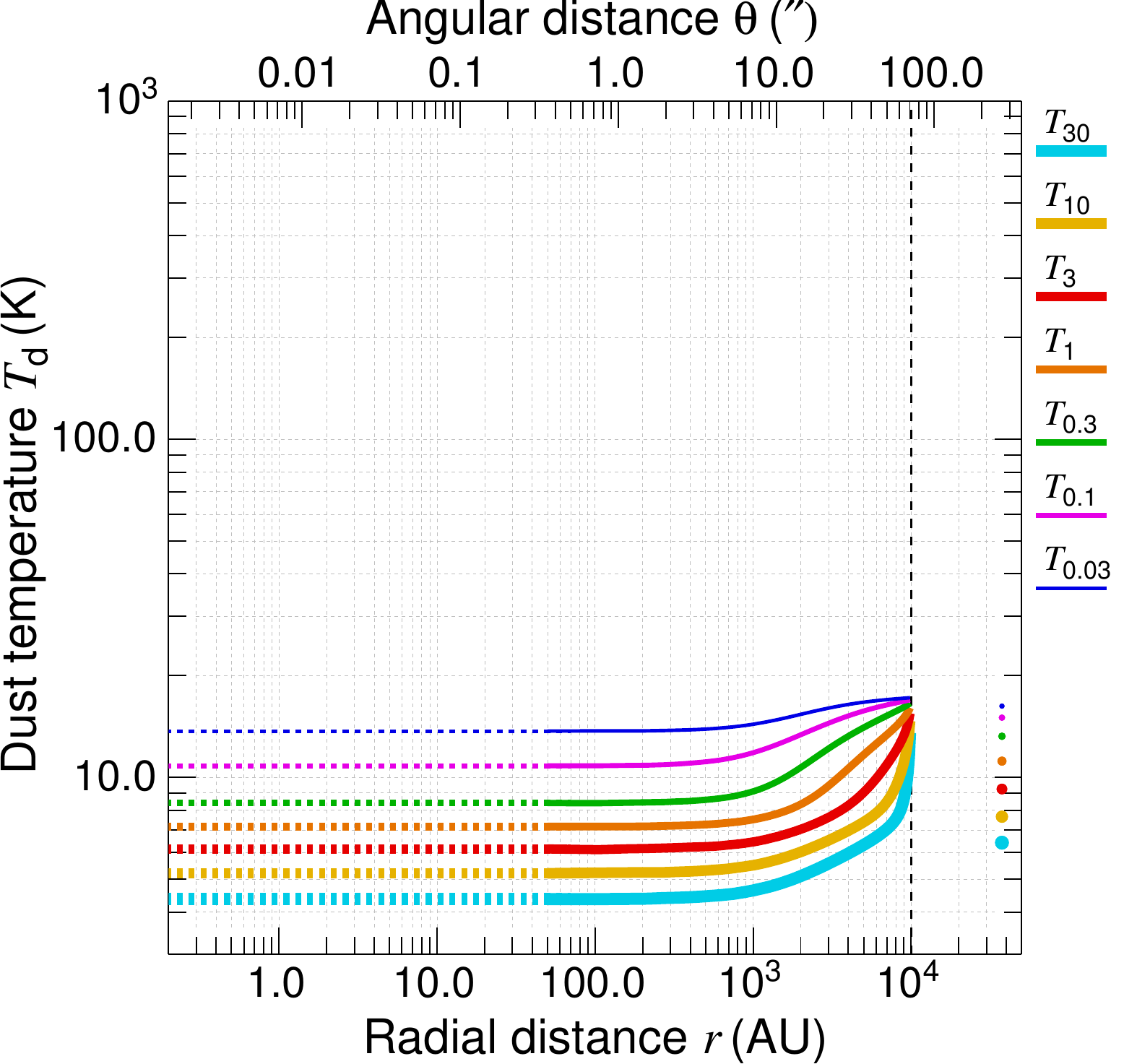}}
            \resizebox{0.3204\hsize}{!}{\includegraphics{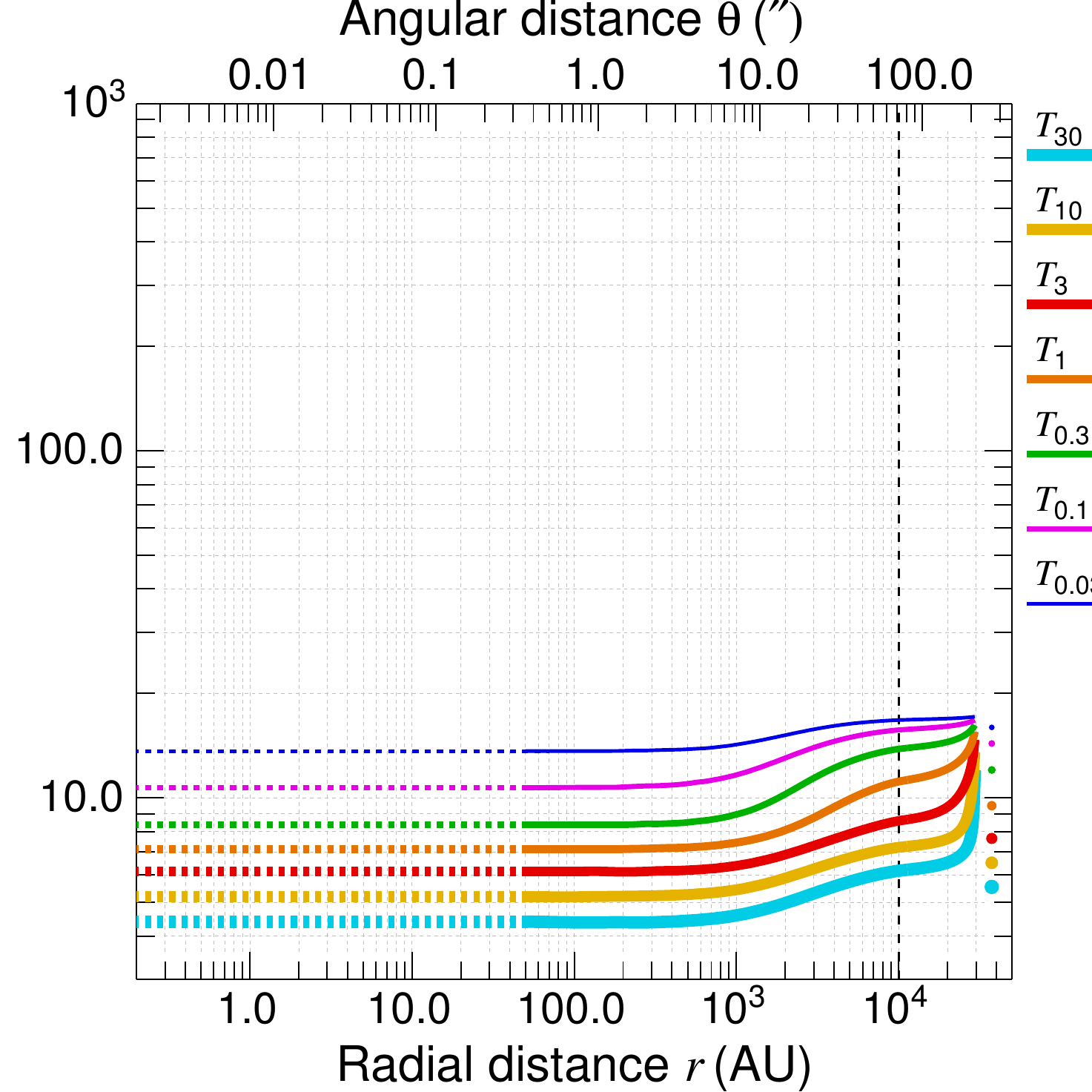}}}
\centerline{\resizebox{0.3386\hsize}{!}{\includegraphics{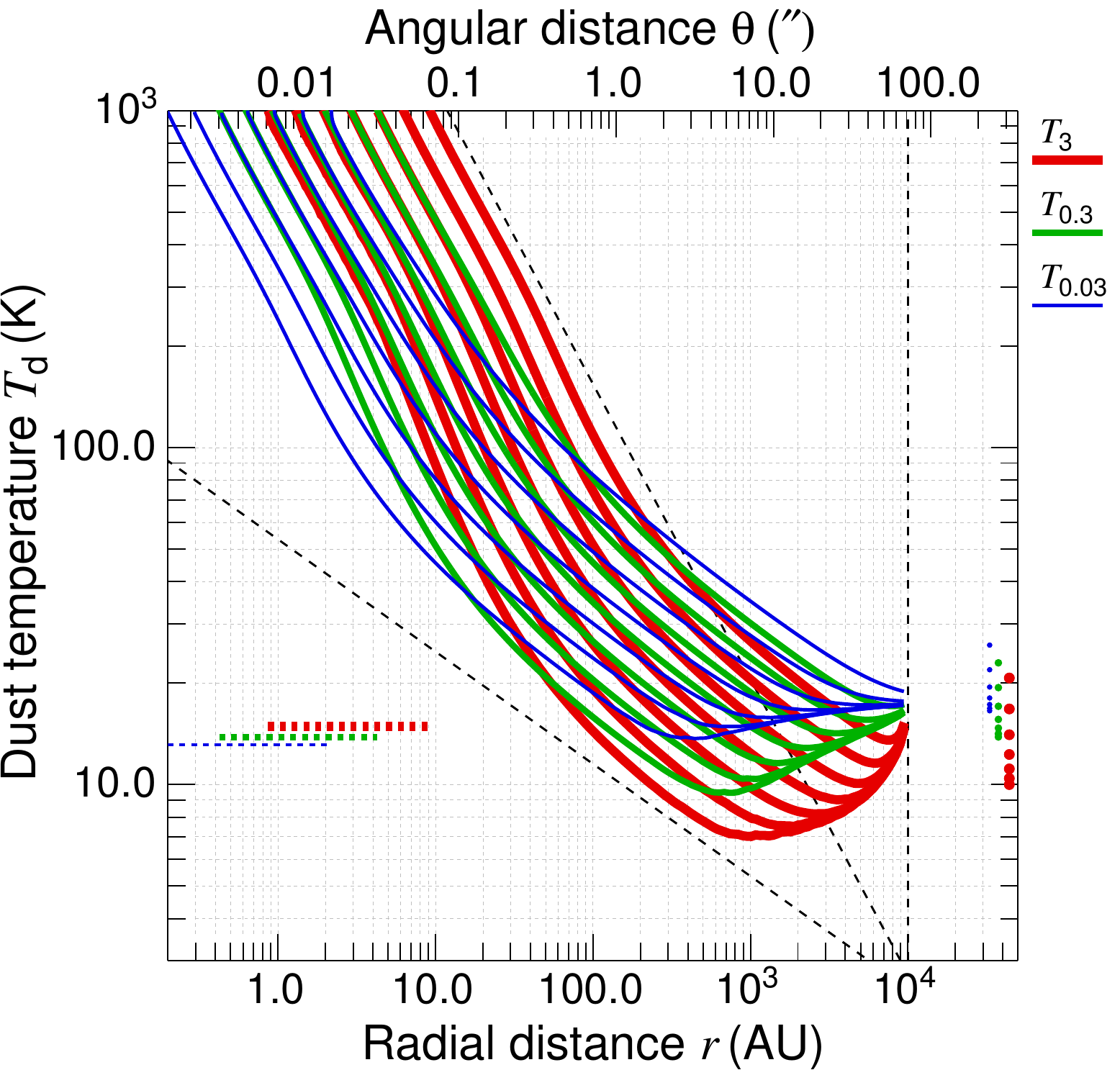}}
            \resizebox{0.3204\hsize}{!}{\includegraphics{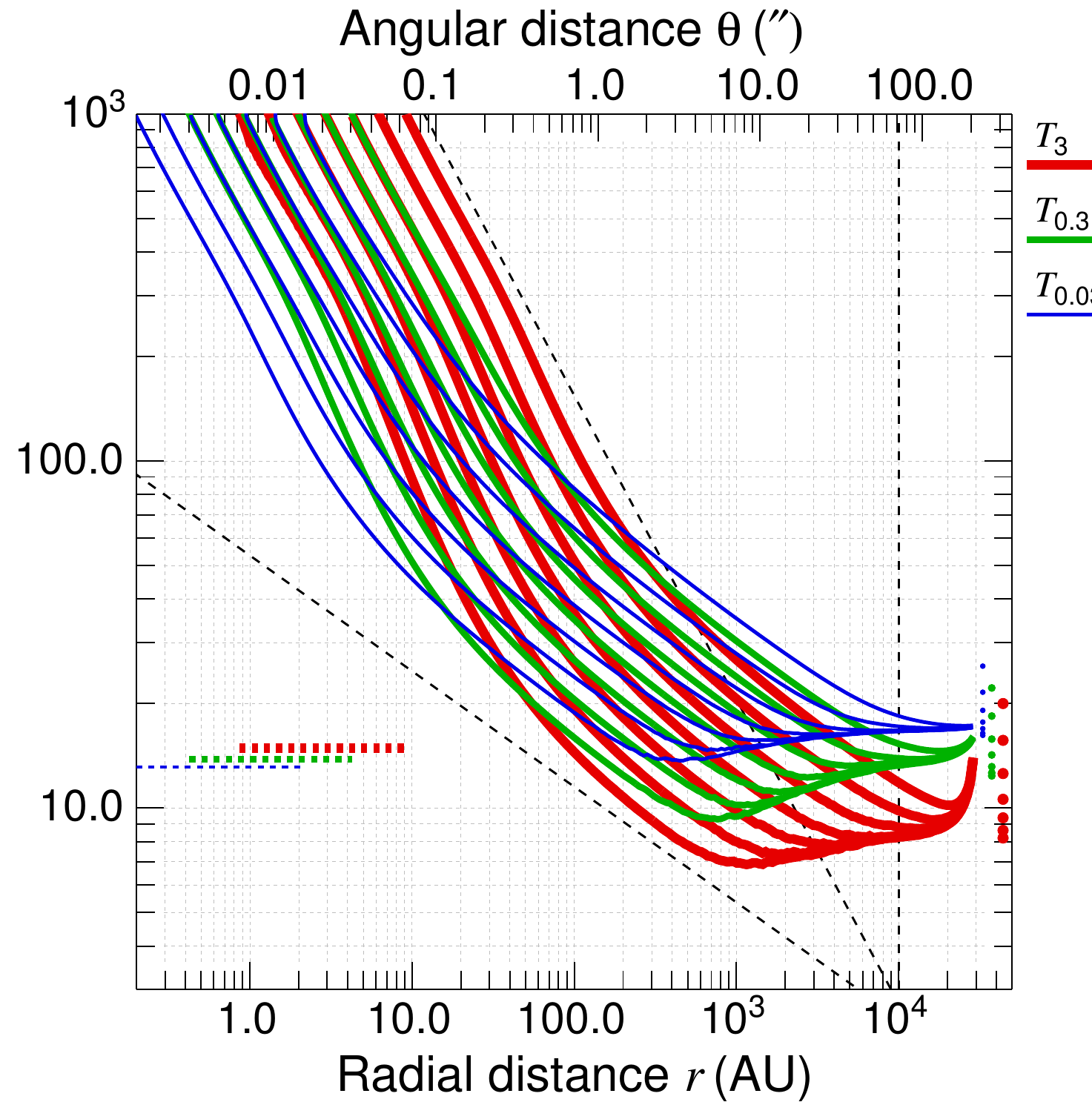}}}
\caption{
Radiative-equilibrium dust temperature profiles of starless cores (\emph{upper}) and protostellar envelopes (\emph{lower}) for the
isolated models (\emph{left}) and their embedded variants (\emph{right}). Subscripts of the curve labels ($T_{30}$ to $T_{0.03}$)
indicate the model mass $M$ (in $M_{\sun}$). The dashed horizontal lines in the upper panels continue the profiles of starless cores
within the innermost radial zone. The dashed vertical line shows the outer boundary radius $R\,{=}\,10^{4}$\,AU of all models. The
maximum temperature and the inner boundary of the dusty protostellar envelopes are defined by the adopted dust sublimation
temperature $T_{\rm S}\,{=}\,10^{3}$\,K. Three dashed horizontal lines in the lower panels indicate the range of radial distances
over which the boundaries $R_{0}$ of the dust-free cavities are located for the model luminosities in the entire range of
$0.03\,{-}\,30$\,$L_{\sun}$ (see Fig.~\ref{drp.bes.pro}). For protostellar envelopes with the same $M$, the ``hotter'' profiles
correspond to higher $L_\star$, larger $R_{0}$ (cf. Eq.~(\ref{inner.boundary})), and lower radial optical depths
$\tau_{\lambda}\,{\propto}\,L_{\star}^{-1/3}$. Two dashed lines bracketing the profiles of protostellar envelopes indicate the
slopes $T_{\rm d}(r)\,{\propto}\,r^{-0.88}$ and ${\propto}\,r^{-1/3}$, the latter describing the temperatures of a
transparent dusty envelope with the adopted grain properties (see also Appendix~\ref{AppendixA}). Vertically aligned filled circles
indicate the mass-averaged temperature $T_{M}$ for each model, defined by Eq.~(\ref{mass.averaged}). The slight wiggling of some
profiles around their minima reflects the discrete nature of the Monte Carlo radiative transfer method.
} 
\label{trp.bes.pro}
\end{figure*}

\subsection{Density distributions}
\label{density.profiles}

The density structure of starless cores was approximated by an isothermal Bonnor-Ebert sphere \citep{Bonnor1956} with a temperature
of $7$\,{K} and a central density of $5.2\,{\times}\,10^{-18}$\,g\,cm$^{-3}$. This somewhat arbitrary choice of $\rho(r)$ gives
just a simple and convenient functional form (Fig.~\ref{drp.bes.pro}) resembling the observed flat-topped density profiles of
starless cores \citep[e.g.,][]{Alves_etal2001,Evans_etal2001}. The issue of the gravitational instability (or stability) of the
model cores is irrelevant for this study of the mass derivation methods. Protostellar envelopes were modeled as infalling spherical
envelopes with the power-law densities $\rho(r)\,{\propto}\,r^{-2}$ \citep[e.g.,][]{Larson1969,Shu_1977} around a central source of
accretion energy (Fig.~\ref{drp.bes.pro}).

Model dust densities were scaled to obtain the desired grid of masses $0.03, 0.1, 0.3, 1, 3, 10$, and $30$\,$M_{\sun}$ using the
standard dust-to-gas mass ratio $\eta\,{=}\,0.01$. The outer boundary of all the models was placed at the same distance of
$R\,{=}\,10^{4}$\,AU, beyond which their density either changed to zero (isolated models) or remained constant until $R_{\rm
E}\,{=}\,3\,{\times}\,R$ (embedded models). The embedding cloud density was set equal to $\rho(R)$ (Fig.~\ref{drp.bes.pro}), which
corresponds to the denser models (i.e., more massive) being formed in a denser environment. Most of the mass of the model starless
cores and protostellar envelopes ($96{\%}$ and $90{\%}$, respectively) is contained in their outer parts ($0.1\,R\,{<}\,r\,{<}\,R$).

For the starless cores, the inner boundary was arbitrarily set to $R_{0}\,{=}\,50$\,AU, as their densities are essentially constant
and hence do not need to be resolved at smaller radii. The inner boundary of the dusty protostellar envelopes is defined by the
dust sublimation temperature $T_{\rm S}\,{\sim}\,10^{3}$\,K. An exact value of $T_{\rm S}$ depends on the chemical composition and
sizes of dust grains and so does the radius $R_{0}$ of the inner dust-free cavity. For the purpose of this study, it is adequate to
adopt a single value $T_{\rm S}\,{=}\,10^{3}$\,K. With the model $\kappa_{\nu}$ and $\rho(r)$ (Sect.~\ref{dust.properties}), the
resulting radiative-equilibrium temperatures (Fig.~\ref{trp.bes.pro}) lead to the inner boundaries of the dusty protostellar
envelopes that are fairly accurately described by a simple formula,
\begin{equation}
R_{0} = 2 \left[\left(M/M_{\sun}\right) \left(L_{\star}/L_{\sun}\right)\right]^{1/3} {\rm AU}.
\label{inner.boundary}
\end{equation}

The model space between the inner and outer boundaries was discretized by nonuniform grids with the relative zone sizes
$\delta\log{r}$ that smoothly varied from $0.6$ to $0.02$ ($100\,{-}\,150$ zones) for starless cores and from $0.002$ to $0.06$
($200\,{-}\,300$ zones) for protostellar envelopes.

\begin{figure*}
\centering
\centerline{\resizebox{0.3327\hsize}{!}{\includegraphics{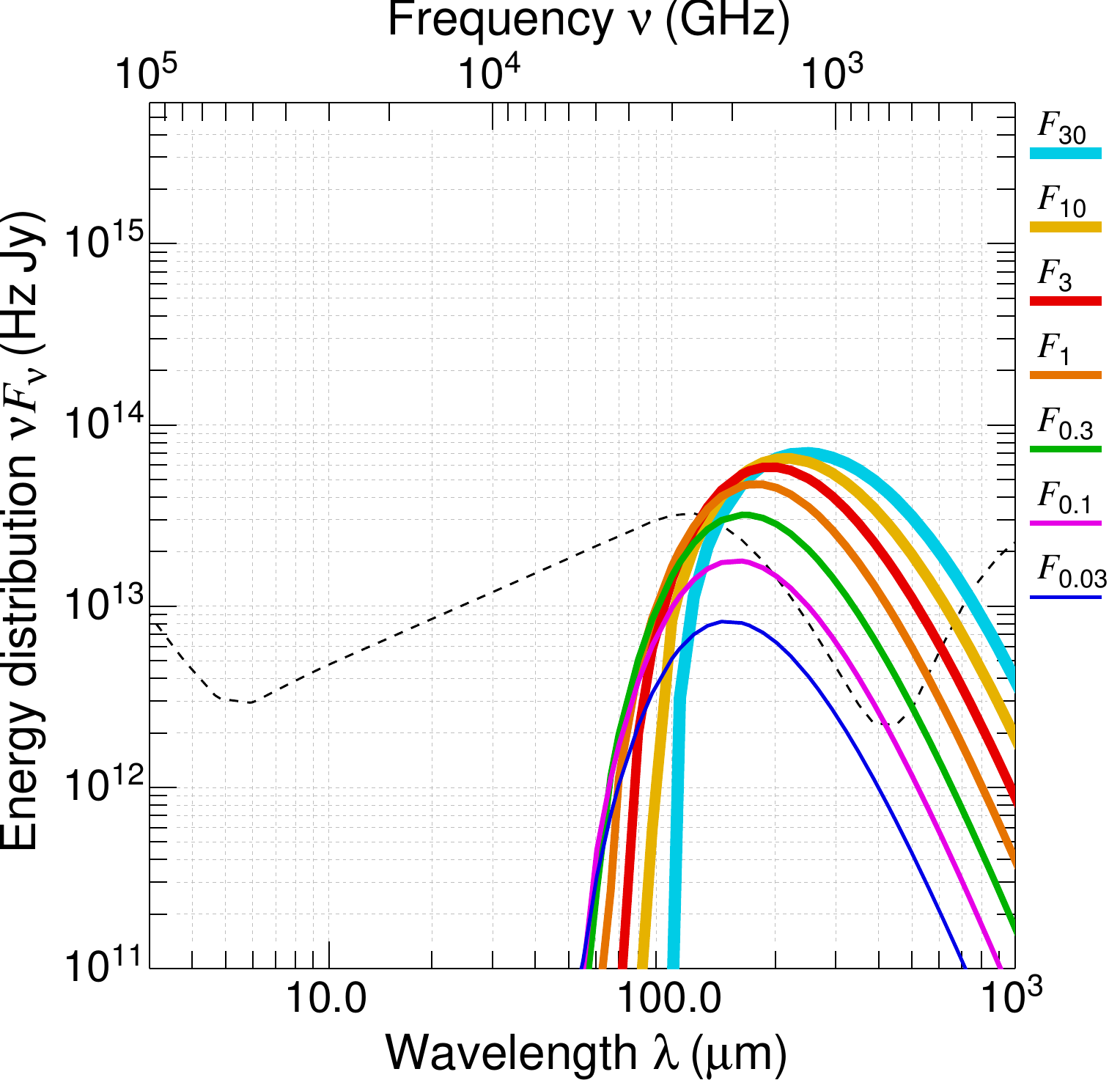}}
            \resizebox{0.3204\hsize}{!}{\includegraphics{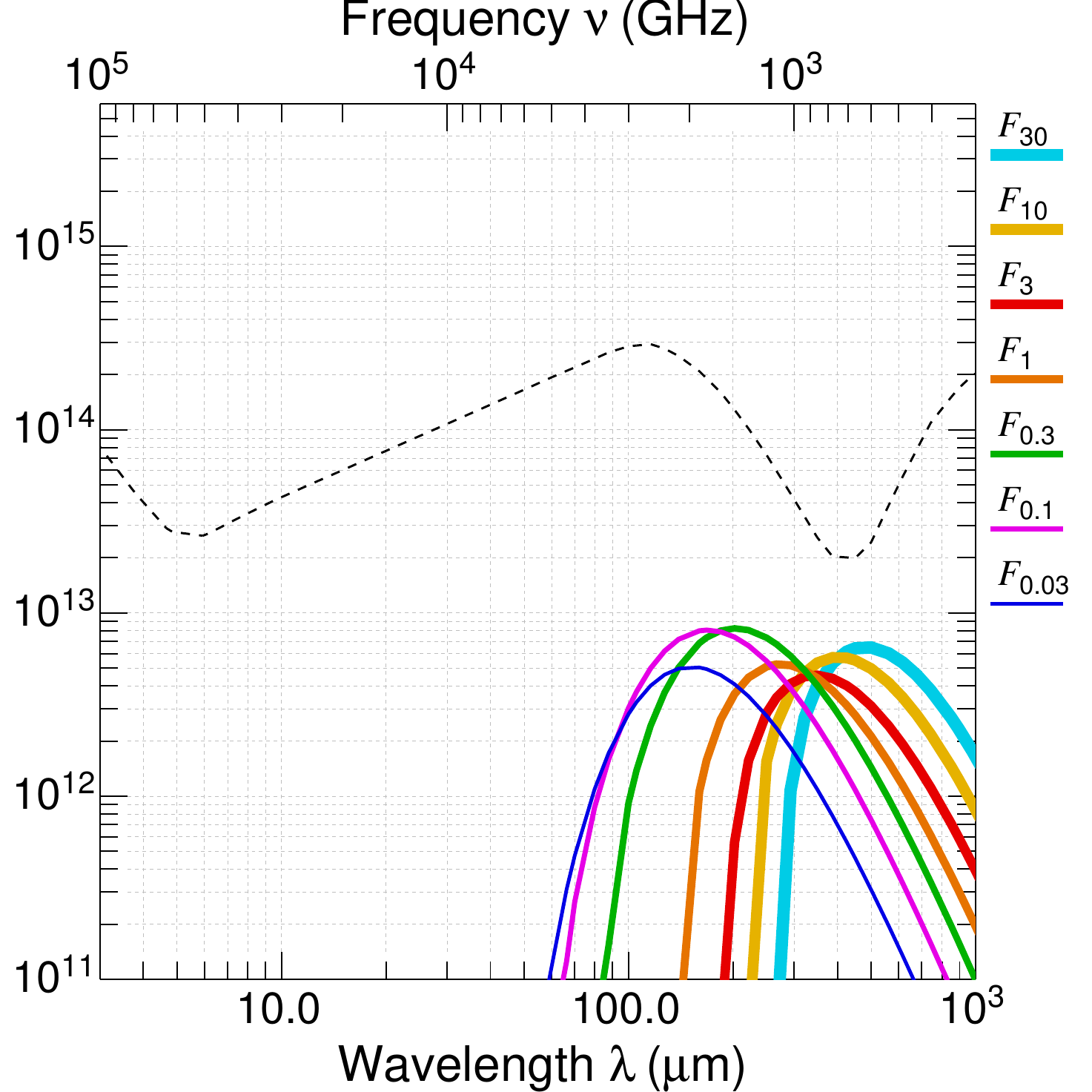}}}
\centerline{\resizebox{0.3327\hsize}{!}{\includegraphics{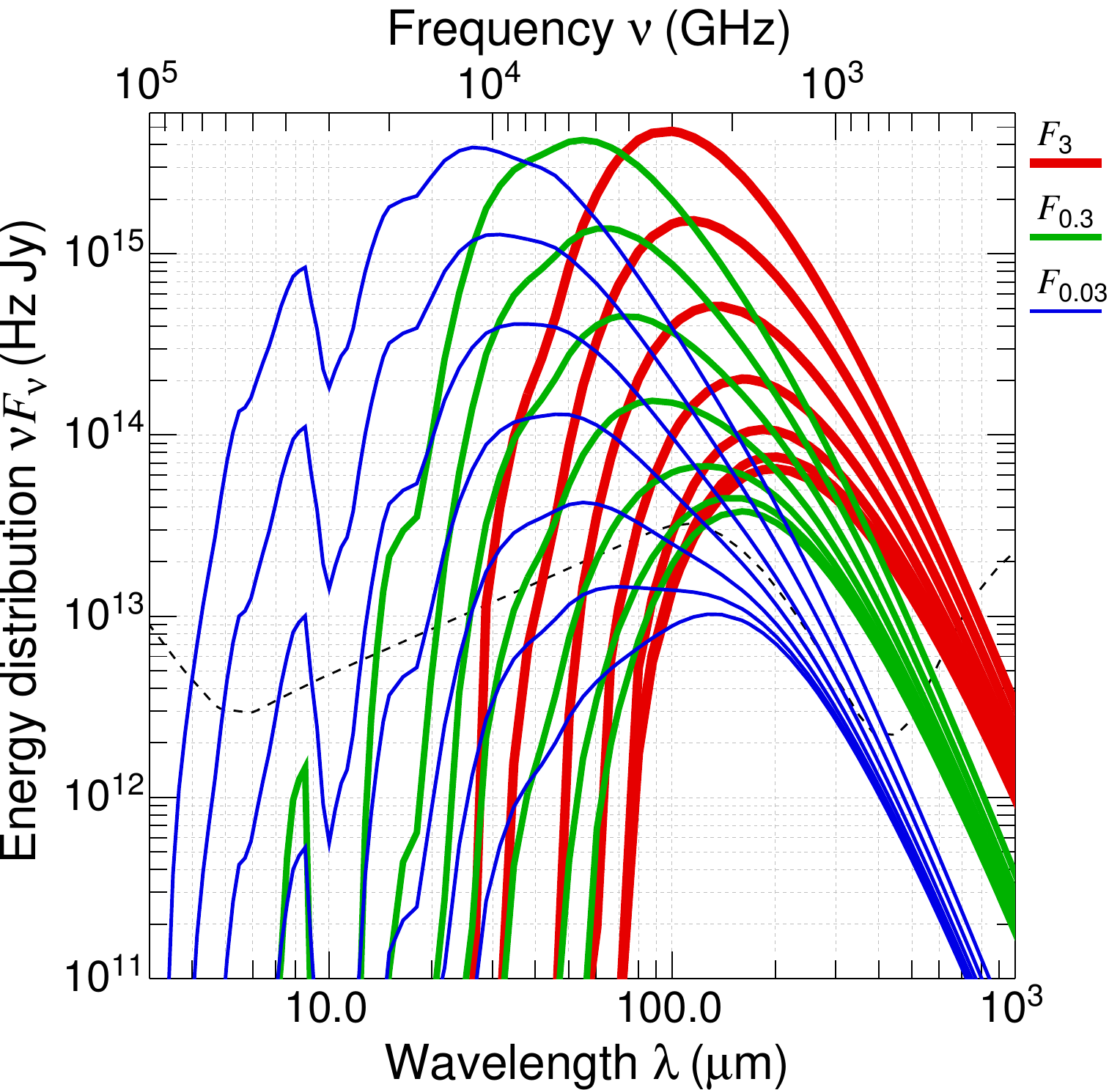}}
            \resizebox{0.3204\hsize}{!}{\includegraphics{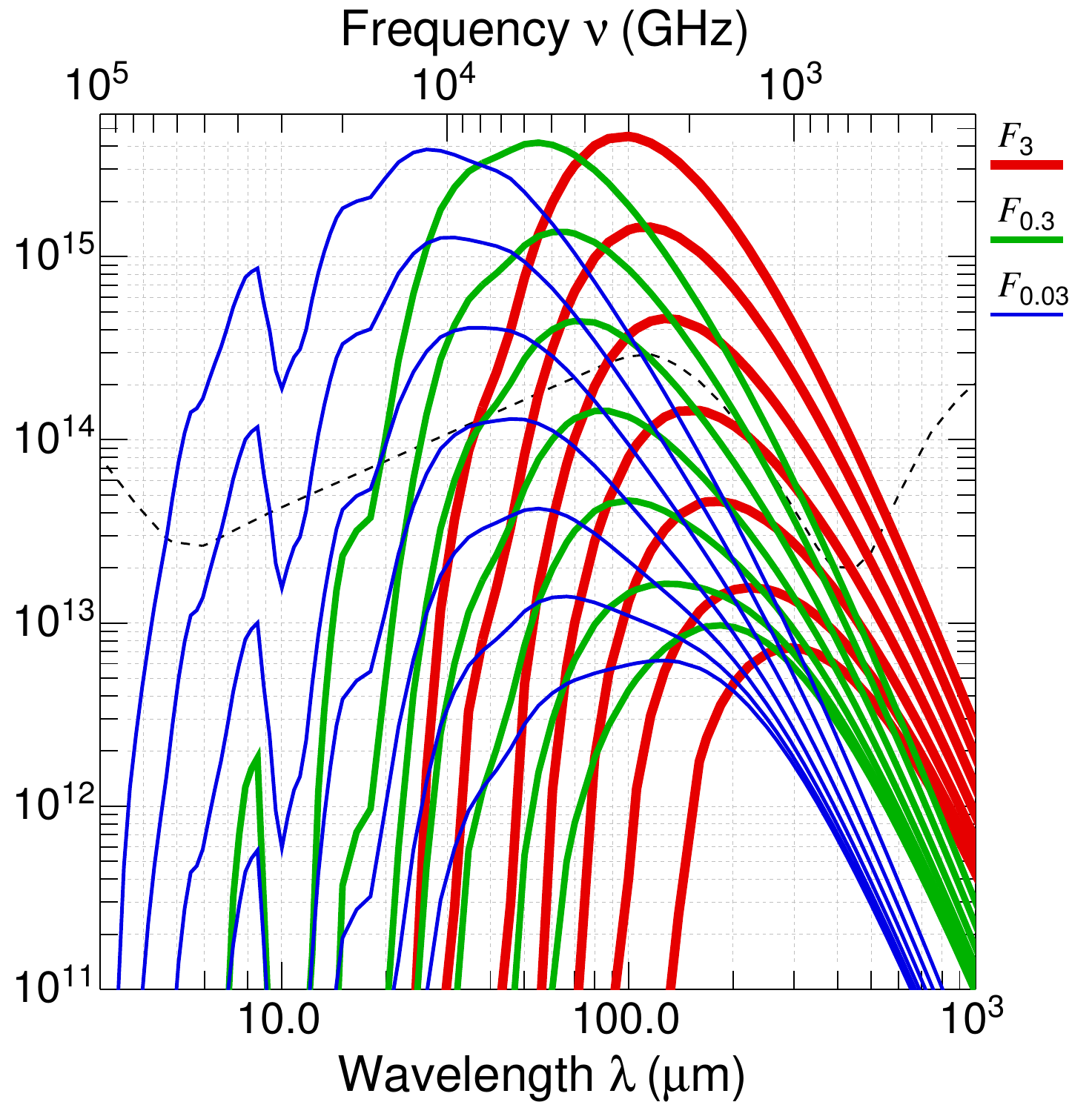}}}
\caption{
Spectral energy distributions starless cores (\emph{upper}) and protostellar envelopes (\emph{lower}). Shown are the
background-subtracted fluxes for the \emph{isolated} models (\emph{left}) and their \emph{embedded} variants (\emph{right}).
Subscripts of the curve labels ($F_{30}$ to $F_{0.03}$) indicate the model mass (in $M_{\sun}$). For protostellar envelopes of the 
same mass, the SEDs with higher fluxes correspond to higher accretion luminosities $L_\star$ (0.03, 0.1, 0.3, 1, 3, 10, 
30\,$L_{\sun}$). Dashed lines indicate the fluxes of the ISRF that were integrated over the projected area of either the isolated 
models or the embedding clouds.
} 
\label{sed.bes.pro}
\end{figure*}

\subsection{Radiation sources and optical depths}
\label{radiation.odepths}

All models were illuminated from the outside by an isotropic interstellar radiation field \citep{Black1994} with the ``strength''
parameter $G_{0}\,{=}\,1$ \citep[e.g.,][]{Parravano_etal2003}. The bolometric luminosity of the interstellar radiation field (ISRF)
entering the isolated models at $R$ amounted to $L_{\rm ISRF}\,{=}\,1$\,$L_{\sun}$, whereas that crossing the boundary of embedding
clouds at $R_{\rm E}$ was $9$\,$L_{\sun}$.

In addition to the external radiation field, the models of protostellar envelopes were assumed to be heated at their centers by a
blackbody source of luminosity $L_{\star}$ of $0.03, 0.1, 0.3, 1, 3, 10$, and $30$\,$L_{\sun}$ with an effective temperature of
$T_{\star}{=}\,5770$\,K. Actual values of $T_{\star}$ are unimportant, as the sources of accretion energy are surrounded by the
completely opaque dusty envelopes reprocessing the hot radiation to $T\,{\la}\,10^{3}$\,K very deep in their interiors.

Distribution of optical depths within dusty envelopes is one of the main parameters (along with the density structure) for the
transfer of radiation and resulting radiative-equilibrium temperatures. All models are quite opaque at visible wavelengths, with
radial optical depths $\tau_{V}\,{\approx}\,3\,{-}\,3\,{\times}\,10^{3}$ for starless cores and
$\tau_{V}\,{\approx}\,500\,{-}\,6\,{\times}\,10^{5}$ for protostellar envelopes of different masses and luminosities
(Fig.~\ref{loo}). At the far-infrared wavelength of $100$\,{${\mu}$m}, starless cores with $M\,{<}\,3$\,$M_{\sun}$ are transparent,
whereas the ones with $M\,{\ga}\,3$\,$M_{\sun}$ are optically thick towards their centers. All protostellar envelopes are optically
thick at $100$\,{${\mu}$m} and some of them (with $M\,{\ga}\,0.3\,M_{\sun}$) are even opaque at $500$\,{${\mu}$m} towards their
centers. Sizes of the dust-free cavities of protostellar envelopes increase with the luminosity of their central energy sources
(cf. Eq.~(\ref{inner.boundary}), Fig.~\ref{drp.bes.pro}), thus the optical depths of the enve\-lopes decrease, approximately as
$\tau_{\lambda}\,{\propto}\,L_{\star}^{-1/3}$.

High far-infrared optical depths of the model starless cores and protostellar envelopes are localized within relatively small
spherical zones around their centers. Angular radii of the opaque dusty zones in the protostellar models can be described (at 
$70\,{-}\,500$\,{${\mu}$m}) by a simple empirical expression
\begin{equation}
\vartheta \approx 1.6{\arcsec} \left(M/M_{\sun}\right) \left(\kappa_{\lambda}/\kappa_{70}\right),
\label{opaque.zone}
\end{equation}
which can also be used (within a factor of $1.5\,{-}\,2$) for the high-mass models of starless cores of $3, 10$, and
$30$\,$M_{\sun}$, in which the opaque zone exists only at ${\lambda}\,{\le}\,70, 160$, and $250$\,{${\mu}$m}, respectively. 

The density profiles $\rho(r)\,{\propto}\,r^{-2}$ of the protostellar envelopes are similar to those of the starless cores for
$r\,{\ga}\,0.1\,R$ (Fig.~\ref{drp.bes.pro}). Therefore, whenever an inner opaque zone exists in the objects, its mass obeys 
$m(\vartheta)\,{\propto}\,\vartheta$, hence the fractional mass is the fractional radius and, using Eq.~(\ref{opaque.zone}), can 
be written as
\begin{equation}
m(\vartheta)/M \approx {\vartheta}D/R \approx 0.022{\arcsec} \left(M/M_{\sun}\right) \left(\kappa_{\lambda}/\kappa_{70}\right),
\label{frac.mass}
\end{equation}
where $\vartheta$, $D$, and $R$ are in units of arcsec, pc, and AU, respectively. At $\lambda \,{\la}\, 100$\,{${\mu}$m}, the
opaque zone of high-mass objects extends over a large fraction of their mass. This means that the standard assumption of the
far-infrared transparency is severely violated for massive objects. Protostellar envelopes with $M\,{<}\,3$\,$M_{\sun}$ have small
opaque zones that contain little mass and thus they cannot substantially affect the standard methods of mass derivation.

\subsection{Temperature distributions}
\label{temperatures}

The models of starless cores and protostellar envelopes acquire radiative-equilibrium dust temperatures $T_{\rm d}(r)$ shown in
Fig.~\ref{trp.bes.pro}. In the adopted isotropic ISRF, the radiative-equilibrium temperature of dust grains with the model
opacities from Fig.~\ref{loo} is $T_{\rm d}\,{=}\,17.4$\,K, the value that the isolated models and embedding clouds acquire at
their outer boundaries in the limit $\tau_{\lambda}\,{\rightarrow}\,0$. The lower mass models of starless cores are transparent and
thus almost isothermal. Their higher mass counterparts develop steeper temperature gradients under the outer boundaries of the
isolated models and embedding clouds and lower temperatures in their interiors (Fig.~\ref{trp.bes.pro}).

Displaying the same behavior under their outer boundaries, protostellar envelopes of all masses develop steep temperature gradients
towards the inner boundary (Fig.~\ref{trp.bes.pro}). Higher accretion luminosities make the dust hotter and thus, with the adopted
dust sublimation temperature $T_{\rm S}\,{=}\,10^{3}$\,K, the boundary of the inner dust-free cavity shifts towards larger radial
distances (cf. Eq.~(\ref{inner.boundary})). An analytical approximation of the profiles $T_{\rm d}(r)$ for protostellar envelopes
can be found in Appendix~\ref{AppendixA}.

Differences between the isolated and embedded models are highlighted by their different temperature distributions at the outer
model boundary (Fig.~\ref{trp.bes.pro}). The temperatures of embedded models at $r\,{=}\,R$ are significantly lower than those of
the isolated models, owing to the absorption of ISRF in the embedding clouds ($R\,{<}\,r\,{\le}\,R_{\rm E}$). The denser the
embedding cloud is, the lower $T_{\rm d}(R)$ is and the greater the contrast to the isolated model (Fig.~\ref{trp.bes.pro}). As the
bulk of the mass of the models is contained in the outer parts, the differences in the temperature profiles between the isolated
and embedded models can greatly affect their observational properties, such as the images and total (integrated) fluxes.

\subsection{Spectral energy distributions}
\label{seds}

After computing the self-consistent radiative-equilibrium dust temperature distributions $T_{\rm d}(r)$ from the radiative transfer
models, observables -- such as the intensity maps $\mathcal{I}_{\nu}$ and total fluxes $F_{\nu}$ -- were obtained by a ray-tracing
algorithm in separate runs of \textsl{RADMC-3D}. Effects of the Monte Carlo noise on $F_{\nu}$, evaluated from the standard
deviations about the azimuthally averaged intensity profiles $I_{\nu}(\vartheta)$, are below $0.003{\%}$ and $3{\%}$ for the
starless cores and protostellar envelopes, respectively, in all models and wavebands.

To emulate the standard observational procedure of flux measurements, $F_{\nu}$ were integrated from the background-subtracted
model images $\mathcal{I}_{\nu}$. The model background $I^{\rm B}_{\nu}$ was evaluated as an average intensity
$\bar{I_{\nu}}(R^{\prime})$ within a ${\delta}R^{\prime}$-wide annulus placed just outside the outer model boundary
($R^{\prime}\,{>}\,R$). In practice, the annulus was one pixel in width (${\delta}R^{\prime}\,{=}\,0.47\arcsec$) and it was
detached from the boundary by one additional pixel. For the isolated models, $I^{\rm B}_{\nu}\,{=}\,I^{\rm ISRF}_{\nu}$ is the
intensity of the isotropic ISRF, whereas for the embedded models, $I^{\rm B}_{\nu}$ is determined by both $I^{\rm ISRF}_{\nu}$ and
the transfer of radiation in the background cloud ($R^{\prime}\,{\le}\,r\,{\le}\,R_{\rm E}$) along the rays passing through the
annulus. Inaccuracies inherent in the standard algorithm of background subtraction are discussed in Sect.~\ref{bg.subtraction} and
Appendix \ref{AppendixB}.

Spectral energy distributions (SEDs) of the models of starless cores and protostars are shown in Fig.~\ref{sed.bes.pro}. The SED
shapes depend on the density and temperature distributions (Figs.~\ref{drp.bes.pro} and \ref{trp.bes.pro}). Large differences
between the SEDs for the isolated and embedded cores are mainly caused by differences in their temperature profiles near the model
boundary. The SEDs of protostellar envelopes are affected by the same effects to a much lesser degree as their density profiles
are centrally peaked and their temperature profiles are dominated by the internal radiation source. The SEDs of the models of
different masses and luminosities show a large variety of shapes in the far-infrared domain (Fig.~\ref{sed.bes.pro}) due to 
varying optical depths and temperatures.

Providing a useful reference in our analysis, additional ray-tracing runs of \textsl{RADMC-3D} computed the fluxes of isothermal
models. These are the same models described above (Fig.~\ref{drp.bes.pro}), in which self-consistent temperature profiles
(Fig.~\ref{trp.bes.pro}) have been replaced with their mass-averaged values:
\begin{equation}
\begin{split}
T_{M} &= M^{\,-1}\!\int{T_{\rm d}(x,y,z)\,\rho(x,y,z)\,\mathrm{d}{x}\,\mathrm{d}{y}\,\mathrm{d}{z}}\\
      &= M^{\,-1}\,4\pi\!\int{T_{\rm d}(r)\,\rho(r)\,r^{2}\,\mathrm{d}{r}}.
\end{split}
\label{mass.averaged}
\end{equation}
The resulting total fluxes of the isothermal models are denoted $F_{\nu}(T_{M})$.


\section{Fitting source fluxes and intensities}
\label{fitting.methods}

In observational studies, after obtaining multiwavelength images $\mathcal{I}_{\nu}$ and integrating background-subtracted (and
deblended) fluxes $F_{\nu}$ of extracted sources, their spectral distributions need to be fitted to derive fundamental physical
parameters, such as the source mass and luminosity.

The standard technique uses the well-known formal solution of the radiative transfer equation that can be written as
\begin{equation}
I_{\nu} = B_{\nu}(T)\left(1 - \exp\left(-\tau_{\nu}\right)\right),
\label{rtsolution}
\end{equation}
where $I_{\nu}$ is the observed specific intensity, $T$ is the homogeneous temperature of an object, $B_{\nu}(T)$ is the blackbody
intensity, and $\tau_{\nu}$ is the optical depth of the object. After obtaining an image
$\mathcal{I}_{\nu}\,{\equiv}\,I_{\nu\,ij}$, the total flux $F_{\nu}\,{=}\,\!\int{\mathcal{I}_{\nu}\,\mathrm{d}{\Omega}}$ can be
integrated over the solid angle $\Omega$ subtended by the object. For constant intensity, it reduces to
$F_{\nu}\,{=}\,I_{\nu}\,\Omega$. A critical assumption used in the derivation of Eq.~(\ref{rtsolution}) is that the object is
homogeneous in temperature, whereas the temperatures of the astrophysical objects are actually nonuniform (cf.
Fig.~\ref{trp.bes.pro}).

Two methods and two fitting models were explored in this work that have been used in observational studies of star formation to
estimate source temperatures and masses.

\subsection{Fitting total fluxes $F_{\nu}$}
\label{fitting.fluxes}


In this method, the total fluxes $F_{\nu}$ are integrated from background-subtracted and deblended images $\mathcal{I}_{\nu}$ of
source intensities and then are fitted to estimate source mass as one of the fitting parameters. With the adopted parameterization
of the power-law opacity $\kappa_{\nu}\,{=}\,\kappa_{0}\left(\nu/\nu_{0}\right)^{\,\beta}$, it is possible to write 
Eq.~(\ref{rtsolution}) in the form
\begin{equation}
F_{\nu} = B_{\nu}(T)\left(1 - \exp\left(-\kappa_{0}\left(\nu/\nu_{0}\right)^{\,\beta\!} \eta M D^{-2} \Omega^{-1}\right)\right) 
\Omega,
\label{modbody}
\end{equation}
where $\eta$ is the dust-to-gas mass ratio and $D$ is the source distance. The fitting model of Eq.~(\ref{modbody}) with five
parameters ($T$, $M$, $\beta$, $D$, $\Omega$) is referred to as \textsl{modbody} in this paper. After fitting $F_{\nu}$ and
estimating the model parameters, the average column density $N_{\rm H_2}$ can be obtained from $M\,{=}\,\mu m_{\rm H} N_{\rm H_2}
D^{2} \Omega$, where $\mu\,{=}\,2.8$ is the mean molecular weight per H$_2$ molecule and $m_{\rm H}$ is the hydrogen mass.

With an additional assumption that measured fluxes $F_{\nu}$ represent optically thin emission\footnote{Far-infrared transparency
is an important assumption that is wrong for high-mass objects (Sect.~\ref{radiation.odepths}).}, Eq.~(\ref{modbody}) can be 
written as
\begin{equation}
F_{\nu} = B_{\nu}(T) \,\kappa_{0}\left(\nu/\nu_{0}\right)^{\,\beta} \eta M D^{-2}.
\label{thinbody}
\end{equation}
The fitting model of Eq.~(\ref{thinbody}) with four parameters ($T$, $M$, $\beta$, $D$) is referred to as \textsl{thinbody} in this
paper. By the definition ($\tau_{\nu}\,{\ll}\,1$), it produces only fits with the modified blackbody shapes
${\kappa_{\nu}\,B_{\nu}(T)}$ that are scaled up or down, depending on $M$. Obviously, the \textsl{modbody} fits with
$\tau_{\nu}\,{\ll}\,1$ produce the same shapes as the \textsl{thinbody} model does, whereas the \textsl{modbody} fits with
$\tau_{\nu}\,{\gg}\,1$ resemble a blackbody $B_{\nu}(T)$. In the intermediate (semi-opaque) cases, the short-wavelength parts of
the fitted curves can be described by $B_{\nu}(T)$ while morphing into ${\kappa_{\nu}\,B_{\nu}(T)}$ at long wavelengths where the
radiation becomes optically thin. With more flexible shapes, \textsl{modbody} can give better fits of the data, but it does not
necessarily lead to good estimates of temperatures and masses.

After fitting fluxes with a \textsl{modbody} or \textsl{thinbody} model, an estimate of $T_{F}$ and the corresponding mass $M_{F}$
are obtained. For the realistic objects with strongly nonuniform temperatures $T_{\rm d}(r)$ (Fig.~\ref{trp.bes.pro}), emerging
fluxes $F_{\nu}$ are heavily distorted from the simple shapes of the fitting models, hence these models are inadequate and an
estimate of $T_{F}$ does not guarantee that $M_{F}$ is close to the true mass $M$. For the purpose of obtaining accurate
$M_{F}\,{\approx}\,M$, it is necessary (but not sufficient) to have $T_{F}\,{\approx}\,T_{M}$, i.e., it is possible to interpret
$T$ in Eq.~(\ref{thinbody}) as the mass-averaged $T_{M}$ from Eq.~(\ref{mass.averaged}). In fact, assuming $\tau_{\nu}\,{\ll}\,1$
in the far-infrared, the observed fluxes contain emission of all dust grains, which is proportional to the mass of dust at 
different $T_{\rm d}$ in the entire volume of an object:
\begin{equation}
F_{\nu} = \kappa_{\nu} \eta D^{-2}\!\int{B_{\nu}(T_{\rm d}(x,y,z))\,\rho(x,y,z)\,\mathrm{d}{x}\,\mathrm{d}{y}\,\mathrm{d}{z}}. 
\label{emission}
\end{equation}
Equations~(\ref{thinbody}) and (\ref{emission}) can immediately be combined into a definition of the mass-averaged intensity
$B_{\nu M}$. Since $B_{\nu}(T)\,{\propto}\,T$ in the Rayleigh-Jeans domain, the equations are also readily converted into $T_{M}$
from Eq.~(\ref{mass.averaged}). In the model objects studied here, differences between $B_{\nu M}$ and $B_{\nu}(T_{M})$ quickly
become negligible beyond the peak wavelength of the latter ($\lambda$ $\ga$ $2\,\lambda_{\rm peak}$). Therefore, $T_{M}$ is fully
consistent with the fitting models at long wavelengths.

\subsection{Fitting image intensities $I_{\nu}$}
\label{fitting.intens}

In this method, it is possible to fit pixel intensity distributions $I_{\nu\,ij}$ of the background-subtracted and deblended images
$\mathcal{I}_{\nu}$ of a source\footnote{In an alternative approach, multiwavelength images of an entire field can be fitted to
derive its $\mathcal{N}_{\rm H_2}$ image, then to identify (extract) the sources and to integrate their masses. Both approaches are
equivalent in this model-based study, hence the alternative method was not used.}, to derive a map of its column densities
$\mathcal{N}_{\rm H_2}\,{\equiv}\,N_{{\rm H_2} ij}$ and then the source mass $M\,{=}\,\mu m_{\rm H} D^{2} \Omega \sum_{}{N_{{\rm
H_2} ij}}$. It is convenient to express the \textsl{modbody} and \textsl{thinbody} models from Eqs.~(\ref{modbody}) and
(\ref{thinbody}) as functions of the pixel column density $N_{\rm H_2}$:
\begin{eqnarray}
I_{\nu}&\!\!\!\!\!=\!\!\!\!\!&B_{\nu}(T)\left(1 - \exp\left(-\kappa_{0}\left(\nu/\nu_0\right)^{\,\beta} \eta \mu m_{\rm H} 
N_{\rm H_2}\right)\right), \label{modbody.intens}\\
I_{\nu}&\!\!\!\!\!=\!\!\!\!\!&B_{\nu}(T) \,\kappa_{0}\left(\nu/\nu_0\right)^{\,\beta} \eta \mu m_{\rm H} 
N_{\rm H_2}. \label{thinbody.intens}
\end{eqnarray}
In this formulation, both models have only three fitting parameters ($T$, $N_{\rm H_2}$, $\beta$) in contrast to the case where
total fluxes are fitted (five parameters for \textsl{modbody} in Eq.~(\ref{modbody}) and four parameters for \textsl{thinbody} in 
Eq.~(\ref{thinbody}); see Sect.~\ref{fitting.fluxes}). Furthermore, limited angular resolutions of real images makes the results of 
fitting $I_{\nu}$ depend sensitively on the degree to which a source is resolved.

For fully resolved sources, such as the model objects used in this work, relatively small pixels sample completely independent
intensities from different rays. For progressively lower angular resolutions, intensities within a beam become increasingly blended
together. Radiation with different temperatures gets mixed not only along the line of sight, but also in the transverse directions,
in the plane of the sky. For unresolved objects with intrinsic temperature gradients, radiation from the entire object becomes
heavily blended, leading to strong distortions of their spectral intensity distributions.

An important assumption used in the derivation of $N_{{\rm H_2} ij}$ is a constant temperature $T_{\rm d}(x,y,z)$ along the lines
of sight within a certain radial distance from a pixel $(i,j)$. The distance depends on the angular resolution of images: for less
resolved sources, temperatures from a larger environment of the pixel contribute to its intensity. With low optical depths
$\tau_{\nu}\,{\ll}\,1$ in the far-infrared, emission is observed from the entire column of dust grains at $(i,j)$ with different
temperatures $T_{\rm d}(z)$ along the line of sight. The reasoning associated with Eq.~(\ref{emission}) can be applied to show that
$T$ in Eq.~(\ref{thinbody.intens}) is consistent with the column-averaged temperature
\begin{equation}
T_{N ij} = N_{{\rm H_2} ij}^{\,-1}\!\int{T_{\rm d}(x,y,z)\,\rho(x,y,z)\,\mathrm{d}{z}}.
\label{coldens.averaged}
\end{equation}
A mass-averaged temperature, equivalent to that from Eq.~(\ref{mass.averaged}), can be obtained as 
$T_{M}\,{=}\,M^{-1} \mu m_{\rm H} D^{2} \Omega \sum_{}{T_{N ij}\,N_{{\rm H_2} ij}}$.

\subsection{Variable and fixed parameters}
\label{variable.fixed}

In most studies, the opacity slope $\beta$ has been kept fixed in the fitting process to reduce the number of free parameters and
improve the robustness of derived parameters. Following this practice, Sect.~\ref{results} presents and discusses only the results
of fitting with a fixed opacity slope. When fitting intensities $I_{\nu}$ with $\beta$ fixed, the number of free variable
parameters becomes $\gamma\,{=}\,2$ for both \textsl{thinbody} and \textsl{modbody} models ($T$, $N_{\rm H_2}$). When fitting
fluxes $F_{\nu}$, distance $D$ is also assigned a fixed value to further reduce the degrees of freedom, although astronomical
distances are poorly known. The number of free variable parameters is thus $\gamma\,{=}\,2$ for \textsl{thinbody} ($T$, $M$) and
$\gamma\,{=}\,3$ for \textsl{modbody} ($T$, $M$, $\Omega$).

In practice, after measuring $F_{\nu}\,{=}\,\!\int{I_{\nu}\,\mathrm{d}{\Omega}}$, the solid angle $\Omega$ over which $I_{\nu}$
were integrated is known\footnote{In real observations, images $\mathcal{I}_{\nu}$ usually have different angular resolutions and
the flux integration area is wavelength dependent.} and its value can be fixed, reducing $\gamma$ for \textsl{modbody} to two free
variables ($T$, $M$). In this model study, one could also keep $\Omega\,{=}\,{\pi\,}(R/D)^{2}$ constant, as the true values of $R$
and $D$ are known; however, \textsl{modbody} would then become completely equivalent to \textsl{thinbody}. Indeed, fixing $\Omega$
of transparent objects at accurate (or even overestimated) values means that the optical depths in Eq.~(\ref{modbody}) are very
small ($\tau_{\nu}\,{=}\,\kappa_{\nu} \eta M D^{-2} \Omega^{-1}{\ll}\,1$), which effectively converts \textsl{modbody} into
\textsl{thinbody}. Only when fixing strongly underestimated values $\Omega\,{\ll}\,{\pi}\,(R/D)^{2}$, the far-infrared $\tau_{\nu}$
become large enough to produce any noticeable differences between \textsl{modbody} and \textsl{thinbody}. This work investigates
qualities of two \emph{different} models, hence $\Omega$ was allowed to vary in all \textsl{modbody} fits of $F_{\nu}$.

\begin{figure*}
\centering
\centerline{\resizebox{0.3400\hsize}{!}{\includegraphics{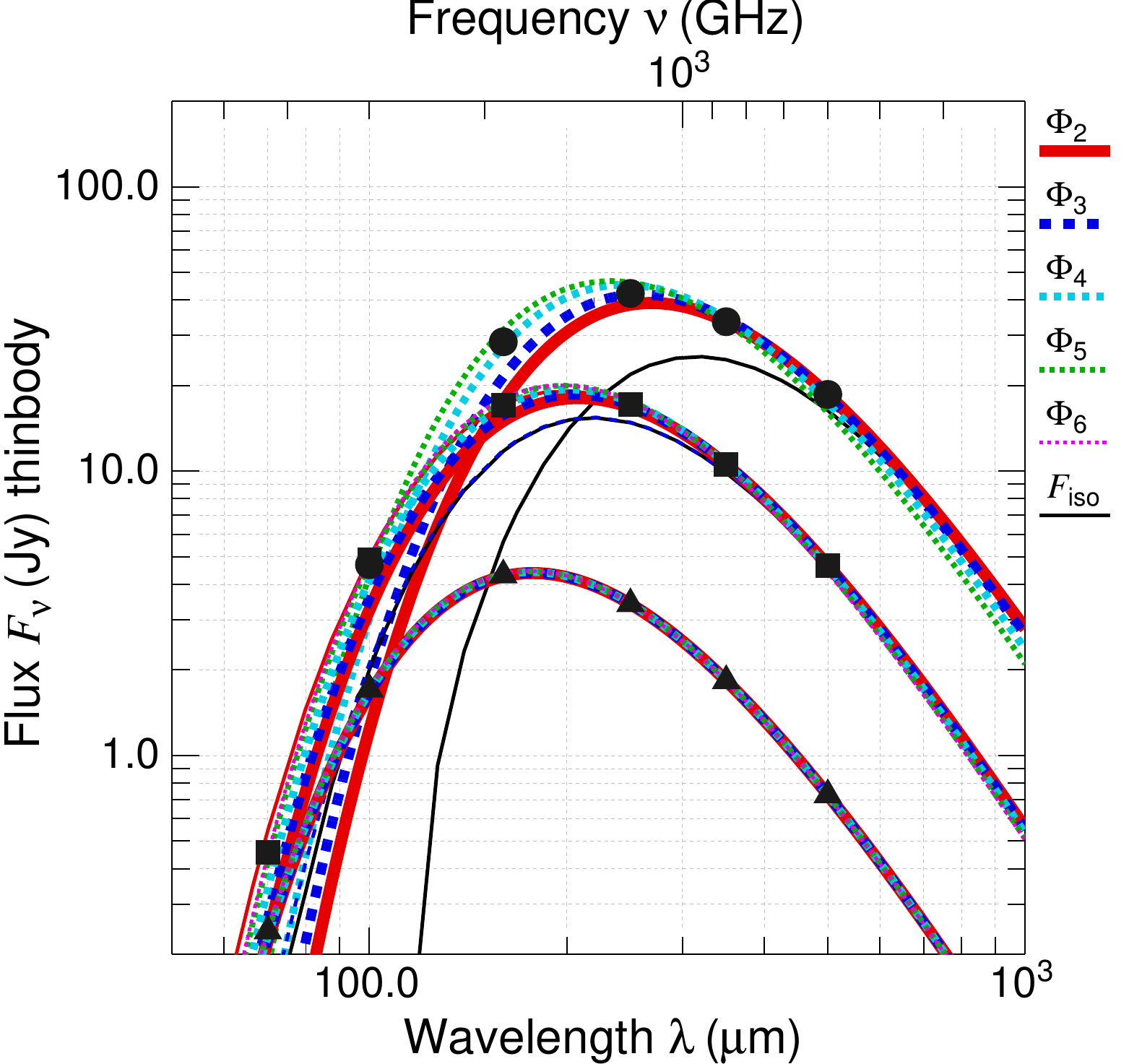}}
            \resizebox{0.3400\hsize}{!}{\includegraphics{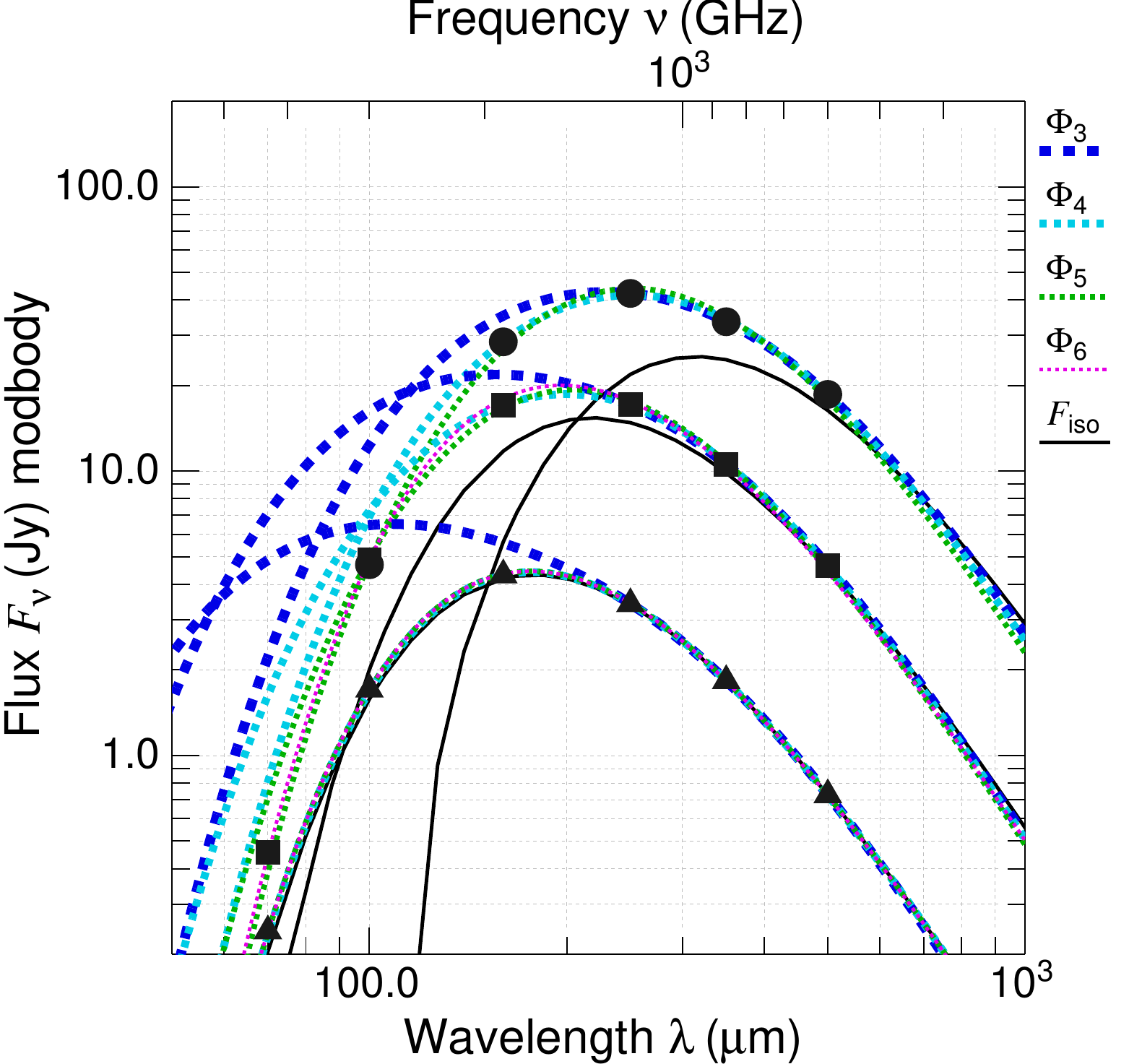}}}
\centerline{\resizebox{0.3400\hsize}{!}{\includegraphics{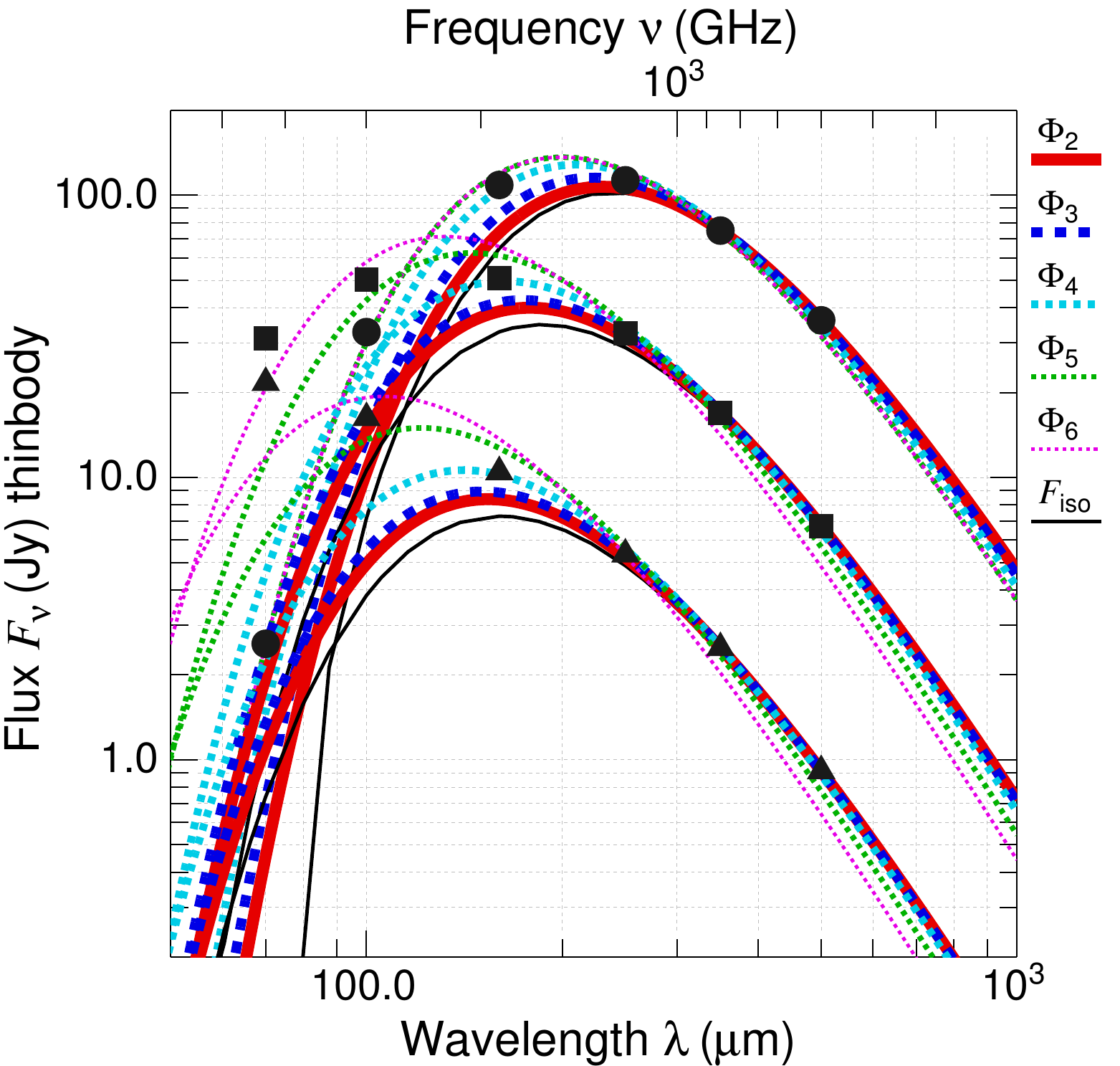}}
            \resizebox{0.3400\hsize}{!}{\includegraphics{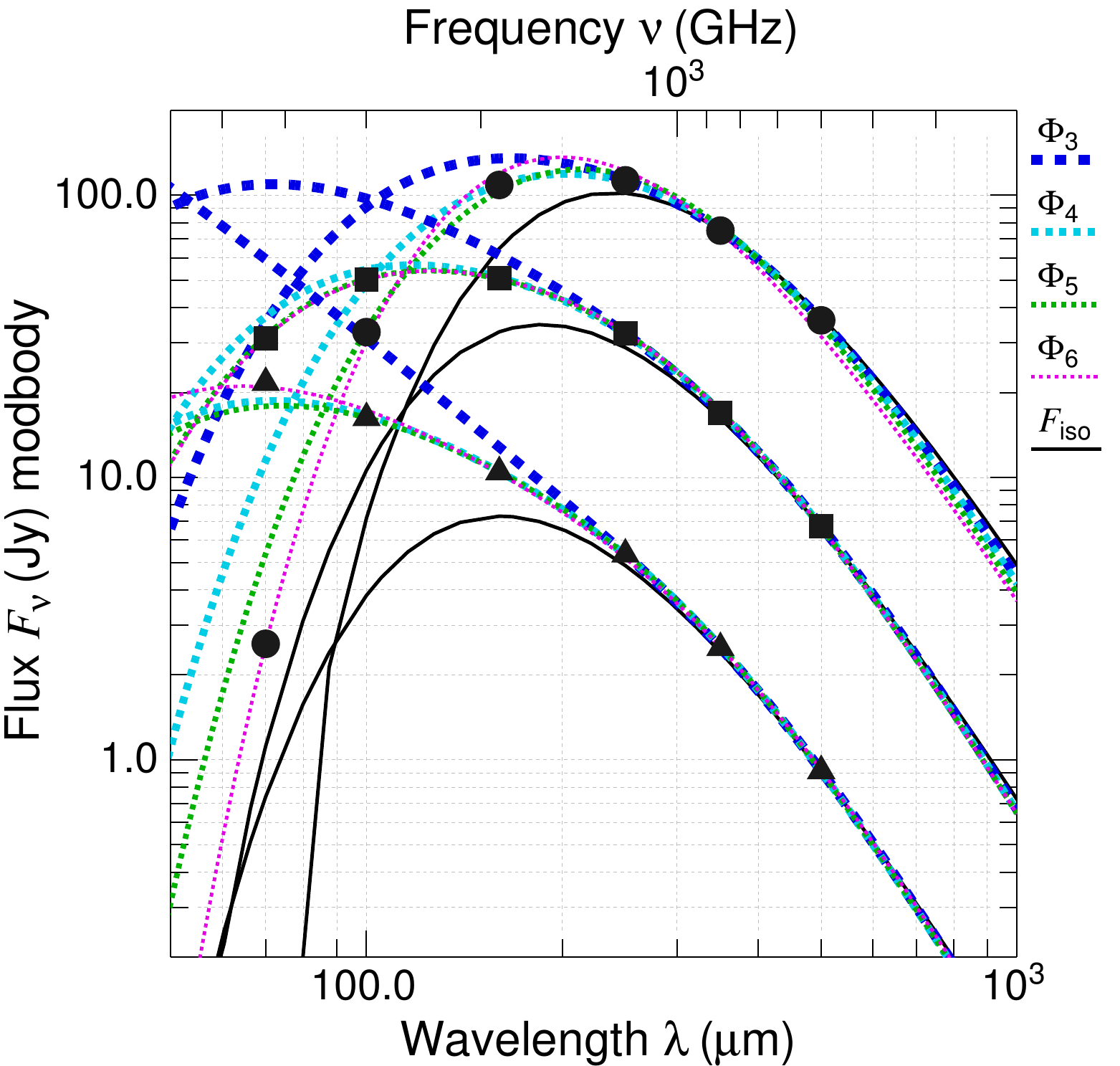}}}
\caption{
Fluxes of the \emph{isolated} starless cores (\emph{upper}) and protostellar envelopes (\emph{lower}) fitted with the
\textsl{thinbody} (\emph{left}) and \textsl{modbody} (\emph{right}) models. The fluxes for the 0.03, 0.3, and 3\,$M_{\sun}$ cores
and envelopes (with $L_{\star}\,{=}\,1\,L_{\sun}$) from Fig.~\ref{sed.bes.pro} are shown as black triangles, squares, and
circles, respectively. Successful fits (Sect.~\ref{data.subsets}) using different subsets $\Phi_{n}$ of fluxes are indicated by
the solid and dashed lines of different widths and colors. The thin black curves show the flux distributions for the
\emph{isothermal} models where the actual model temperatures (Fig.~\ref{trp.bes.pro}) were replaced with their mass-averaged 
values: $T_{\rm d}(r)\,{=}\,T_{M}$ ($16.3, 13.2, 9.2$\,K for starless cores and $18.1, 15.6, 12.3$\,K for protostellar envelopes).
} 
\label{fits.examples}
\end{figure*}

When fitting pixel intensities $I_{\nu}$ instead of $F_{\nu}$, the far-infrared $\tau_{\nu\,ij}$ within most of the image pixels
$(i,j)$ are small, even for perfectly resolved sources. For poorly resolved or unresolved sources, radiation within the beams gets
diluted and maximum values of $\tau_{\nu\,ij}$ in the images become smaller. All models of starless cores and protostellar
envelopes contain $96{\%}$ and $90{\%}$ of their masses, respectively, in their outer parts ($r\,{\ga}\,0.1\,R$,
Fig.~\ref{drp.bes.pro}). Intensities in the outer parts of the source images come mostly from the pixel columns of dust with
$\tau_{\nu\,ij}{\,\ll\,}1$ in the far-infrared. Only in the models with $M\,{\ga}\,3$\,$M_{\sun}$ do they become substantially
affected by the radiation from the central opaque zone (Sect.~\ref{radiation.odepths}). As a result, the masses derived from
fitting $I_{\nu}$ are almost the same (within ${\sim}\,20{\%}$) for both fitting models and hence only the \textsl{thinbody}
results are presented for this method.

\subsection{Data points and their subsets}
\label{data.subsets}

Fitting was executed for a set of the total model fluxes $\left\{ F_{\lambda_{i}} \right\}$ or pixel intensities $\left\{
I_{\lambda_{i}} \right\}$ $(i\,{=}\,1, 2, \dots, 6)$ at the \emph{Herschel} wavelengths $\lambda_{i}$ of 70, 100, 160, 250, 350,
and 500\,{${\mu}$m}. In this model-based study, the intensities and fluxes of numerical models have essentially no measurement
errors. It makes sense, however, to make their uncertainties resemble typical observational values, hence to get an idea of
realistic inaccuracies of the estimated parameters (masses, temperatures). Before the fitting, the model intensities and fluxes
were assigned an additional (optimistic) uncertainty of $15{\%}$, a value similar to the levels of calibration errors in real
observations (e.g., with \emph{Herschel}). The above uncertainties were associated with the exact data points to see how typical
data uncertainties translate into the resulting error bars of the derived parameters. Extra uncertainties come from the fact that
the dust-to-gas ratio $\eta$, reference opacity $\kappa_0$, and distance $D$, which are used in the fitting models
(Sects.~\ref{fitting.fluxes}, \ref{fitting.intens}) but held constant, are actually poorly known. Conservatively assuming that the
quantities have random and independent uncertainties of $20{\%}$, the latter were added in quadrature to those of the derived
masses, for the same purpose of obtaining the total resulting mass uncertainties.

To isolate the effects of temperature gradients in starless cores and protostellar envelopes (Fig.~\ref{trp.bes.pro}), the fitting
was done for several subsets of data, removing some (or none) of the shortest-wavelength points from the fitting process. The data
subsets are denoted $\Phi_{n}\,{=}\,\left\{Y_{\lambda_{i}}\right\}$, where $Y_{\lambda_{i}}$ is either $F_{\lambda_{i}}$ or
$I_{\lambda_{i}}$ and $n$ is the number of the longest wavelengths used in the fitting\footnote{ $\Phi_{6}\,{=}\,\{Y_{70}, Y_{100},
Y_{160}, Y_{250}, Y_{350}, Y_{500}\}$, $\Phi_{5}\,{=}\,\{Y_{100}, Y_{160}, Y_{250}, Y_{350}, Y_{500}\}$, $\Phi_{4}\,{=}\,\{Y_{160},
Y_{250}, Y_{350}, Y_{500}\}$, $\Phi_{3}\,{=}\,\{Y_{250}, Y_{350}, Y_{500}\}$, $\Phi_{2}\,{=}\,\{Y_{350}, Y_{500}\}$.}. Fits of
total fluxes were considered successful (acceptable) and their results are shown below, if $\chi^{2}\,{\le}\,{n-\gamma+1}$, with
the last term added to allow testing $\chi^{2}$ for zero degrees of freedom ($n\,{=}\,\gamma$). Fits of image intensities were
considered successful, if the same goodness condition was fulfilled in \emph{all} pixels within an object. These results, as well
as the somewhat less reliable results with ${n-\gamma+1}\,{<}\,\chi^{2}\,{<}\,10$ in \emph{some} pixels are presented below.
Details of the fitting algorithm can be found in Appendix \ref{AppendixC}.


\section{Results}
\label{results}

This section describes derived parameters for both starless cores and protostellar envelopes, obtained from acceptable fits for all
subsets $\Phi_{n}$ (Sect.~\ref{data.subsets}) for both \textsl{modbody} and \textsl{thinbody} (Sect.~\ref{fitting.fluxes}). Results
for the isothermal models are presented in Appendix \ref{AppendixD}. To evaluate the effects of the un\-certain far-infrared
opacity slope, results are shown for $\beta\,{=}\,2$ used in the radiative transfer modeling and for two other $\beta$ values
($1.67$, $2.4$), differing from the true value by a factor of $1.2$. Results obtained with variable fitting parameter $\beta$ are
described in Appendix~\ref{AppendixE}.

Masses derived from fitting images $\mathcal{I}_{\nu}$ of objects with temperature gradients must depend on their angular
resolutions (Sect.~\ref{fitting.intens}). To investigate this effect, the model images with pixels of $0.47{\arcsec}$ were
convolved with Gaussian beams of $1$, $36$, and $144{\arcsec}$ (FWHM) and then resampled to $1$, $12$, and $48{\arcsec}$ pixels,
respectively. For the objects with diameters of $142{\arcsec}$ ($2\,{\times}\,10^{4}$\,AU, Fig.~\ref{drp.bes.pro}), the three
variants represent resolved, partially resolved, and unresolved cases. 

In this paper, the term \emph{uncertainties} refers to the error bars of measured or derived quantities, the term
\emph{inaccuracies} (sometimes simply \emph{errors}) refers to the deviations of the derived quantities from their model values,
and the term \emph{biases} denotes variable systematic dependences of inaccuracies across the ranges of model parameters ($M$, $L$,
$T_{M}$).

\subsection{Selected examples}
\label{selected.examples}

Examples of the fits of $F_{\nu}$ for the isolated starless cores and protostellar envelopes with masses of $0.03$, $0.3$, and
$3\,M_{\sun}$ are shown in Fig.~\ref{fits.examples}. Although the qualitatively similar plots for embedded models are not
presented, their derived parameters and uncertainties are described in Sect.~\ref{derived.properties}.

\begin{figure*}
\centering
\centerline{\resizebox{0.3327\hsize}{!}{\includegraphics{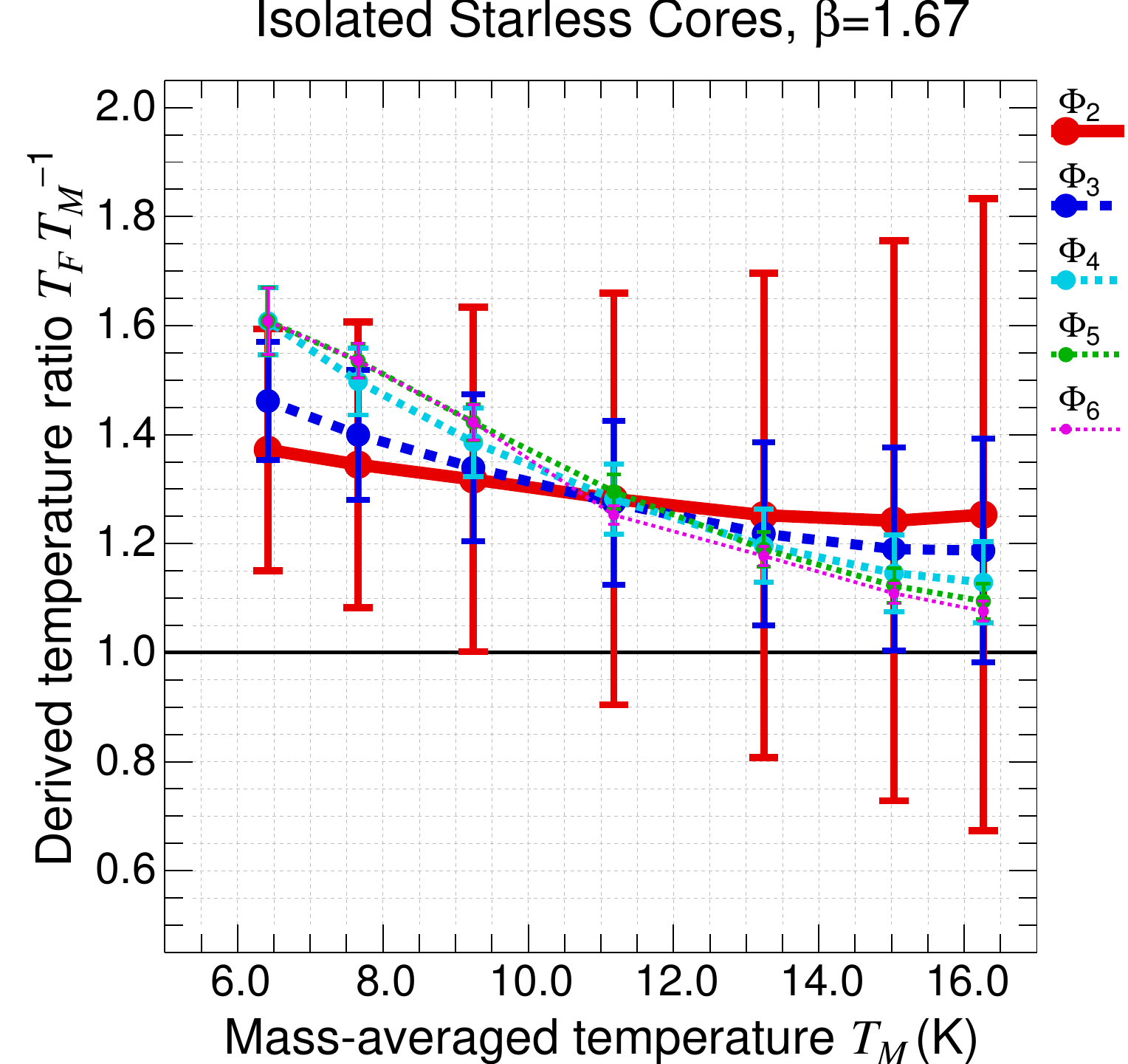}}
            \resizebox{0.3204\hsize}{!}{\includegraphics{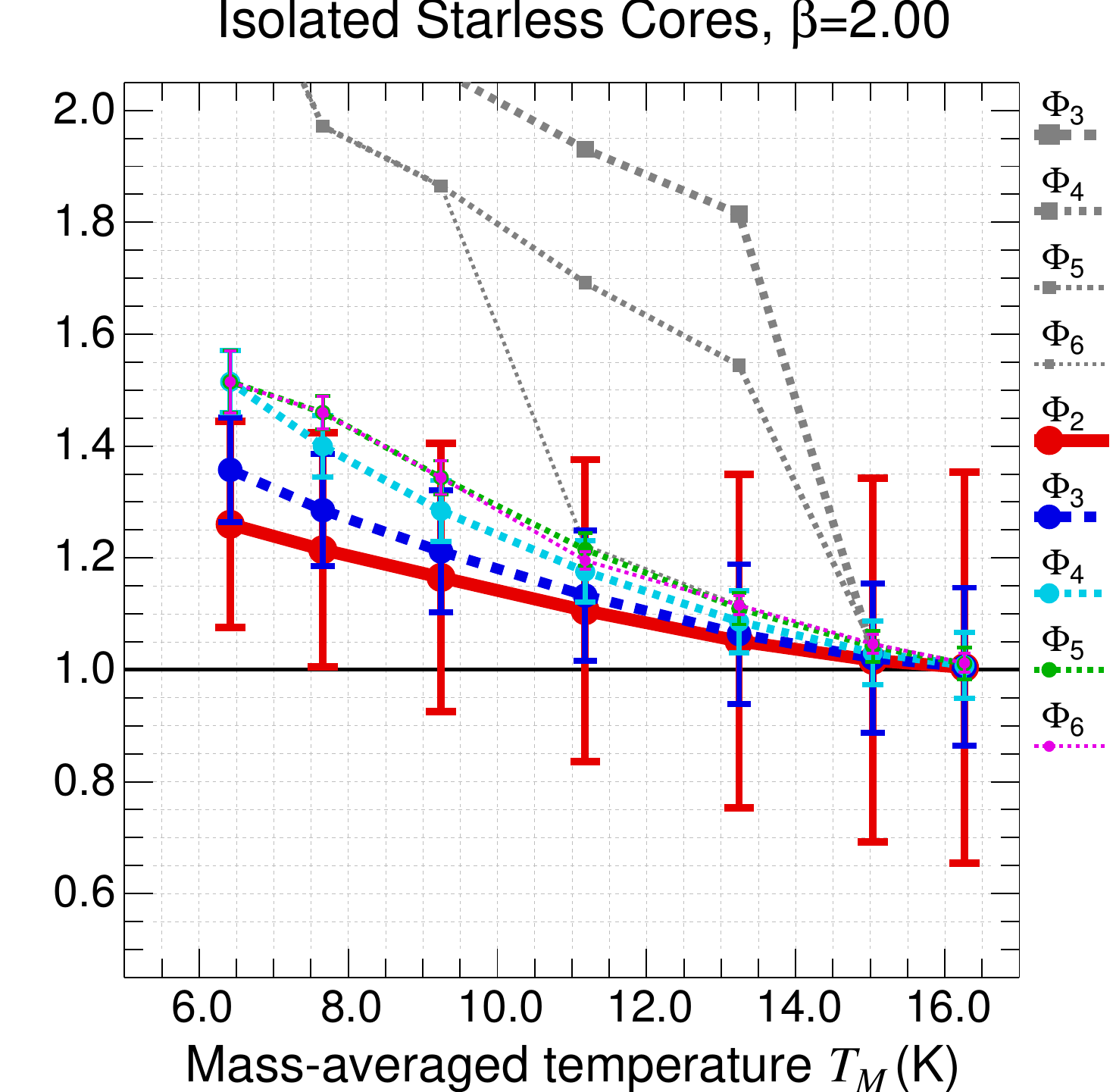}}
            \resizebox{0.3204\hsize}{!}{\includegraphics{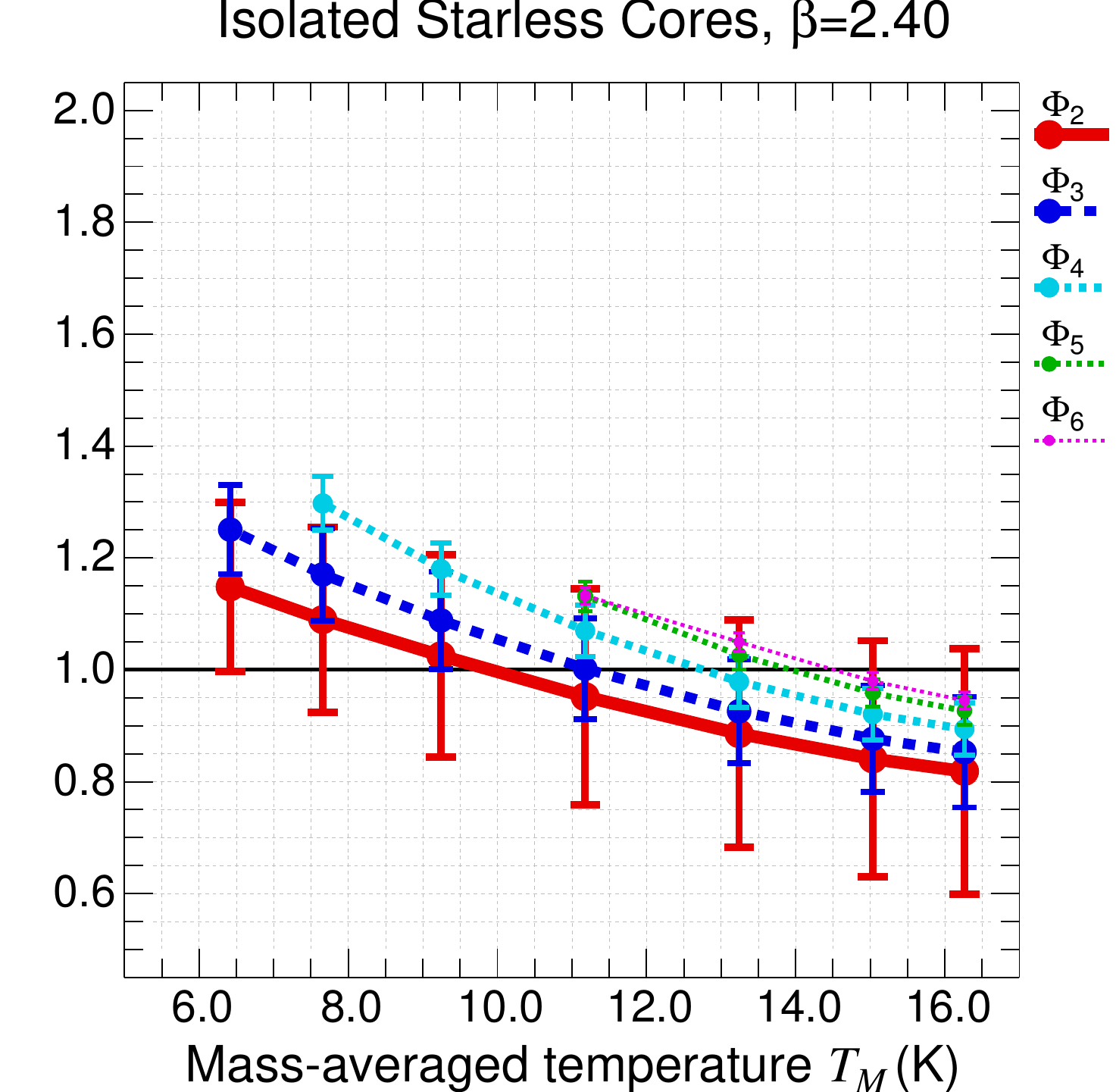}}}
\centerline{\resizebox{0.3327\hsize}{!}{\includegraphics{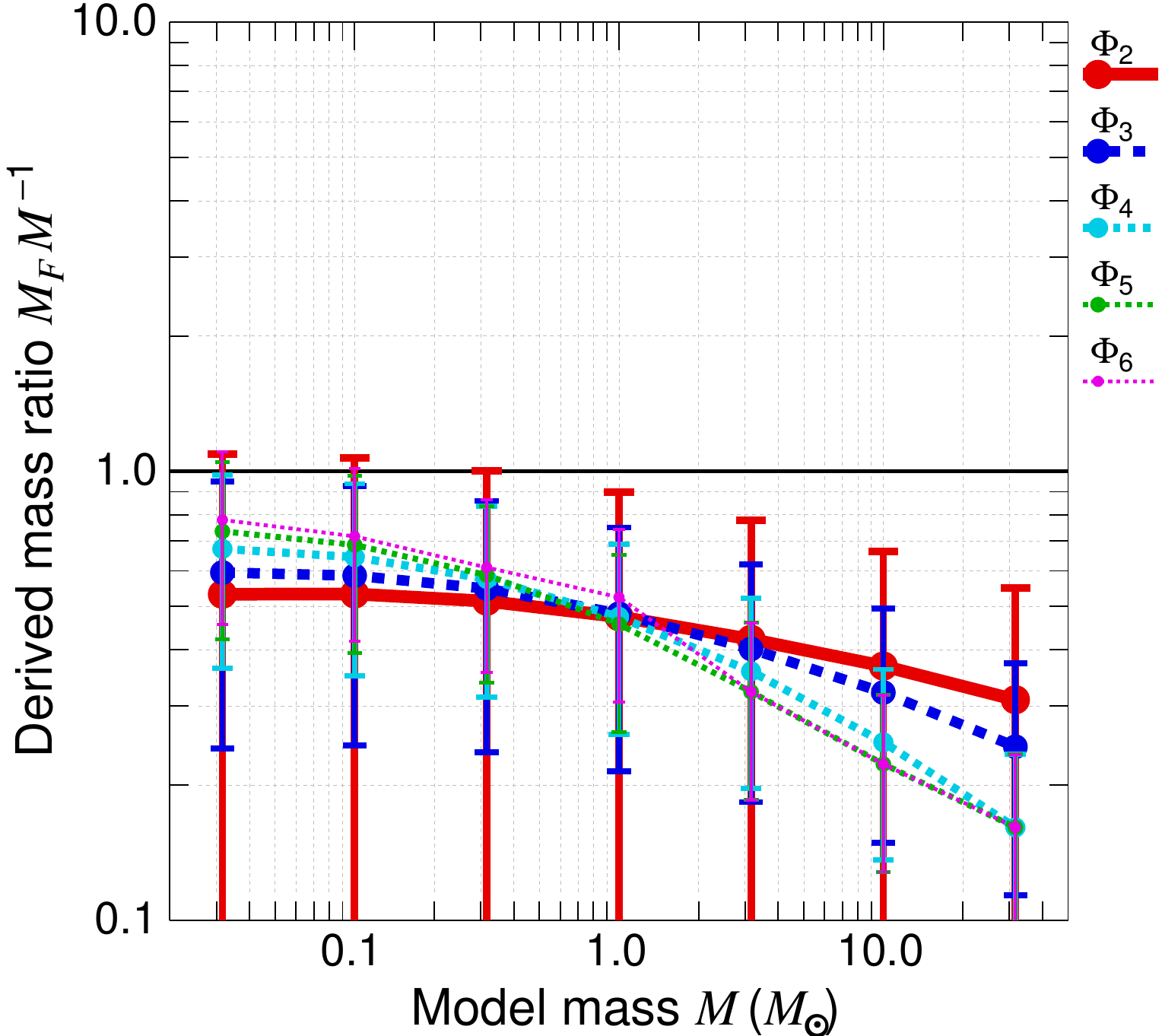}}
            \resizebox{0.3204\hsize}{!}{\includegraphics{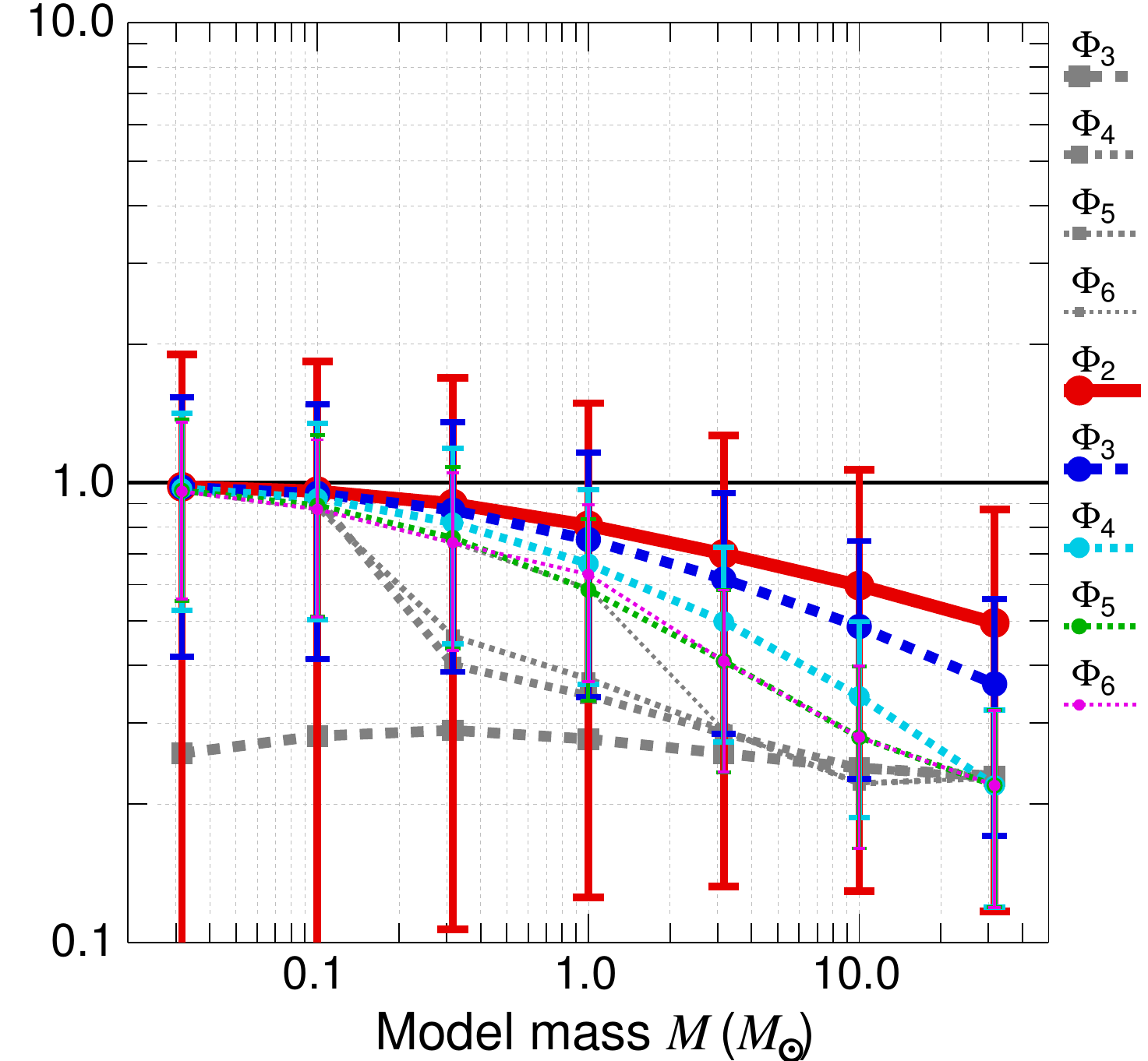}}
            \resizebox{0.3204\hsize}{!}{\includegraphics{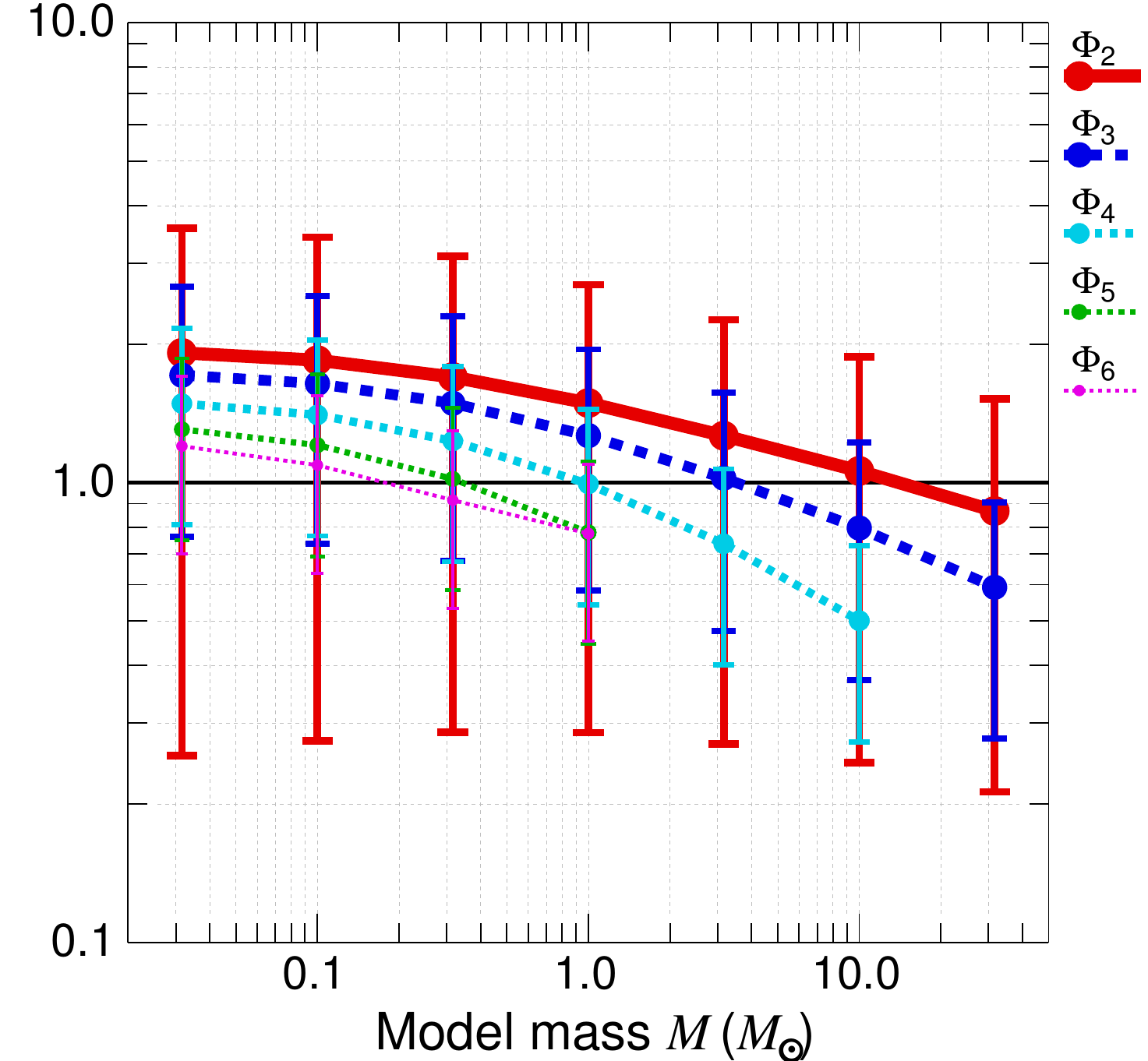}}}
\centerline{\resizebox{0.3327\hsize}{!}{\includegraphics{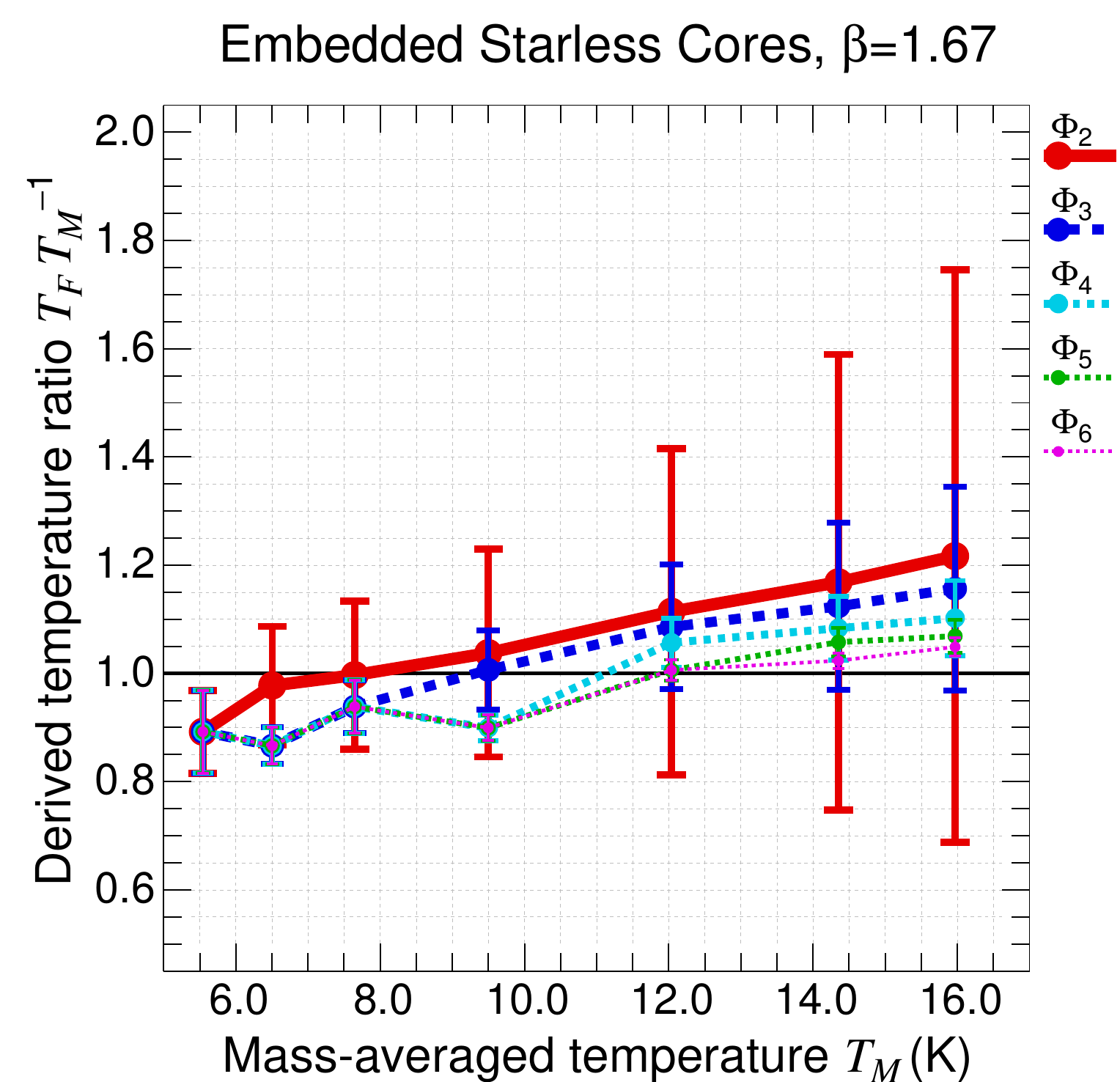}}
            \resizebox{0.3204\hsize}{!}{\includegraphics{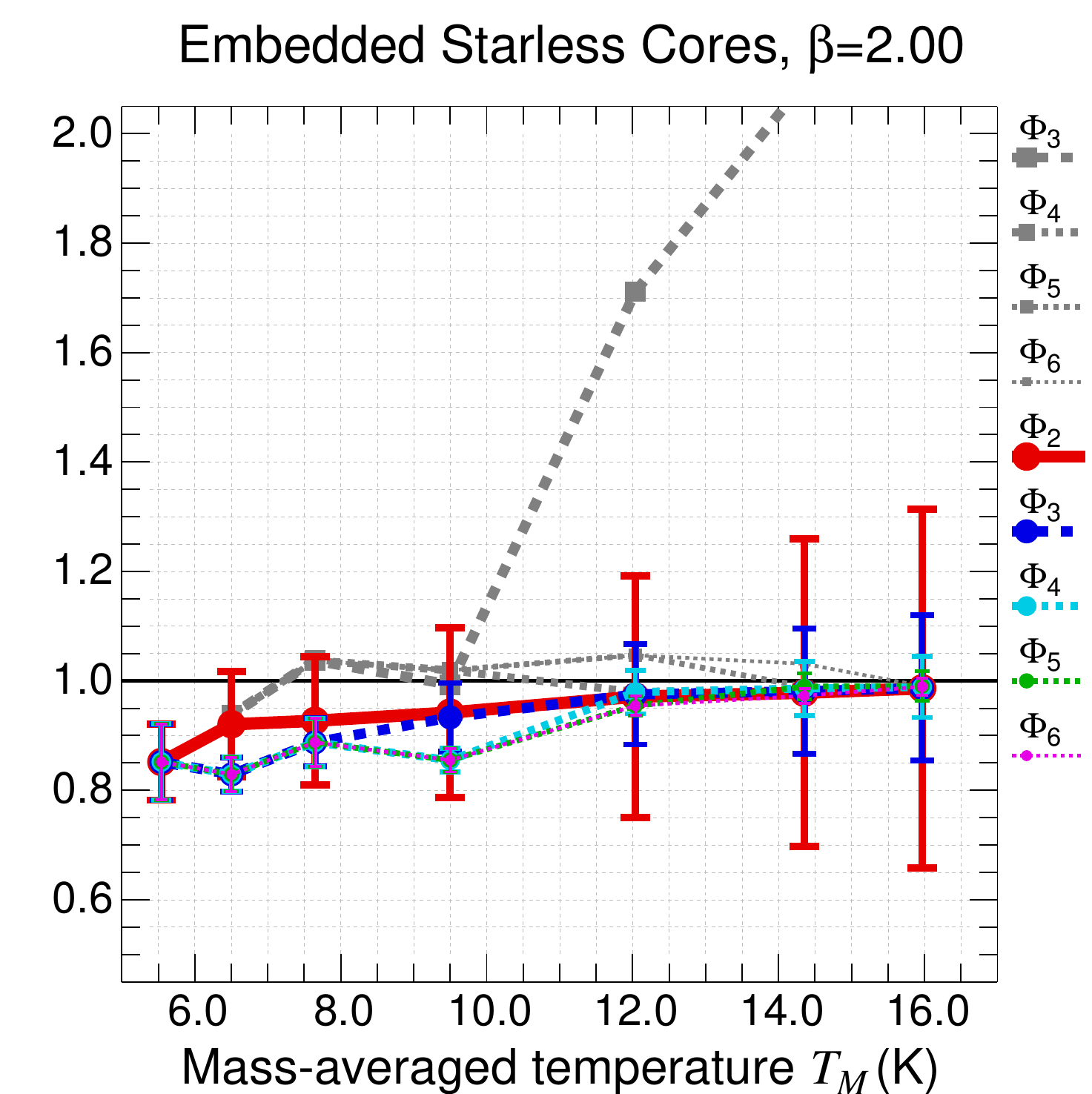}}
            \resizebox{0.3204\hsize}{!}{\includegraphics{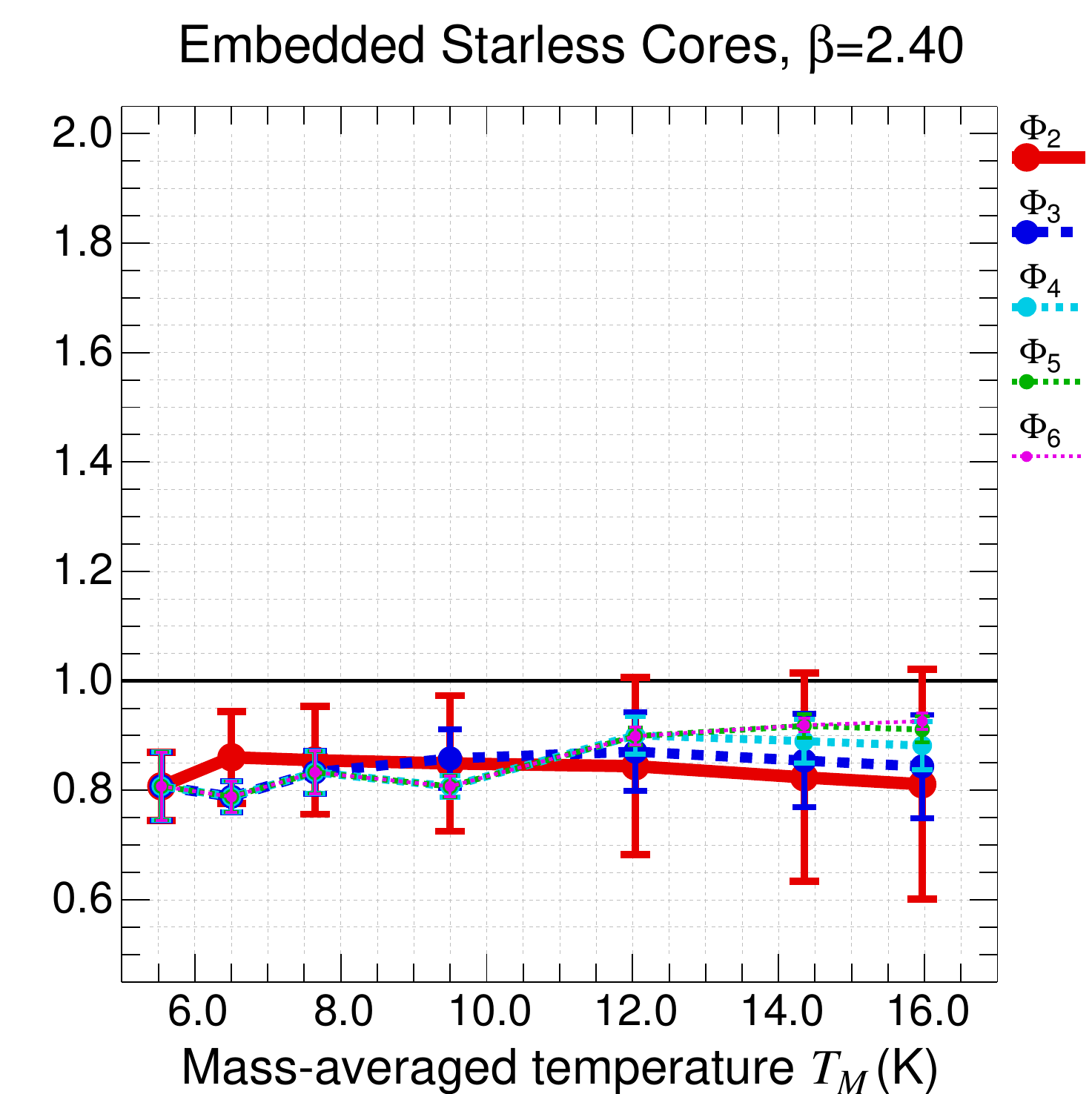}}}
\centerline{\resizebox{0.3327\hsize}{!}{\includegraphics{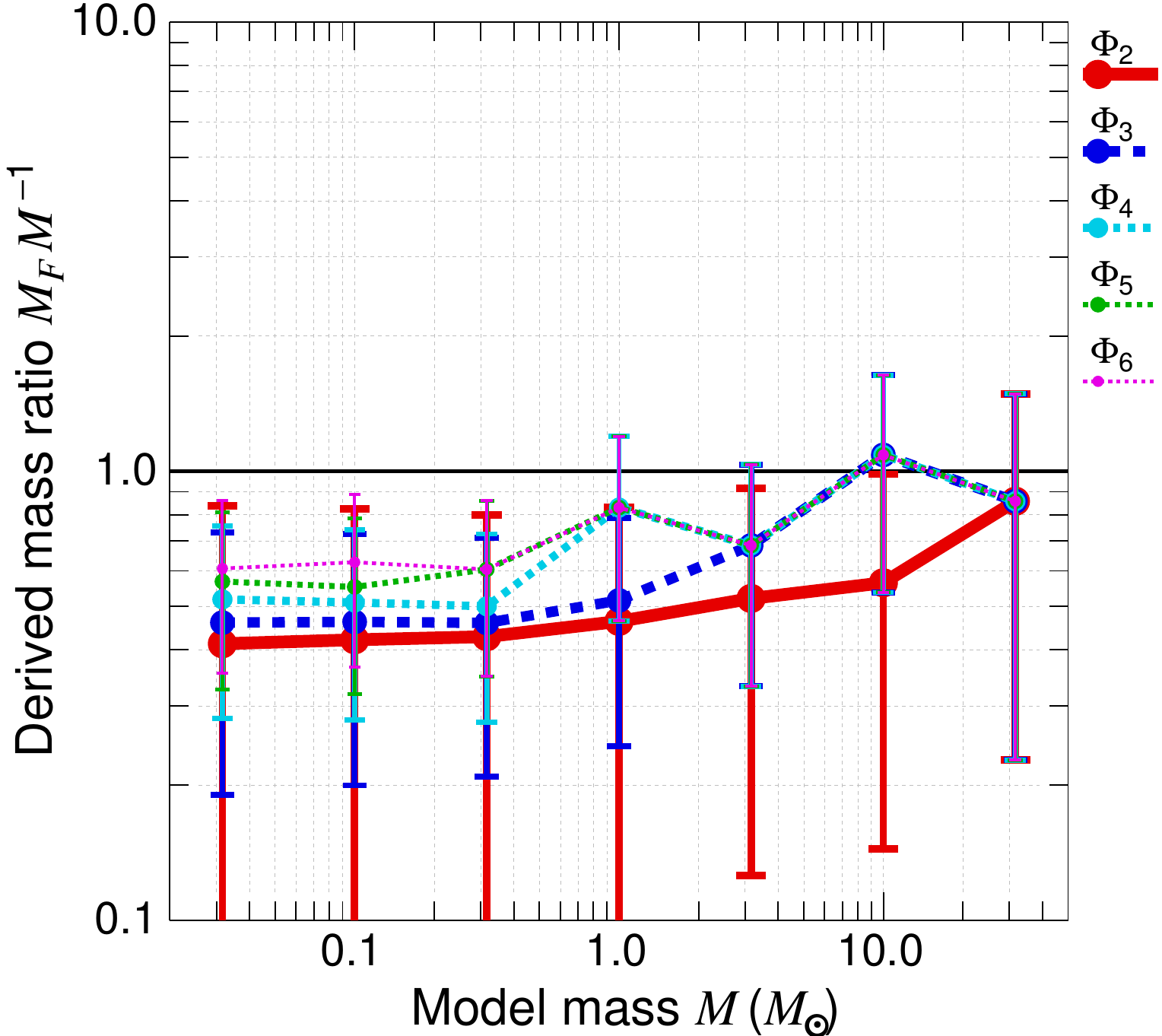}}
            \resizebox{0.3204\hsize}{!}{\includegraphics{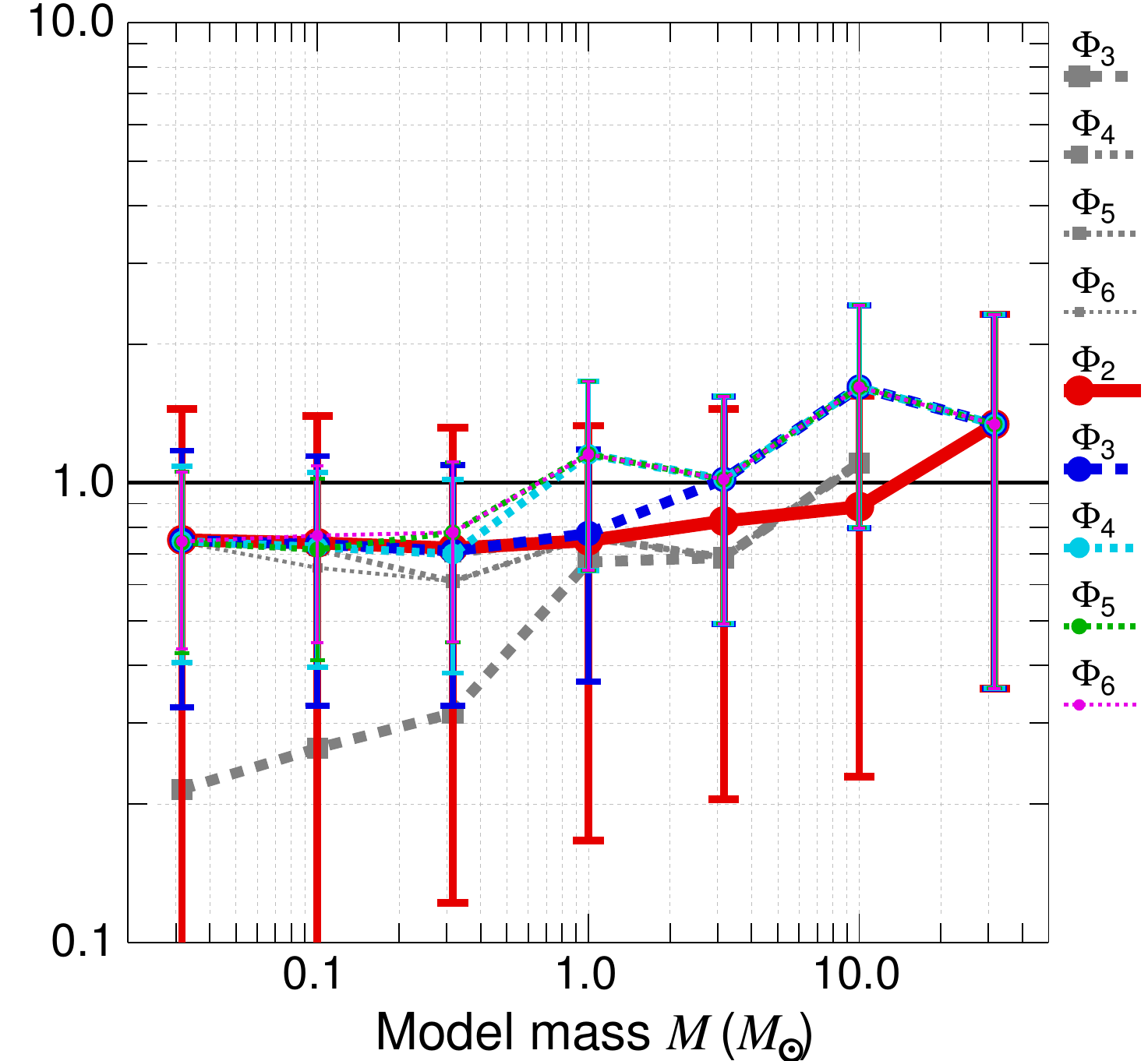}}
            \resizebox{0.3204\hsize}{!}{\includegraphics{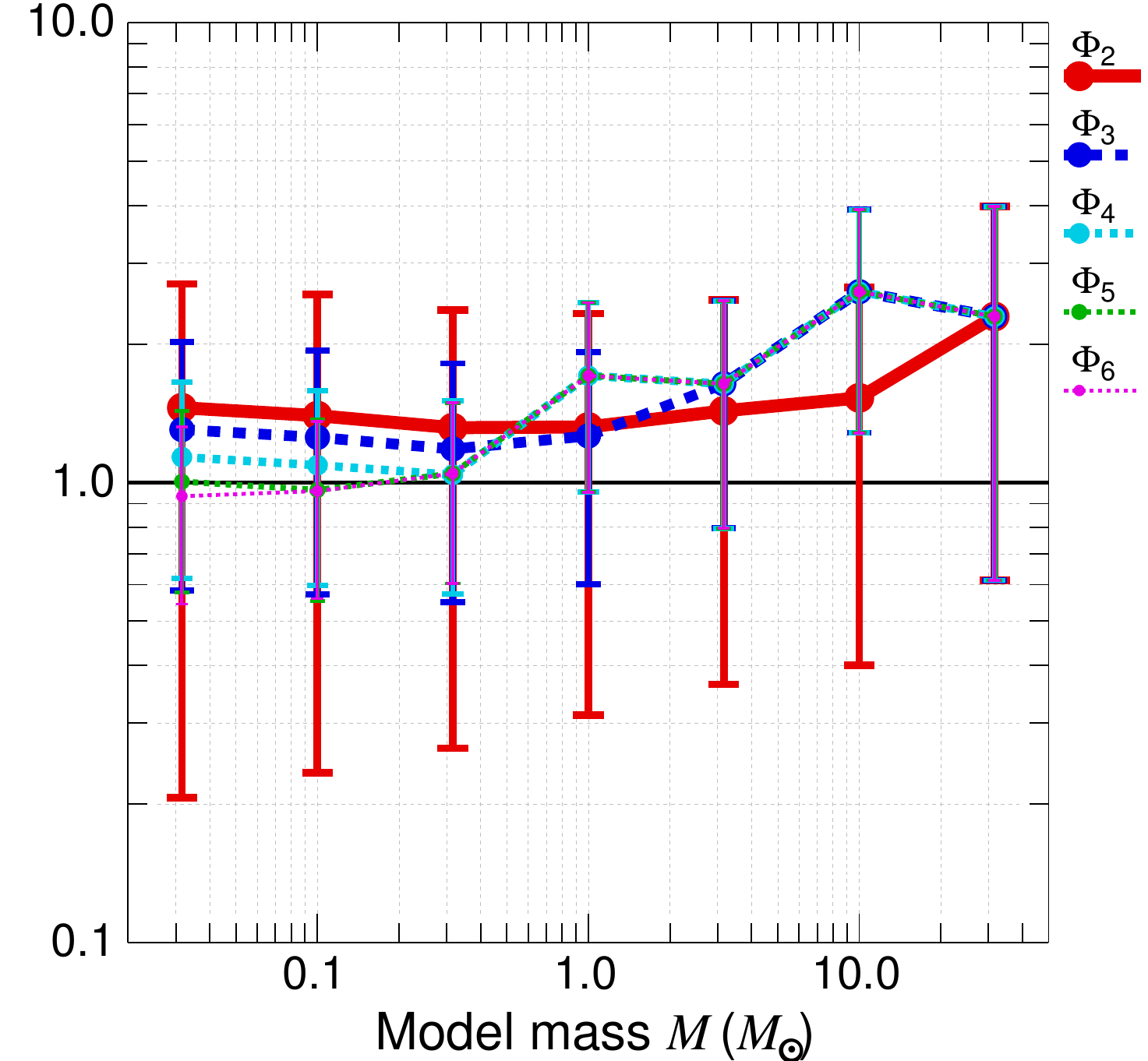}}}
\caption{
Temperatures $T_{F}$ and masses $M_{F}$ derived from fitting $F_{\nu}$ of both \emph{isolated} and \emph{embedded} starless cores 
vs. the true model values of $T_{M}$ and $M$ for three $\beta$ values (1.67, 2, 2.4). For various subsets $\Phi_{n}$ of fluxes, 
results from successful \textsl{thinbody} and \textsl{modbody} fits (Sect.~\ref{data.subsets}) are displayed by the colored and gray 
lines, respectively. Error bars represent the $1\,{\times}\,\sigma$ uncertainties of the derived parameters returned by the fitting 
routine combined with the assumed $\pm\,20{\%}$ uncertainties of $\eta$, $\kappa_{0}$, and $D$ (Sect.~\ref{data.subsets}). The black 
solid lines show the locations where $T_{F}$ and $M_{F}$ are equal to the true values. To preserve clarity of the plots, \emph{much} 
less accurate \textsl{modbody} results are displayed only for correct $\beta\,{=}\,2$ and without error bars.
} 
\label{temp.mass.bes}
\end{figure*}

\begin{figure*}
\centering
\centerline{\resizebox{0.3327\hsize}{!}{\includegraphics{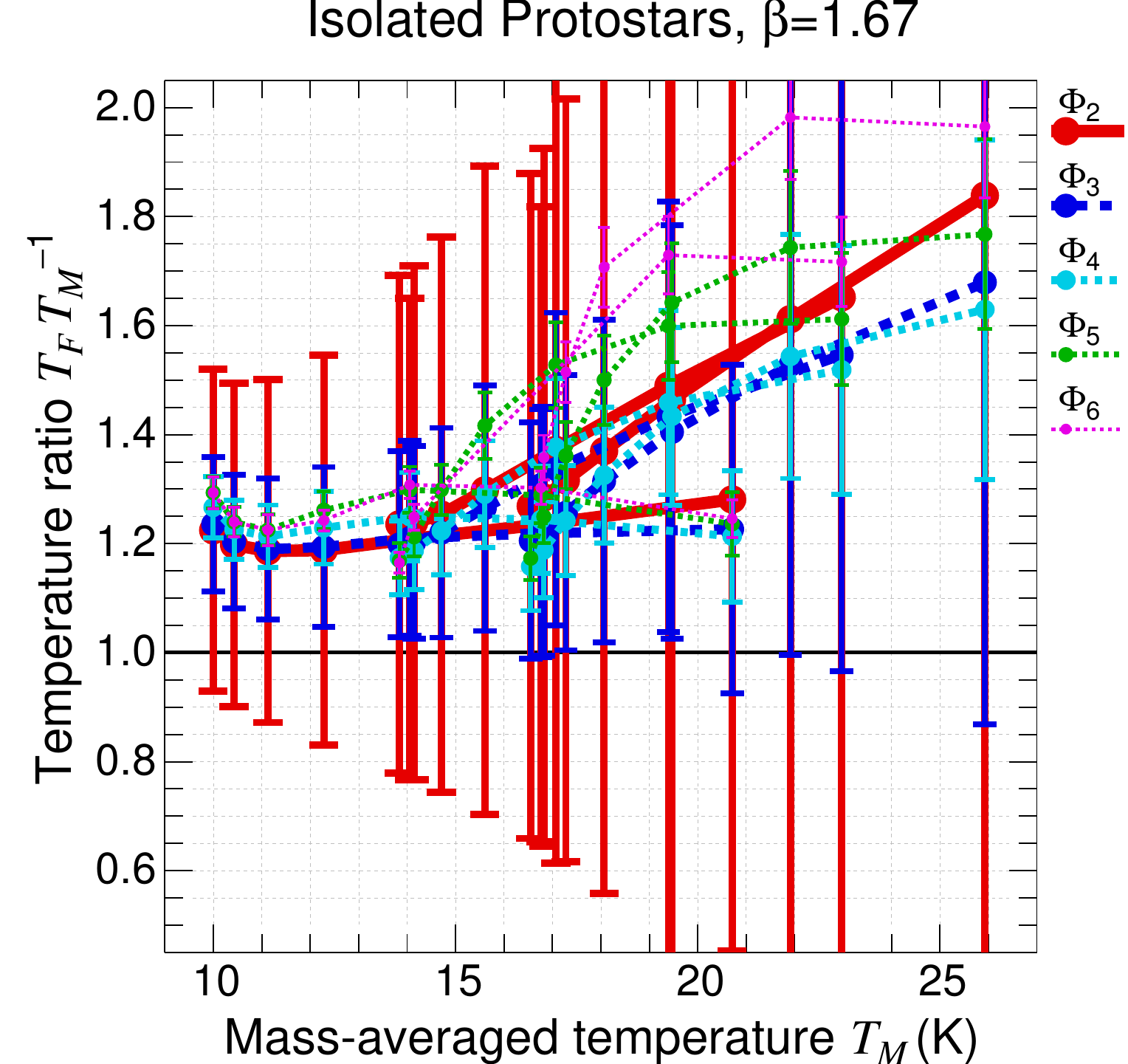}}
            \resizebox{0.3204\hsize}{!}{\includegraphics{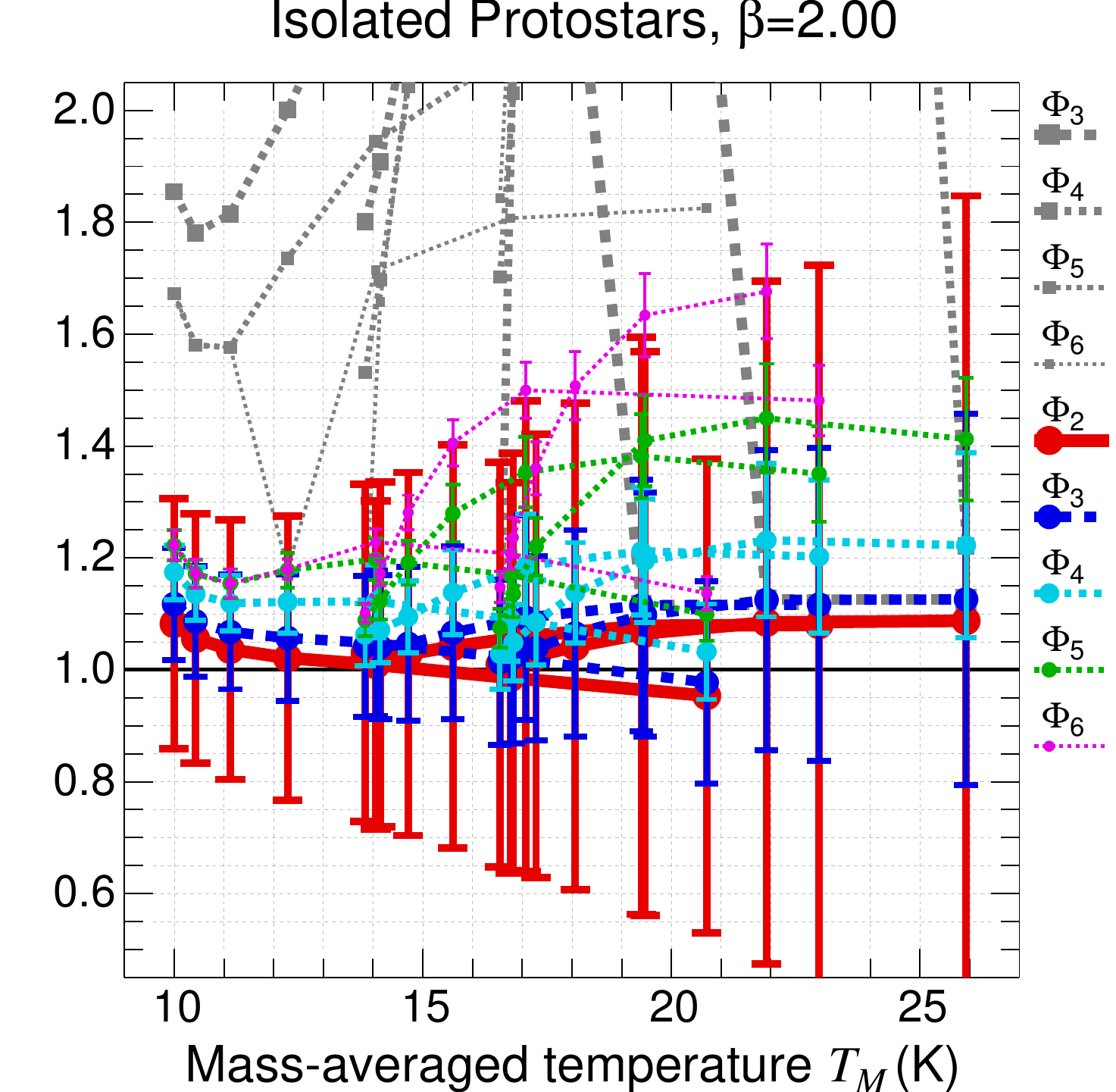}}
            \resizebox{0.3204\hsize}{!}{\includegraphics{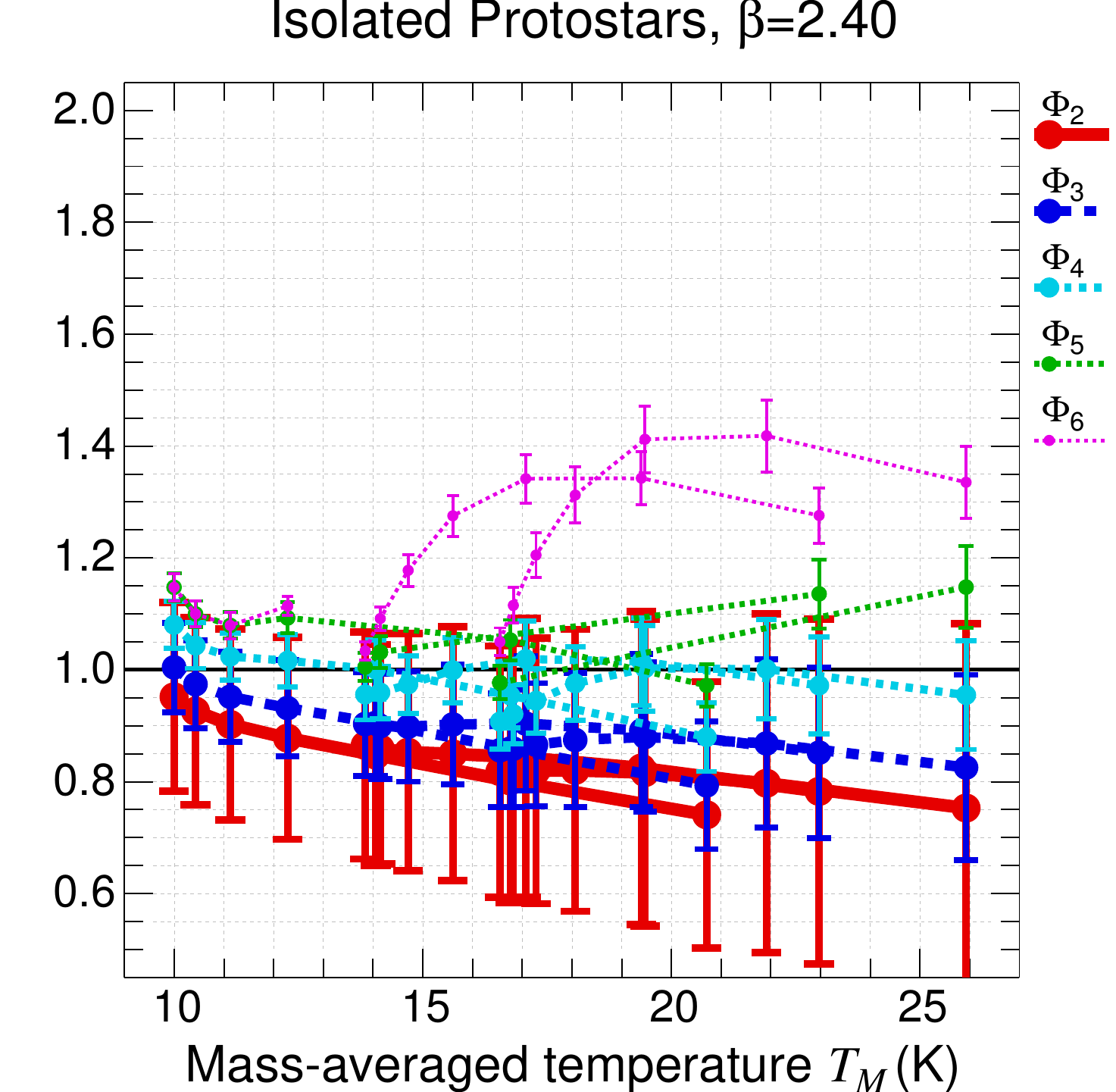}}}
\centerline{\resizebox{0.3327\hsize}{!}{\includegraphics{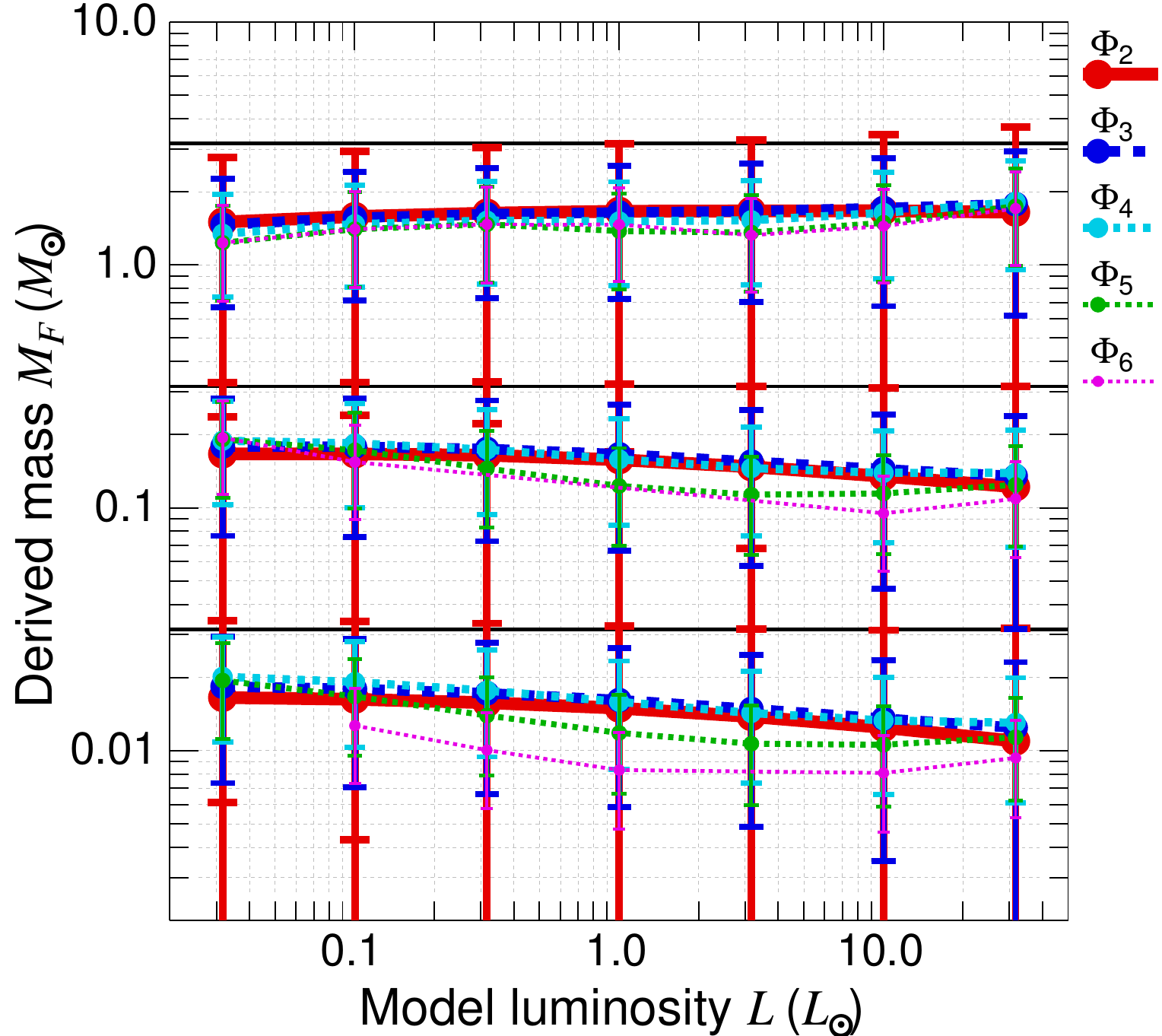}}
            \resizebox{0.3204\hsize}{!}{\includegraphics{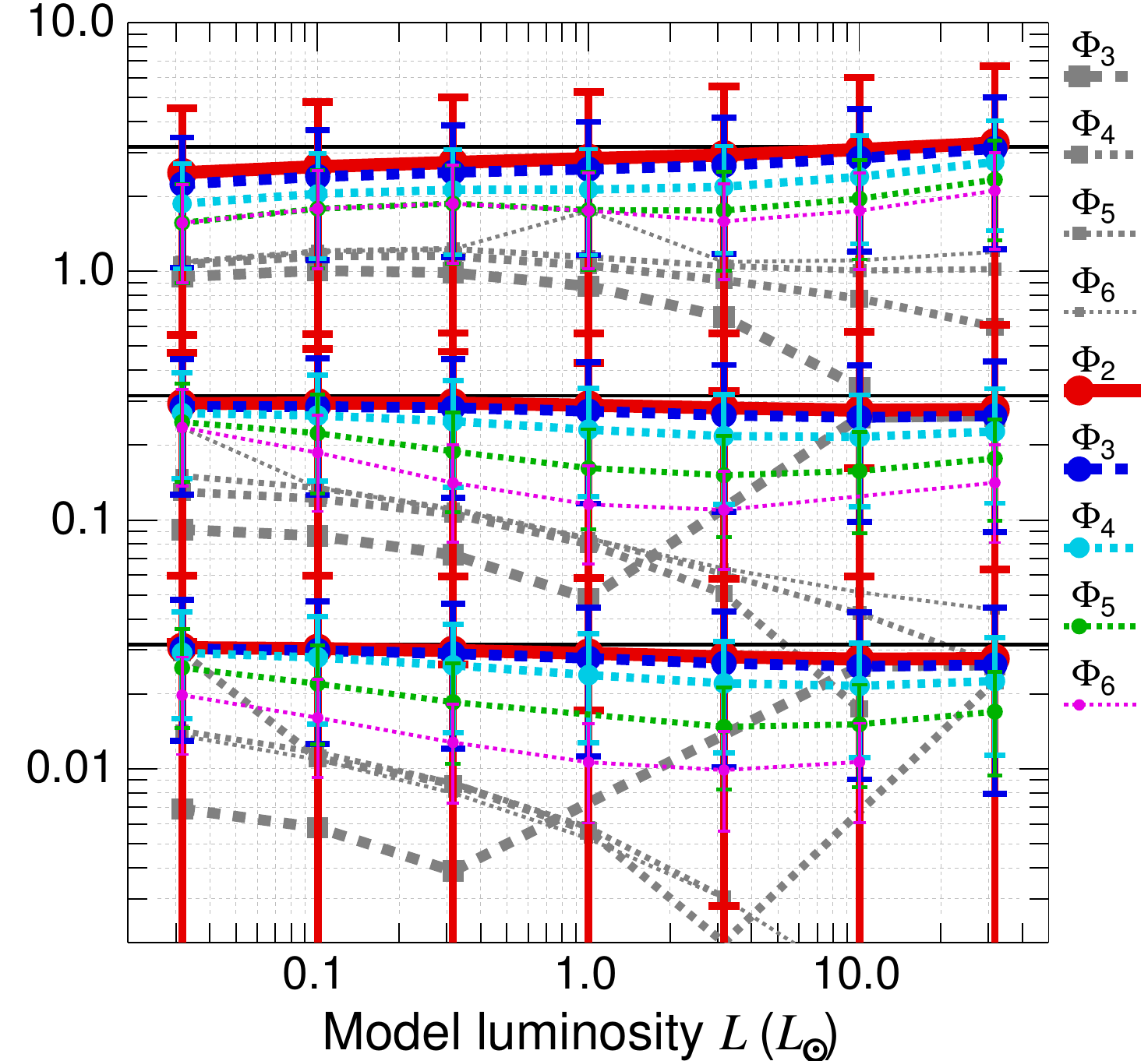}}
            \resizebox{0.3204\hsize}{!}{\includegraphics{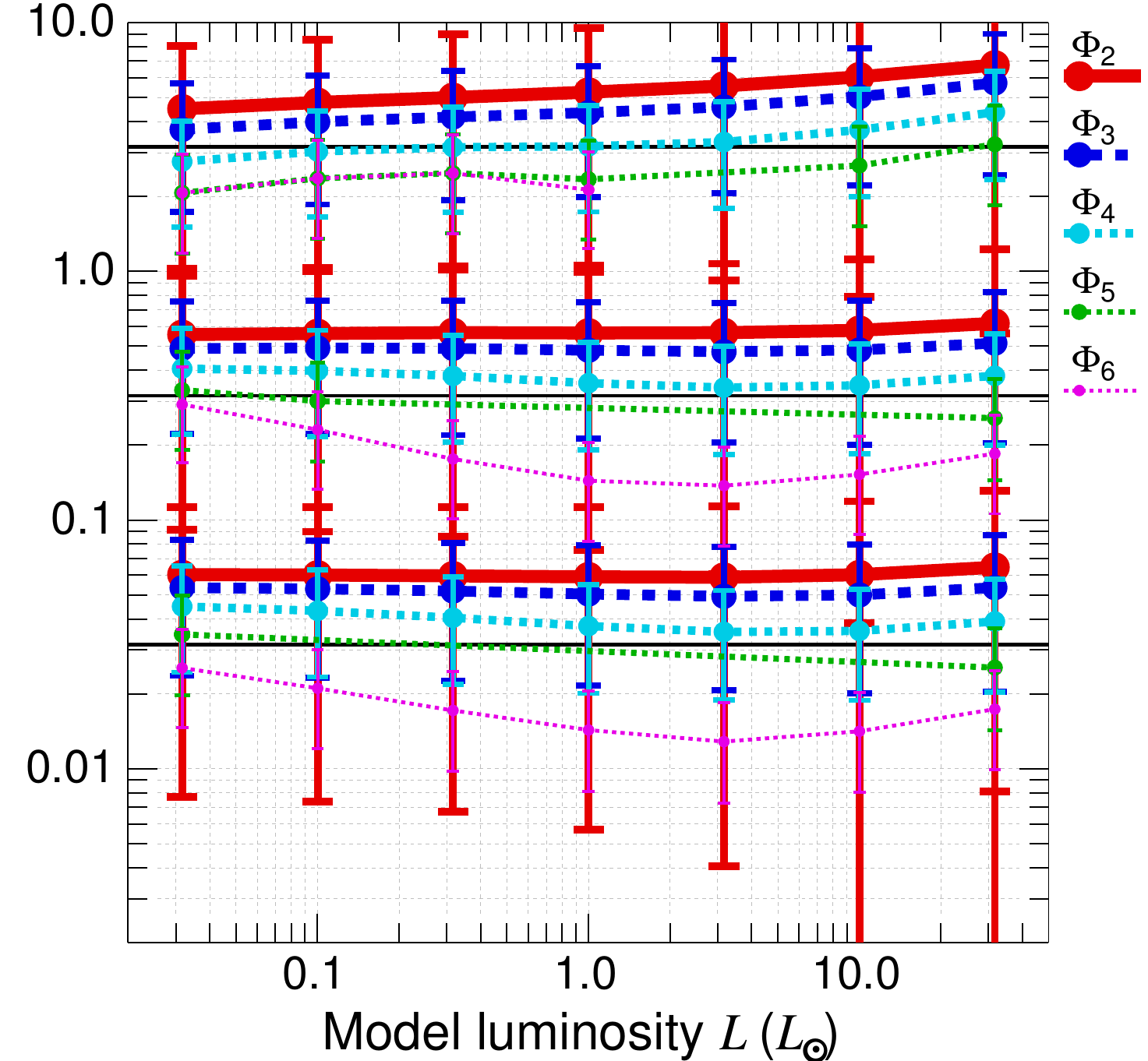}}}
\centerline{\resizebox{0.3327\hsize}{!}{\includegraphics{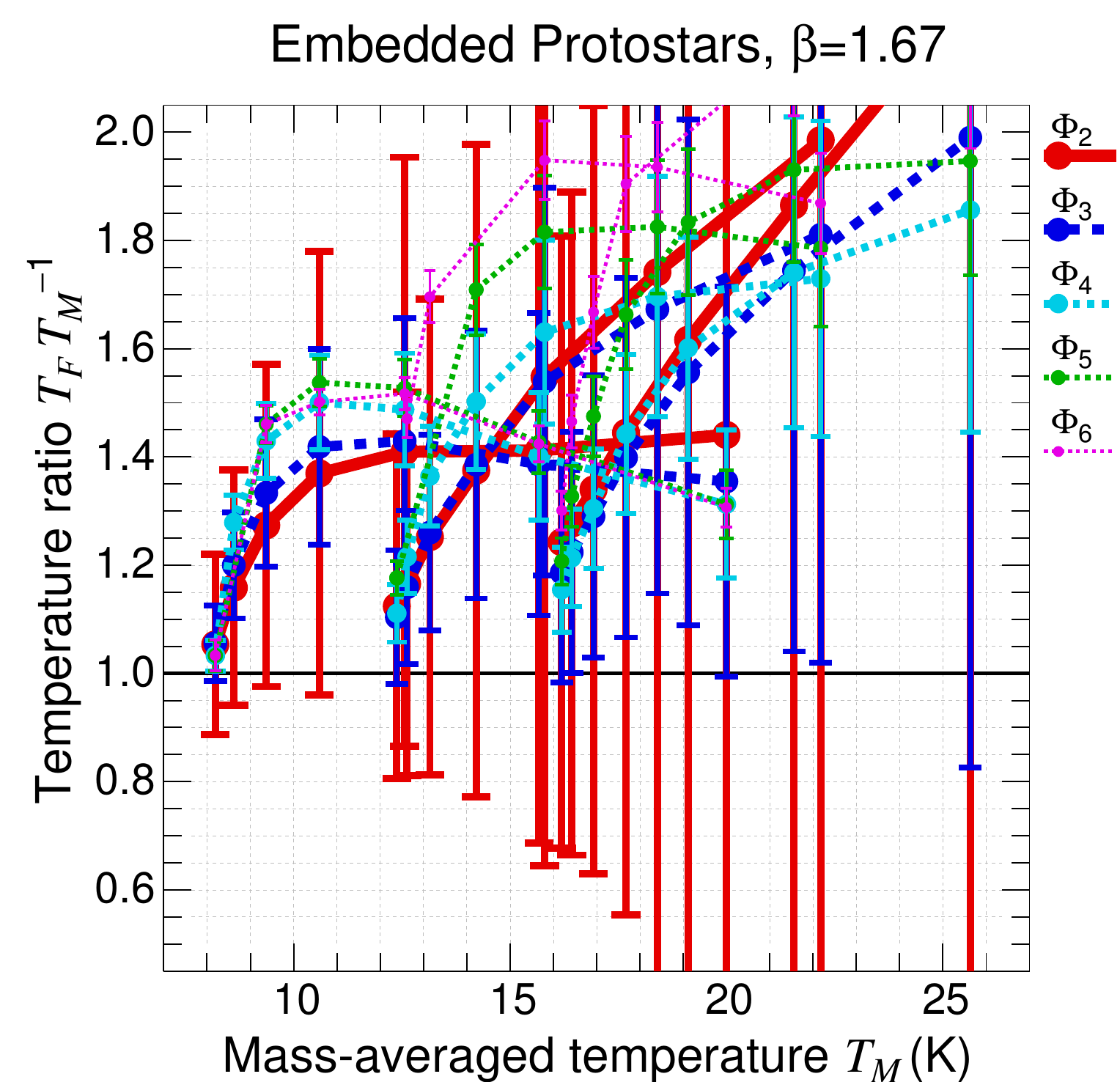}}
            \resizebox{0.3204\hsize}{!}{\includegraphics{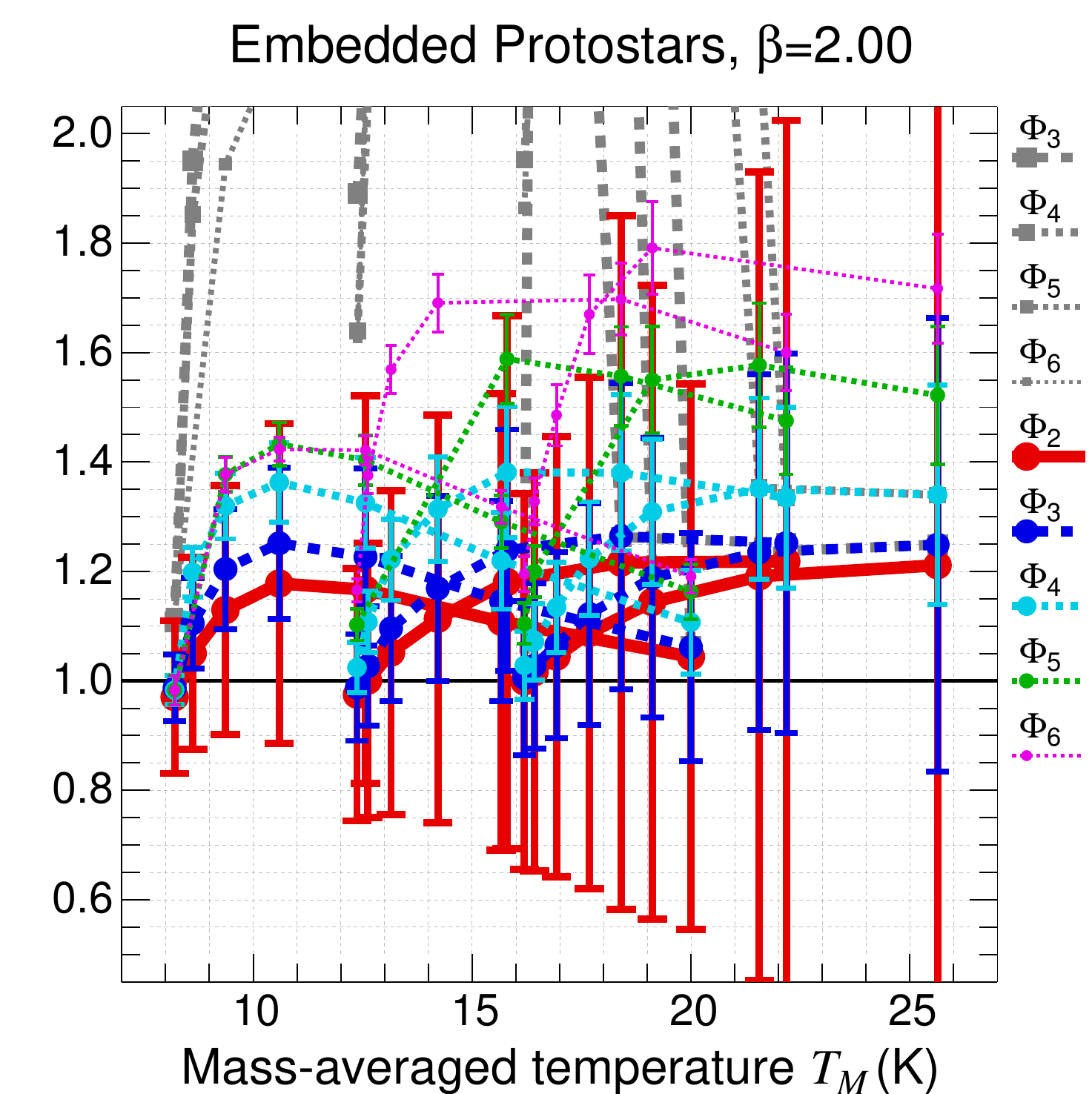}}
            \resizebox{0.3204\hsize}{!}{\includegraphics{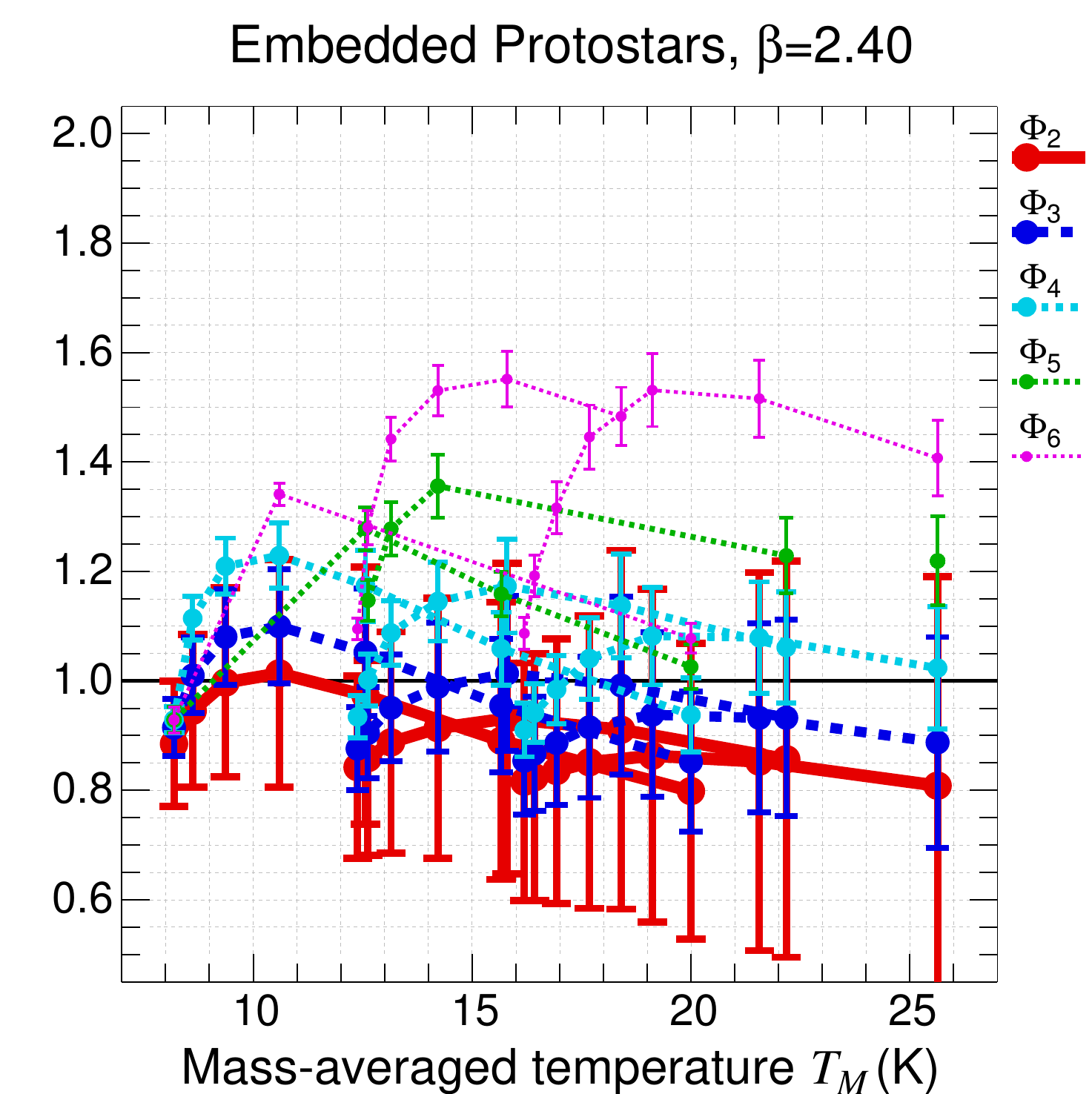}}}
\centerline{\resizebox{0.3327\hsize}{!}{\includegraphics{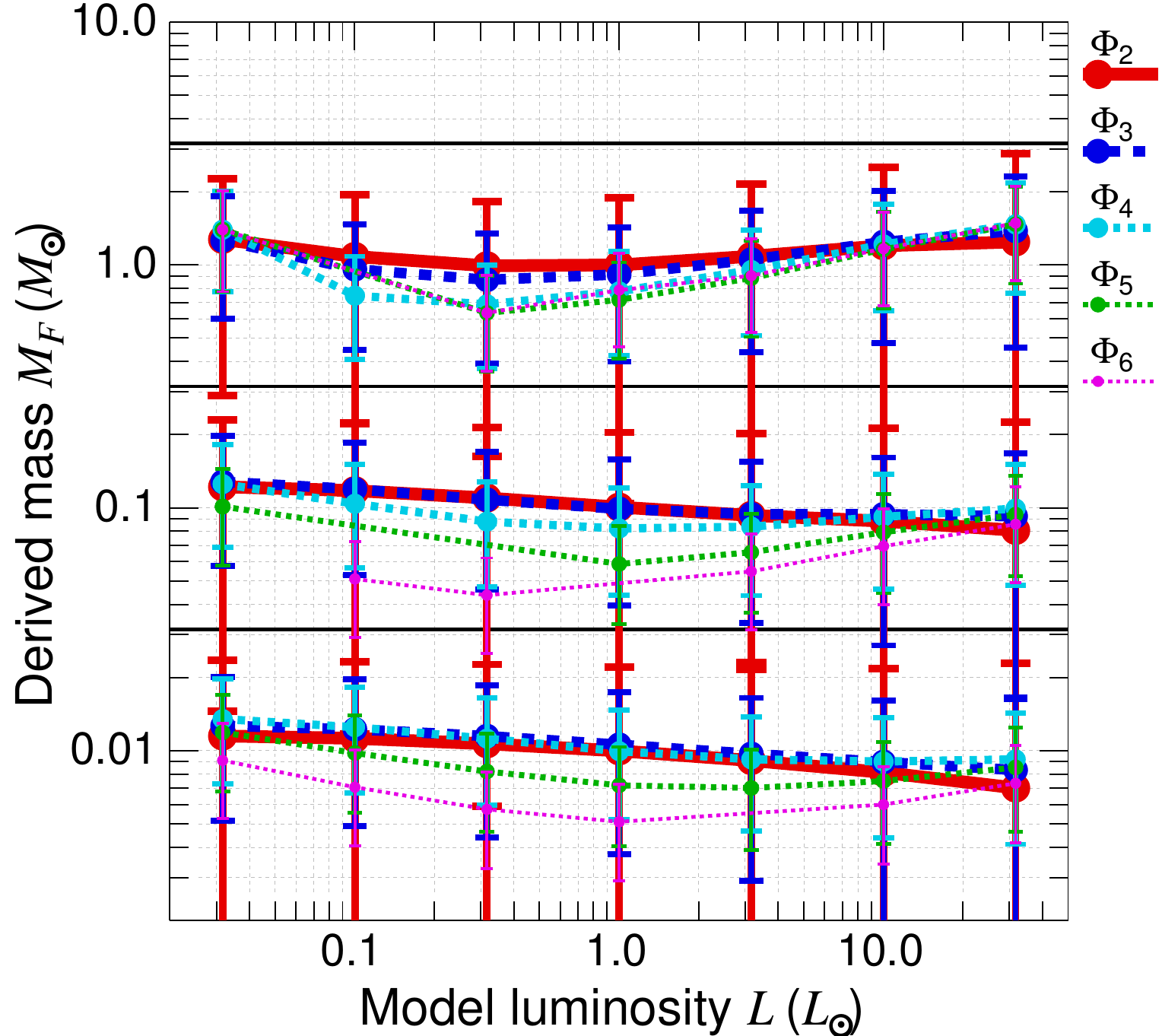}}
            \resizebox{0.3204\hsize}{!}{\includegraphics{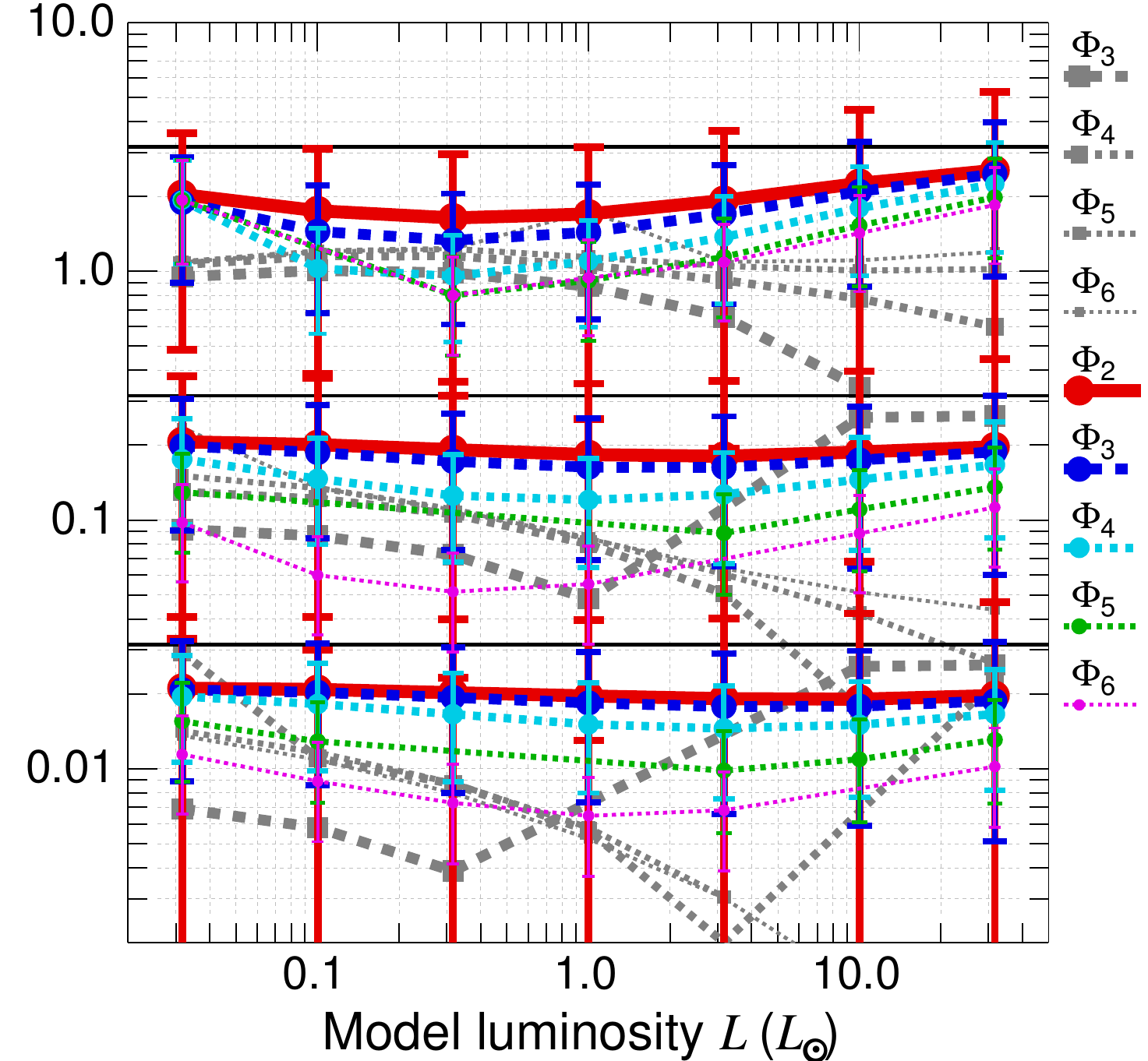}}
            \resizebox{0.3204\hsize}{!}{\includegraphics{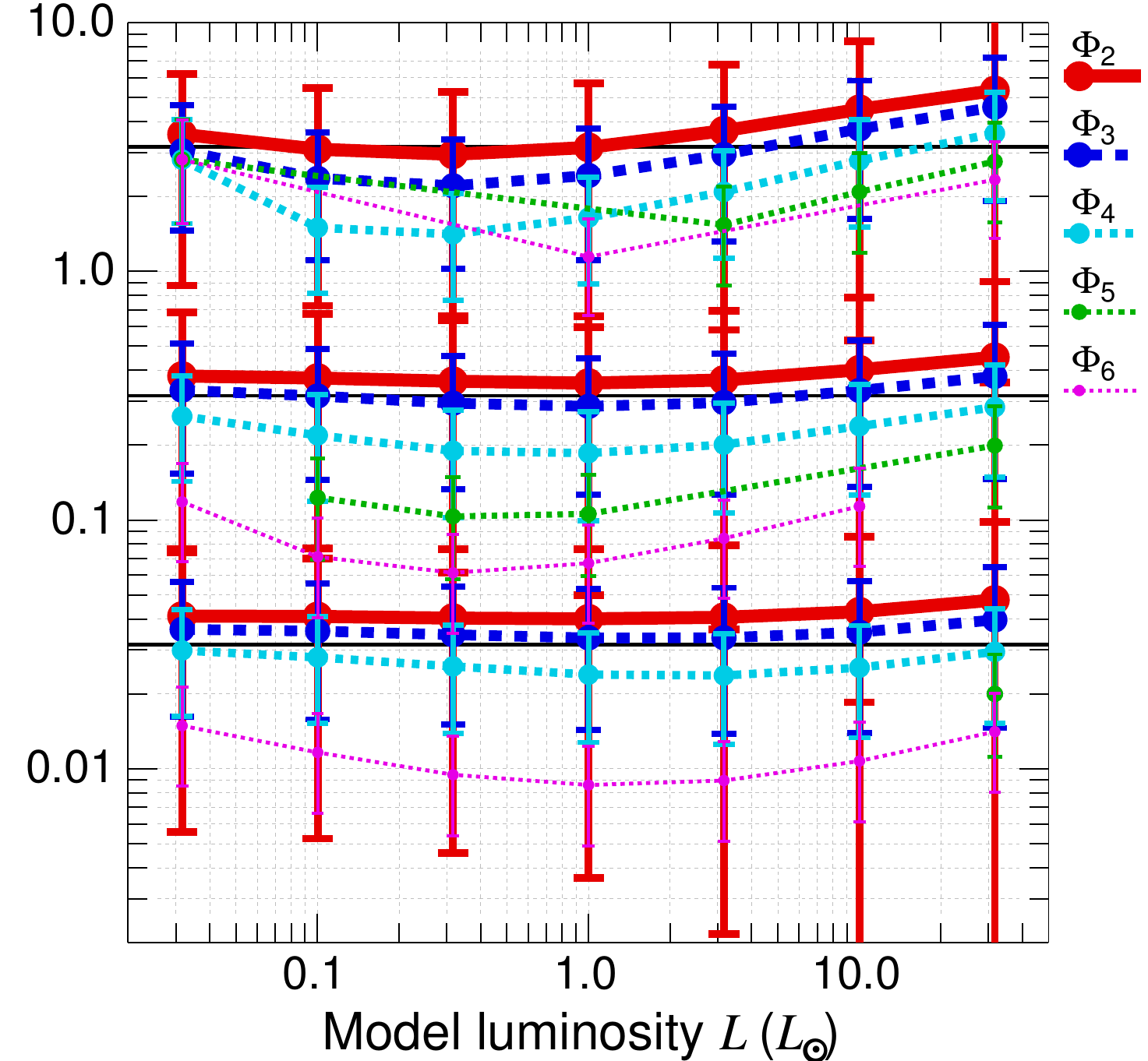}}}
\caption{
Temperatures $T_{F}$ and masses $M_{F}$ derived from fitting $F_{\nu}$ of both \emph{isolated} and \emph{embedded} protostellar
envelopes (with the true masses $M$ of 0.03, 0.3, and 3$\,M_{\sun}$) vs. the true model values of $T_{M}$ and $L$ for three
$\beta$ values (1.67, 2, 2.4). Results from successful \textsl{thinbody} and \textsl{modbody} fits for various subsets $\Phi_{n}$ 
of fluxes (Sect.~\ref{data.subsets}) are displayed by the colored and gray lines, respectively. See Fig.~\ref{temp.mass.bes} for 
more details.
} 
\label{temp.mass.pro}
\end{figure*}

Flux distributions of the isolated starless cores (Fig.~\ref{sed.bes.pro}) are similar to those of the modified blackbodies
$\kappa_{\nu}\,B_{\nu}(T_{M})$. The fits for a low-mass core with $M\,{=}\,0.03\,M_{\sun}$ shown in Fig.~\ref{fits.examples} are
identical for all subsets $\Phi_{n}$ since the core is nearly isothermal, with $T_{\rm d}(r)$ very similar to its
$T_{M}\,{=}\,16.3$\,K. Fluxes of the higher-mass cores of $0.3$ and $3\,M_{\sun}$ display larger deviations from the fluxes
$F_{\nu}(T_{M})$ of isothermal models for larger subsets $\Phi_{n}$ ($n\,{=}\,3\,{\rightarrow}\,6$). The shapes of $F_{\nu}$ become
``hotter'' because of the steeper temperature profiles (Sect.~\ref{temperatures}) at the outer boundary (Fig.~\ref{trp.bes.pro}).
For a massive core of $3\,M_{\sun}$, discrepancies between $F_{\nu}$ and $F_{\nu}(T_{M})$ at $\lambda\,{\la}\,160$\,{${\mu}$m}
reach factors ${\ga}\,5$.

Flux distributions of an isolated protostellar envelope with $M\,{=}\,0.03\,M_{\sun}$ (Fig.~\ref{sed.bes.pro}) display various
shapes that are quite different from those of $\kappa_{\nu}\,B_{\nu}(T_{M})$, whereas for a more opaque envelope of $3\,M_{\sun}$
they become similar to the modified blackbody shapes. The protostellar fits (Fig.~\ref{fits.examples}) show greater deviations for
larger subsets $\Phi_{n}$ ($n\,{=}\,3\,{\rightarrow}\,6$), much larger than those of starless cores. Differences between $F_{\nu}$
and $F_{\nu}(T_{M})$ reach orders of magnitude at $\lambda\,{\la}\,100$\,{${\mu}$m}. The shapes appear much ``hotter'' owing to
$T_{\rm d}\,{\sim}\,100{-}10^{3}$\,K (Sect.~\ref{temperatures}) deep inside the envelopes (Fig.~\ref{trp.bes.pro}). The
lower-mass protostellar envelopes are more transparent and the hot emission greatly distorts $F_{\nu}$ at
$\lambda\,{\la}\,250$\,{${\mu}$m}.

\subsection{Properties derived from fitting fluxes $F_{\nu}$}
\label{derived.properties}

Isolated starless cores, $\beta\,{=}\,2$ (Fig.~\ref{temp.mass.bes}). For the low-mass, transparent cores
($M\,{\rightarrow}\,0.03\,M_{\sun}$), quite accurate values $T_{F}\,{\approx}\,T_{M}$ and $M_{F}\,{\approx}\,M$ are derived for all
subsets $\Phi_{n}$. For the denser, more opaque cores ($M\,{\rightarrow}\,30\,M_{\sun}$), derived $T_{F}$ and $M_{F}$ become more
over- and underestimated, respectively, as the spectral shapes of $F_{\nu}$ become much wider and distorted towards shorter
wavelengths (Fig.~\ref{sed.bes.pro}). The biases and inaccuracy of the estimates depend on the subset $\Phi_{n}$, with the least
inaccurate $T_{F}$ and $M_{F}$ obtained for the \textsl{thinbody} fits of $\Phi_{2}$. However, the biases of the parameters across
the entire mass range remains fairly strong. Derived masses of the starless cores are underestimated within a factor of $2$ for
$1\,{<}\,M\,{\le}\,3\,M_{\sun}$ and factor of $5$ for $3\,{<}\,M\,{\le}\,30\,M_{\sun}$.

Embedded starless cores, $\beta\,{=}\,2$ (Fig.~\ref{temp.mass.bes}). For the low-mass, transparent cores
($M\,{\rightarrow}\,0.03\,M_{\sun}$), $M_{F}$ are underestimated by a factor of $1.35$ for all subsets $\Phi_{n}$, although $T_{F}$
are quite accurate because the standard observational procedure of background subtraction ignores the fact that embedding
backgrounds tend to be rim-brightened at their outer boundary $R$ (Appendix \ref{AppendixB}, Sect.~\ref{bg.subtraction}). The
embedded cores have $T_{\rm d}(r)$ that are quite flat across their boundary for all masses (Fig.~\ref{trp.bes.pro}). Having no
flux distortions caused by nonuniform temperatures (Fig.~\ref{sed.bes.pro}), the $F_{\nu}$ peaks of the most massive cores
($M\,{\rightarrow}\,30\,M_{\sun}$) move towards the longest wavelength ($\lambda_{6}\,{=}\,500$\,{${\mu}$m}), which leads to
$T_{F}$ and $M_{F}$ that are under- and overestimated, respectively.

Isolated protostellar envelopes, $\beta\,{=}\,2$ (Fig.~\ref{temp.mass.pro}). Emission of the hot dust with $T_{\rm
d}{\sim}\,100{-}10^{3}$\,K greatly skews their $F_{\nu}$ towards shorter wavelengths (Fig.~\ref{sed.bes.pro}). This becomes
especially significant for the lower mass, more transparent envelopes ($M\,{\rightarrow}\,0.03\,M_{\sun}$,
$L_{\star}\,{\rightarrow}\,30\,L_{\sun}$) that produce hotter dust over a much larger volume (Fig.~\ref{trp.bes.pro}). The
\textsl{thinbody} fits of larger subsets $\Phi_{n}$ ($n\,{=}\,3\,{\rightarrow}\,6$), lead to errors in $T_{F}$ and $M_{F}$ that
reach factors of $1.6$ and $3$, respectively. The smallest subset $\Phi_{2}$ is unaffected by the hot emission and it produces
fairly accurate \textsl{thinbody} estimates of $T_{F}$ and $M_{F}$ (for all $M$ and $L_{\star}$) within factors of $1.1$ and
$1.3$, respectively.

Embedded protostellar envelopes, $\beta\,{=}\,2$ (Fig.~\ref{temp.mass.pro}). Results are qualitatively similar to those of the
isolated envelopes, although with larger inaccuracies. Derived $M_{F}$ are underestimated by at least a factor of $1.5$, mostly due
to over-subtraction of the rim-brightened embedding background (Appendix \ref{AppendixB}, Sect.~\ref{bg.subtraction}). Although the
envelopes have $T_{\rm d}(r)$ that are quite flat across their boundaries (Fig.~\ref{trp.bes.pro}), their derived parameters are
greatly affected by the skewed $F_{\nu}$ owing to the hot dust deep in their interiors. The \textsl{thinbody} fits of large subsets
$\Phi_{n}$ ($n\,{=}\,3\,{\rightarrow}\,6$) lead to inaccuracies in $T_{F}$ and $M_{F}$ as large as factors of $1.4{-}1.8$ and
$3{-}5$, respectively. The most accurate $T_{F}$ and $M_{F}$, obtained for the smallest subset $\Phi_{2}$, are underestimated
within factors of $1.2$ and $2$.

Effects of the adopted opacity slope $\beta$ on the estimated parameters are similar for both starless cores
(Fig.~\ref{temp.mass.bes}) and protostellar envelopes (Fig.~\ref{temp.mass.pro}). Although detailed behavior of the differences
with respect to the above results for true $\beta\,{=}\,2$ depends on the subset $\Phi_{n}$, clear general trends can be seen.
Shallower slopes ($\beta\,{=}\,1.67$) lead to an increase in $T_{F}$ and thus $M_{F}$ becomes smaller, whereas steeper slopes
($\beta\,{=}\,2.4$) lead to a decrease in $T_{F}$ and hence $M_{F}$ becomes larger, in both cases by a factor of approximately $2$.

The \textsl{thinbody} fitting model produces much better overall results than \textsl{modbody} does. Parameters estimated with
\textsl{modbody} become so incorrect that they may be considered completely unusable. The importance of estimating accurate
mass-averaged temperatures $T_{M}$ for deriving correct masses $M_{F}$ is illustrated by the isothermal models presented in
Appendix \ref{AppendixD}.

\subsection{Properties derived from fitting images $\mathcal{I}_{\nu}$}
\label{coldens.properties}

This section presents results for both starless cores and protostellar envelopes, obtained from successful fits of the
background-subtracted $\mathcal{I}_{\nu}$ for all subsets $\Phi_{n}$, for only the \textsl{thinbody} fitting model. Derived
\textsl{modbody} masses are practically the same as the \textsl{thinbody} masses, because the bulk of the model mass is in
optically thin regions (Sect.~\ref{variable.fixed}). Effects of the adopted far-infrared opacity slopes are the same as when
fitting $F_{\nu}$ (Sect.~\ref{derived.properties}): under- or overestimating $\beta$ by a factor of $1.2$ gives masses
$M_{\mathcal{I}}$ that are systematically under- or overestimated by a factor of $2$. The method of fitting images
$\mathcal{I}_{\nu}$, thereby deriving $\mathcal{N}_{\rm H_2}$, and afterwards integrating source mass $M_{\mathcal{I}}$ brings
clear benefits for well-resolved starless cores with nonuniform temperatures, compared to the other method (Sect.
\ref{derived.properties}) of first integrating total fluxes $F_{\nu}$ from $\mathcal{I}_{\nu}$ (losing all spatial information) and
then estimating $M_{F}$ from the fitting model.

\begin{figure*}
\centering
\centerline{\resizebox{0.3327\hsize}{!}{\includegraphics{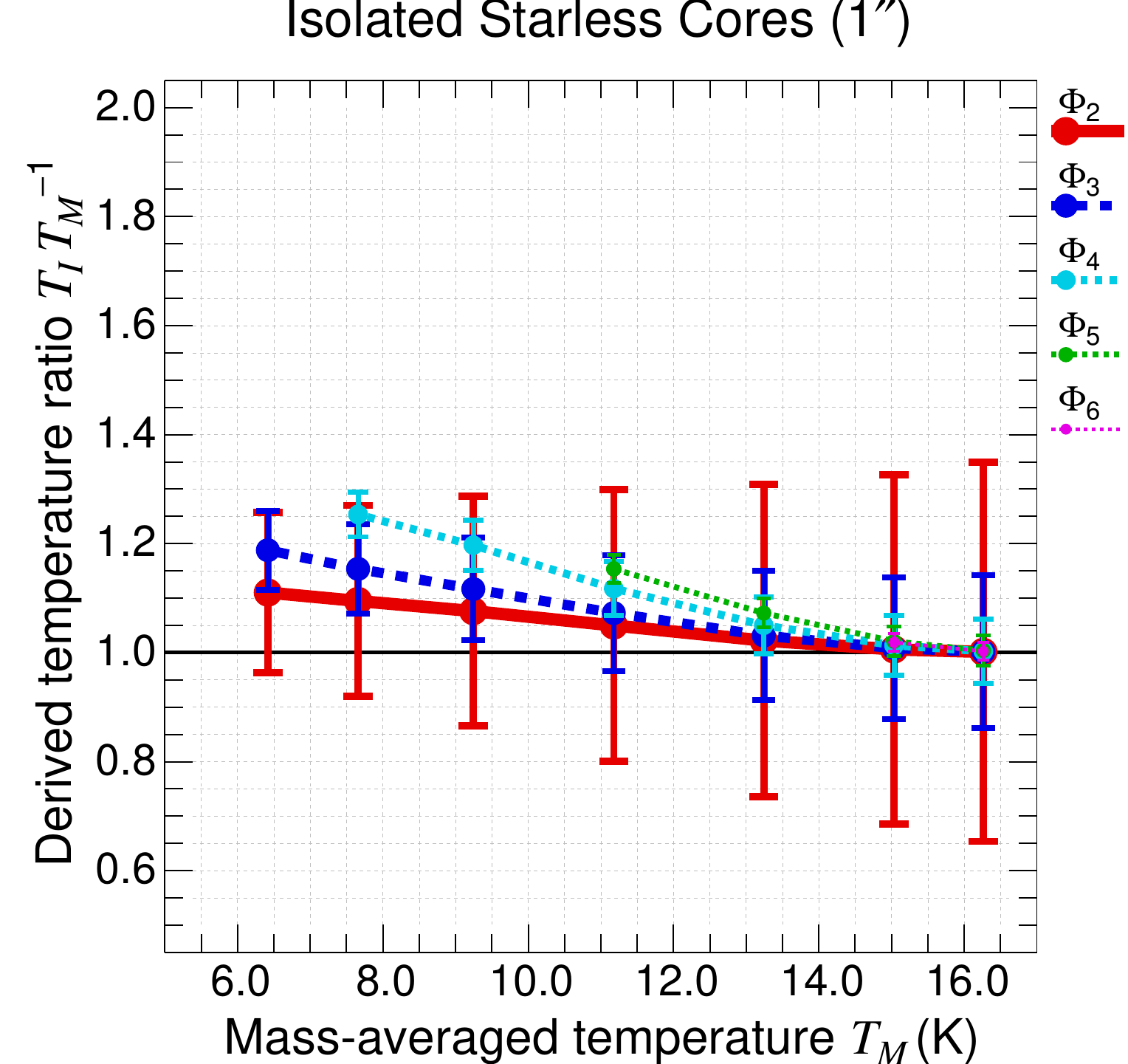}}
            \resizebox{0.3204\hsize}{!}{\includegraphics{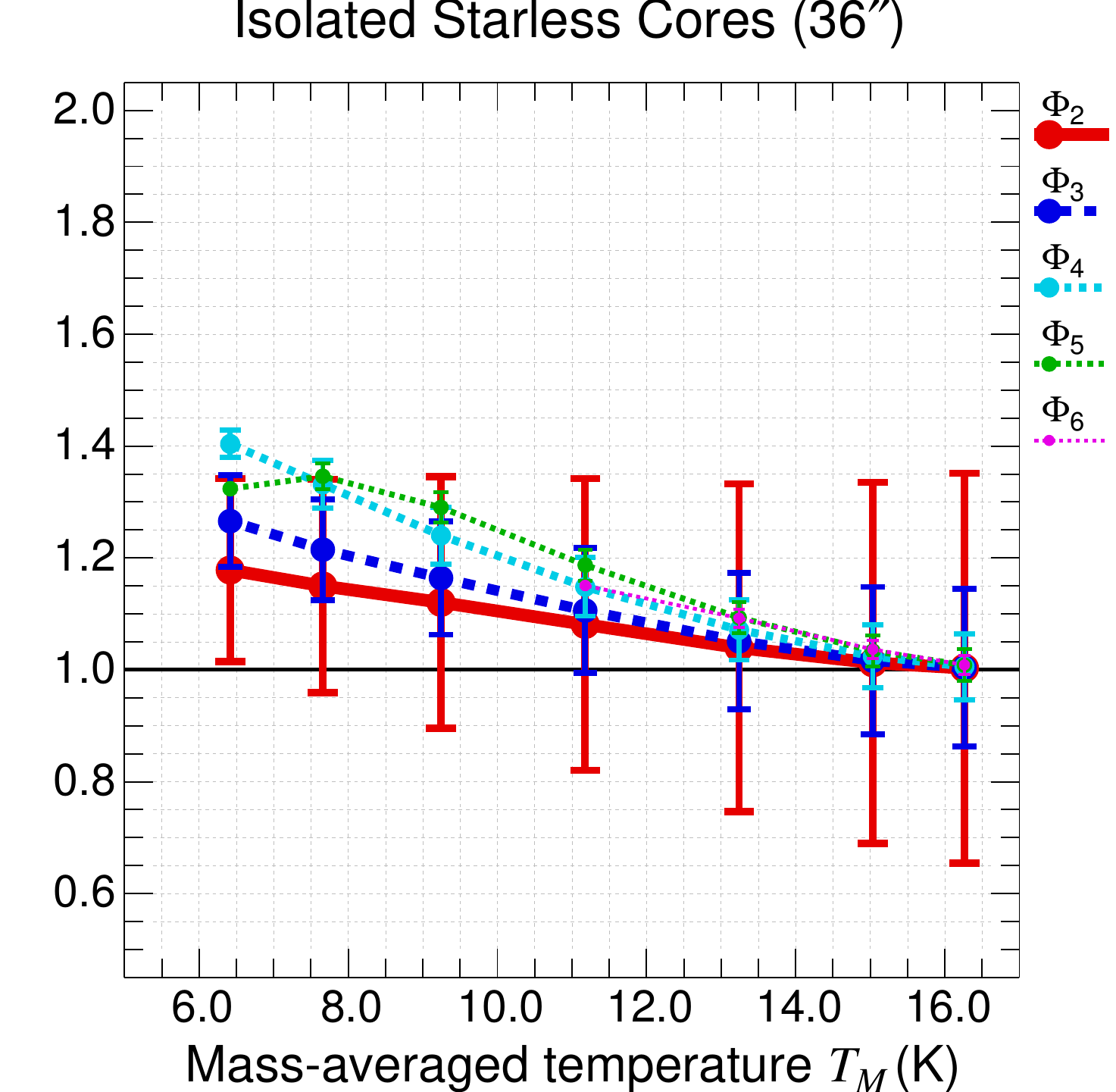}}
            \resizebox{0.3204\hsize}{!}{\includegraphics{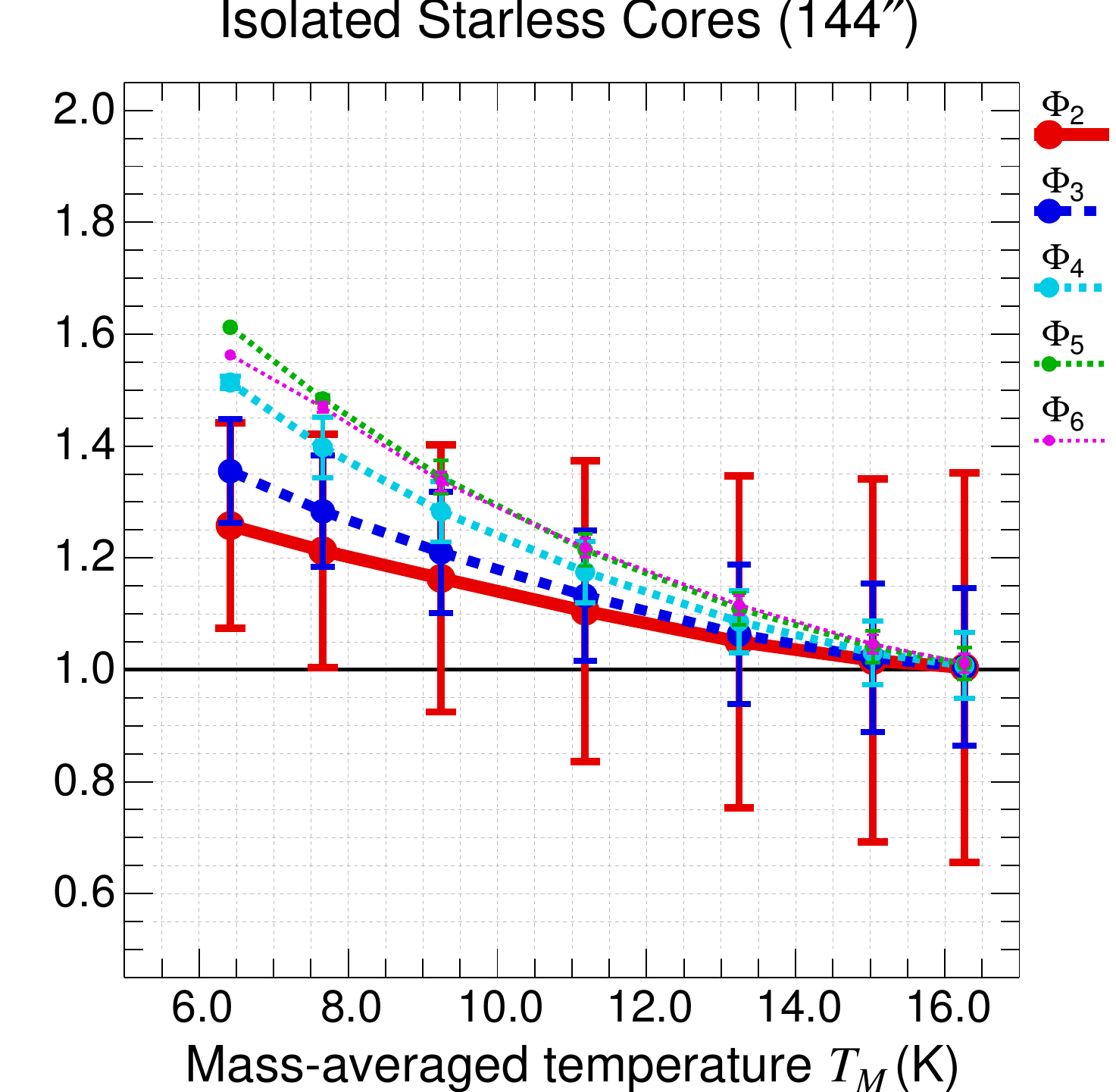}}}
\centerline{\resizebox{0.3327\hsize}{!}{\includegraphics{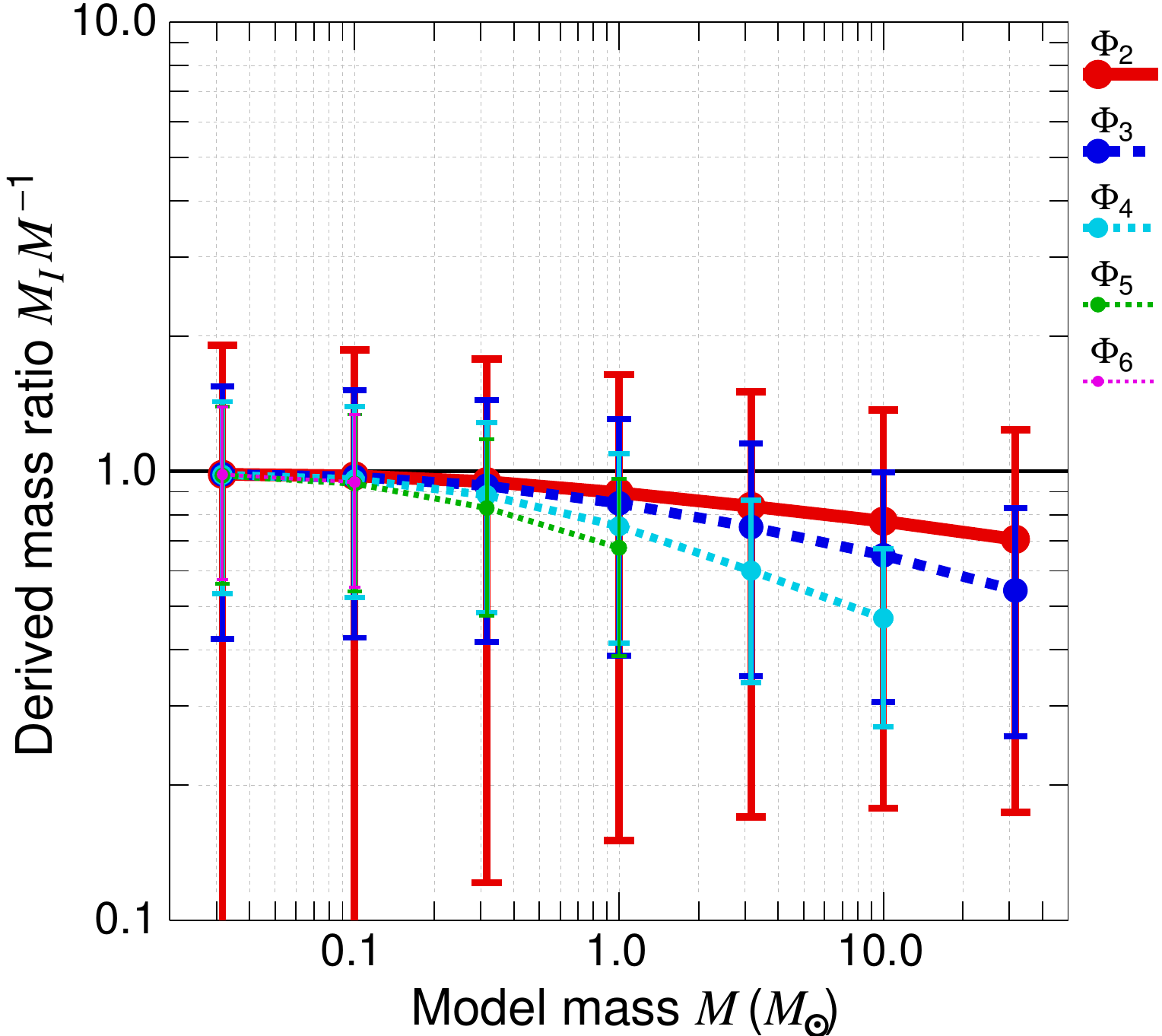}}
            \resizebox{0.3204\hsize}{!}{\includegraphics{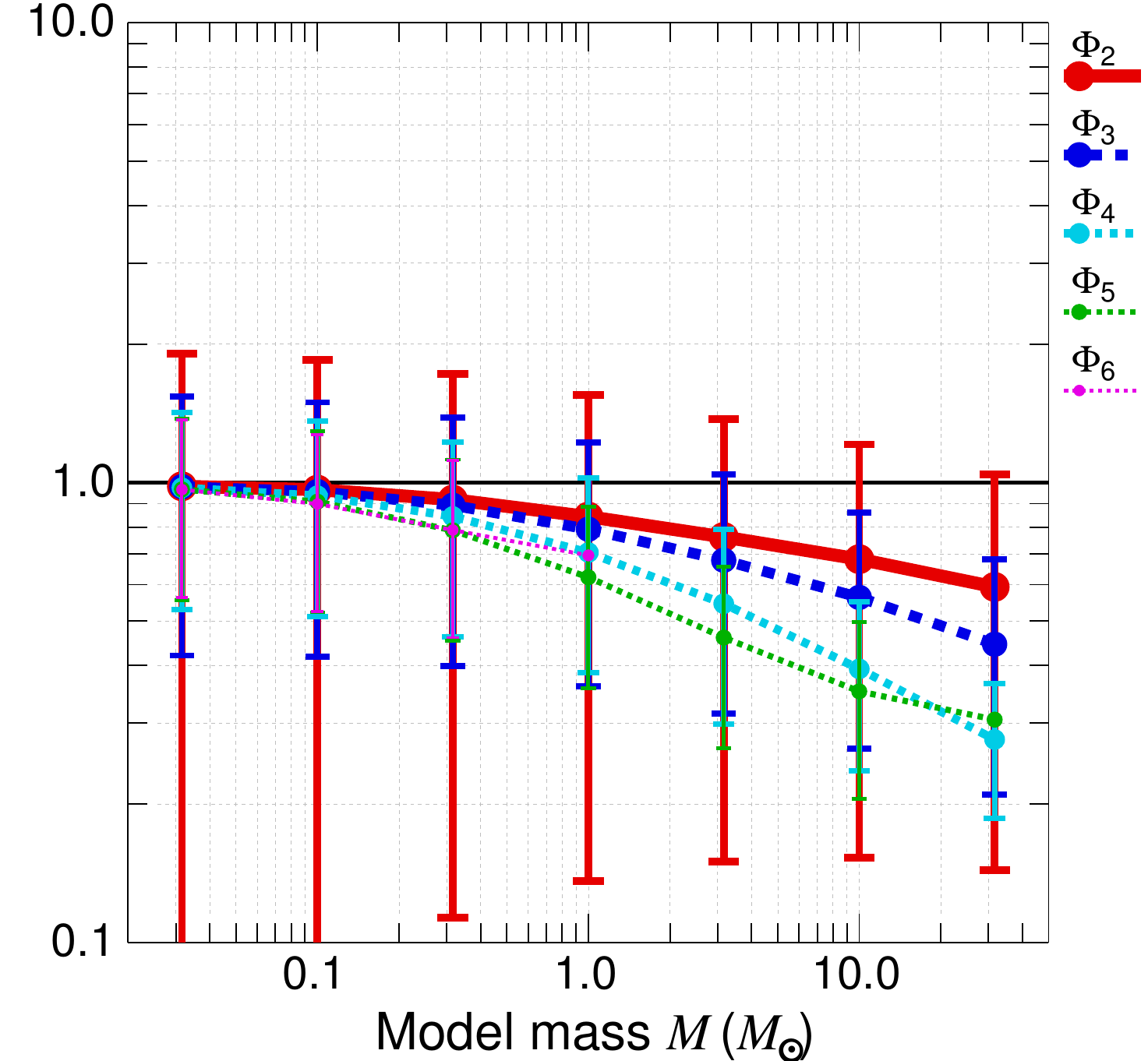}}
            \resizebox{0.3204\hsize}{!}{\includegraphics{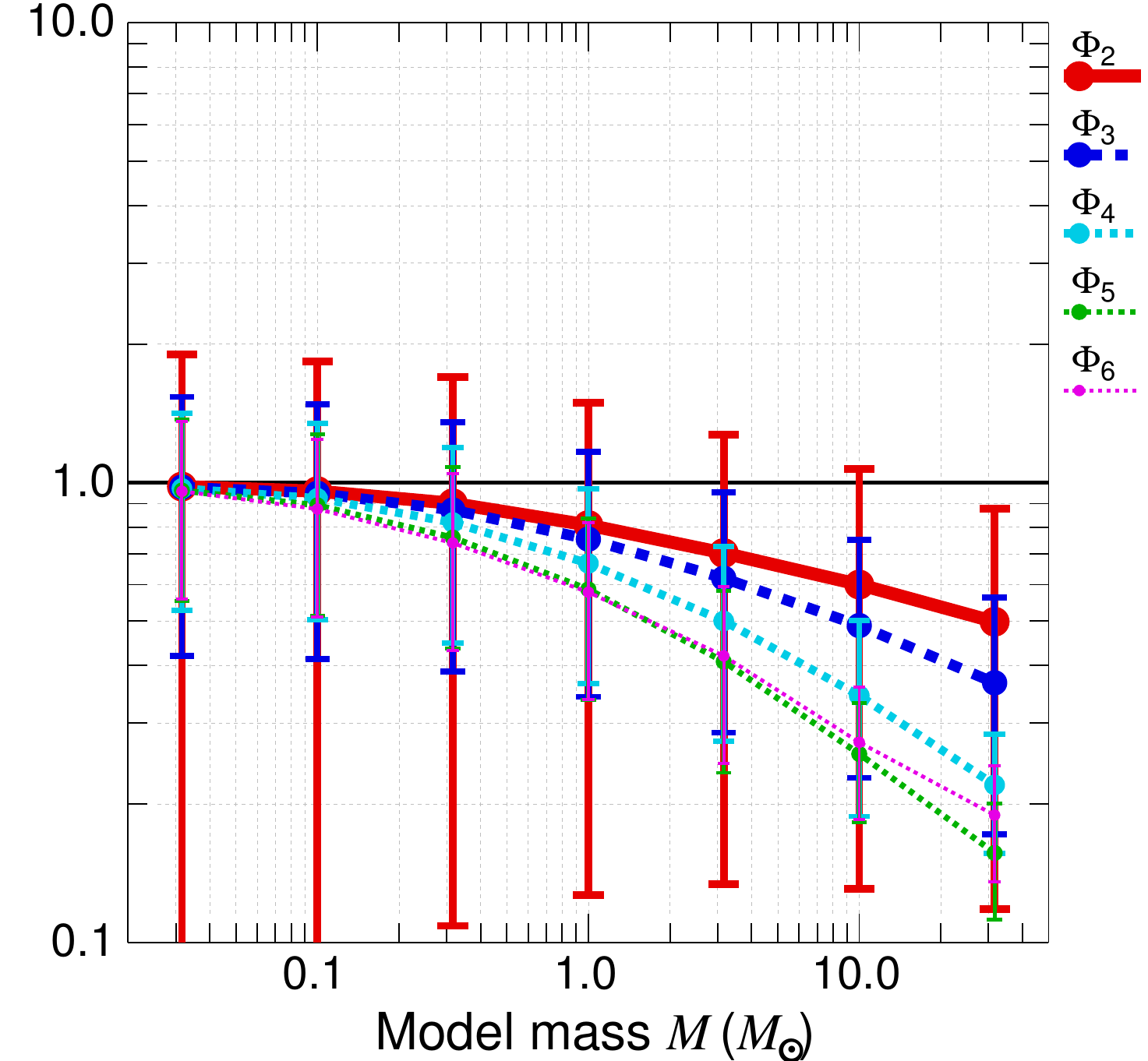}}}
\centerline{\resizebox{0.3327\hsize}{!}{\includegraphics{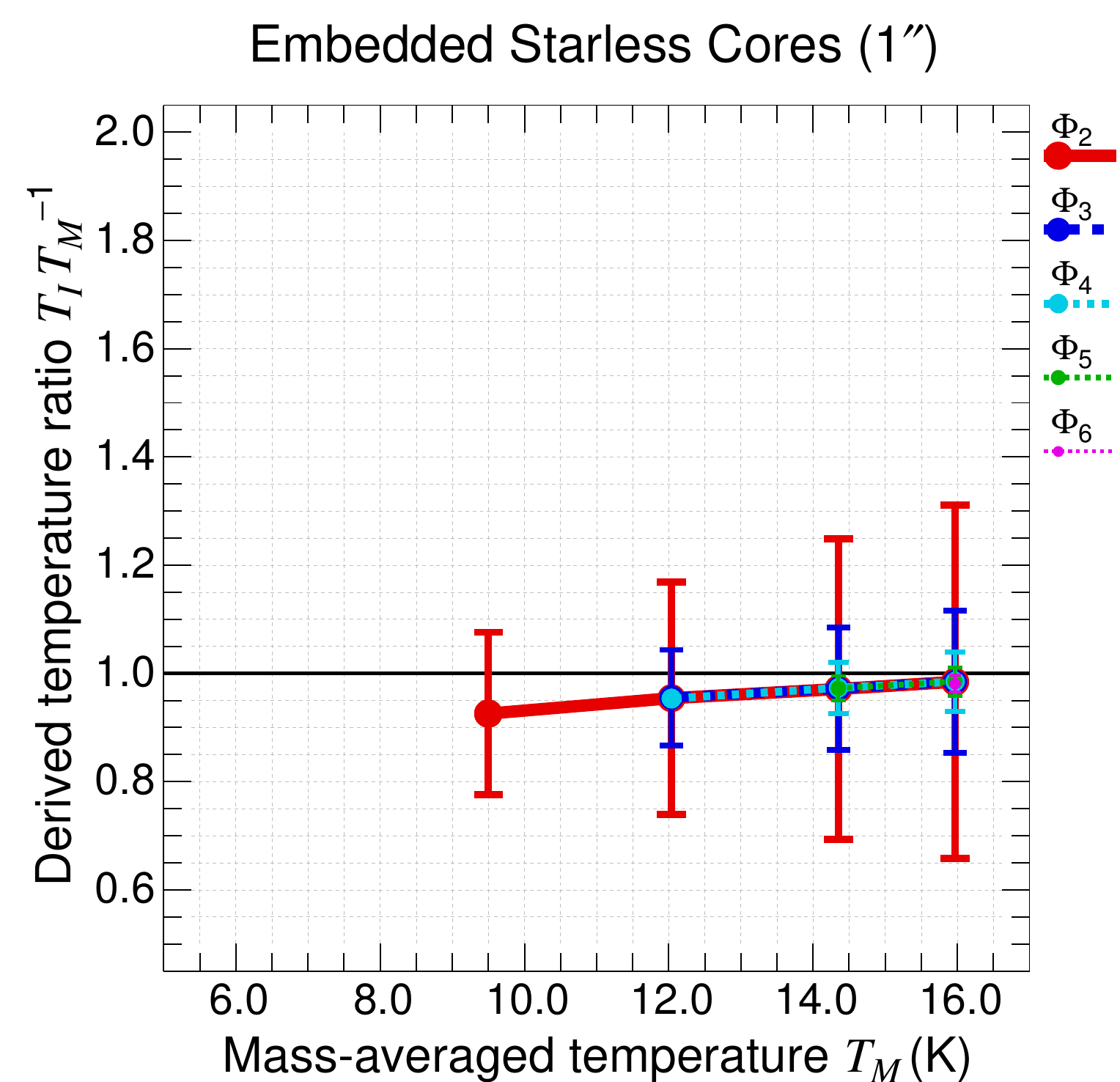}}
            \resizebox{0.3204\hsize}{!}{\includegraphics{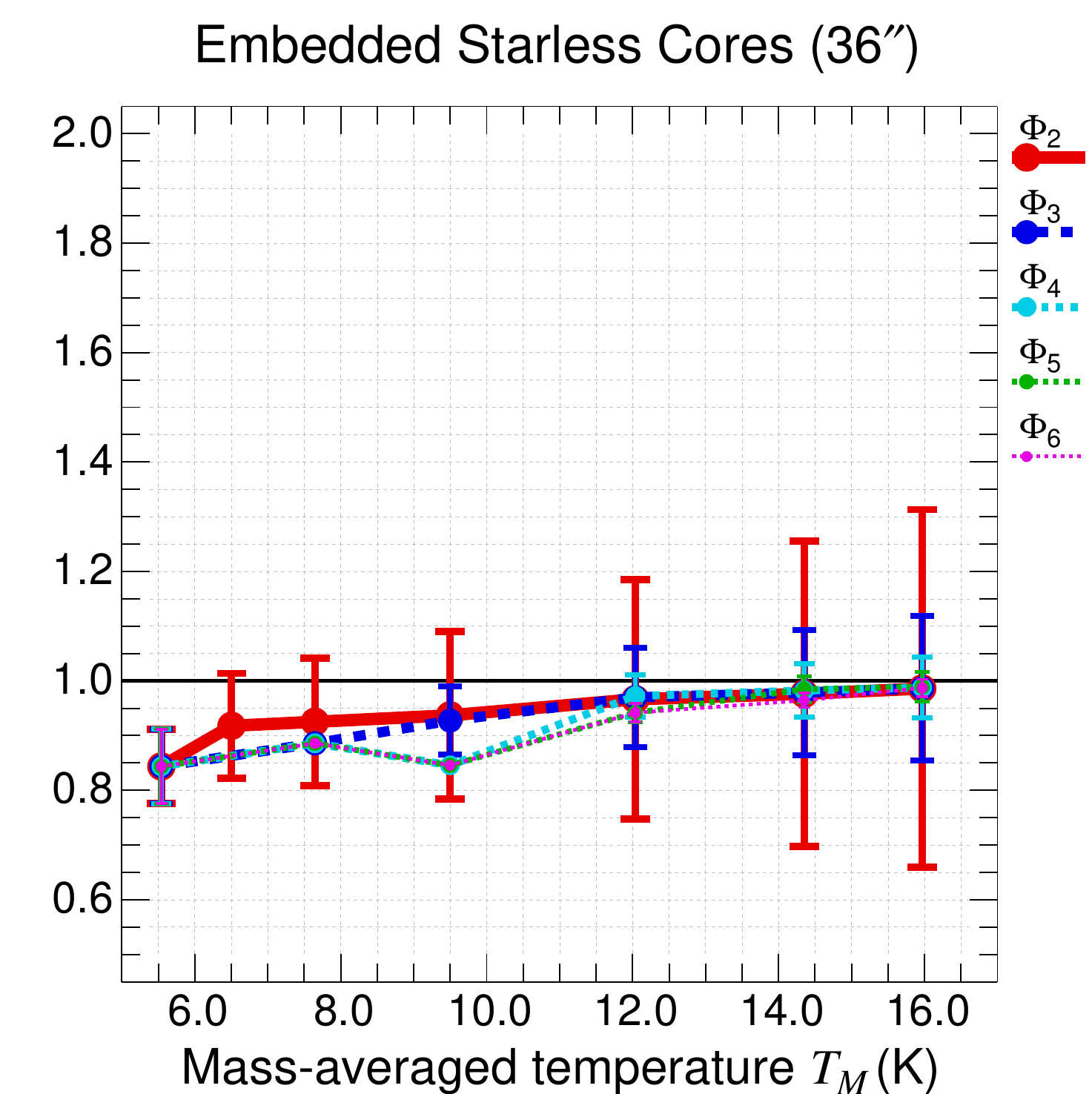}}
            \resizebox{0.3204\hsize}{!}{\includegraphics{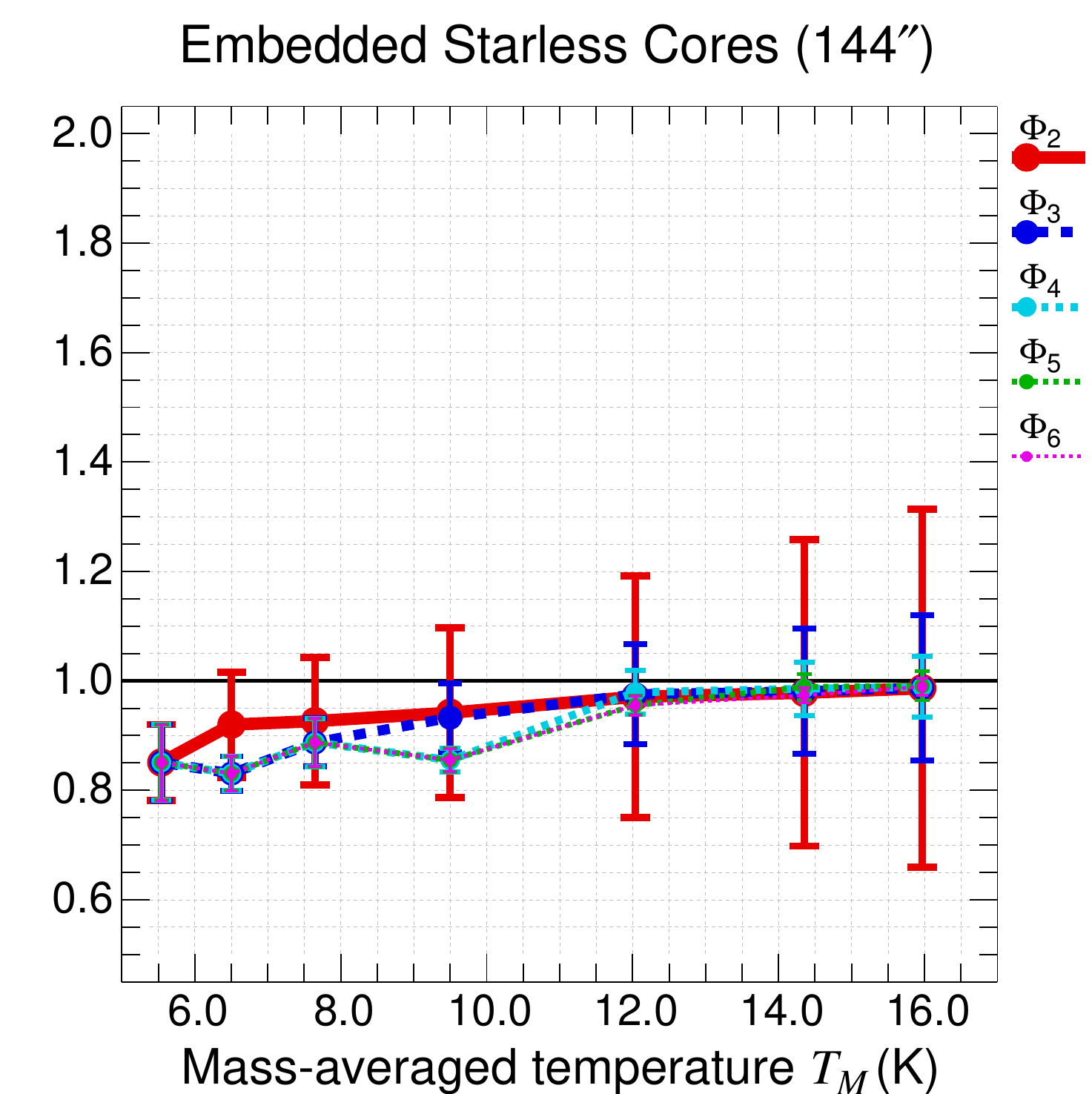}}}
\centerline{\resizebox{0.3327\hsize}{!}{\includegraphics{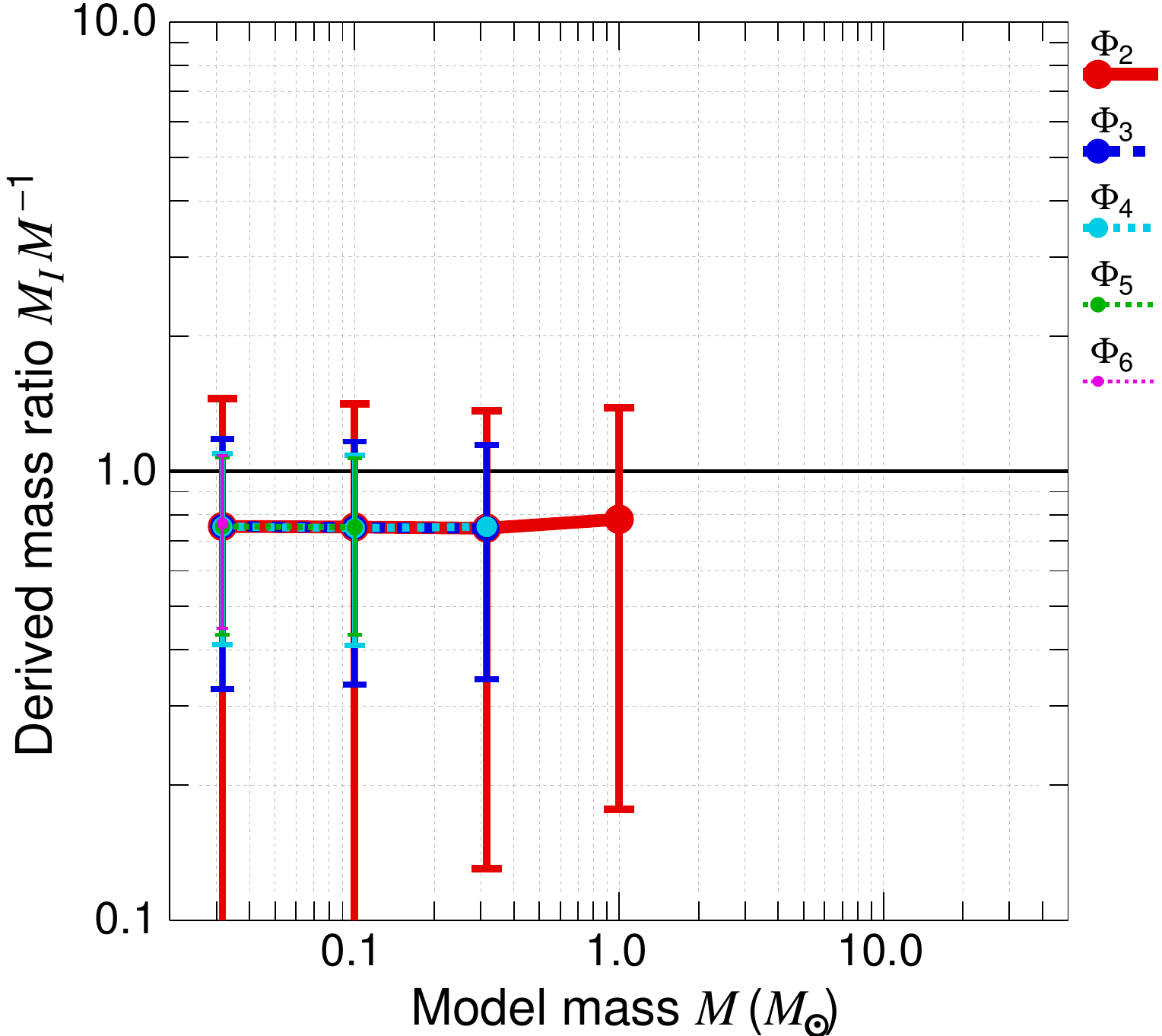}}
            \resizebox{0.3204\hsize}{!}{\includegraphics{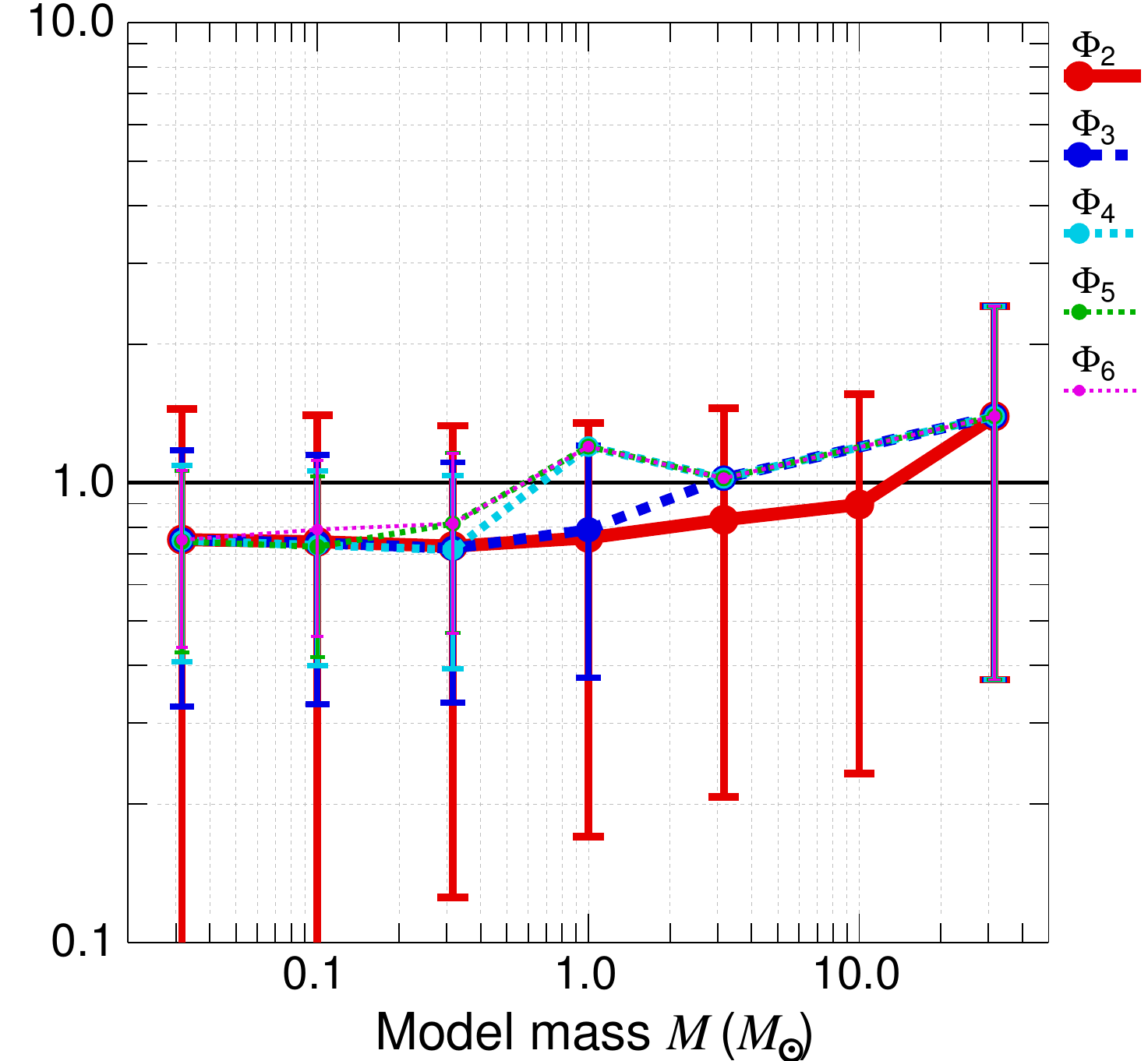}}
            \resizebox{0.3204\hsize}{!}{\includegraphics{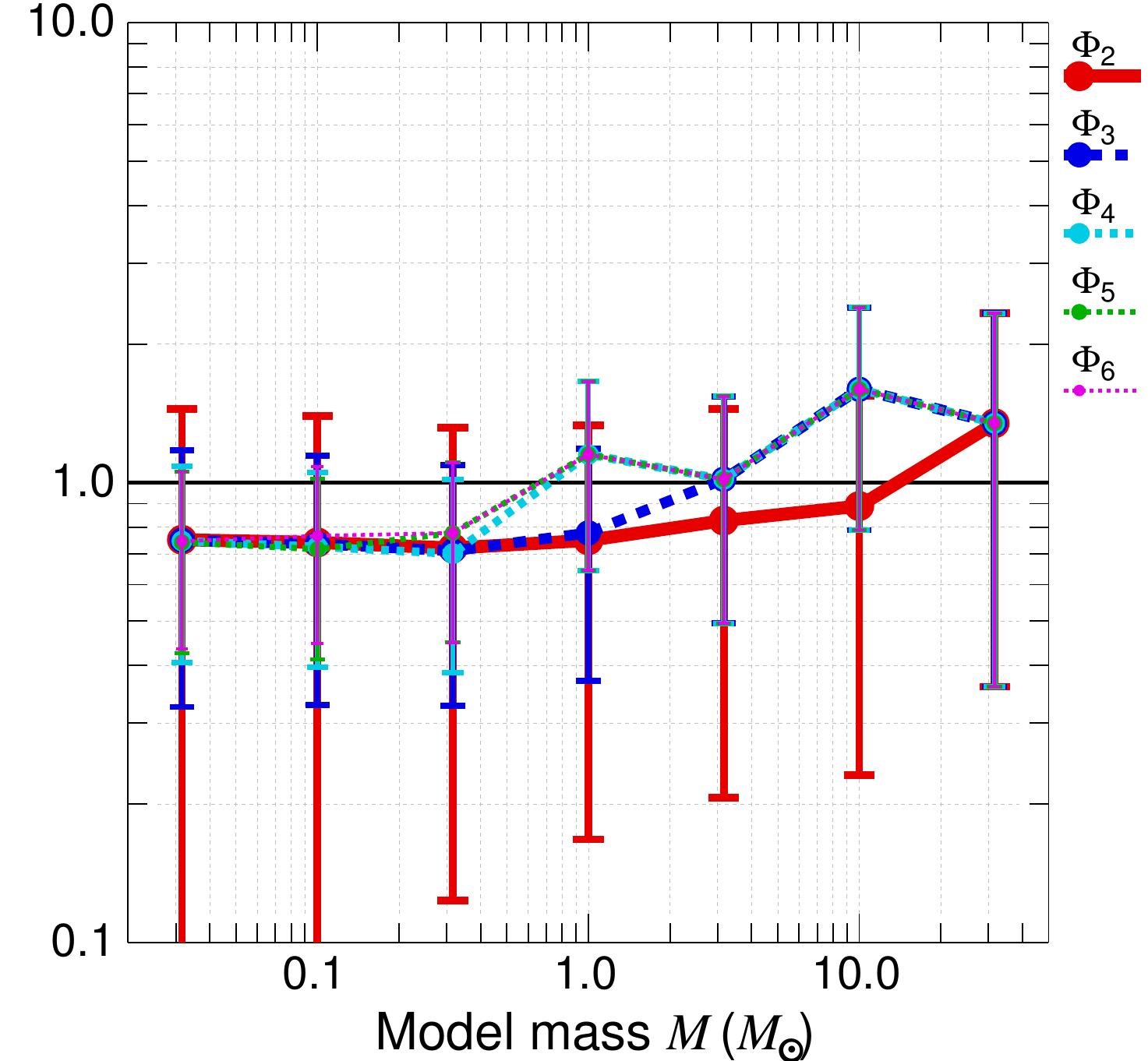}}}
\caption{
Temperatures $T_{\mathcal{I}}$ and masses $M_{\mathcal{I}}$ derived from fitting images $\mathcal{I}_{\nu}$ of both \emph{isolated} 
and \emph{embedded} starless cores vs. the true model values of $T_{M}$ and $M$ for correct $\beta\,{=}\,2$. The three columns of 
panels present results for three angular resolutions (resolved, partially resolved, and unresolved cases) and for various subsets 
$\Phi_{n}$ of pixel intensities. Error bars represent the $1\,{\times}\,\sigma$ uncertainties of the derived $T_{\mathcal{I}}$ and 
$M_{\mathcal{I}}$ (computed over all pixels as the $N_{\rm H_2}$-averaged errors of $T_{N ij}$ and integrated errors of 
$N_{\rm H_2}$, correspondingly), combined with the assumed $\pm\,20{\%}$ uncertainties of $\eta$, $\kappa_{0}$, and $D$ 
(Sect.~\ref{data.subsets}). Less reliable results (${n-\gamma+1}\,{<}\,\chi^{2}\,{<}\,10$ in some pixels) are shown without 
error bars. See Fig.~\ref{temp.mass.bes} for more details.
}
\label{coldens.bes}
\end{figure*}

Isolated starless cores, $\beta\,{=}\,2$ (Fig.~\ref{coldens.bes}). For the fully resolved models, derived $T_{\mathcal{I}}$ and
$M_{\mathcal{I}}$ have fairly good accuracy and little bias for acceptable fits, although the range of the latter for larger
$\Phi_{n}$ ($n\,{=}\,3\,{\rightarrow}\,6$) shrinks to the lowest masses. As the transparent low-mass cores
($M\,{\rightarrow}\,0.03\,M_{\sun}$) are almost isothermal, derived $T_{\mathcal{I}}$ and $M_{\mathcal{I}}$ perfectly agree with
$T_{M}$ and $M$ for any subset $\Phi_{n}$. Massive cores with more variable $T_{\rm d}(r)$ (Fig.~\ref{trp.bes.pro}) also have
significant variations of $T_{\rm d}(z)$ along the line of sight at pixel $(i,j)$. Emission of hot dust skews the spectral shapes
of $I_{\nu\,ij}$ towards shorter wavelengths, even more so at the high-mass end. The most accurate masses are obtained for
$\Phi_{2}$, whereas larger $\Phi_{n}$ ($n\,{=}\,3\,{\rightarrow}\,6$) give increasingly incorrect $T_{\mathcal{I}}$ and
$M_{\mathcal{I}}$. With degrading angular resolutions, the inaccuracies and biases increase, especially for
$M\,{\rightarrow}\,30\,M_{\sun}$ and larger $\Phi_{n}$ ($n\,{=}\,3\,{\rightarrow}\,6$). As expected, in the limiting case of
unresolved objects the results approach those obtained with the method of fitting fluxes $F_{\nu}$ (Fig.~\ref{temp.mass.bes}).

Embedded starless cores, $\beta\,{=}\,2$ (Fig.~\ref{coldens.bes}). For the fully resolved models, derived $T_{\mathcal{I}}$ have
fairly good accuracy and little bias for the acceptable fits, although the range of the latter for larger $\Phi_{n}$
($n\,{=}\,3\,{\rightarrow}\,6$) shrinks to even lower masses than for the isolated models. Showing no particularly large bias over
almost the entire range of model masses, $M_{\mathcal{I}}$ are underestimated by a factor of $1.3$ owing to the standard
observational procedure of background subtraction (Appendix \ref{AppendixB}, Sect.~\ref{bg.subtraction}). Derived parameters of the
models do not depend on angular resolutions, as they have relatively flat $T_{\rm d}(r)$ across their boundaries
(Fig.~\ref{trp.bes.pro}), hence the spectral distortions of $I_{\nu\,ij}$ are negligible.

\begin{figure*}
\centering
\centerline{\resizebox{0.3327\hsize}{!}{\includegraphics{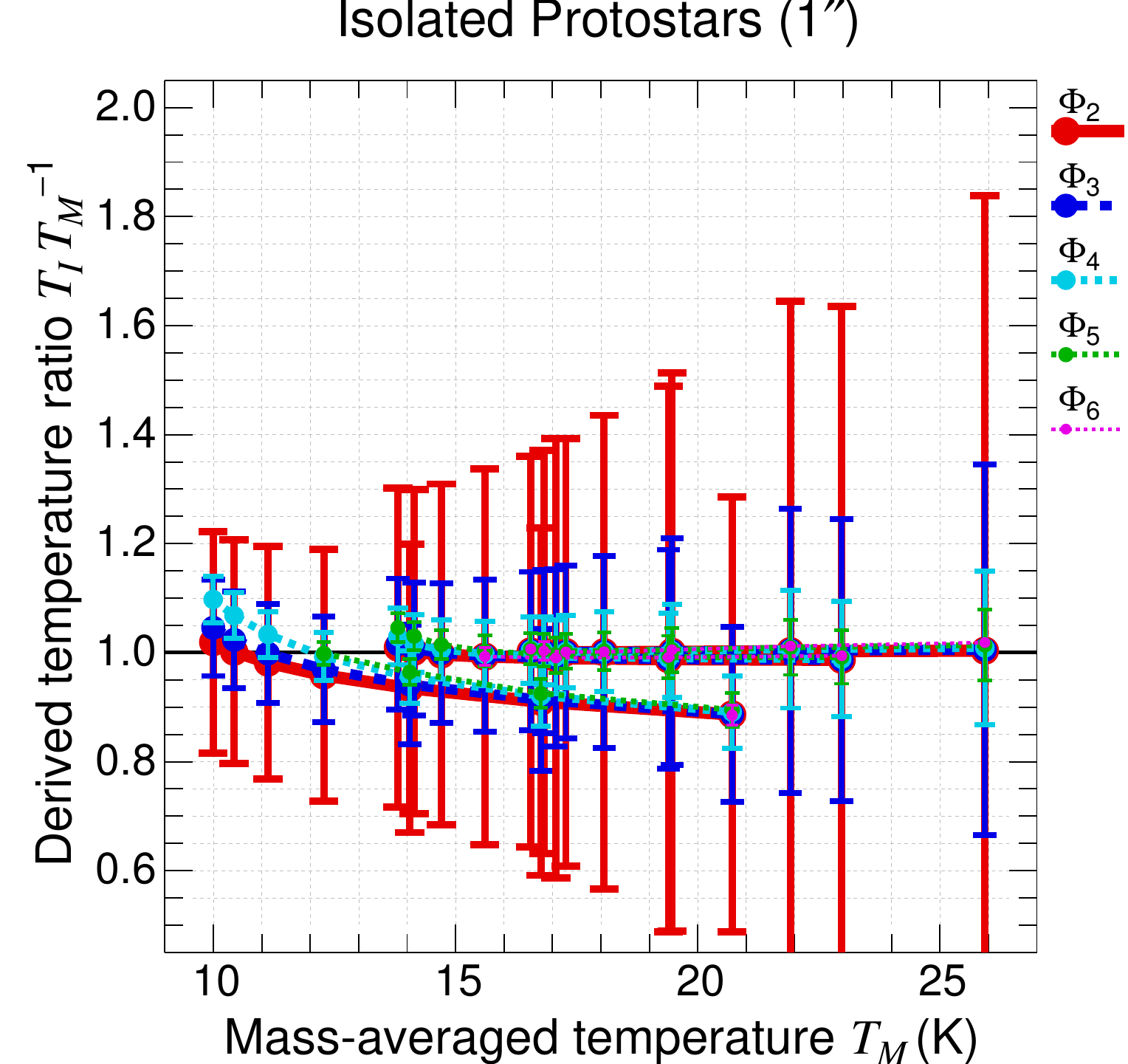}}
            \resizebox{0.3204\hsize}{!}{\includegraphics{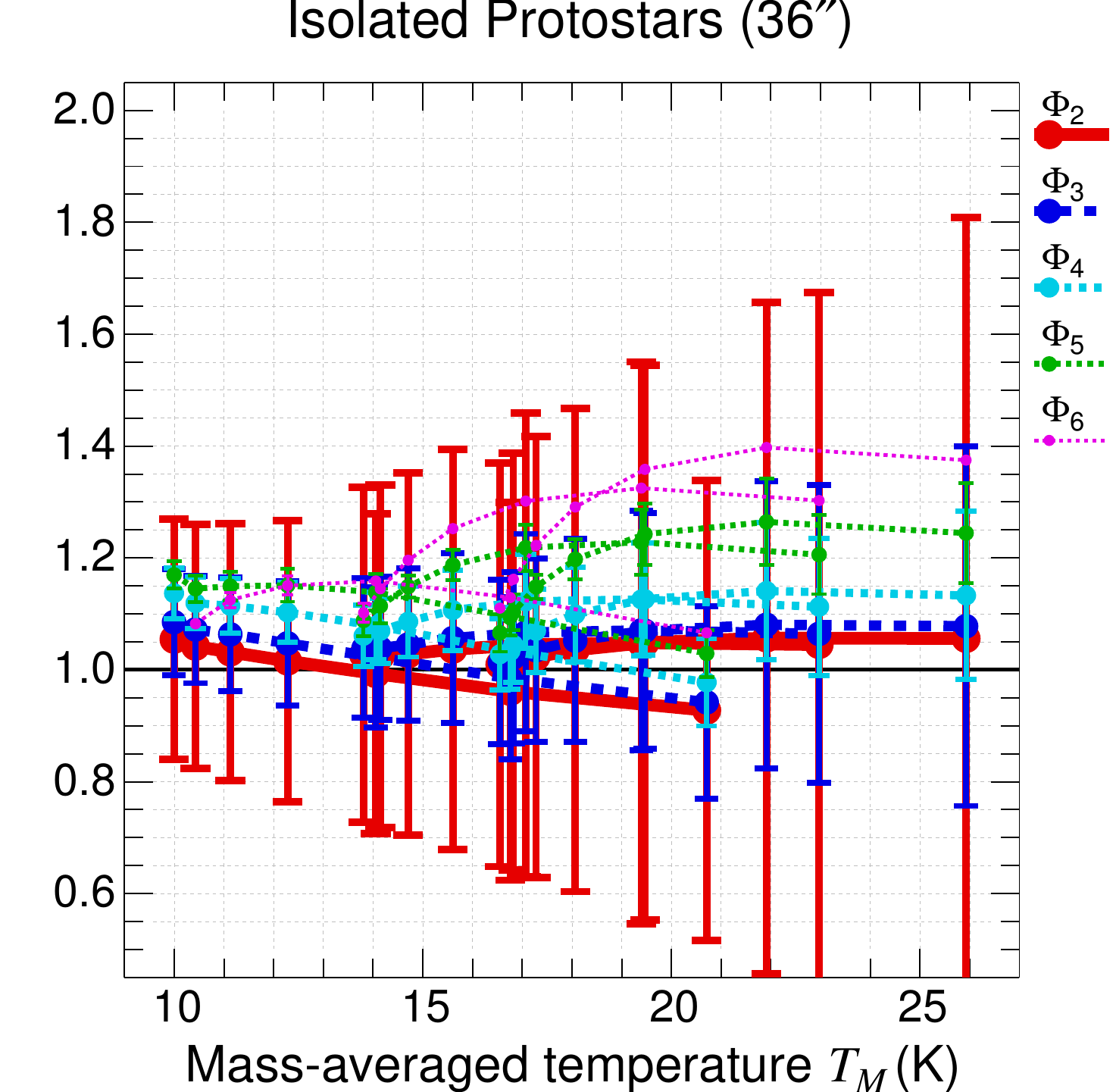}}
            \resizebox{0.3204\hsize}{!}{\includegraphics{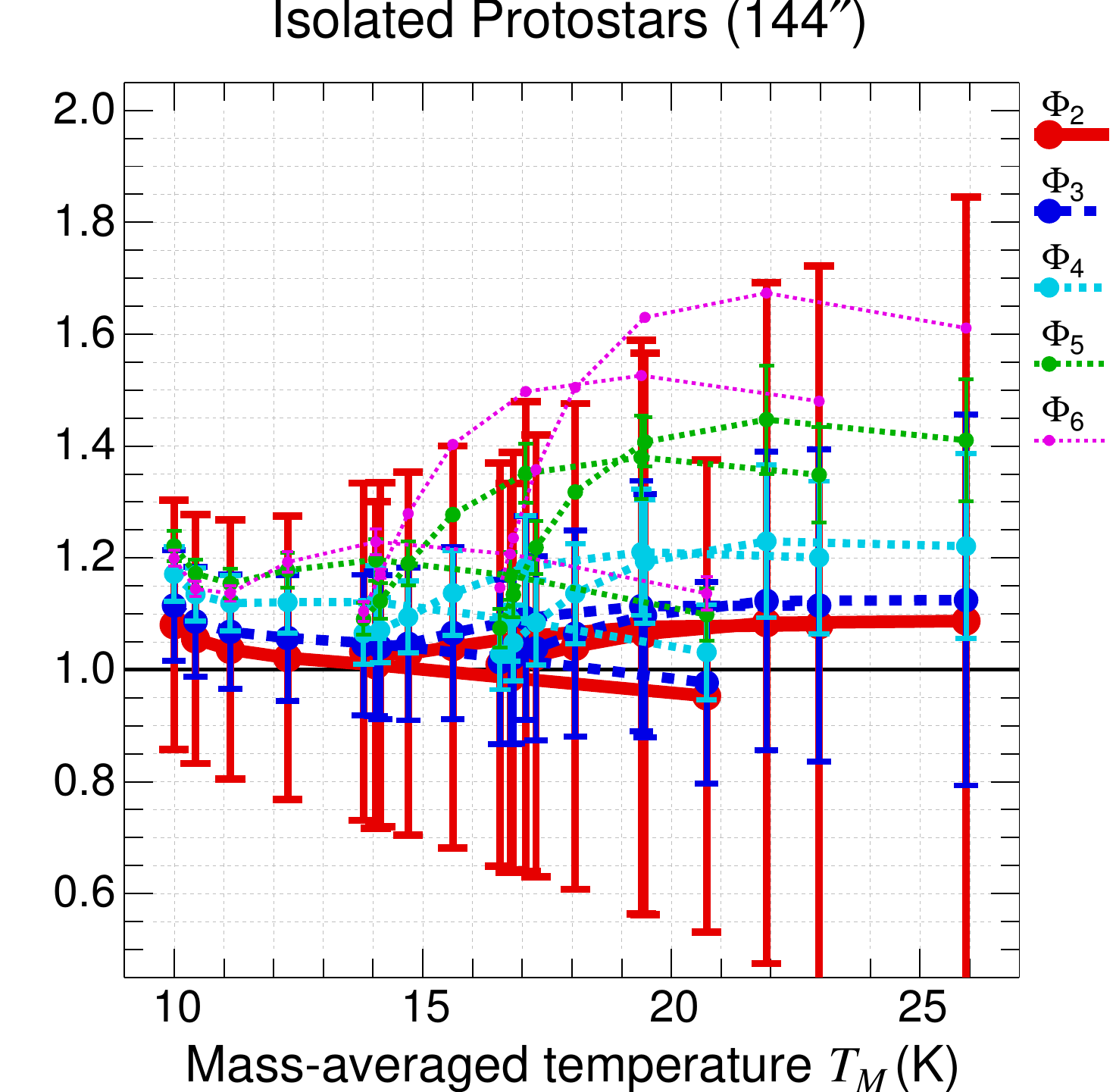}}}
\centerline{\resizebox{0.3327\hsize}{!}{\includegraphics{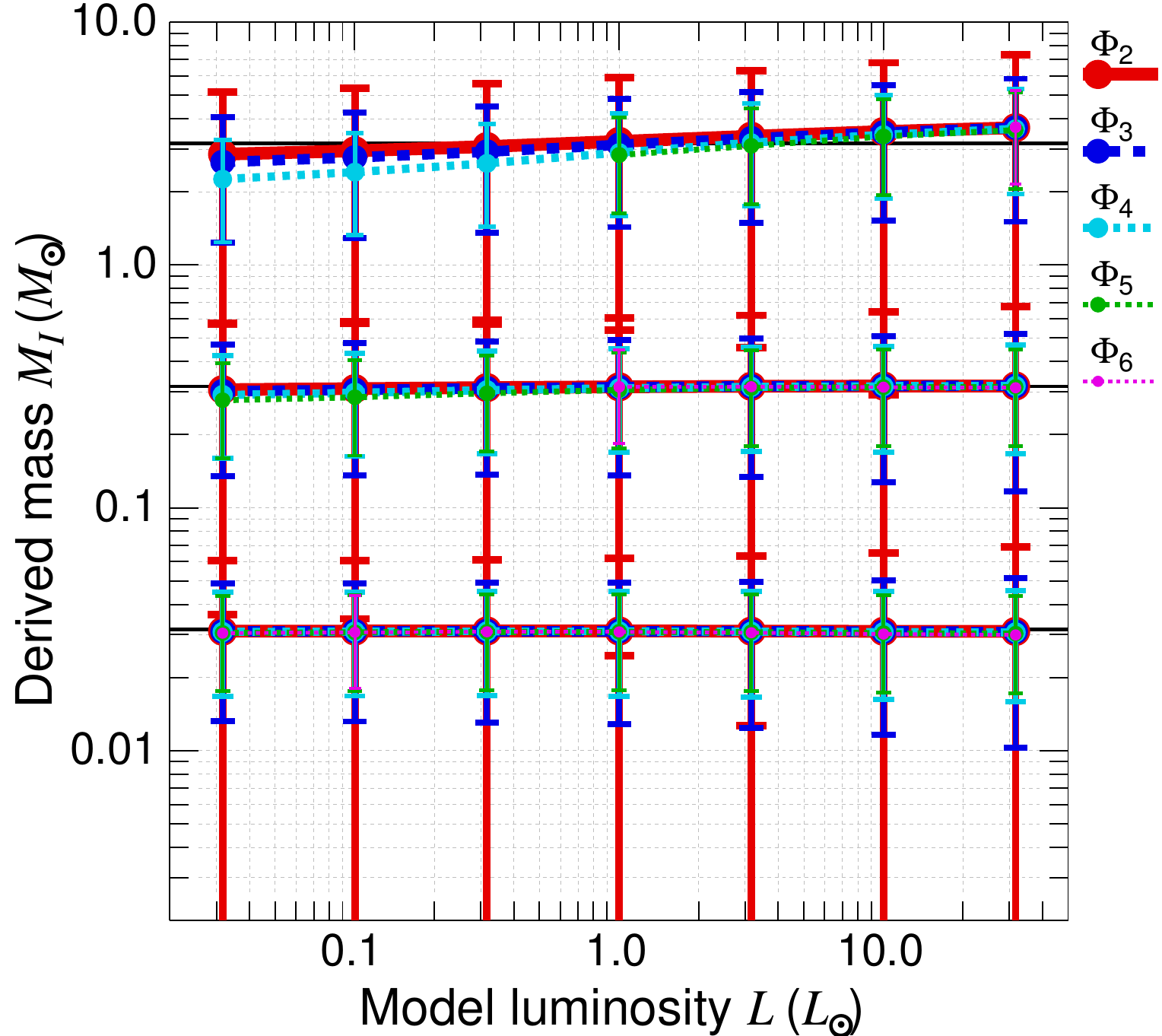}}
            \resizebox{0.3204\hsize}{!}{\includegraphics{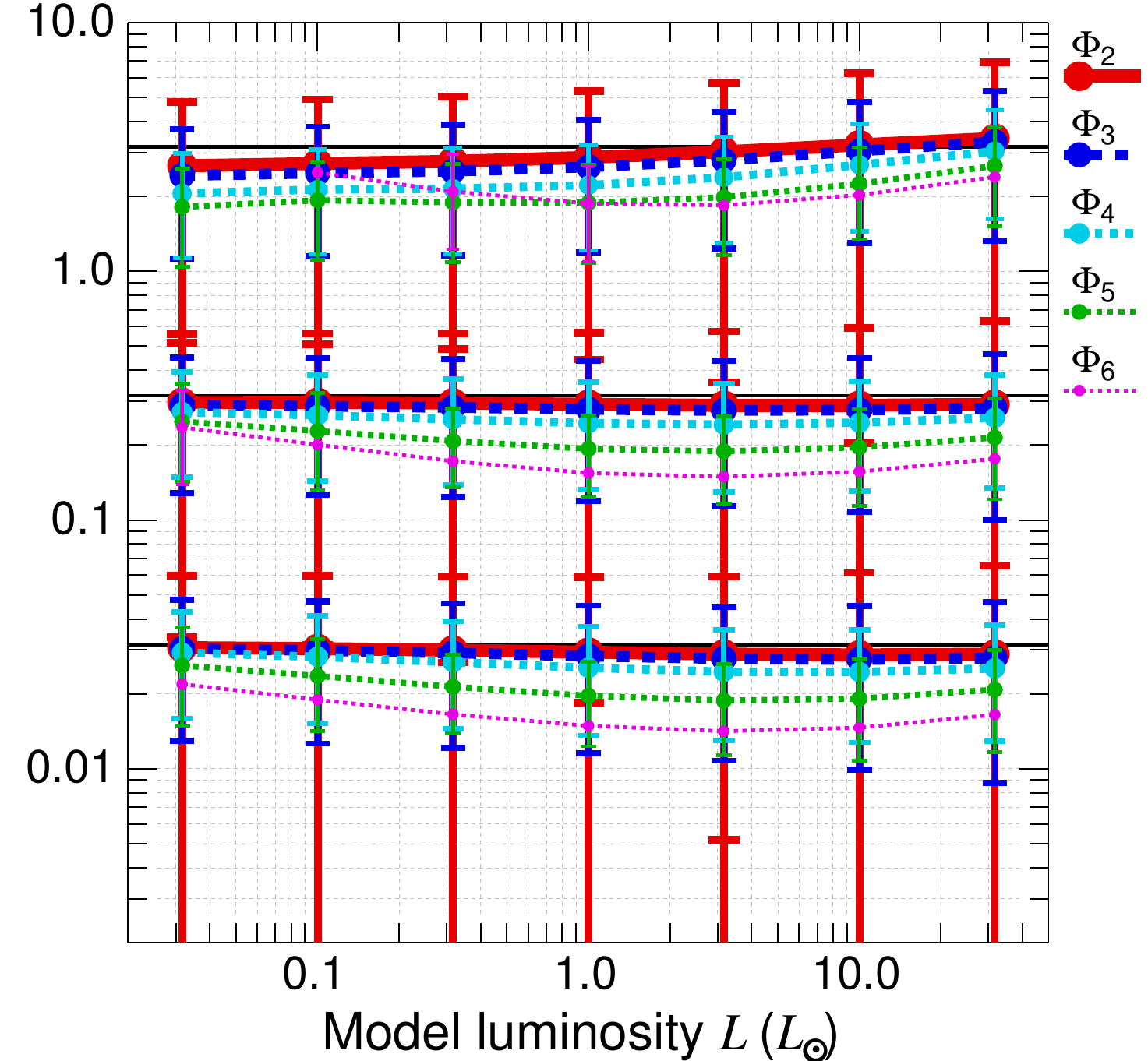}}
            \resizebox{0.3204\hsize}{!}{\includegraphics{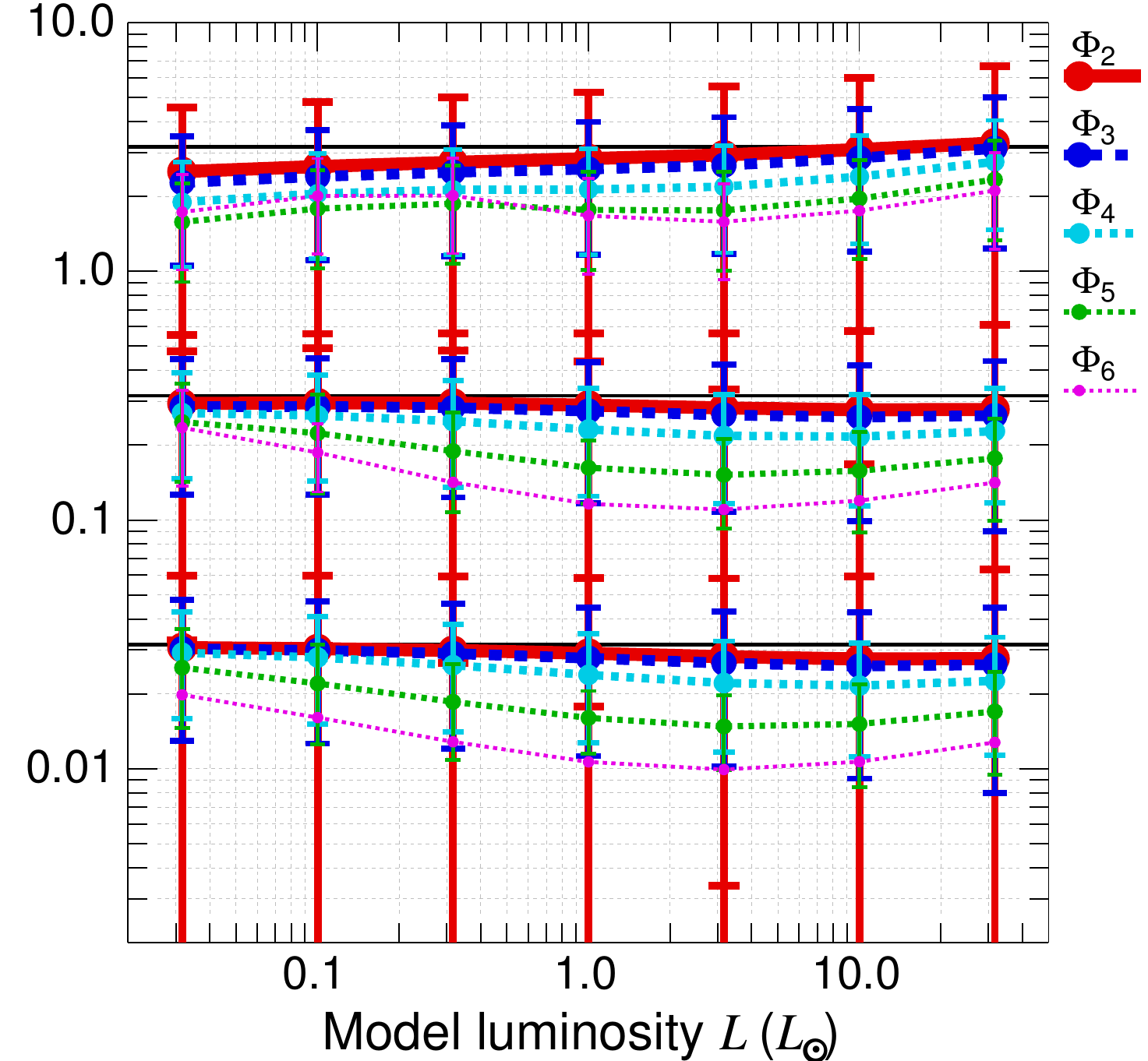}}}
\centerline{\resizebox{0.3327\hsize}{!}{\includegraphics{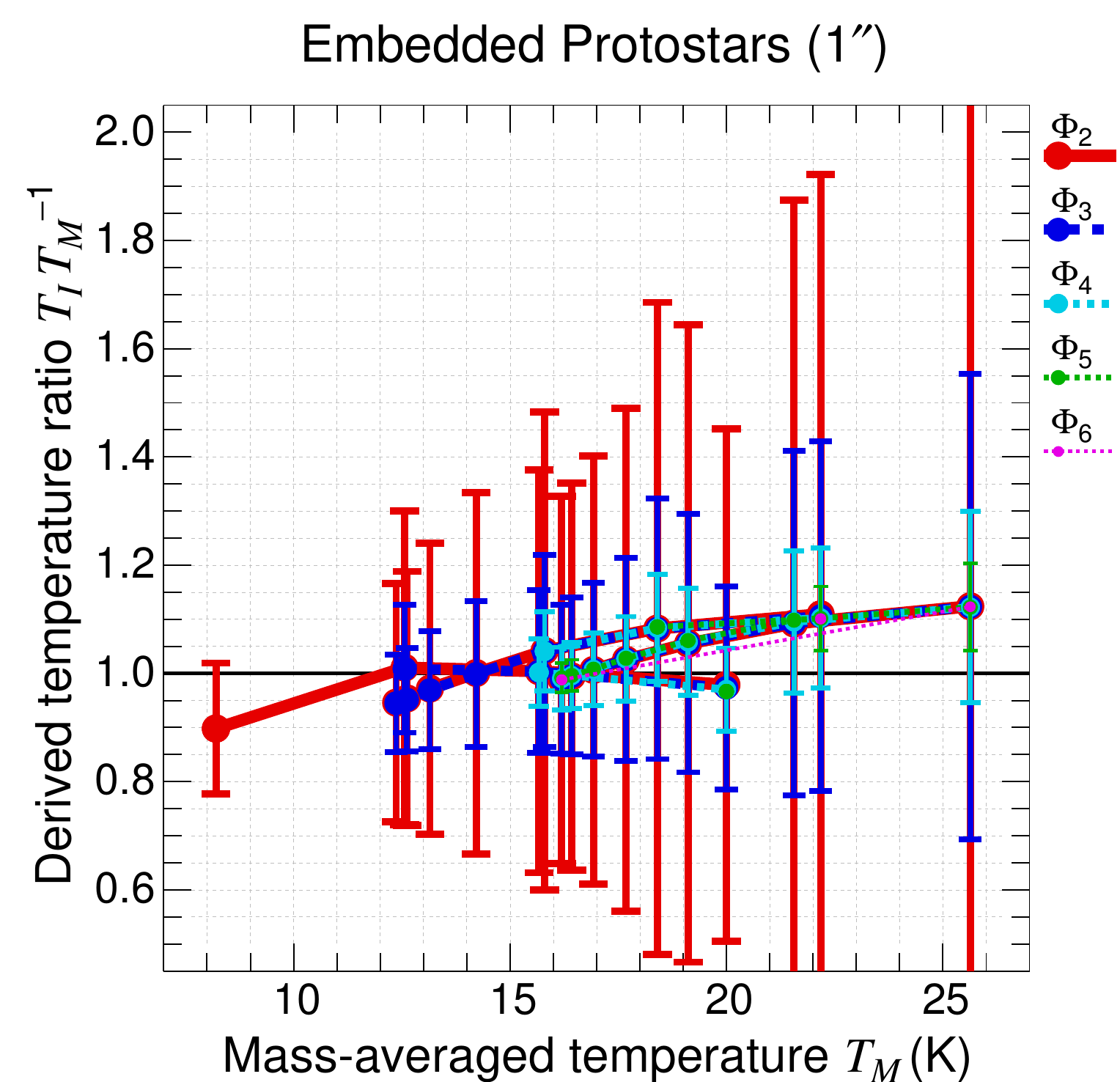}}
            \resizebox{0.3204\hsize}{!}{\includegraphics{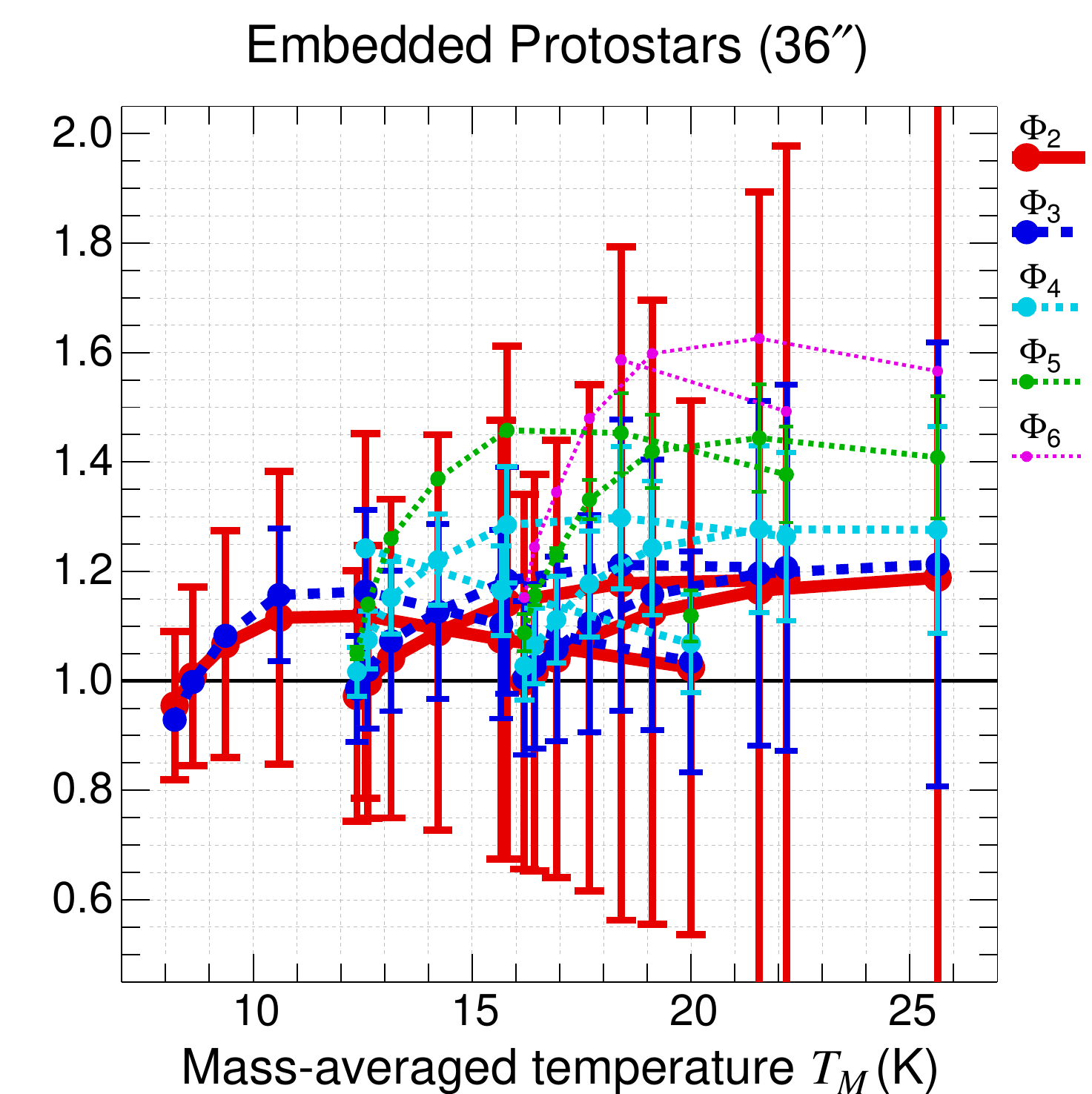}}
            \resizebox{0.3204\hsize}{!}{\includegraphics{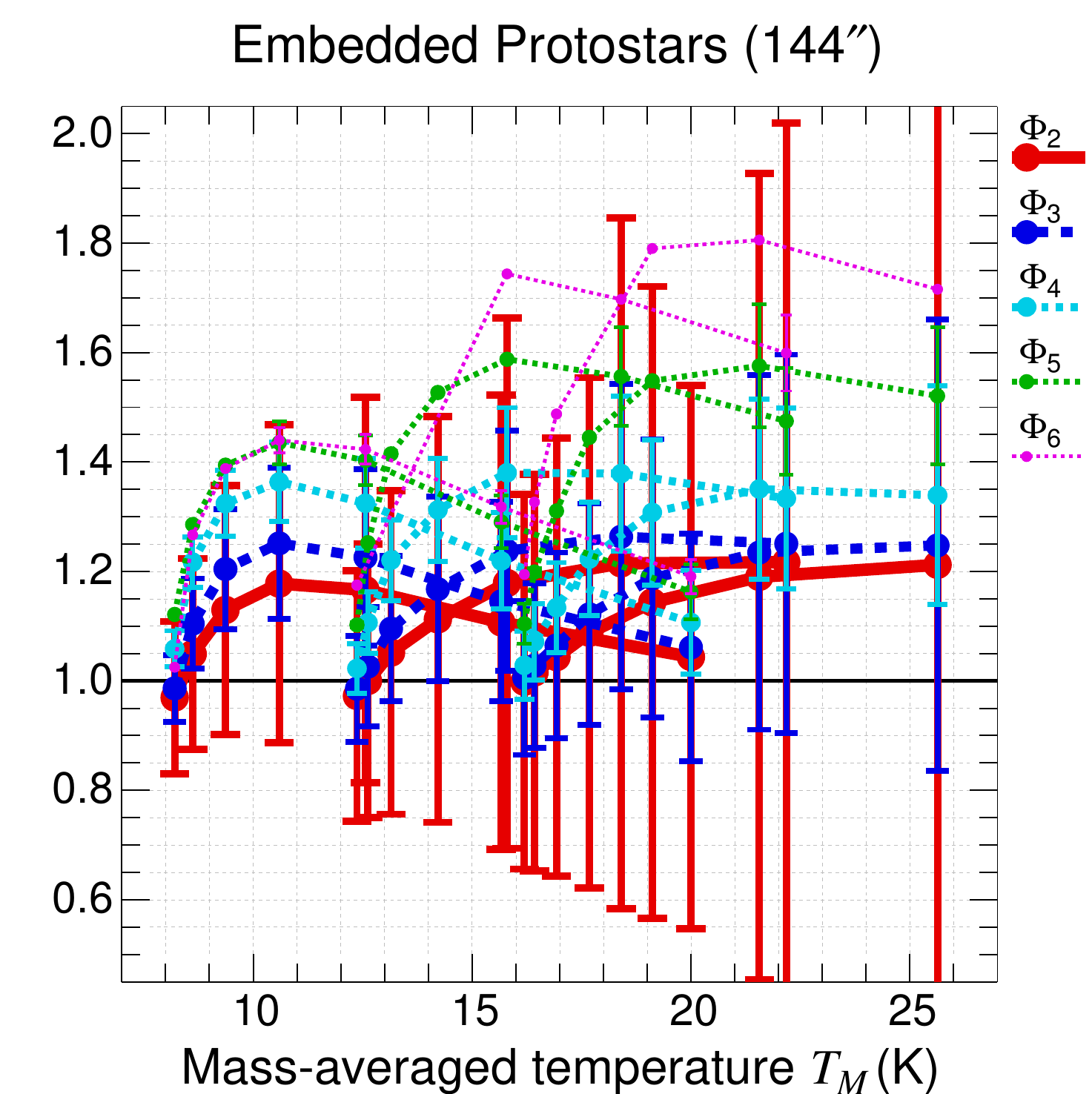}}}
\centerline{\resizebox{0.3327\hsize}{!}{\includegraphics{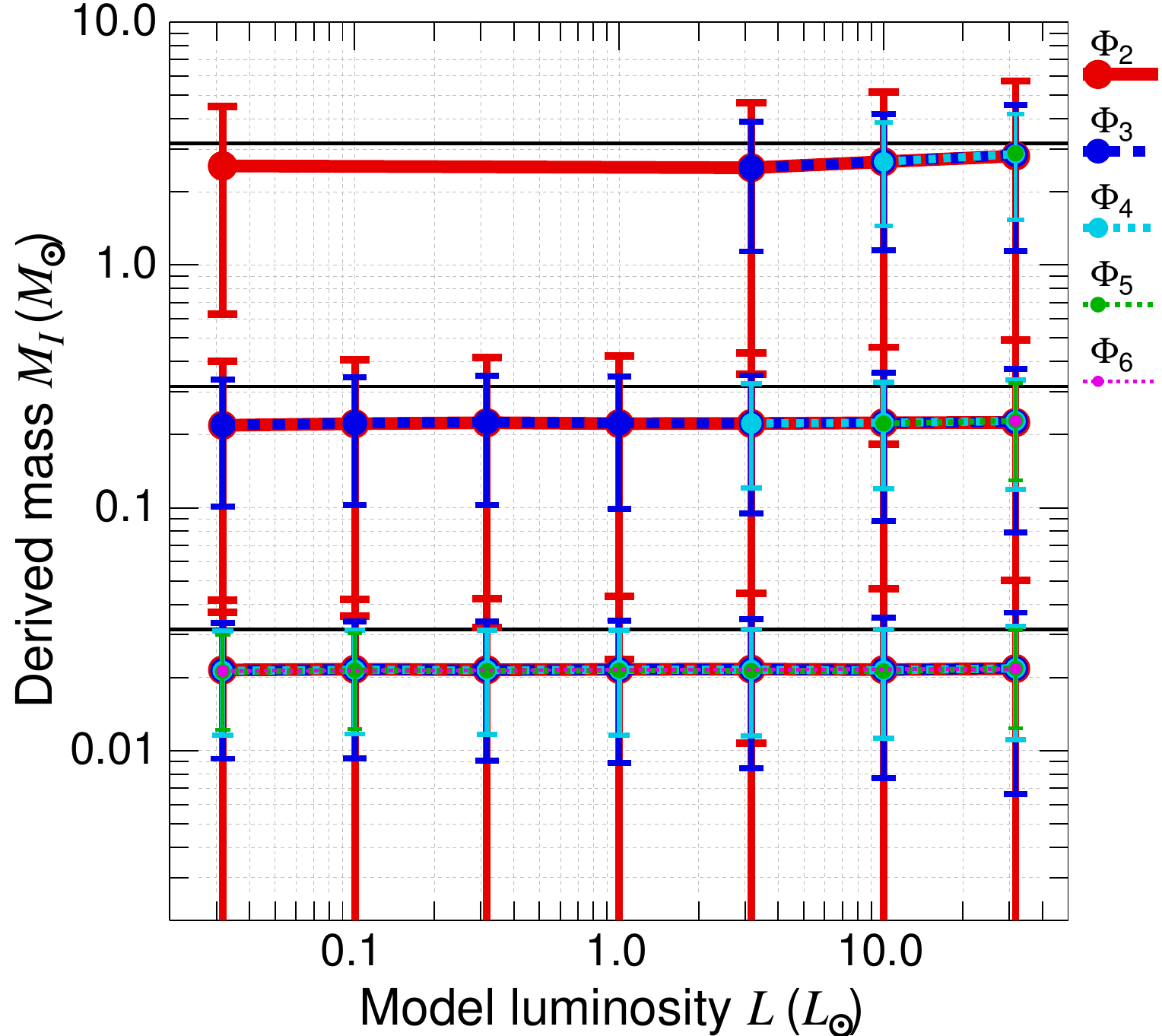}}
            \resizebox{0.3204\hsize}{!}{\includegraphics{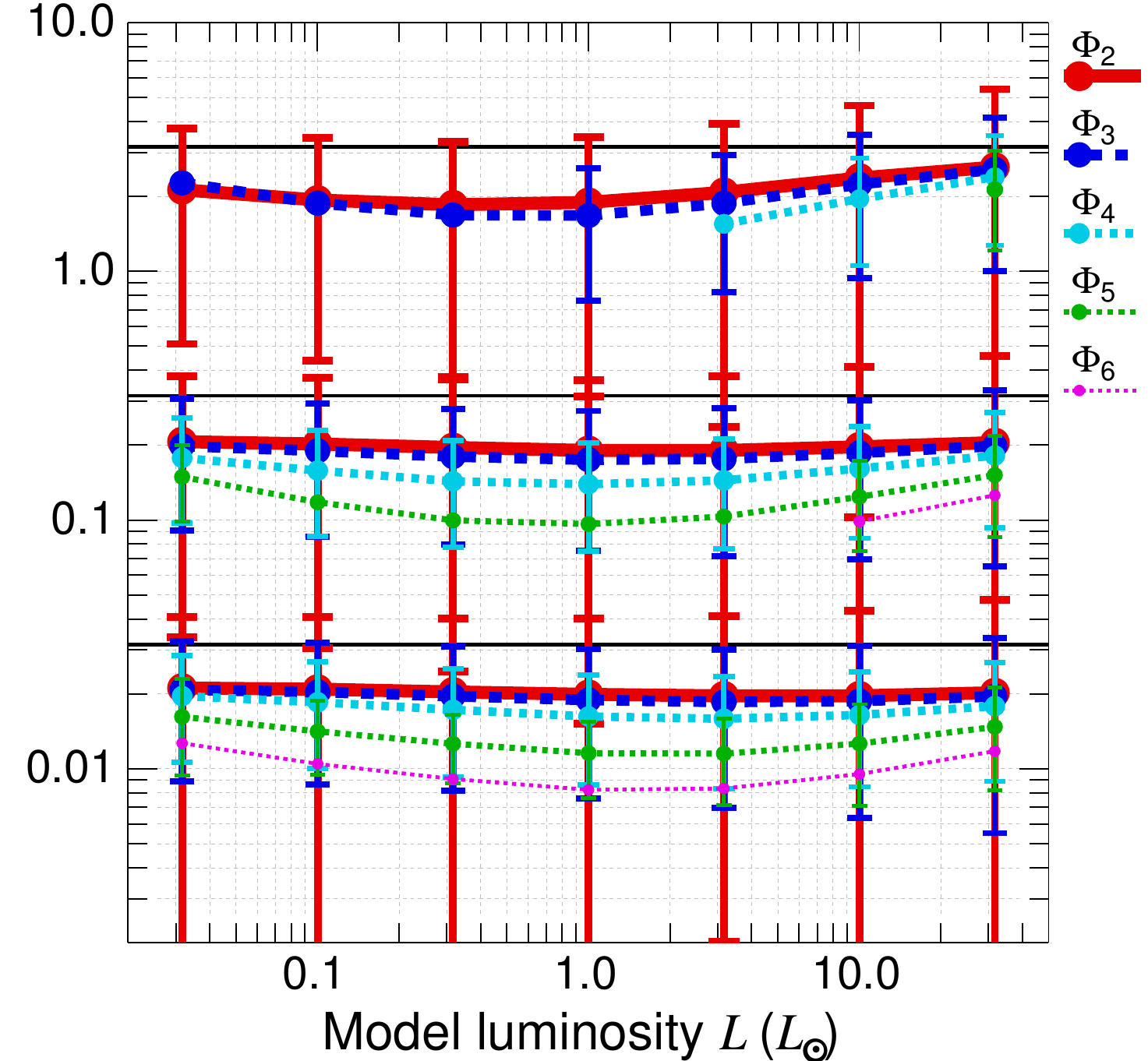}}
            \resizebox{0.3204\hsize}{!}{\includegraphics{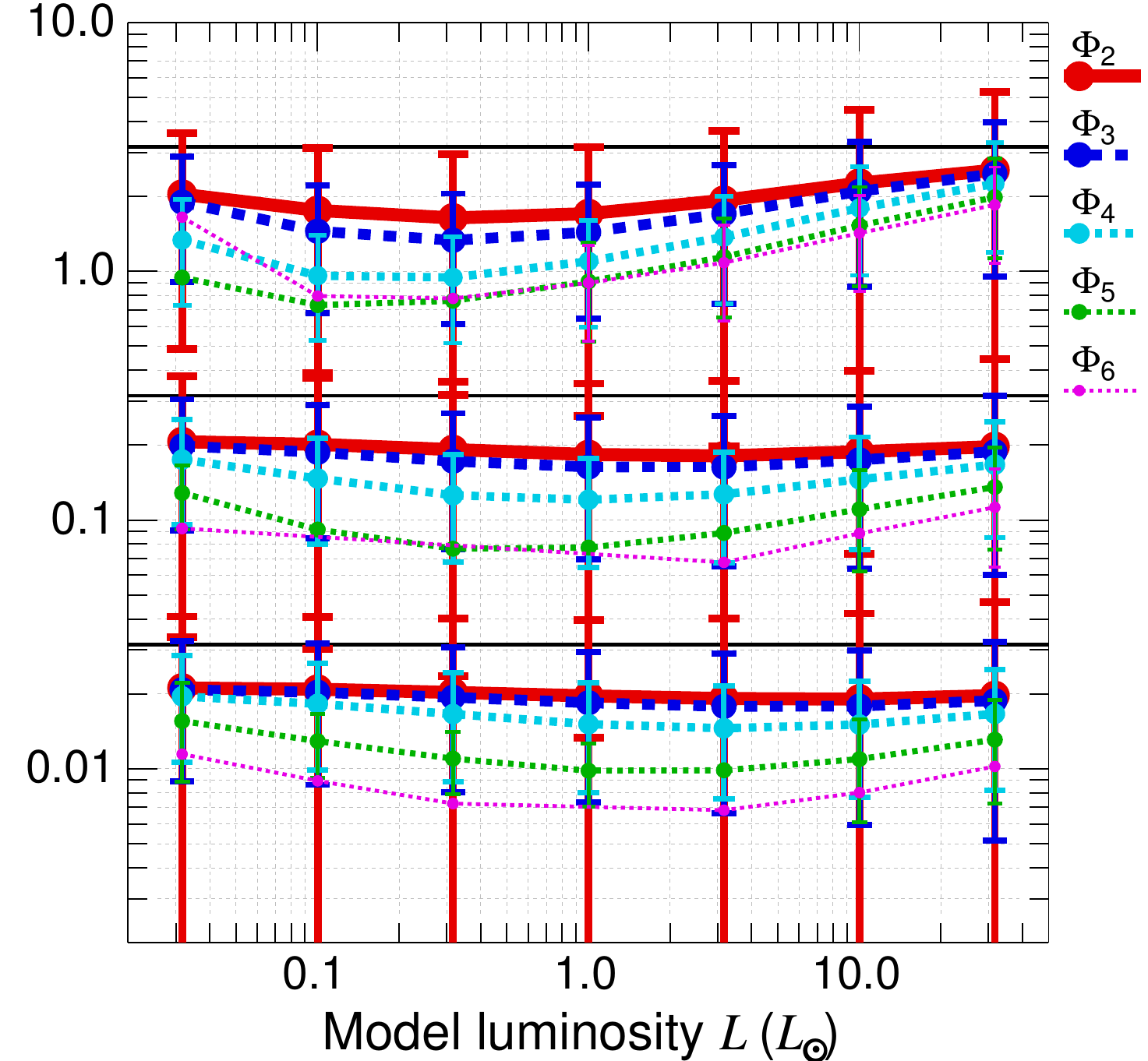}}}
\caption{
Temperatures $T_{\mathcal{I}}$ and masses $M_{\mathcal{I}}$ derived from fitting images $\mathcal{I}_{\nu}$ of both \emph{isolated} 
and \emph{embedded} protostellar envelopes (with the true masses $M$ of $0.03, 0.3$, and $3\,M_{\sun}$) vs. the true model values 
of $T_{M}$ and $L$ for correct $\beta\,{=}\,2$. The three columns of panels present results for three angular resolutions indicated 
(resolved, partially resolved, and unresolved cases). See Fig.~\ref{coldens.bes} for more details.
} 
\label{coldens.pro}
\end{figure*}

Isolated protostellar envelopes, $\beta\,{=}\,2$ (Fig.~\ref{coldens.pro}). For the fully resolved models, derived $T_{\mathcal{I}}$
and $M_{\mathcal{I}}$ are very accurate across all masses and luminosities. With degrading angular resolutions and for larger
$\Phi_{n}$ ($n\,{=}\,3\,{\rightarrow}\,6$) the inaccuracies and biases increase quite considerably. The accretion energy released
in the envelopes centers heats the dust to $T_{\rm S}\,{\sim}\,10^{3}$\,K, making $T_{\rm d}(r)$ strongly nonuniform. For the lines
of sight passing through the inner radial zones, the hot emission skews the $I_{\nu\,ij}$ shapes towards shorter wavelengths. For
the unresolved envelopes, the results become similar to those obtained with the method of fitting fluxes $F_{\nu}$
(Fig.~\ref{temp.mass.pro}).
                                                 
Embedded protostellar envelopes, $\beta\,{=}\,2$ (Fig.~\ref{coldens.pro}). For the fully-re\-solved models, derived
$T_{\mathcal{I}}$ are slightly less accurate for the acceptable fits than in the case of the isolated envelopes. The range of the
latter in more massive models for larger $\Phi_{n}$ ($n\,{=}\,3\,{\rightarrow}\,6$) shrinks towards higher $L_{\star}$. The most
accurate $M_{\mathcal{I}}$, obtained for $\Phi_{2}$, are underestimated by a factor of $1.45$, mostly because of the
over-subtraction of the rim-brightened background (Appendix \ref{AppendixB}, Sect.~\ref{bg.subtraction}). For the
partially resolved and unresolved envelopes, the most accurate $M_{\mathcal{I}}$ (for $\Phi_{2}$) are underestimated by factors of
$1.5\,{-}\,2$, whereas fitting larger $\Phi_{n}$ ($n\,{=}\,3\,{\rightarrow}\,6$) leads to errors by factors of $4\,{-}\,5$.


\section{Discussion}
\label{discussion}

Spectral flux and intensity distributions of the radiative transfer models of the starless cores and protostellar envelopes
($0.03$\,{--}\,$30\,M_{\sun}, L_{\sun}$) were fitted using the \textsl{modbody} and \textsl{thinbody} models. Derived values of the
fitting parameters were then compared to their true values to quantify the qualities of the mass derivation methods, fitting
models, and various sources of errors.

As shown in Sect.~\ref{results}, large \emph{intrinsic} inaccuracies and biases need to be taken into account when applying the
methods of mass derivation to the observed sources. In addition to being affected by nonuniform temperatures, estimated masses are
also affected by the adopted value of $\beta$ and subset of data points $\Phi_{n}$, as well as by the removal algorithm of the
background emission of an embedding cloud. In the method of fitting fluxes $F_{\nu}$, the masses depend on the fitting model,
whereas in the method of fitting images $\mathcal{I}_{\nu}$, they depend on the angular resolution.

The results of this purely model-based work discussed below may be directly applicable \emph{only} to sources with \emph{very}
accurate measurements (with negligible errors). Real observations deal with images of relatively faint, crowded sources on strong
and variable backgrounds, obtained with quite different angular resolutions, and thus they carry much larger measurement errors.
Observations are substantially affected by various statistical and systematic errors, depending on the adopted source extraction
method \citep[e.g.,][]{Men'shchikov_etal2012, Men'shchikov2013} and especially the treatment of background subtraction and
deblending. Implications for the real-life studies are considered below, whenever possible.

\subsection{Mass derivation methods}
\label{the.methods}

In the first method, source fluxes $F_{\nu}$ are integrated from the images $\mathcal{I}_{\nu}$, their spectral distribution is
fitted, and source mass $M_{F}$ is estimated from the fitting model. In the second method, the pixel spectral shapes $I_{\nu}$ of
the images $\mathcal{I}_{\nu}$ are fitted and the source mass $M_{\mathcal{I}}$ is integrated from the resulting image
$\mathcal{N}_{\rm H_2}$ of column densities. For unresolved sources and the \textsl{thinbody} fitting model, the methods give very
similar levels of inaccuracy, whereas for resolved images, the methods differ quite substantially.

When fitting $F_{\nu}$, the observed source emission from its entire volume is blended in the spatially integrated fluxes that
retain no spatial information. For the models with strongly nonuniform $T_{\rm d}(r)$ (Fig.~\ref{trp.bes.pro}), resulting heavy
distortions of the spectral shapes of $F_{\nu}$ (Fig.~\ref{sed.bes.pro}) from those of the fitting models lead to large systematic
errors in estimated parameters (Figs.~\ref{temp.mass.bes} and \ref{temp.mass.pro}).

When fitting $\mathcal{I}_{\nu}$, it is very beneficial to have a higher angular resolution. For fully resolved objects, pixels
$(i,j)$ sample independent $I_{\nu\,ij}$ from different columns of dust. For the transparent lower mass models
($M\,{\la}\,0.3\,M_{\sun}$), derived $M_{\mathcal{I}}$ are quite accurate (Figs.~\ref{coldens.bes},\,\ref{coldens.pro}). For lower
resolutions, the intensity of each pixel $(i,j)$ blends with that of its larger surroundings within the beam, not only along the
line of sight. The contamination of $I_{\nu\,ij}$ by the more distant areas, leads to a substantial degradation of
$T_{\mathcal{I}}$ and $M_{\mathcal{I}}$, especially when fitting large $\Phi_{n}$ ($n\,{=}\,3\,{\rightarrow}\,6$). Thus, the
benefits of this method are vanishing with decreasing angular resolutions.

Multiwavelength \emph{Herschel} images have been used to reconstruct radial temperature and density profiles of well-resolved
sources \citep[][]{Roy_etal2014}. Whenever such reconstructed densities are accurate enough, they can be used to obtain masses of
the nearby sources. Results of this study demonstrate, however, that the simple method of fitting images $\mathcal{I}_{\nu}$ is
able to deliver accurate masses for spatially resolved sources (Sect.~\ref{coldens.properties}).

\subsection{Background subtraction}
\label{bg.subtraction}

Stars form in the densest parts of interstellar clouds, hence the embedded models of starless cores and protostellar envelopes must
be more realistic than the isolated models. Although the spherical uniform-density embedding clouds are idealized, in a first
approximation they account for the absorption and re-emission of ISRF, leading to realistic temperature profiles within the model
objects. However, the presence of surrounding material makes it necessary to subtract its contribution to study the properties of
the starless cores and protostellar envelopes alone. In observational practice, backgrounds are estimated by an average intensity
in a narrow annulus placed just outside a source (cf. Sect.~\ref{seds}). Subtraction of such a flat background is not quite
accurate as a transparent embedding cloud around any object always tends to be rim-brightened and resembles a crater, in contrast
to a distant, physically unrelated back- or foreground. This effect is discussed in detail in Appendix \ref{AppendixB}.

The actual observable depths of the background craters may be shallower, when the local (filamentary) background itself is embedded
in a less dense but more extended cloud or is seen in projection onto a distant, physically unrelated back- or foreground. The
rim-brightening effect gets diluted, if the column densities of the source-embedding background and of the other unrelated clouds
are similar. Poorer angular resolutions also tend to smear out the effect for less resolved sources. Realistic temperature
gradients within the embedding backgrounds (Fig.~\ref{trp.bes.pro}) can either reduce or increase the crater depths by
${\sim}\,10{\%}$ for starless cores and protostellar envelopes, respectively (Appendix \ref{AppendixB}).

For unresolved sources, the observational algorithm of background subtraction is likely to overestimate fluxes as stars are born
within the gravitationally unstable densest peaks of the parent clouds. Large beams blend the object's emission with that of its
mountain-like environment, spreading the mix downhill, towards the valleys of lower cloud densities. The real background under an
unresolved source must be hill-like, whereas the background values from an annulus tend to come from more distant valleys. The
problem is aggravated in crowded regions, where no local source-free annuli around overlapping sources can be found and where one
needs to deblend sources. Angular resolution degrades with wavelengths, hence the degree of flux overestimation becomes strongly
biased towards longer wavelengths.

\subsection{Nonuniform temperatures}
\label{nonuniform.temps}
                                        
Both fitting models make a sensitive assumption that the objects have a uniform temperature $T$, which seems to make them
inadequate for the applications to starless cores and protostellar envelopes with nonuniform $T_{\rm d}(r)$. For the purpose of the
derivation of accurate masses, however, the uniform $T$ can be interpreted as an appropriate average quantity. In the methods of
fitting $F_{\nu}$ and $\mathcal{I}_{\nu}$, the temperature is consistent with $T_{M}$ and $T_{N ij}$ from
Eqs.~(\ref{mass.averaged}) and (\ref{coldens.averaged}), respectively (cf. Sects.~\ref{fitting.fluxes} and \ref{fitting.intens}).
In other words, to estimate masses $M_{F}$ or $M_{\mathcal{I}}$ that are accurate (${\approx}\,M$), it is necessary that the
fitting returns $T_{F}$ or $T_{\mathcal{I} ij}$ as close as possible to the average values $T_{M}$ or $T_{N ij}$, respectively.
This is clearly demonstrated in Appendix \ref{AppendixD} by the accurate masses obtained for the isothermal models with $T_{\rm
d}(r)\,{=}\,T_{M}$.

The inhomogeneous temperatures tend to distort the spectral shapes of $F_{\nu}$ and $I_{\nu\,ij}$ of the objects towards shorter
wavelengths (Figs.~\ref{sed.bes.pro} and \ref{fits.examples}). With a strong dependence of the dust emission peak on temperature
(${\kappa_{\nu_{\rm P}} B_{\nu_{\rm P}}(T)}\,{\propto}\,T^{5}$), the radial zones with higher $T_{\rm d}(r)$ make a much greater
contribution to the observed spectral shapes. Therefore, the shapes are skewed mainly owing to the emission of those parts of the
objects that have $T_{\rm d}(r)\,{>}\,T_{M}$ or $T_{\rm d}(r)\,{>}\,T_{N ij}$. In other words, distortions of the spectral shapes
are caused by the dust with \emph{excess} temperatures above the average values.

This is further demonstrated by additional ray-tracing observations of the models, in which the excess temperatures were removed:
$T_{\rm d}(r){\rightarrow}\min\left\{T_{\rm d}(r),T_{M}\right\}$. Derived masses of these mostly isothermal models (not shown) are
almost as accurate as those of the fully isothermal models (Appendix \ref{AppendixD}), only within a few percent lower. As is
expected, there is almost no dependence on the subsets $\Phi_{n}$ ($n{\,=\,}2{\,\rightarrow\,}6$), which indicates that the
spectral shapes are indeed not distorted.

\subsection{Fitting models}
\label{choice.model}

When fitting images $\mathcal{I}_{\nu}$, both fitting models are equivalent and estimated parameters are indistinguishable
(Sect.~\ref{variable.fixed}). When fitting fluxes $F_{\nu}$, the results of this work show that \textsl{thinbody} generally returns
far more accurate masses than \textsl{modbody} does for both isolated and embedded variants of starless cores and protostellar
envelopes (Figs.~\ref{temp.mass.bes} and \ref{temp.mass.pro}).

Although the \textsl{modbody} fits often look better (i.e., they have smaller $\chi^{2}$ values), they generally bring parameters
that are much more inaccurate. Indeed, the spectral shapes of $F_{\nu}$ are skewed towards short wavelengths by emission from their
hotter parts. With more free parameters, \textsl{modbody} describes more flexible shapes, between $B_{\nu}(T)$ and
${\kappa_{\nu}\,B_{\nu}(T)}$. It is able to produce better fits of the distorted spectral shapes of objects with nonuniform $T_{\rm
d}(r)$ and hence it always tends to produce significantly over- and underestimated $T_{F}$ and $M_{F}$, respectively
(Sect.~\ref{derived.properties}). Furthermore, most of the \textsl{modbody} fits have $\tau_{\nu}\,{\sim}\,1$ even in the
far-infrared, which is fundamentally inconsistent with the radiative transfer models whose fluxes $F_{\nu}$ represent optically
thin emission ($\tau_{\nu}\,{\ll}\,1$).

The \textsl{thinbody} model produces the best overall results and smallest biases and inaccuracies in derived $T_{F}$ and $M_{F}$
for the isolated and embedded starless cores and protostellar envelopes (Figs.~\ref{temp.mass.bes}\,{--}\,\ref{coldens.pro}). The
\textsl{thinbody} fits are, by definition, optically thin in the far-infrared and thus consistent with the radiative transfer
models. Only two variable fitting parameters of \textsl{thinbody} contribute to better robustness of $T_{F}$ and $M_{F}$, compared
to \textsl{modbody} with one extra free parameter.

Contrary to what is usually assumed in observational studies, the results show that it must be counterproductive to aim at precise
fitting of the peaks and shorter wavelength parts of $I_{\nu}$ and $F_{\nu}$. When the distorted shapes are reproduced more
accurately, the estimates of the temperatures and masses are less accurate.

\subsection{Opacity slopes}
\label{choice.beta}

The standard methods of mass derivation ignore the presence of very small stochastically heated dust particles, assuming just a
simple power-law opacity across all bands, and so do the radiative transfer models in this study. Emission of such very small
grains within the real objects could enhance fluxes at $70$ and $100\,{\mu}$m and, in effect, skew their spectral shapes farther
towards short wavelengths, leading to more heavily overestimated temperatures and underestimated masses.

Various compositional and structural properties of real cosmic dust grains in different environments may lead to far-infrared
opacity slopes that are different from $\beta\,{\approx}\,2$ (expected for small compact spherical grains) and even to
wavelength-de\-pendent $\beta_{\lambda}$. This study explored three constant values ($1.67$, $2.0$, $2.4$) to probe their influence
on the accuracy of derived masses. Fixing $\beta$ in the fitting process reduces the number of free parameters and improves the
consistency (reduces biases) of derived parameters for objects with different physical properties ($M$, $L_{\star}$).

With the correct $\beta$ value, masses derived with the \textsl{thinbody} fits are generally off the true mass $M$, the magnitude
of discrepancy depending on how much and in what direction derived temperature deviates from $T_{M}$. When $\beta$ is over- or
underestimated by a factor of $1.2$, derived masses become over- or underestimated within a factor of $2$ with respect to the
masses obtained using the true value $\beta{\,=\,}2$. This is a direct consequence of the temperatures being under- or
overestimated, correspondingly, a behavior that is easy to understand. In contrast to the \textsl{thinbody} fits, no clear trends
with respect to the inaccuracies in the adopted $\beta$ value can be found for \textsl{modbody}, except that it generally returns
greatly over- and underestimated $T_{F}$ and $M_{F}$.
 
To quantify the effects of freedom in this fitting parameter, additional fits with variable $\beta$ were performed
(Appendix~\ref{AppendixE}). As is expected, they showed much greater biases and inaccuracies in derived parameters
(Figs.~\ref{beta.bes.pro} and \ref{beta.bes.pro.coldens}), as the extra degree of freedom also makes the resulting $\beta$ values
incorrect (Fig.~\ref{beta.beta}), the magnitude of error depending on the true values of $M$ and $L_{\star}$. It is possible to
compare these results with those obtained in previous studies focused on the relationships between the derived $\beta_{F}$ and
$T_{F}$ \citep[][and references therein]{Shetty_etal2009a,Shetty_etal2009b,JuvelaYsard2012a}. The present models of starless cores
and protostellar envelopes show that the correlations of the two quantities may be both positive and negative
(Fig.~\ref{tem.beta.bes.pro}), with almost no correlation in the case of isothermal models. They must be induced by deviations of
the spectral shapes of $F_{\nu}$ (Fig.~\ref{sed.bes.pro}) from $F_{\nu}(T_{M})$ (Fig.~\ref{sed.tmav.bes.pro}), caused by the
nonuniform $T_{\rm d}(r)$ (Fig.~\ref{trp.bes.pro}). For the protostellar envelopes, the correlations are non-monotonic and they may
either be strongly negative or positive, depending on the luminosity.

\subsection{Data subsets}
\label{choice.subset}

For nonuniform profiles $T_{\rm d}(r)$ of starless cores and protostellar envelopes (Fig.~\ref{trp.bes.pro}), better parameters are
estimated with \textsl{thinbody} when using smaller subsets of data ($n\,{=}\,6\,{\rightarrow}\,2$) as the latter are less affected
by the skewed spectral shapes. The most accurate masses are obtained by fitting just two of the longest wavelength data points; in
most cases, however, a subset $\Phi_{3}$ produces very similar results. Larger subsets $\Phi_{n}$ ($n\,{=}\,2\,{\rightarrow}\,6$)
may give slightly better $M_{F}$ only when fixing an incorrect $\beta$ value for the lower-mass starless cores
($M\,{\la}\,1\,M_{\sun}$, Fig.~\ref{temp.mass.bes}). Using the inadequate fitting model with an incorrect $\beta$, larger
$\Phi_{n}$ can constrain $T_{F}$ to better resemble $T_{M}$. For overestimated $\beta$, derived $M_{F}$ always shift to higher
values (Fig.~\ref{temp.mass.pro}), which offsets the general opposite trend to underestimate $M_{F}$ and thus may give more
accurate results.

Inaccuracies of the data points in real observations are usually more substantial than those assumed in this work, aggravated by
the systematic uncertainties that may lead to both over- and underestimated $F_{\nu}$ (Sect.~\ref{bg.subtraction}).
Background-subtracted and deblended $I_{\nu}$ at each wavelength with different angular resolutions have independent and different
systematic errors. The latter must be large and uncertain on the bright and structured backgrounds in star-forming regions.
Moreover, unresolved sources are likely to include emission from \emph{clusters} of objects.

Results of this model study are directly relevant to real observations \emph{only} in the simplest case (which is rare) of accurate
measurements with negligible errors. A blind application of the findings to real complex images may lead to incorrect results if
the above caution is ignored and small subsets $\Phi_{2}$ of data points with large and independent measurement errors are fitted.
For such data, it would be safer and more appropriate to fit a larger subset of the longest wavelength data ($\Phi_{3}$ or
$\Phi_{4}$, depending on the quality of measurements). Distortions of the observed spectral shapes towards shorter wavelengths is
an intrinsic property of both starless cores and protostellar envelopes, affecting all sources in star-forming regions,
independently of the level of measurement errors.

It beyond the scope of this model-based work to give general recipes to observers on how to select data points to fit. This study
highlights the intrinsic behavior of the mass derivation methods by eliminating the ``observational layer'' (with all its
complications and uncertainties) between the objects and the observer. It is important to realize that the peak and shorter
wavelength shapes of $I_{\nu}$ and $F_{\nu}$ are most skewed by the temperature excesses within objects
(Sect.~\ref{nonuniform.temps}) and their influence has to be minimized to obtain accurate results. In view of the strong dependence
of the results on $\Phi_{n}$, it is advisable to examine fits of \emph{all} subsets of data points for each observed source to
estimate the robustness of the results and to possibly choose the fits giving the best mass estimate.

\subsection{Mass uncertainties}
\label{uncert.masses}

To make a bridge between this purely model-based study with no measurement errors and actual observational studies and see how
typical statistical errors in the input data would translate into those of the derived masses, this work assigned (fairly
optimistically) $\pm\,15{\%}$ errors to the model intensities and fluxes, and adopted $\pm\,20{\%}$ errors in $\eta$, 
$\kappa_0$, and $D$.

The uncertainties in derived masses returned by the fitting algorithm are $40\,{-}\,70{\%}$, depending on the subset $\Phi_{n}$ of
fluxes (Figs.~\ref{temp.mass.bes}\,{--}\,\ref{coldens.pro}, \ref{mass.bes.pro.tmav}, \ref{coldens.bes.pro.tmav}). For the
acceptable fits of larger subsets ($n{\,=\,}2{\,\rightarrow\,}6$), the derived mass uncertainty is dominated by the $20{\%}$ errors
of the parameters $\eta$, $\kappa_0$, and $D$, because the effect of the $15{\%}$ measurement errors becomes smaller for the fits
constrained by a larger number of independent data points. For smaller subsets ($n{\,=\,}6{\,\rightarrow\,}2$), the fits are less
constrained, hence the contribution of the $15{\%}$ error bars to the derived mass uncertainty becomes larger. Different subsets
$\Phi_{n}$ give very similar results only for fully resolved sources with the method of fitting images $\mathcal{I}_{\nu}$
(Figs.~\ref{coldens.bes} and \ref{coldens.pro}).

In real observations, statistical measurement uncertainties in $I_{\nu}$ and $F_{\nu}$ are larger than the $\pm\,15{\%}$ errors
assumed in this study. Furthermore, it would be more realistic to adopt uncertainties of $\eta$, $\kappa_0$, and $D$ of at least
$\pm\,50{\%}$, which would raise the derived mass uncertainties well beyond $100{\%}$. By including the mass inaccuracies (of a
factor of $2$) induced by a $20{\%}$ uncertainty in $\beta$ and systematic errors (of factors of at least $2$) caused by the
nonuniform temperatures within the observed sources, it is clear that the absolute values of masses derived from fitting are
inaccurate and uncertain (within a factor of at least $2\,{-}\,3$). It is possible to neglect the uncertainties in $\eta$,
$\kappa_0$, and $D$, if the focus is on studying relative properties of a population of objects all at roughly the same distance
within a certain star-forming cloud with homogeneous dust properties. Apart from this, however, one has to derive accurate
\emph{absolute} values of the most fundamental parameters to make correct and physically meaningful conclusions.

It is quite important to carefully estimate mass uncertainties: without realistic error bars, derived masses are meaningless and
correct conclusions are unlikely. To go one step further and obtain an idea of the \emph{actual} errors of derived masses, it is
possible to construct radiative transfer models of the observed population of sources, distribute the model sources over the
observed images and extract them, and finally derive their masses. Comparing derived masses with the fully known model properties,
reasonable estimates of the actual errors in derived masses are obtained.


\section{Conclusions}
\label{conclusions}

This paper presented a model-based study of the uncertainties and biases of the standard methods of mass derivation (fitting fluxes
$F_{\nu}$ and images $\mathcal{I}_{\nu}$), widely applied in observational studies of the low- and intermediate star formation. To
focus on the intrinsic effects related to the physical objects, all observational complications leading to additional flux or
intensity errors (filamentary and fluctuating backgrounds, instrumental noise, calibration errors, different resolutions, blending
with nearby sources, etc.) were assumed to be nonexistent. As a consequence, results of this work are directly relevant \emph{only}
for the simplest case of bright isolated sources on faint backgrounds with negligible measurement errors. The real mass
uncertainties for starless cores and protostellar envelopes are likely to be larger than those found in this work.
 
Background subtraction. Embedding backgrounds of physical objects are rim-brightened (i.e., they tend to resemble craters), their
depths depend on the sizes of the object and embedding cloud. The standard observational procedure of flat background subtraction
may give systematically underestimated $I_{\nu}$ and $F_{\nu}$, and hence masses for resolved sources. Poorer angular resolutions
at longer wavelengths tend to systematically overestimate $I_{\nu}$ and $F_{\nu}$, and hence masses for unresolved objects, as
their emission gets blended with that of the mountain-like background and possibly with other objects within the same beam.

Nonuniform temperatures. Temperature excesses above average values $T_{M}$ and $T_{N ij}$ is the primary reason for the skewness of
the spectral shapes of $F_{\nu}$ and $I_{\nu\,ij}$ towards shorter wavelengths. Depending on $M$, $L_{\star}$, $\beta$, $\Phi_{n}$,
fitting model, and angular resolution, they lead to overestimated temperatures and various biases. With the method of fitting
$F_{\nu}$, masses become underestimated by factors $2\,{-}\,5$. When fitting $\mathcal{I}_{\nu}$, similarly large inaccuracies are
found only for unresolved objects, whereas with better angular resolutions they decrease and become very small for well-resolved
objects.

Fitting models. When fitting $\mathcal{I}_{\nu}$, both models are equivalent and estimated $M_{\mathcal{I}}$ are indistinguishable.
When fitting $F_{\nu}$, \textsl{thinbody} gives far more accurate $M_{F}$ than \textsl{modbody} does. The latter causes such great
biases and inaccuracies in $T_{F}$ and $M_{F}$ that \textsl{modbody} must be considered unusable.

Opacity slopes. Fixing $\beta$ reduces biases in derived parameters. When $\beta$ is too high or low by a factor of $1.2$, derived
masses become over- or underestimated by a factor of $2$ with respect to those obtained using the true $\beta\,{=}\,2$.
Qualitatively, this behavior is caused by the natural tendency of steeper $\beta$ to produce lower temperatures, hence higher
masses. Quantitatively, the factors are approximate and they may depend on some of the assumptions used in this study. Mass
derivation with a free variable $\beta$ should be avoided, as it tends to lead to very strong biases and erroneous masses.

Data subsets. Derived masses strongly depend on the subsets $\Phi_{n}$ of data points, except when fitting images
$\mathcal{I}_{\nu}$ of fully resolved sources. Given the nonuniform $T_{\rm d}(r)$ of the model objects, the most accurate masses
are estimated with \textsl{thinbody} using subsets that are as small as possible ($n\,{=}\,6\,{\rightarrow}\,2$). In real
observations with substantial independent errors in different wavebands, it should be much safer and more accurate to fit slightly
larger subsets ($\Phi_{3}$ or even $\Phi_{4}$). Those data points that are on the peak of their spectral distribution or on the
short-wavelength side should be ignored, whenever possible, to improve the accuracy of derived masses. In practice, it is advisable
to investigate fits of \emph{all} subsets of data for each observed source, to verify robustness of the results and to possibly
choose the best mass estimate.

Derived masses. Dividing the mass range of $0.03\,{-}\,30\,M_{\sun}$ at $1\,M_{\sun}$ into the low- and high-mass objects and
considering unresolved or poorly resolved sources with $\beta\,{=}\,2$, the following conclusions can be drawn. Masses of the
isolated low- and high-mass starless cores are underestimated by factors $1\,{-}\,1.3$ and $1.3\,{-}\,4$, respectively. The mass
inaccuracies increase towards the high-mass end and for larger subsets $\Phi_{n}$ ($n\,{=}\,2\,{\rightarrow}\,6$). Masses of the
embedded low-mass cores are underestimated by a factor of $1.4$. They are more biased towards the high-mass end, changing from
under- to overestimated within a similar factor. Masses of the protostellar envelopes are considerably biased over the range of
$0.03\,{-}\,30\,L_{\sun}$ and their inaccuracies strongly increase for larger subsets of $\Phi_{n}$
($n\,{=}\,2\,{\rightarrow}\,6$). Masses of the isolated and embedded envelopes become underestimated by factors $2\,{-}\,3$ and
$3\,{-}\,5$, respectively. Masses of the low-mass starless cores are likely to be determined much more accurately than those of
protostellar envelopes.

Mass uncertainties. Adopting statistical errors of $15{\%}$ for model intensities (fluxes) and optimistically assuming that $\eta$,
$\kappa_0$, and $D$ were known to within $20{\%}$, typical mass uncertainties returned by the fitting algorithm are
$40\,{-}\,70{\%}$, depending on $\Phi_{n}$. If more realistic statistical errors in the measurements and parameters of at least
$50{\%}$ are adopted, the mass uncertainties increase well beyond $100{\%}$. Larger subsets $\Phi_{n}$
($n\,{=}\,2\,{\rightarrow}\,6$) of independent data points are beneficial in somewhat reducing the resulting mass uncertainties. On
the other hand, the larger subsets are also highly undesirable, because they escalate the systematic mass inaccuracies by at least
a factor of $2$ as a result of nonuniform temperatures. Smaller subsets $\Phi_{n}$ ($n\,{=}\,6\,{\rightarrow}\,2$) are able to
minimize the systematic errors caused by the temperature variations, but they increase the chances of getting incorrect masses in
the case of inaccurate data measurements in real observations.

Global inaccuracies. Without extremely accurate flux measurements and knowledge of the free parameters ($\eta, \kappa_0, \beta,
D$), and without radiative transfer simulations to have an idea of the actual mass errors, it would be reasonable to assume that
the \emph{absolute} values of masses of the unresolved or poorly resolved objects are inaccurate to within \emph{at least} a factor
of $2\,{-}\,3$. This may be less problematic, if the relative properties are studied of a population of objects within a
star-forming cloud, hopefully with the same distance and dust opacities. Ultimately, however, accurate absolute masses are
necessary to make correct, physically meaningful conclusions.
                       
Accuracy is paramount. There are several ways to improve mass estimates: (1) using a multiwavelength source extraction method
measuring the most accurate, least biased background-subtracted and deblended fluxes across all wavebands; (2) selecting the best
sources from the extraction catalogs, with the most accurately and consistently measured fluxes over at least three longest
wavelengths; (3) using the \textsl{thinbody} fitting model for the purposes of temperature or mass derivation; (4) estimating the
model parameters $\eta$, $\kappa_0$, $\beta$, and $D$ as accurately as possible and always performing fitting with $\beta$ fixed;
(5) for resolved sources, fitting their background-subtracted (and deblended) images and integrating source masses from column
densities; (6) fitting all subsets of data points for each source and choosing the smallest possible subset ($\Phi_{3}$ or
$\Phi_{4}$) that gives the most accurate temperatures and masses; (7) using radiative transfer models to simulate observed images,
extracting the model sources, deriving their masses, and comparing them to the true model values to have an idea of the
\emph{actual} errors for the derived masses of observed sources.


\begin{acknowledgements}
This study employed \textsl{SAOImage DS9} (by William Joye) developed at the Smithsonian Astrophysical Observatory (USA), the
\textsl{CFITSIO} library (by William D. Pence) developed at HEASARC NASA (USA), and \textsl{SWarp} (by Emmanuel Bertin) developed
at Institut d'Astrophysique de Paris (France). Radiative transfer code \textsl{MC3D-sph} version 3.12 \citep[by Sebastian 
Wolf,][]{Wolf2003} was used to compute the first generation of the models in this work. The \textsl{plot} utility and
\textsl{ps12d} library used in this work to draw figures directly in the \textsl{PostScript} language were written by the author
using the \textsl{PSPLOT} library (by Kevin E. Kohler) developed at Nova Southeastern University Oceanographic Center (USA) and the
plotting subroutines from the \textsl{AZEuS} MHD code (by David A. Clarke and the author) developed at Saint Mary's University
(Canada). Collaborative work within the \emph{Herschel} Gould Belt and HOBYS key projects was very beneficial. Useful comments on a
draft made by Pierre Didelon, Arabindo Roy, Alana Rivera-Ingraham, Sarah Sadavoy, Philippe Andr{\'e}, and by the anonymous referee 
helped improve this paper.
\end{acknowledgements}


\begin{appendix}

\section{Protostellar envelopes temperatures}
\label{AppendixA}

With the adopted $\kappa_{\nu}$ and $\rho(r)$ (Sects.~\ref{dust.properties} and \ref{density.profiles}), the radial temperature
profiles of protostellar envelopes (Fig.~\ref{trp.bes.pro}) can be approximated by a combination of two power laws:
\begin{equation}
T_{\rm d}(r) = A\,r^{-1.44} + B\,r^{-1/3},
\label{approximation}
\end{equation}
where $r$ is in AU and parameters $A$ and $B$ depend on mass and accretion luminosity:
\begin{eqnarray*}
A&\!\!\!\!\!=\!\!\!\!\!&2.93^{C\,} 900 \left(M/M_{\sun}\right)^{1/2},\\
B&\!\!\!\!\!=\!\!\!\!\!&1.63^{C\,} 84 \left(1.33+0.22\log{\left(M/M_{\sun}\right)}\right)^{-1},\\
C&\!\!\!\!\!=\!\!\!\!\!&1.5+\log{\left(L_{\star}/L_{\sun}\right)}.
\end{eqnarray*}
Equation (\ref{approximation}) describes the temperatures induced by the central accretion energy source (ignoring ISRF), valid for
$T_{\rm d}\,{\la}\,300$\,K. The first term in Eq.~(\ref{approximation}) approximates the steepest profiles in the inner semi-opaque
region, the second term represents temperatures in the transparent outer part of the envelopes. An approximate borderline between 
the two regimes can be estimated directly from Fig.~\ref{trp.bes.pro} as
\begin{equation}
\begin{split}
\hat{T}\,&{=}\,0.65\,(M/M_{\sun})^{-0.606}\,\hat{R}^{\,0.714},\\
\hat{R}\,&{=}\,175\left(M/M_{\sun}\right)^{1/2}\left(L_{\star}/L_{\sun}\right)^{1/5}\,{\rm AU},\\
\hat{T}\,&{=}\,26\left(M/M_{\sun}\right)^{-1/4}\left(L_{\star}/L_{\sun}\right)^{1/7}\,{\rm K}.
\end{split}
\label{boundary}
\end{equation}

\section{Rim-brightened backgrounds}
\label{AppendixB}

\begin{figure*}
\centering
\centerline{\resizebox{0.3250\hsize}{!}{\includegraphics{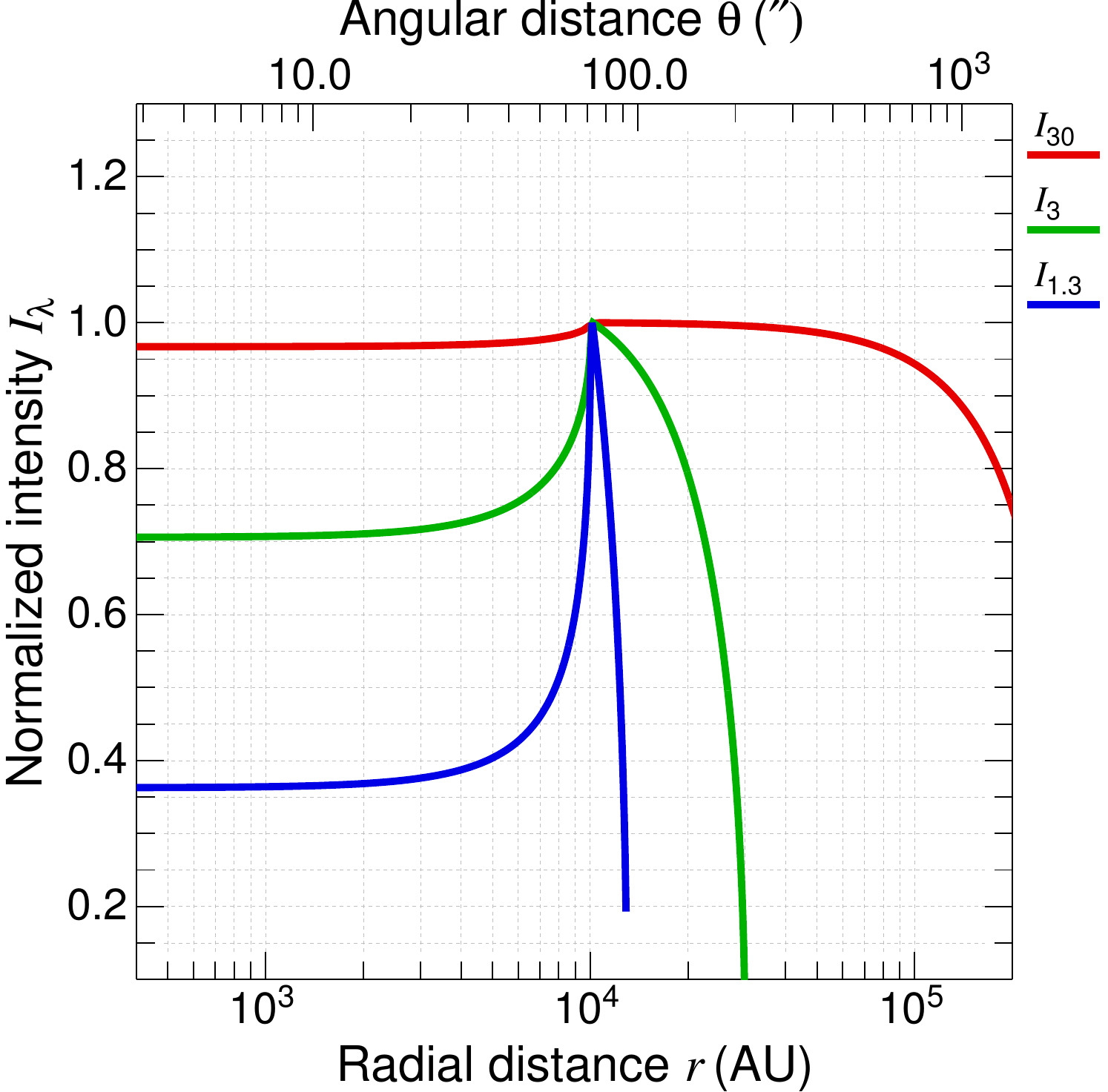}}
            \resizebox{0.3204\hsize}{!}{\includegraphics{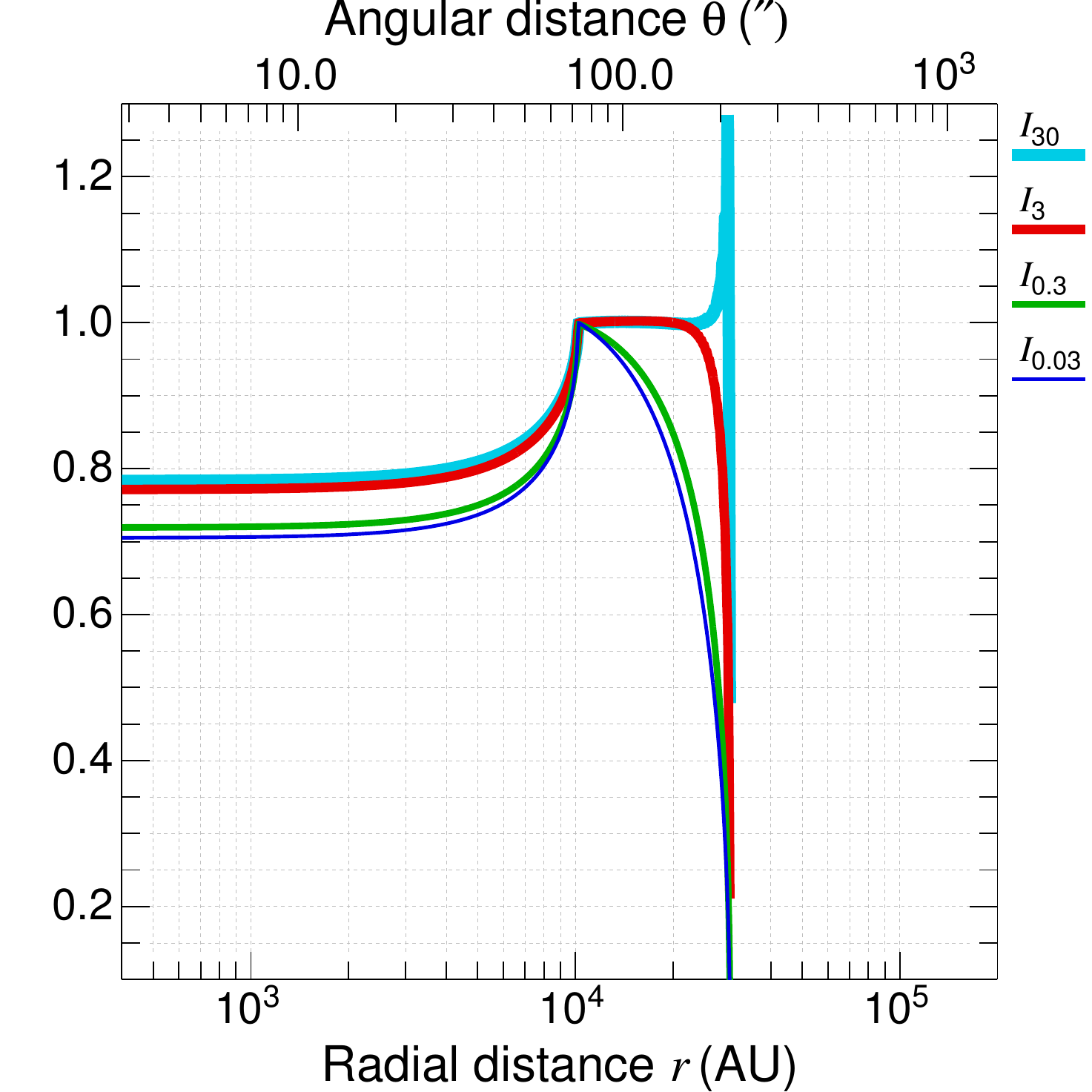}}
            \resizebox{0.3204\hsize}{!}{\includegraphics{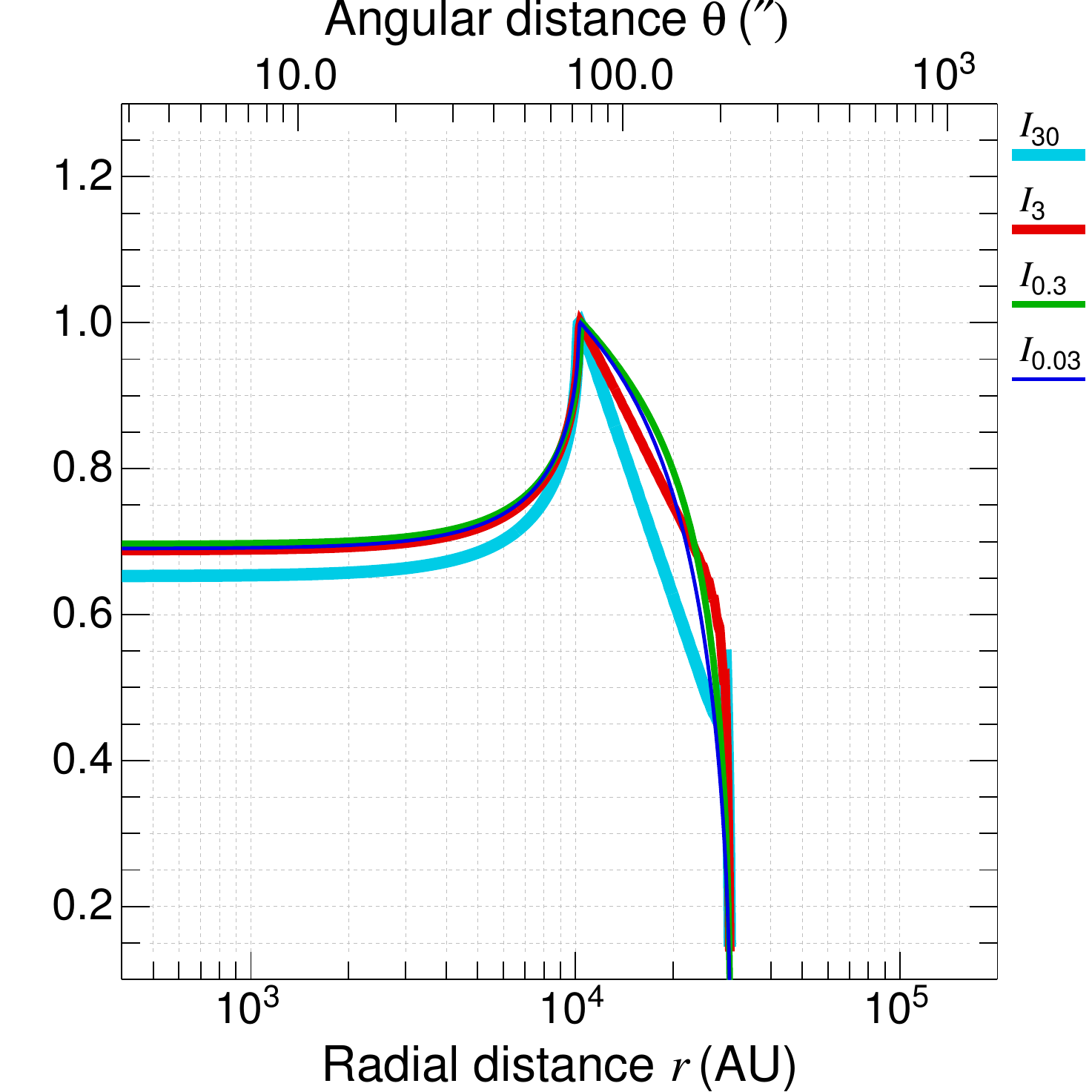}}}
\caption{
Background rim brightening in the spherical uniform-density embedding shells, either isothermal (\emph{left}) or with the
radiative-equilibrium $T_{\rm d}(r)$ from the models (Sect.~\ref{rtmodels}) of embedded starless cores (\emph{middle}) and
protostellar envelopes with $L_{\star}\,{=}\,30\,L_{\sun}$ (\emph{right}). The intensity profiles on the left are labeled by the
size ratios $R_{\rm E}/R\,{=}\,\{30, 3, 1.3\}$ of the embedding cloud and the central cavity, whereas the intensity profiles at
$500\,{\mu}$m in the middle and on the right indicate the mass (in $M_{\sun}$) of the embedded models. For uniform temperatures 
(\emph{left}), the brightening factors $f_{\rm S}$ are $1.03, 1.41$, and $2.77$ are the values given by Eq.~(\ref{background}). For 
nonuniform temperatures (and $R_{\rm E}/R\,{=}\,3$, $R\,{=}\,10^{4}$ AU), they range from $1.27$ to $1.40$ (\emph{middle}) and
from $1.44$ to $1.53$ (\emph{right}).
} 
\label{rim.bright}
\end{figure*}

In contrast to the emission of the distant and physically unrelated backgrounds or foregrounds, embedding backgrounds resemble
craters. The central spherical region occupied by an object ($r\,{\le}\,R$) does not belong to the embedding cloud
($R\,{<}\,r\,{\le}\,R_{\rm E}$) and thus must be considered empty when determining the object's background. Rim brightening for
uniform-density transparent isothermal clouds with such a cavity depends only on their relative radial thickness $(R_{\rm
E}\,{-}\,R)/R$. It can be quantified by the ratio $f_{\rm S}$ of intensities (or column densities) along the lines of sight passing
through the rim and the center of the cavity:
\begin{equation}
f_{\rm S} = \left(1 + \frac{2\,R}{R_{\rm E} - R}\right)^{1/2}.
\label{background}
\end{equation}

According to Eq.~(\ref{background}), the background under embedded objects can be overestimated from a few percent to a factor of
several, hence background-subtracted values and masses may become substantially underestimated. For the present models with
$R\,{=}\,10^{4}\,$AU and $R_{\rm E}/R\,{=}\,3$, the background and masses are over- and underestimated by $f_{\rm S}\,{=}\,1.41$,
respectively. The value is the discrepancy of derived masses seen for the embedded starless cores and protostellar envelopes
(Figs.~\ref{temp.mass.bes}\,{--}\,\ref{coldens.pro}, \ref{mass.bes.pro.tmav}, \ref{coldens.bes.pro.tmav}). The effect becomes much
stronger for very thin shell-like embedding clouds, whereas it vanishes for extended background clouds. For the size ratios $R_{\rm
E}/R$ of $1.3$ and $30$, the factor $f_{\rm S}$ takes the values of $2.77$ and $1.03$, respectively. Numerical examples of the
rim-brightened backgrounds for both isothermal and non-isothermal spherical embedding clouds are shown in Fig.~\ref{rim.bright}.

Realistic temperature profiles in the embedding clouds bring only minor quantitative changes, not altering qualitatively the rim 
brightening effect (Fig.~\ref{rim.bright}). Steep positive or negative temperature gradients in the dense shells around embedded
starless cores and protostellar envelopes (Fig.~\ref{trp.bes.pro}) tend to slightly reduce or increase the brightening effect (by
${\sim\,}10{\%}$), respectively.

Actual geometry of the real background clouds is of minor importance, the only relevant assumption is that the embedded object
(hence, its background cavity) has a convex shape. For instance, assuming a plane-parallel embedding cloud with thickness 
$2\,R_{\rm E}$ along the line of sight, it is possible to obtain a slightly different expression than Eq.~(\ref{background}) for 
the rim brightening factor:
\begin{equation}
f_{\rm P} = \frac{1}{2} \left(f_{\rm S} + \frac{R_{\rm E}}{R_{\rm E} - R}\right).
\label{planeparallel}
\end{equation}
Plane-parallel geometry makes the brightening factors $f_{\rm P}$ somewhat larger than $f_{\rm S}$ from Eq.~(\ref{background}), 
with the difference being stronger for thinner shell-like clouds. For instance, the size ratios $R_{\rm E}/R$ of $30, 3$, and 
$1.3$, correspond to the brightening factors $f_{\rm P}$ of $1.03, 1.46$, and $3.55$, respectively. 

Depths of the background craters may be quite dissimilar for different objects in real observations. The observations show that the
interstellar medium is strongly filamentary and that stars tend to form in narrow, very dense filaments
\citep[e.g.,][]{Men'shchikov_etal2010,Andre_etal2014}. For an object embedded in a cylindrical filament of radius $R_{\rm E}$ in
the plane of the sky, the brightening factor $f_{\rm C}$ is intermediate between $f_{\rm S}$ and $f_{\rm P}$, depending on the
position angle of the radius-vector from the center of the object to its outer boundary $R$. It is easy to see that $f_{\rm
C}\,{=}\,f_{\rm P}$ along the filament's axis, whereas $f_{\rm C}\,{=}\,f_{\rm S}$ in the orthogonal direction, across the filament.

The widths of the embedding filaments appear to be similar to the sizes of embedded objects \citep[][]{Men'shchikov_etal2010}.
Assuming their cylindrical geometry, the embedding filaments of starless cores and protostellar envelopes are likely to have
$R_{\rm E}$ only a factor of about $2\,{-}\,3$ larger than $R$. For such narrow filamentary backgrounds of resolved objects, the
rim brightening effect is quantified by factors $f_{\rm C}\,{\approx}\,1.8{-}1.4$.

Background rim brightening may be observable only when imaging the nearby resolved sources unaffected by other distant back- or
foregrounds. With this effect in action, the standard observational algorithm of background subtraction underestimates $F_{\nu}$ by
factors similar to $f_{\rm S}$ or $f_{\rm P}$. With poorer angular resolutions, the rim of the background crater gets smeared out
and thus the brightening effect eventually vanishes for unresolved sources which also have an opposite trend to produce
underestimated background and hence overestimated $I_{\nu}$ and $F_{\nu}$ (cf. Sect.~\ref{bg.subtraction}).

\section{Fitting procedure}
\label{AppendixC}

The nonlinear least-squares fitting algorithm used in this work employs the Le\-ven\-berg-Marquardt method \citep{Press_etal1992}
that minimizes $\chi^{2}$ residuals between the model and data. The method requires a user to provide initial guesses for model
parameters. Tests have shown that an arbitrary choice of the initial values of the fitting models (Sects.~\ref{fitting.fluxes},
\ref{fitting.intens}) does not guarantee convergence to the global $\chi^{2}$ minimum. A fully automated fitting procedure has been
designed to overcome this problem and avoid any need to make arbitrary initial guesses.

The algorithm explores the multidimensional parameter space of the model with a large number of trial fittings of the spectral
distributions of data points. The parameter space is discretized in logarithmically equidistant steps $\delta\log{p}$, covering all
relevant initial values of temperature $T$ ($4\,{-}\,10^{3}$\,K), mass $M$
($3\,{\times}\,10^{-3}\,{-}\,3\,{\times}\,10^{2}\,M_{\sun}$), column density $N_{\rm H_2}$ ($10^{17}\,{-}\,10^{27}$\,cm$^{-2}$),
and solid angle $\Omega$ ($1.85\,{\times}\,10^{-13}\,{-}\,1.85\,{\times}\,10^{-5}$\,sr). Large initial discretization steps
$\delta\log{p}\,{\sim}\,10$ and the above ranges of parameters are adaptively refined in an iterative binary search down to
$\delta\log{p}\,{\approx}\,0.1$, accelerating the algorithm in finding the global minimum. Data points are fitted using all
combinations of the model parameters\footnote{Although the number of trial fittings for a spectral shape of $F_{\nu}$ or
$I_{\nu\,ij}$ may reach ${\sim\,}10^{3}$ in some cases, computation time is never an issue as all of the fits are completed within
a second. Fitting of an entire set of six images with $10^{6}$ pixels each may take a couple of hours.} in the adaptively refined
parameter space and their initial values are found that converge to the globally smallest $\chi^{2}$ in a fully automated procedure.

Algorithms described in this paper were written as a versatile and robust FORTRAN utility \textsl{fitfluxes} that efficiently
estimates \textsl{modbody} or \textsl{thinbody} parameters from fitting either total fluxes of cataloged sources or intensities
of multiwavelength images. The code is easy to install and use and it is freely available from the author upon request.

\section{Results for isothermal models}
\label{AppendixD}

\begin{figure*}
\centering
\centerline{\resizebox{0.3327\hsize}{!}{\includegraphics{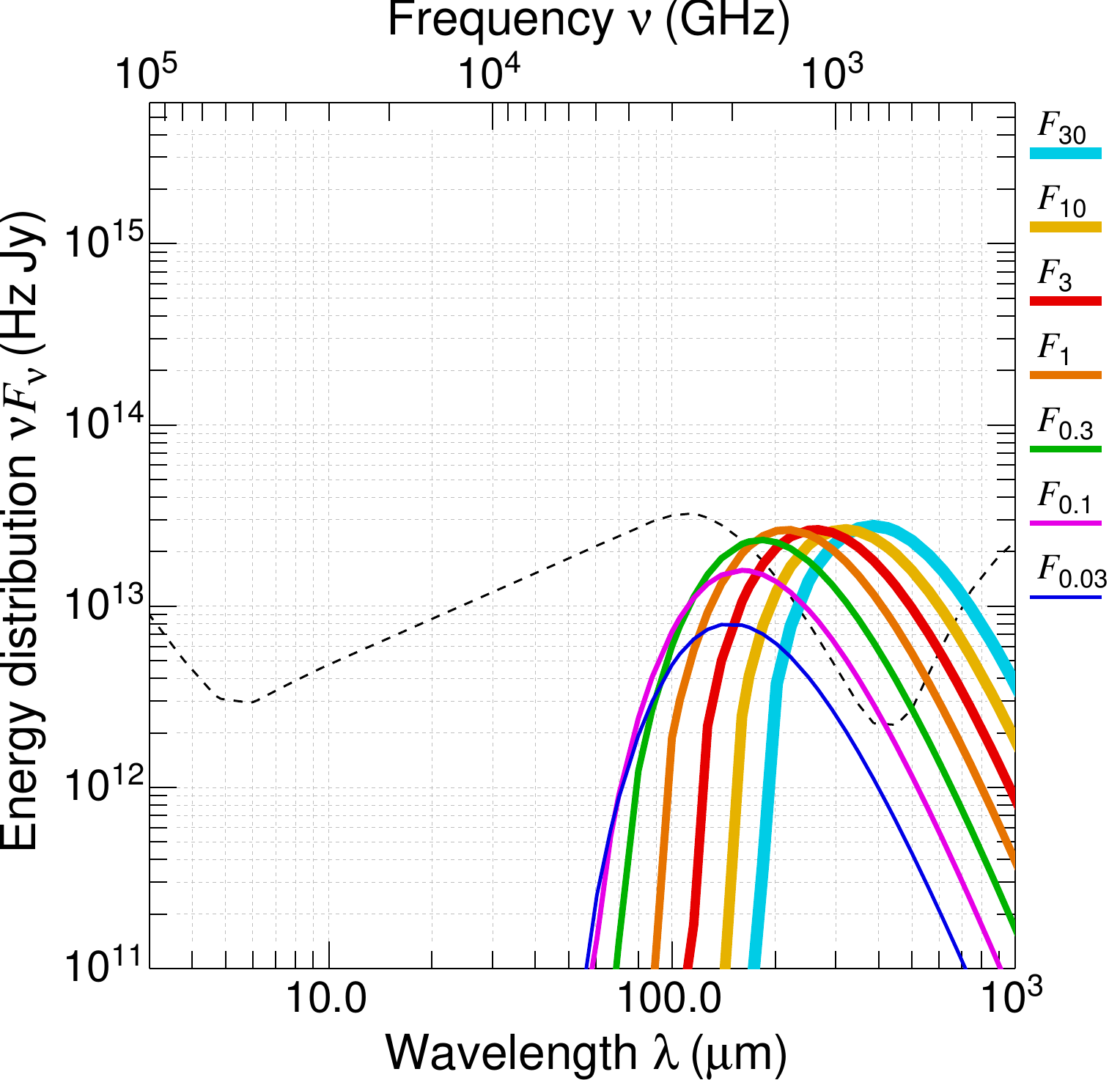}}
            \resizebox{0.3204\hsize}{!}{\includegraphics{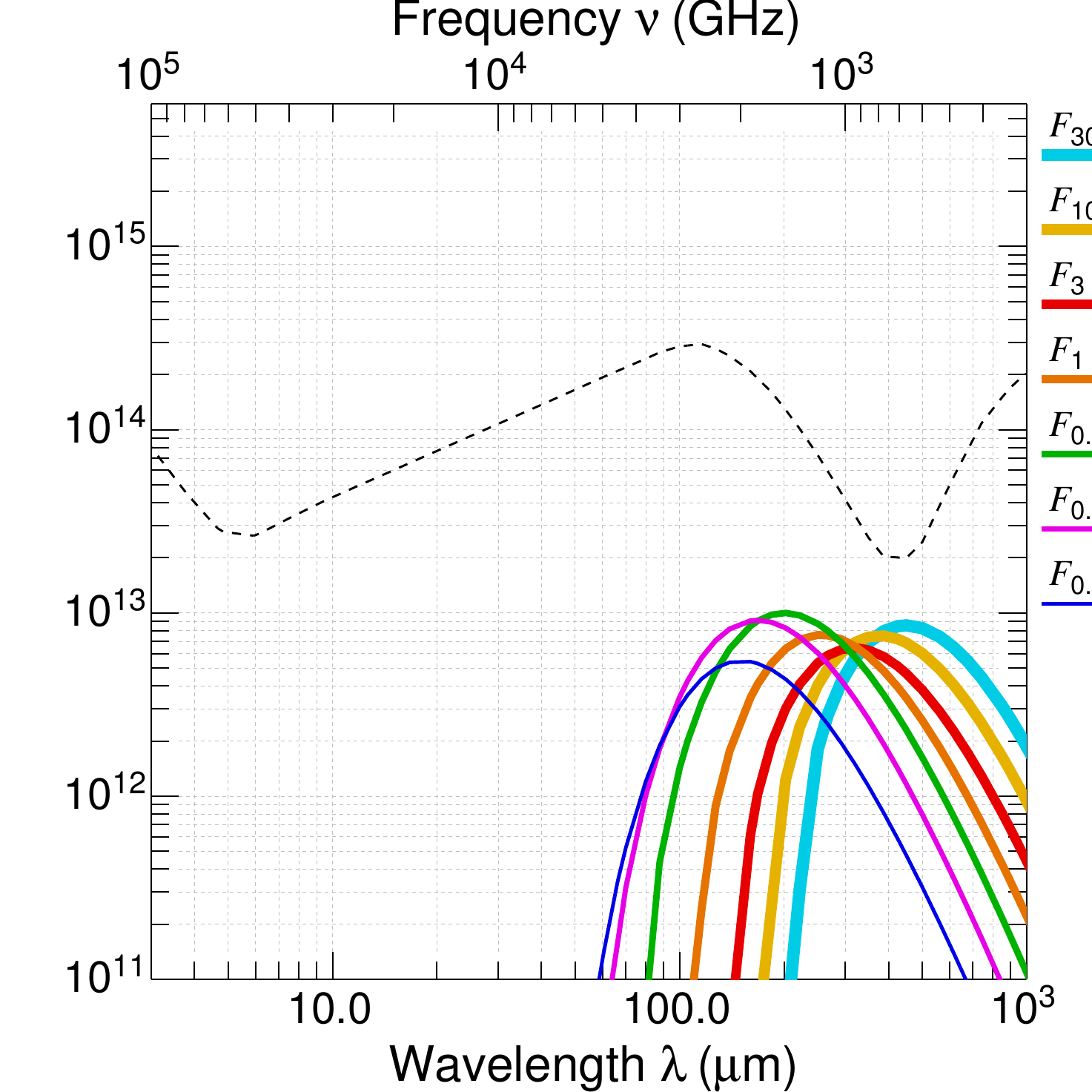}}}
\centerline{\resizebox{0.3327\hsize}{!}{\includegraphics{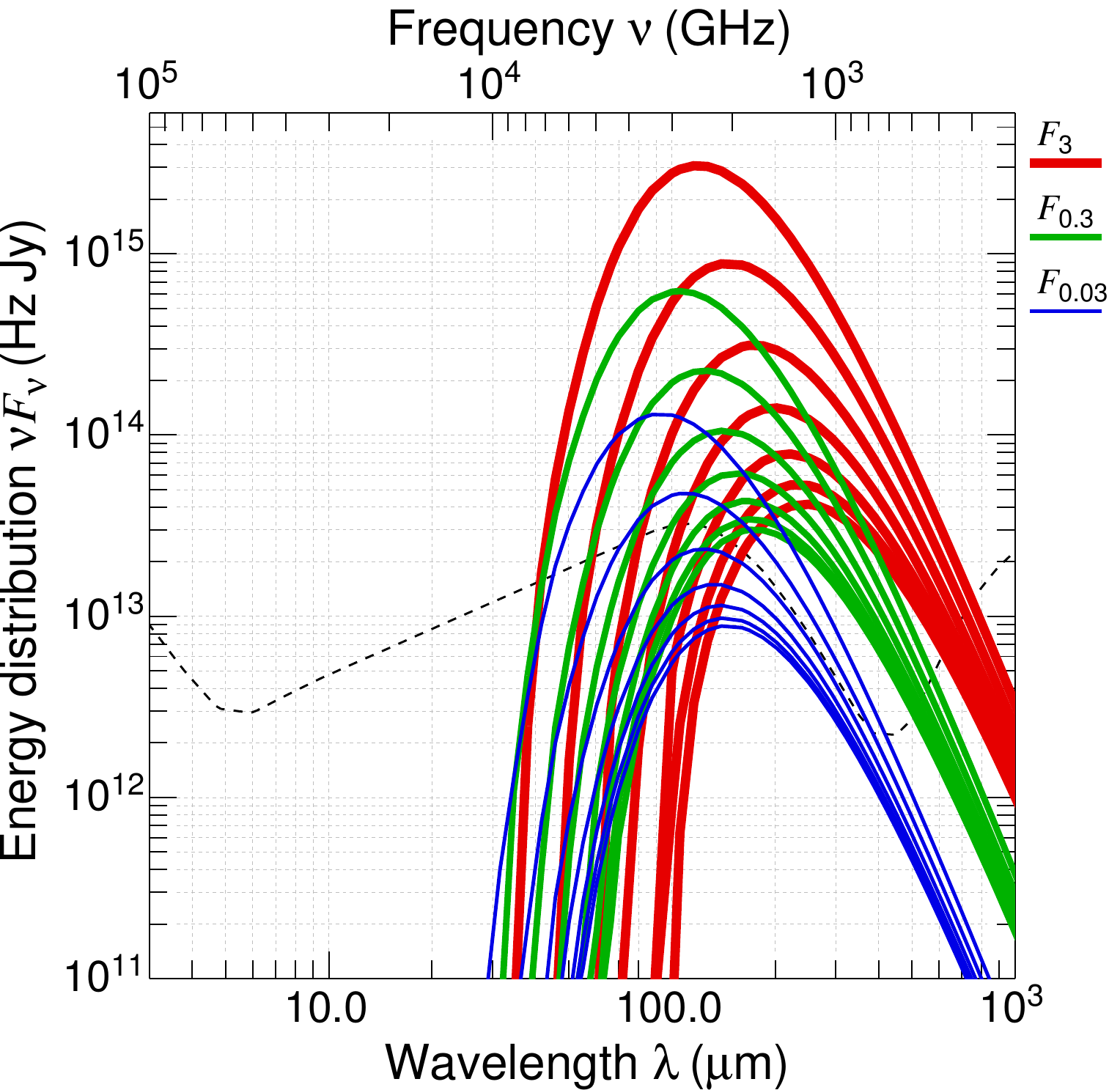}}
            \resizebox{0.3204\hsize}{!}{\includegraphics{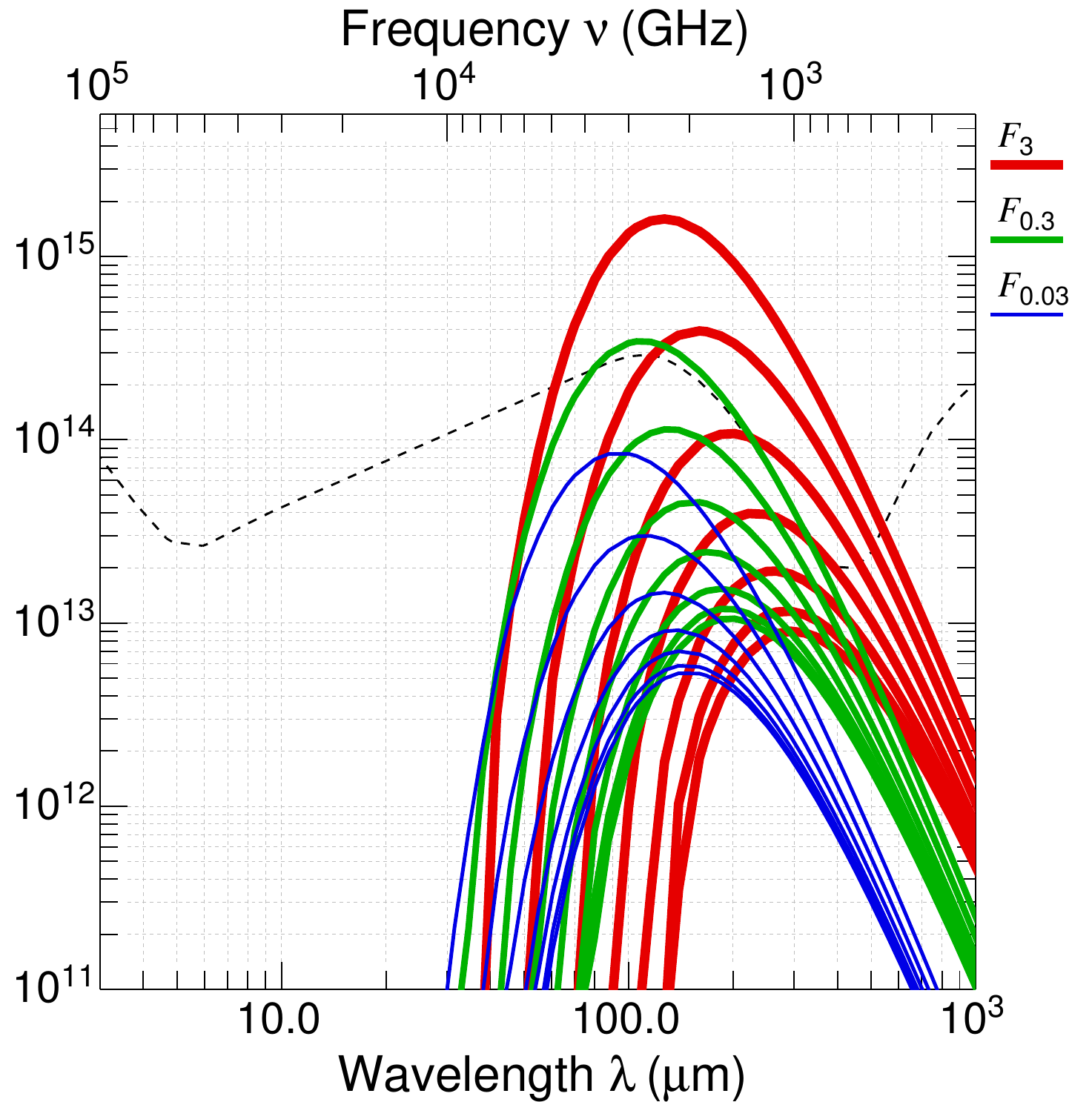}}}
\caption{
Spectral energy distributions of the \emph{isothermal} models of starless cores (\emph{upper}) and protostellar envelopes
(\emph{lower}). Shown are the background-subtracted fluxes of the \emph{isolated} models (\emph{left}) and of their \emph{embedded}
variants (\emph{right}). For the cores of increasing masses, $T_{M}\,{=}\,$\{16.3, 15.0, 13.2, 11.2, 9.25, 7.66, 6.42 K\}. For the
envelopes of increasing luminosities and masses, $T_{M}\,{=}\,$\{16.6, 16.8, 17.3, 18.1, 19.5, 21.9, 25.9 K\},
$T_{M}\,{=}\,$\{13.8, 14.1, 14.7, 15.6 17.1, 19.4, 23.0 K\}, $T_{M}\,{=}\,$\{10.0, 10.4, 11.1, 12.3, 14.1, 16.8, 20.7 K\}. See
Fig.~\ref{sed.bes.pro} for more details.
} 
\label{sed.tmav.bes.pro}
\end{figure*}

\begin{figure*}
\centering
\centerline{\resizebox{0.3327\hsize}{!}{\includegraphics{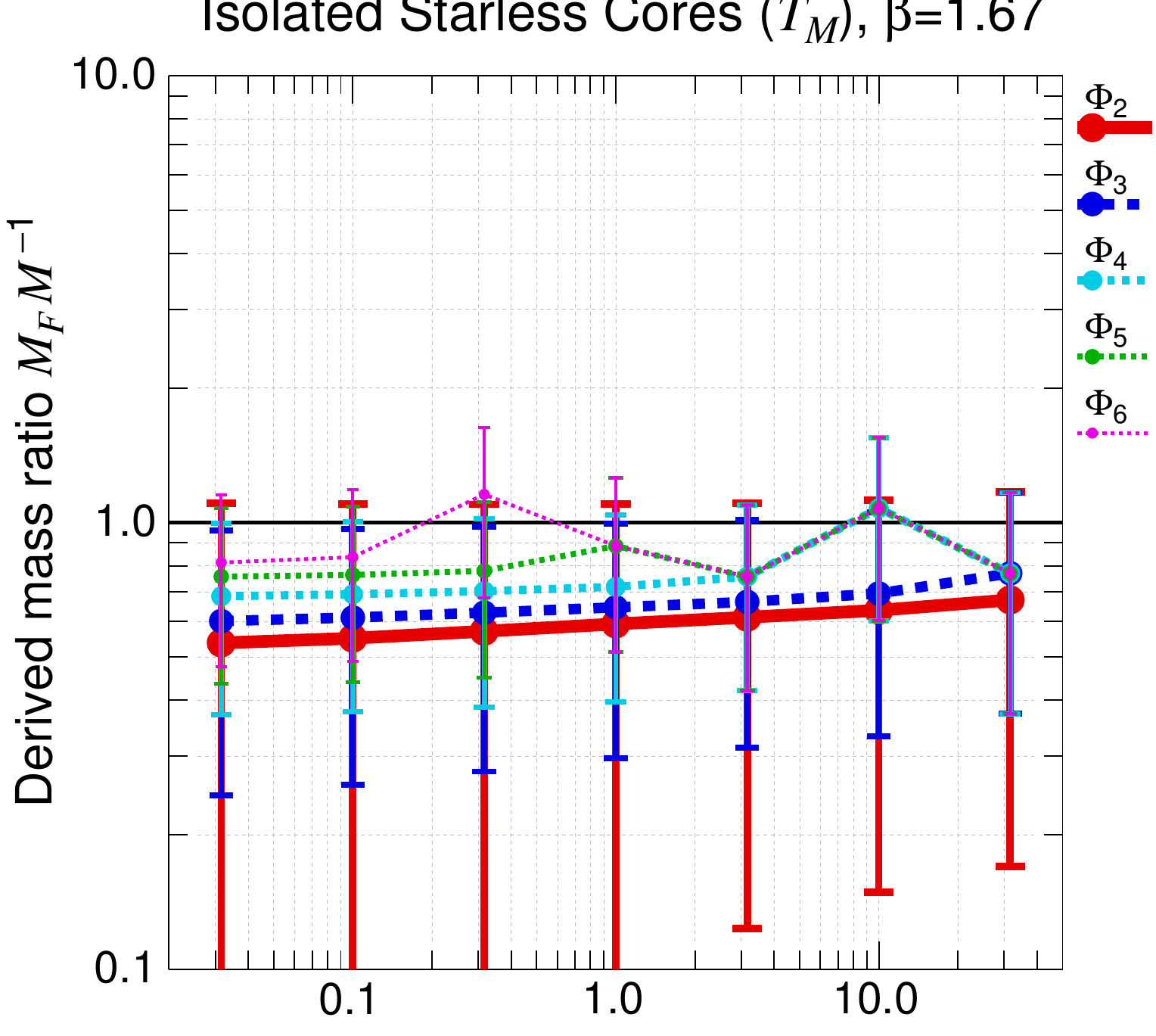}}
            \resizebox{0.3204\hsize}{!}{\includegraphics{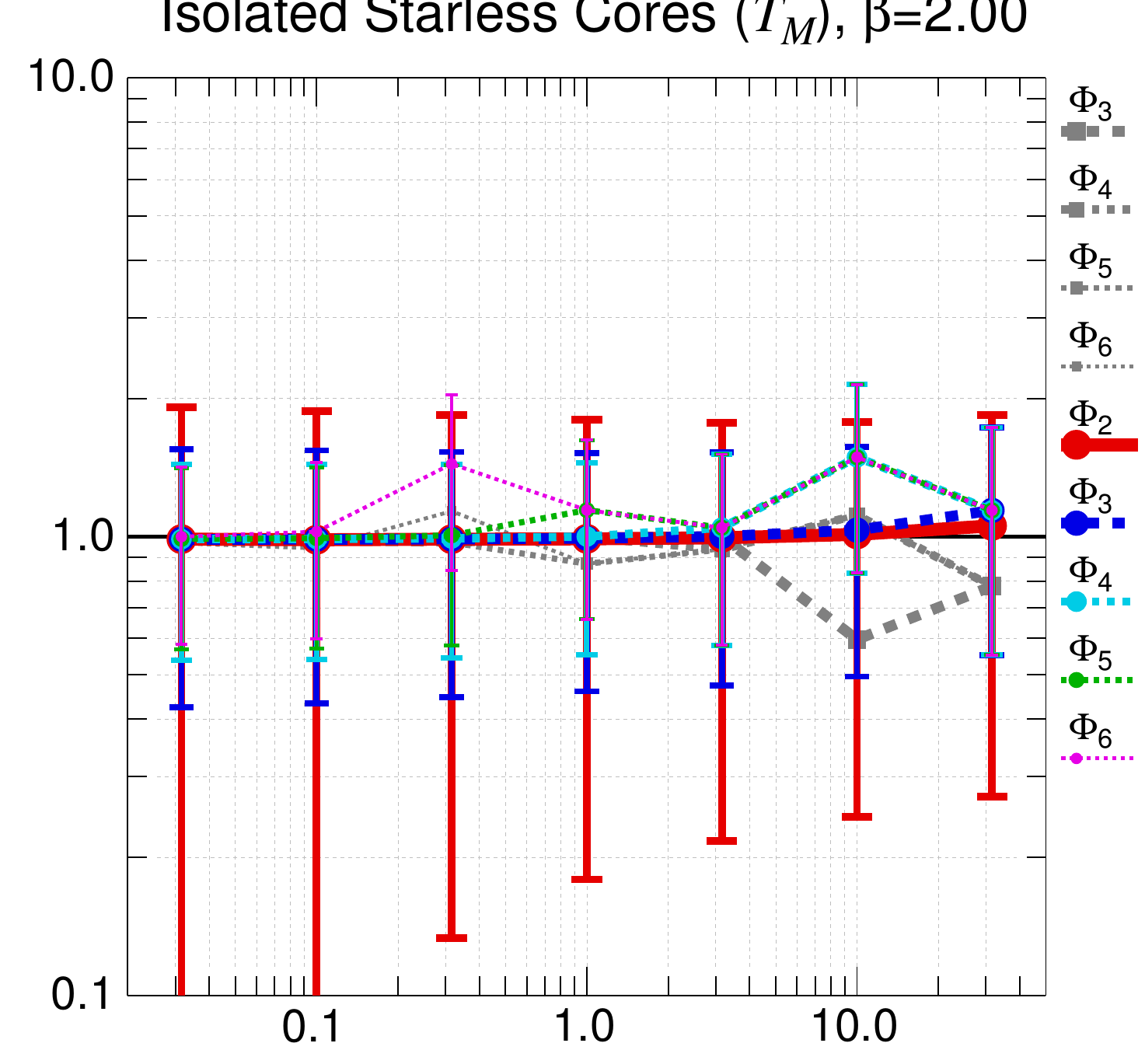}}
            \resizebox{0.3204\hsize}{!}{\includegraphics{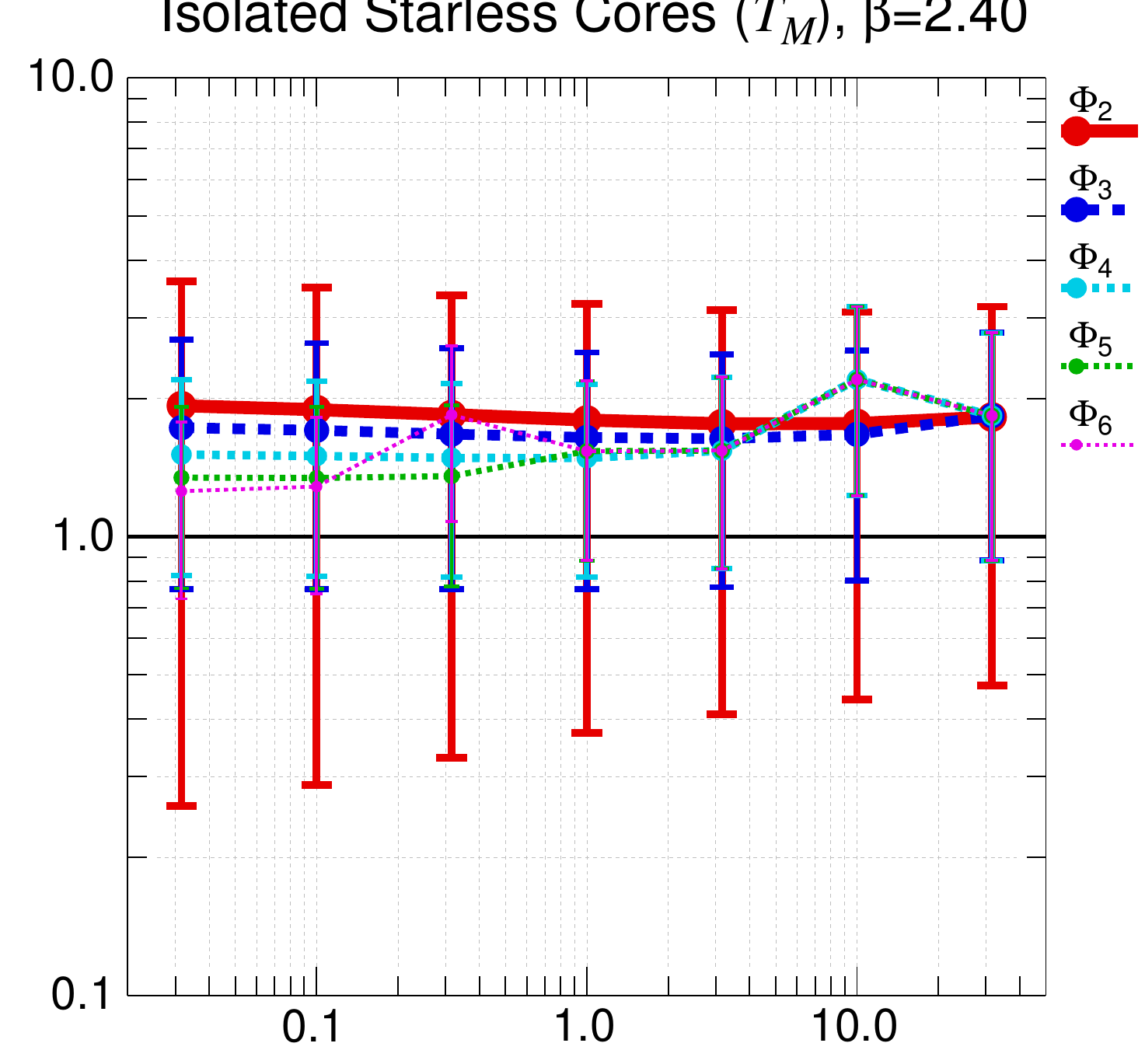}}}
\centerline{\resizebox{0.3327\hsize}{!}{\includegraphics{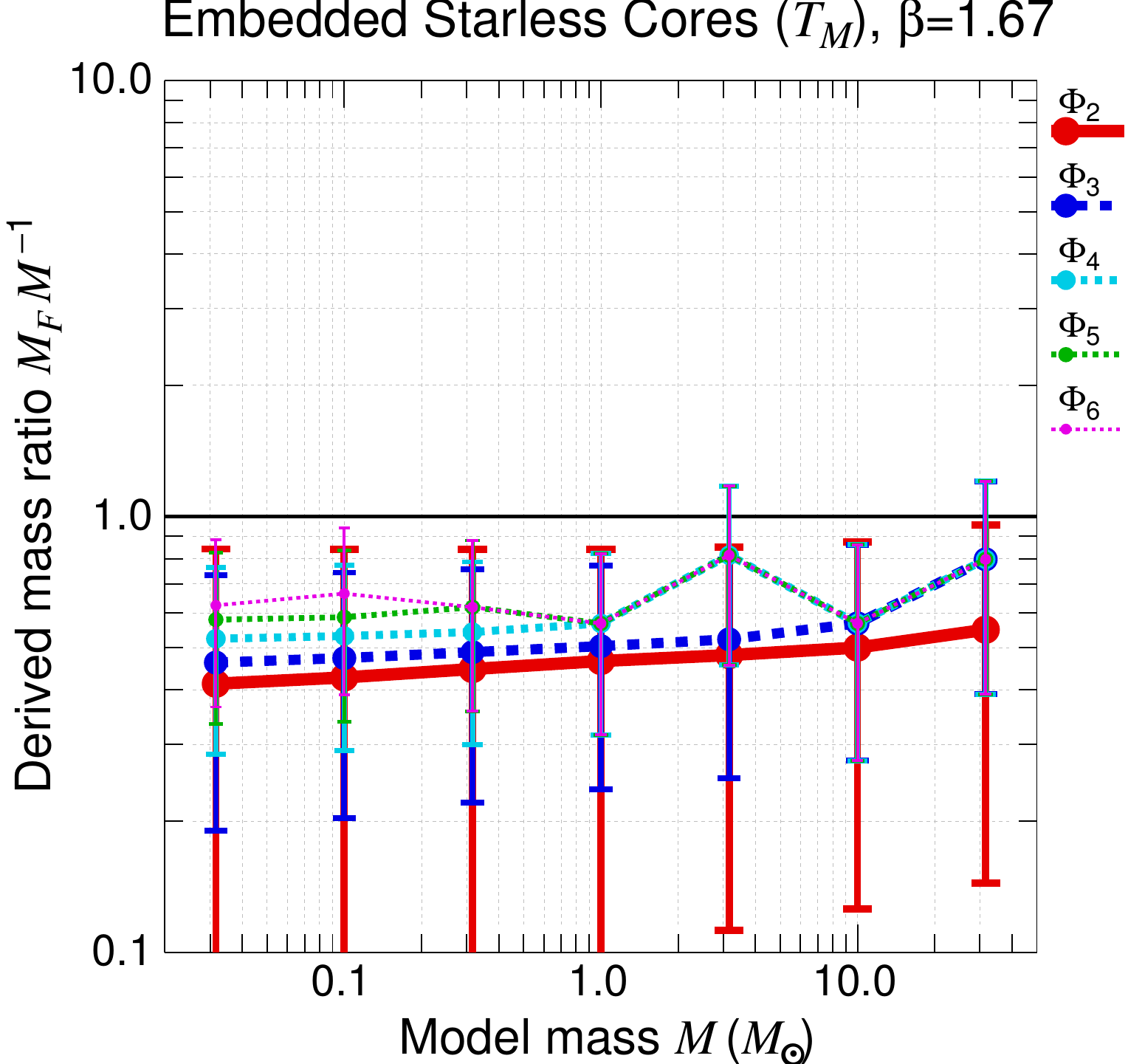}}
            \resizebox{0.3204\hsize}{!}{\includegraphics{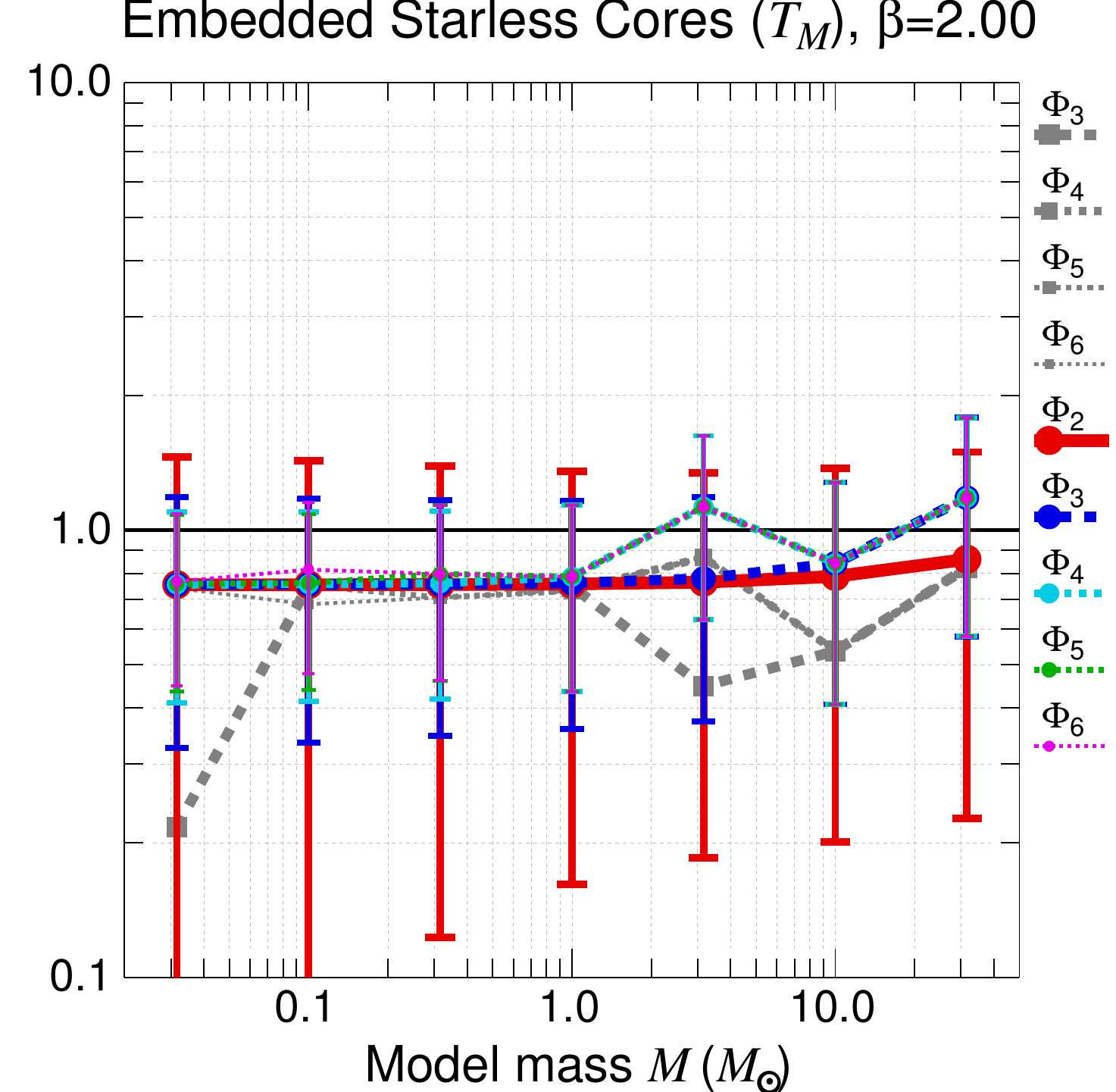}}
            \resizebox{0.3204\hsize}{!}{\includegraphics{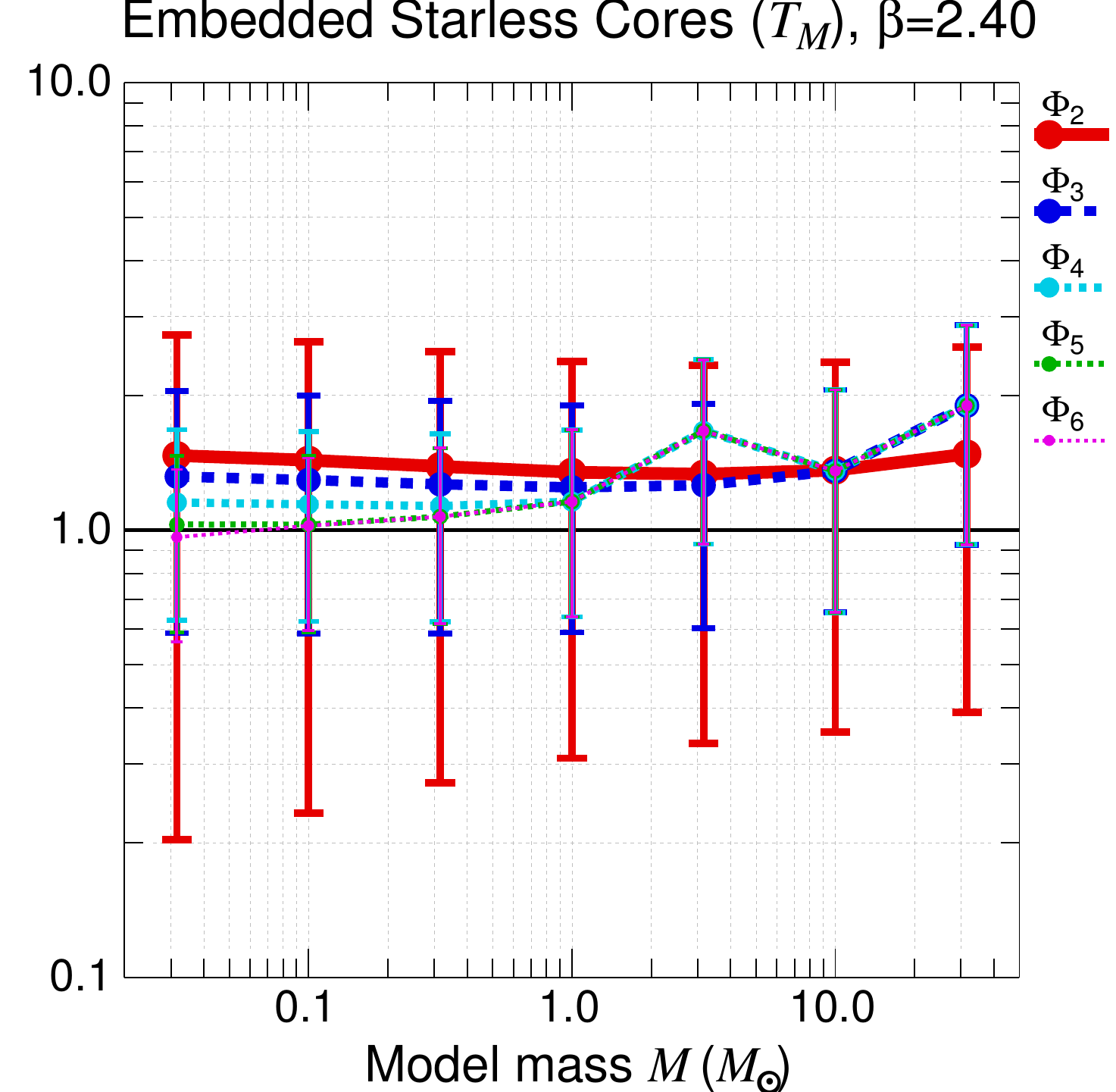}}}
\centerline{\resizebox{0.3327\hsize}{!}{\includegraphics{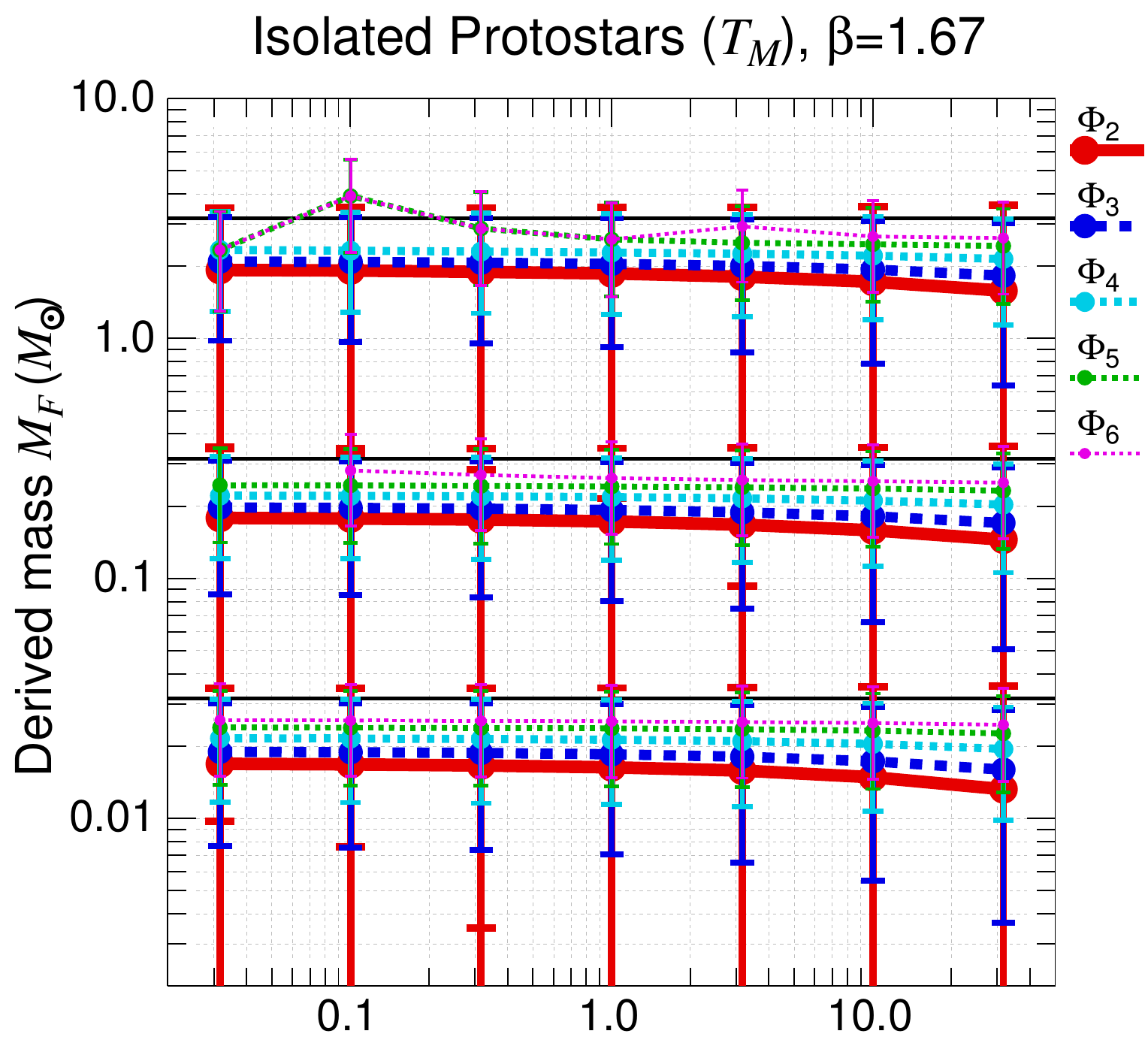}}
            \resizebox{0.3204\hsize}{!}{\includegraphics{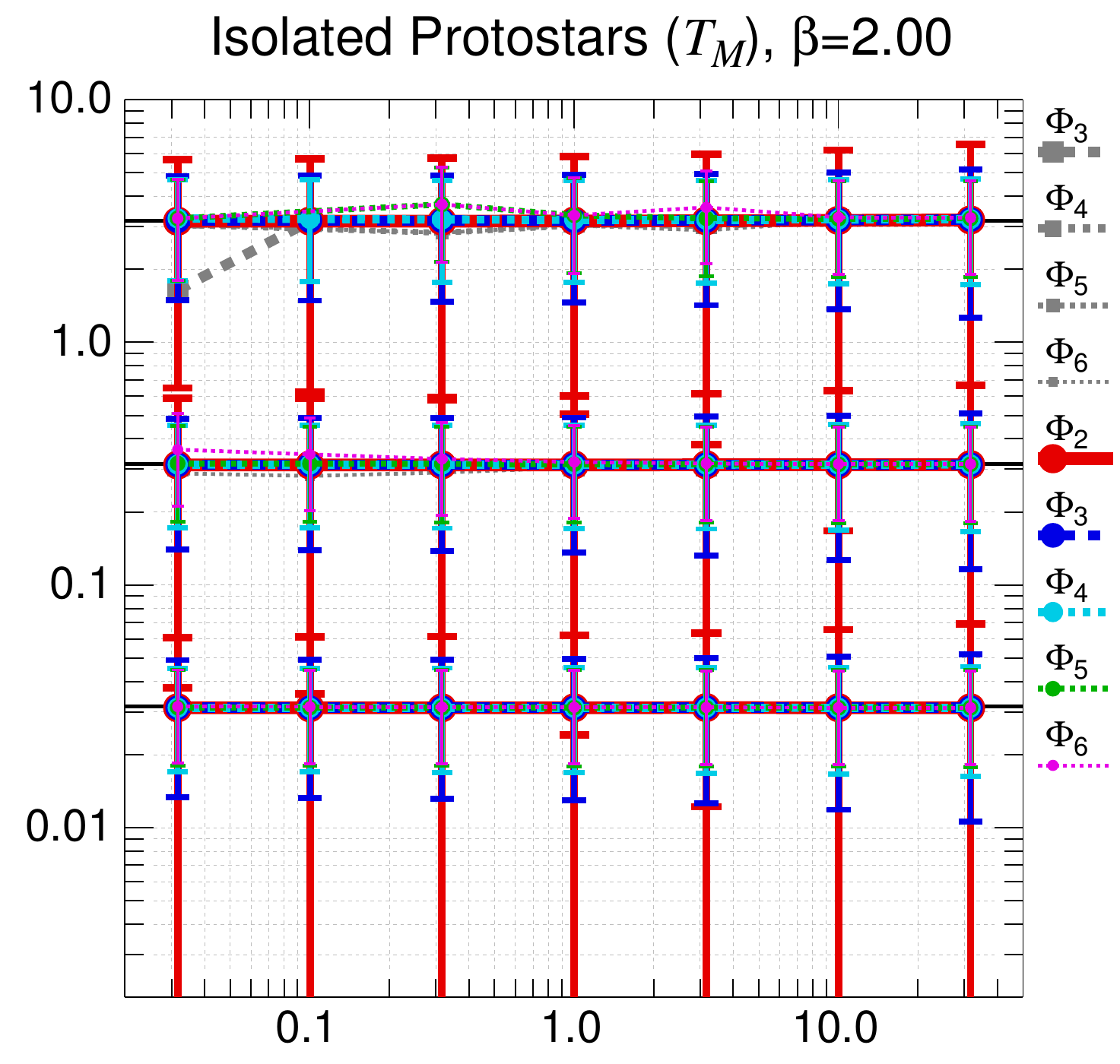}}
            \resizebox{0.3204\hsize}{!}{\includegraphics{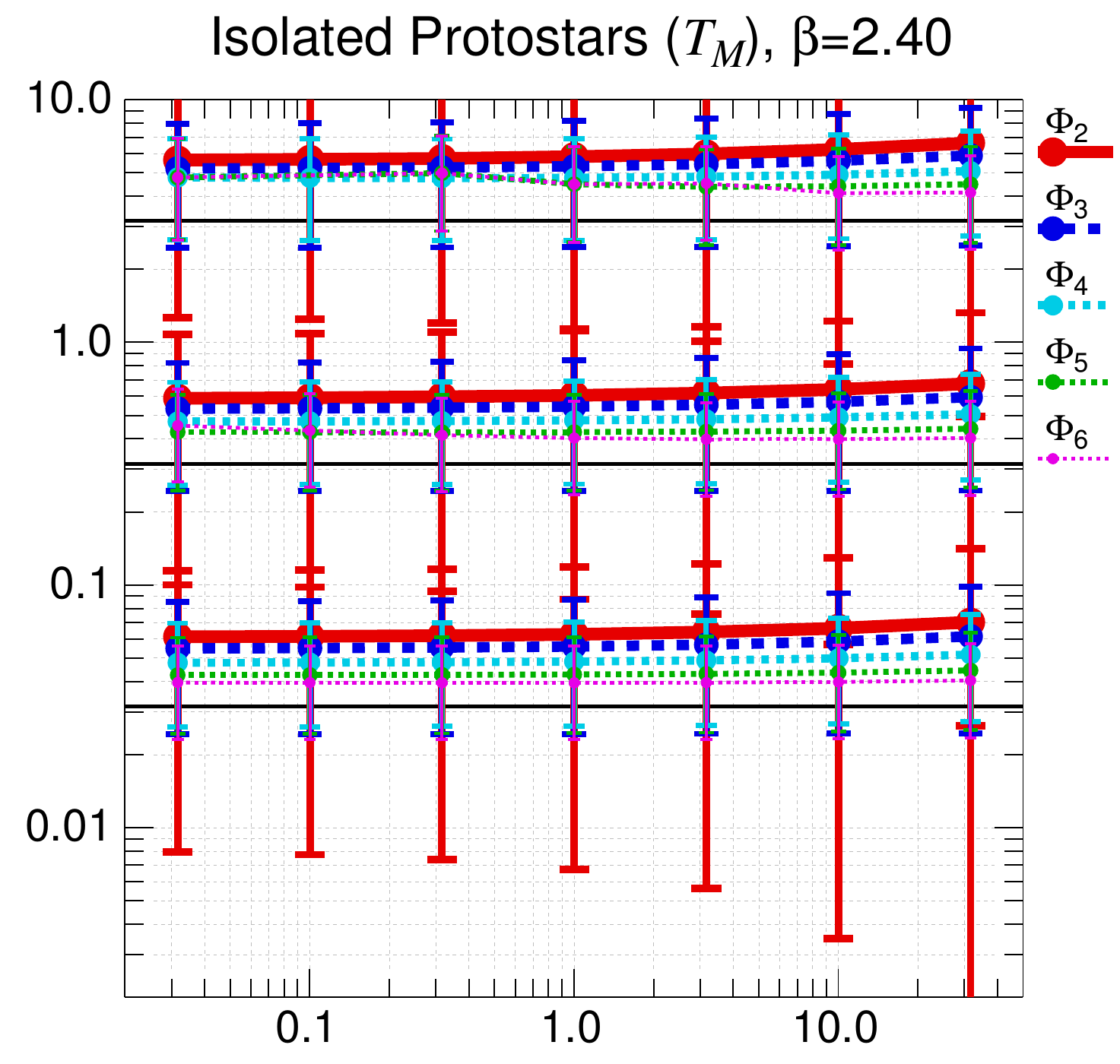}}}
\centerline{\resizebox{0.3327\hsize}{!}{\includegraphics{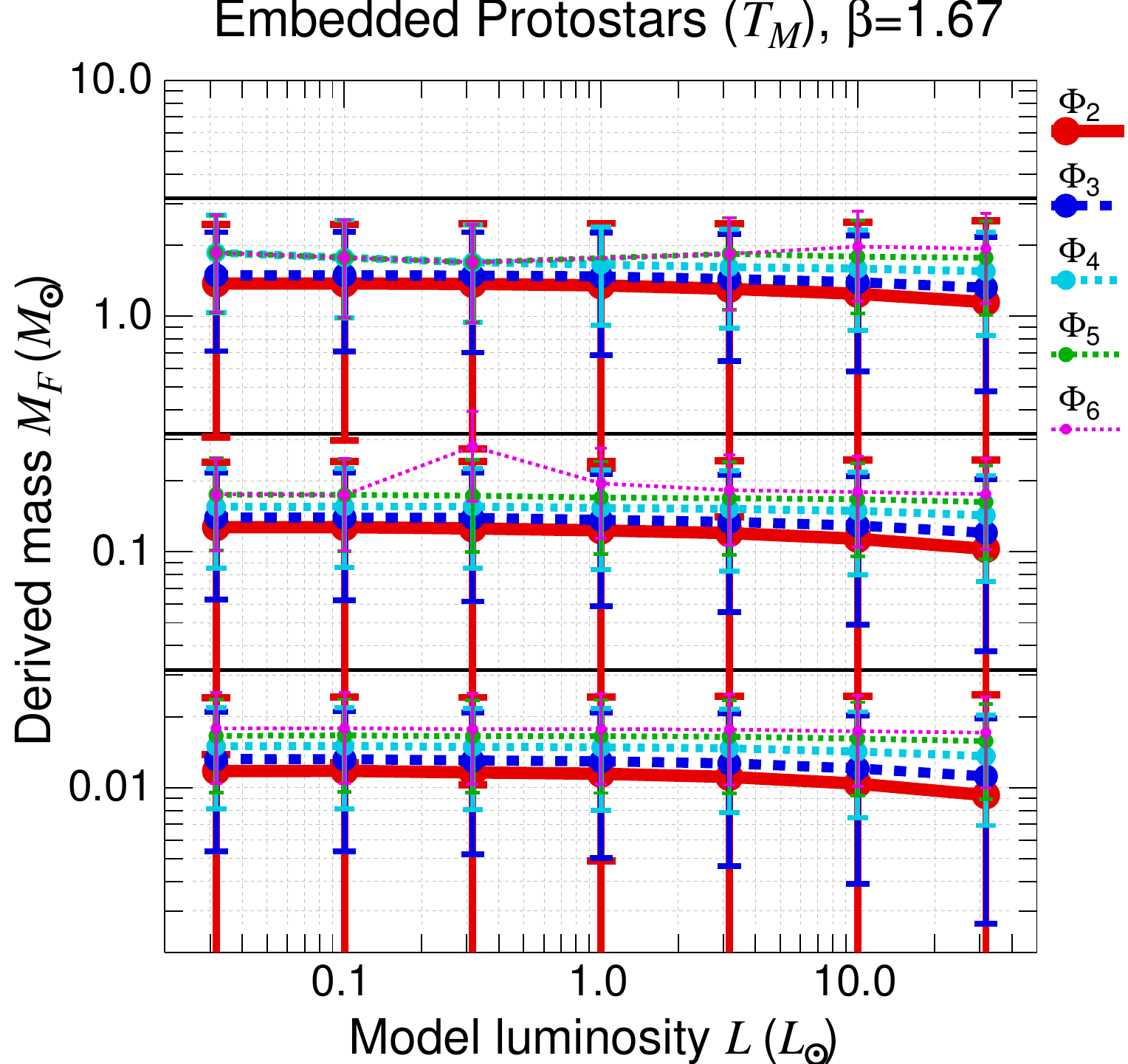}}
            \resizebox{0.3204\hsize}{!}{\includegraphics{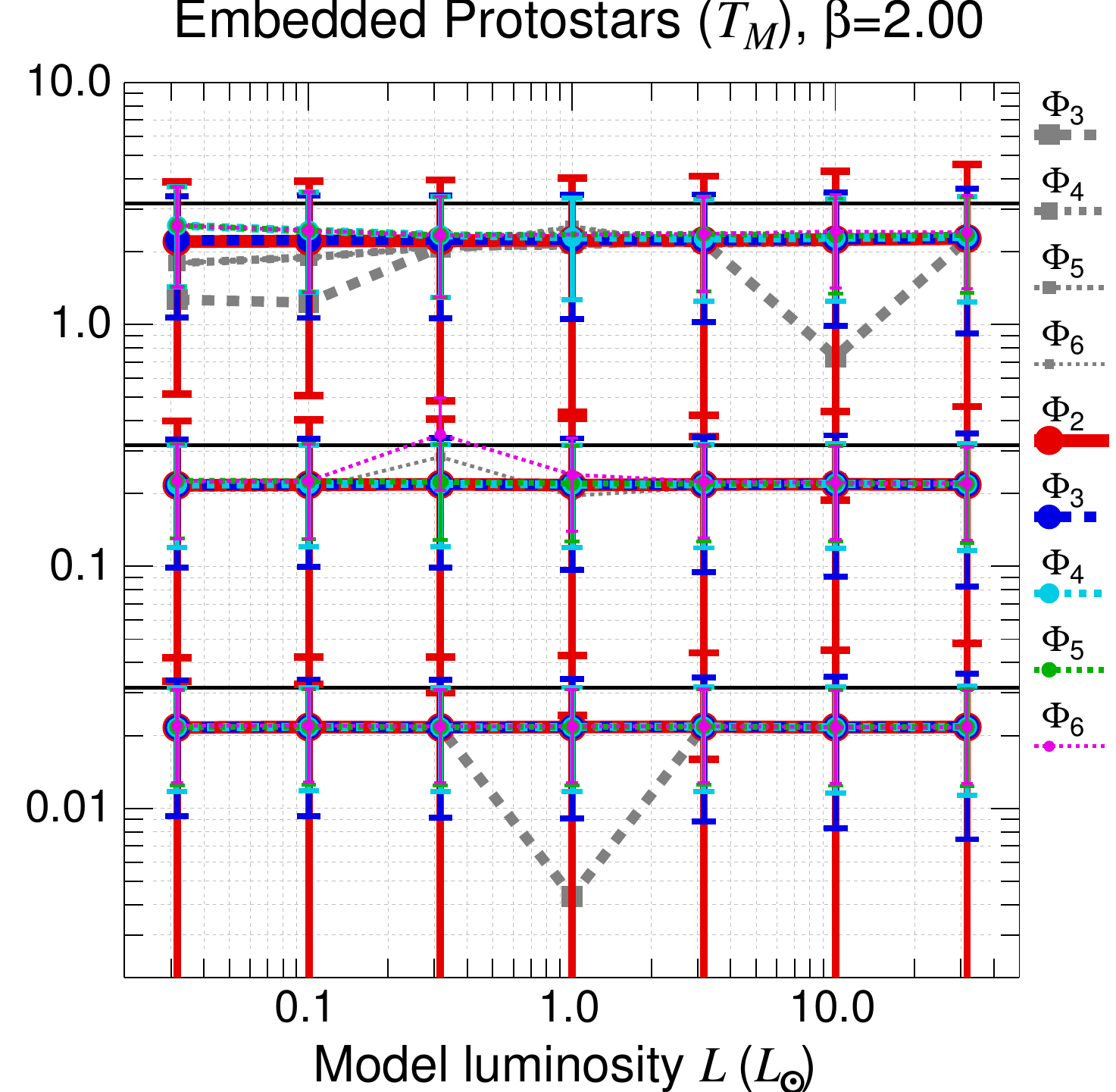}}
            \resizebox{0.3204\hsize}{!}{\includegraphics{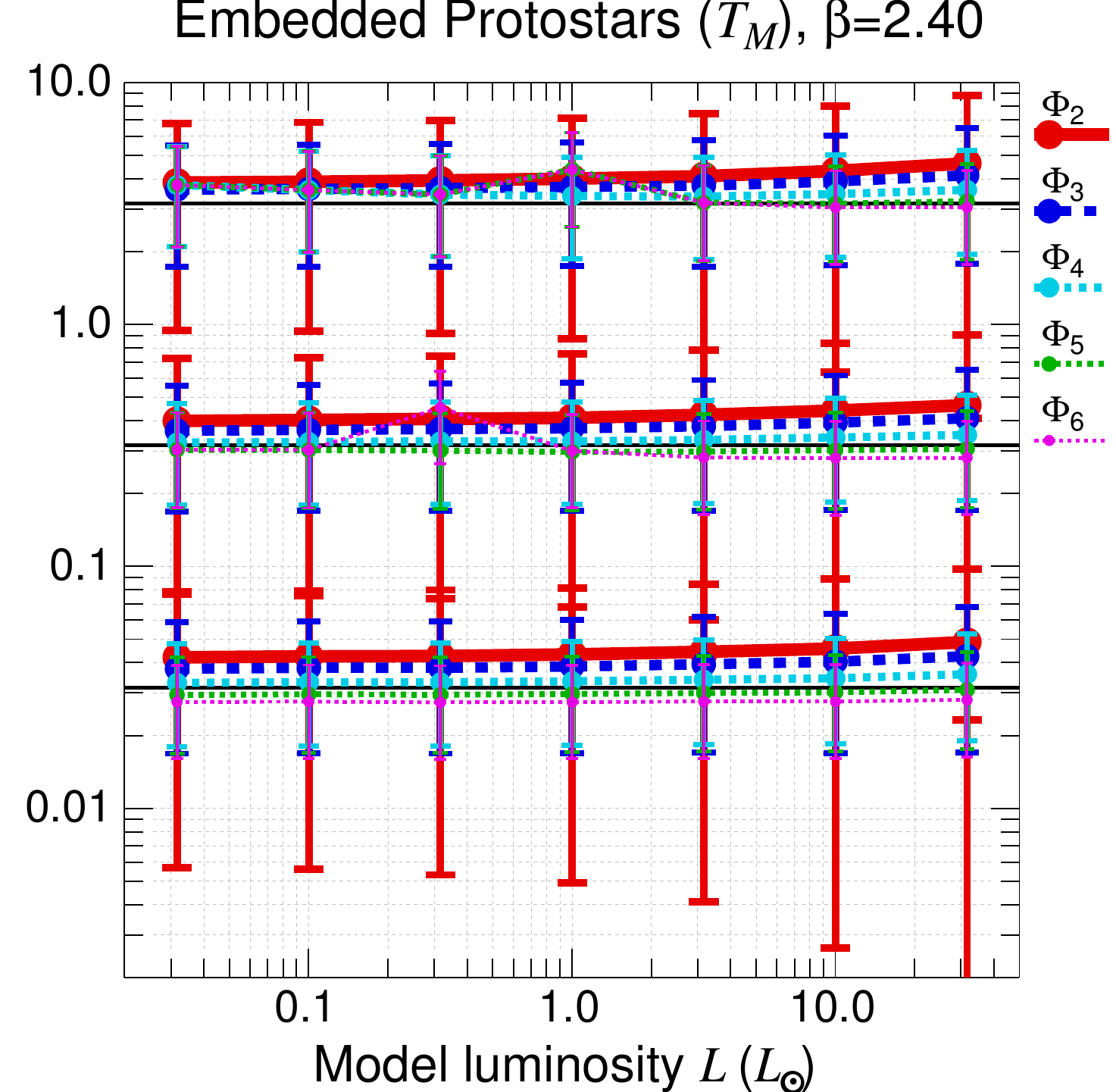}}}
\caption{
Masses $M_{F}$ derived from fitting $F_{\nu}$ for the \emph{isothermal} models of both \emph{isolated} and \emph{embedded} starless
cores and protostellar envelopes. Compared to the results for isolated models, all derived masses of embedded models are
underestimated by approximately a factor of $1.3$, owing to background over-subtraction (Sect.~\ref{bg.subtraction}). See 
Fig.~\ref{sed.tmav.bes.pro} for SEDs and Fig.~\ref{temp.mass.bes} for more details.
} 
\label{mass.bes.pro.tmav}
\end{figure*}

The importance of estimating $T_{F}$ that approaches the mass-ave\-raged temperature $T_{M}$ from Eq.~(\ref{mass.averaged}) for
deriving accurate masses is shown by isothermal models, those described in Sect.~\ref{rtmodels} and used throughout this paper
where self-consistent (radiative-equilibrium) profiles $T_{\rm d}(r)$ were replaced with $T_{M}$. The isothermal models were then
observed and imaged in a ray-tracing run of the radiative transfer code. The resulting SEDs (Fig.~\ref{sed.tmav.bes.pro}) are
essentially the modified blackbody shapes $\kappa_{\nu}\,B_{\nu}(T_{M})\,{\nu}$ and the same is true for the spectral shapes of
image pixels. The temperature excesses above $T_{M}$ (Sect.~\ref{nonuniform.temps}) that greatly distorted the model SEDs
(Fig.~\ref{sed.bes.pro}) towards shorter wavelengths do not exist in the isothermal models. Consequently, the isothermal shapes of
$F_{\nu}$ and $I_{\nu\,ij}$ bring much more expected and accurate results.

\begin{figure*}
\centering
\centerline{\resizebox{0.3327\hsize}{!}{\includegraphics{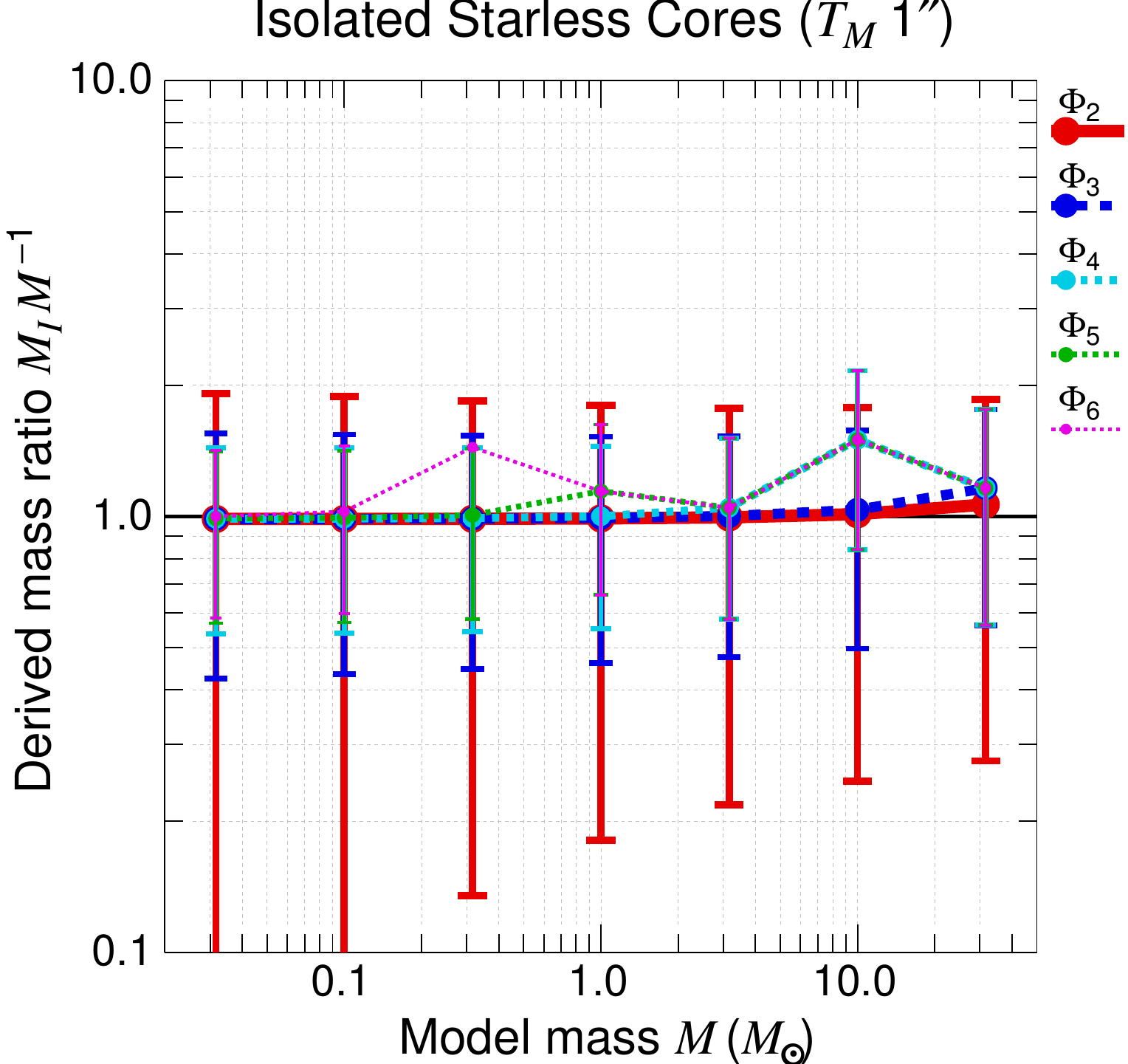}}
            \resizebox{0.3204\hsize}{!}{\includegraphics{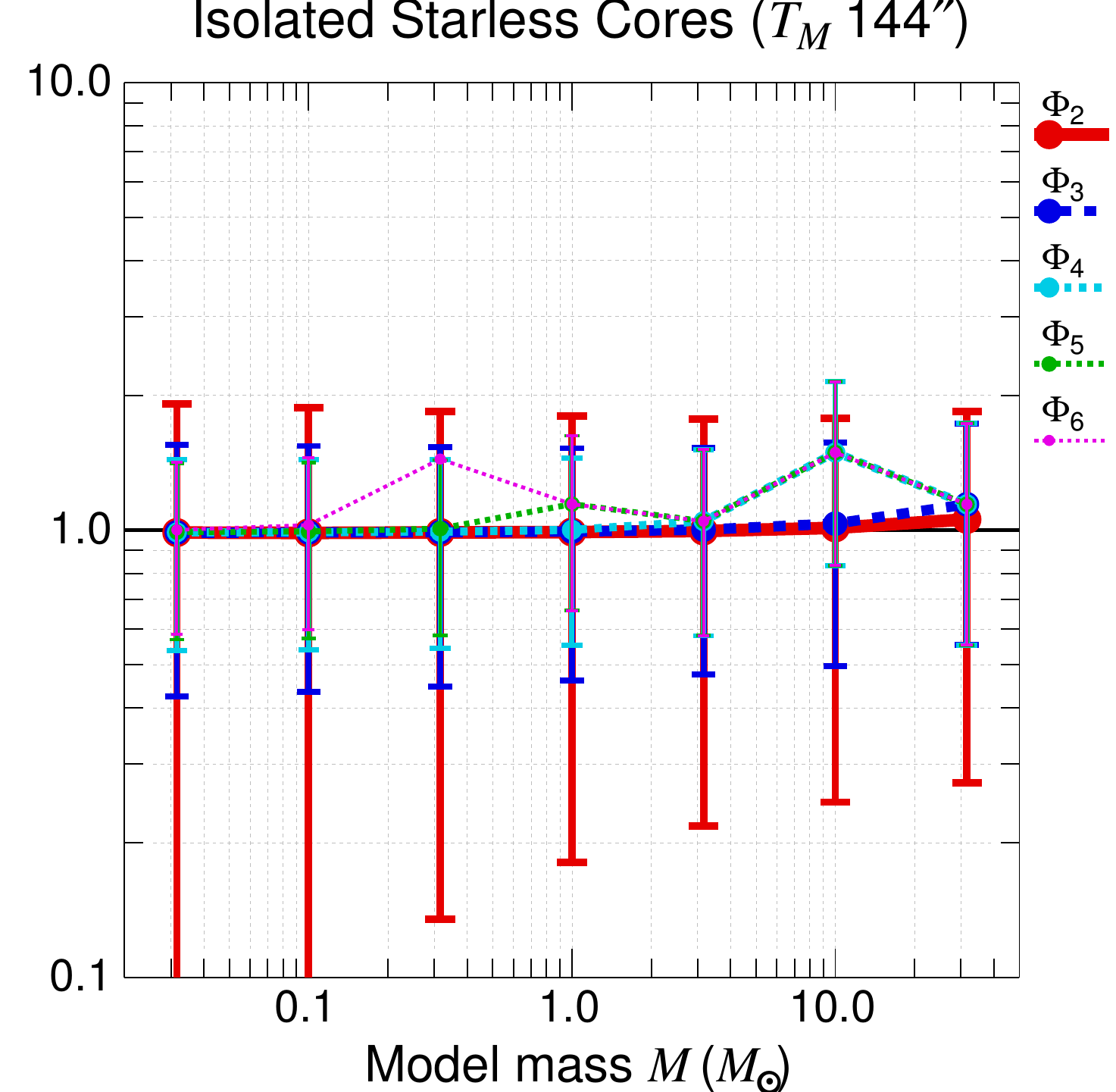}}
            \resizebox{0.3204\hsize}{!}{\includegraphics{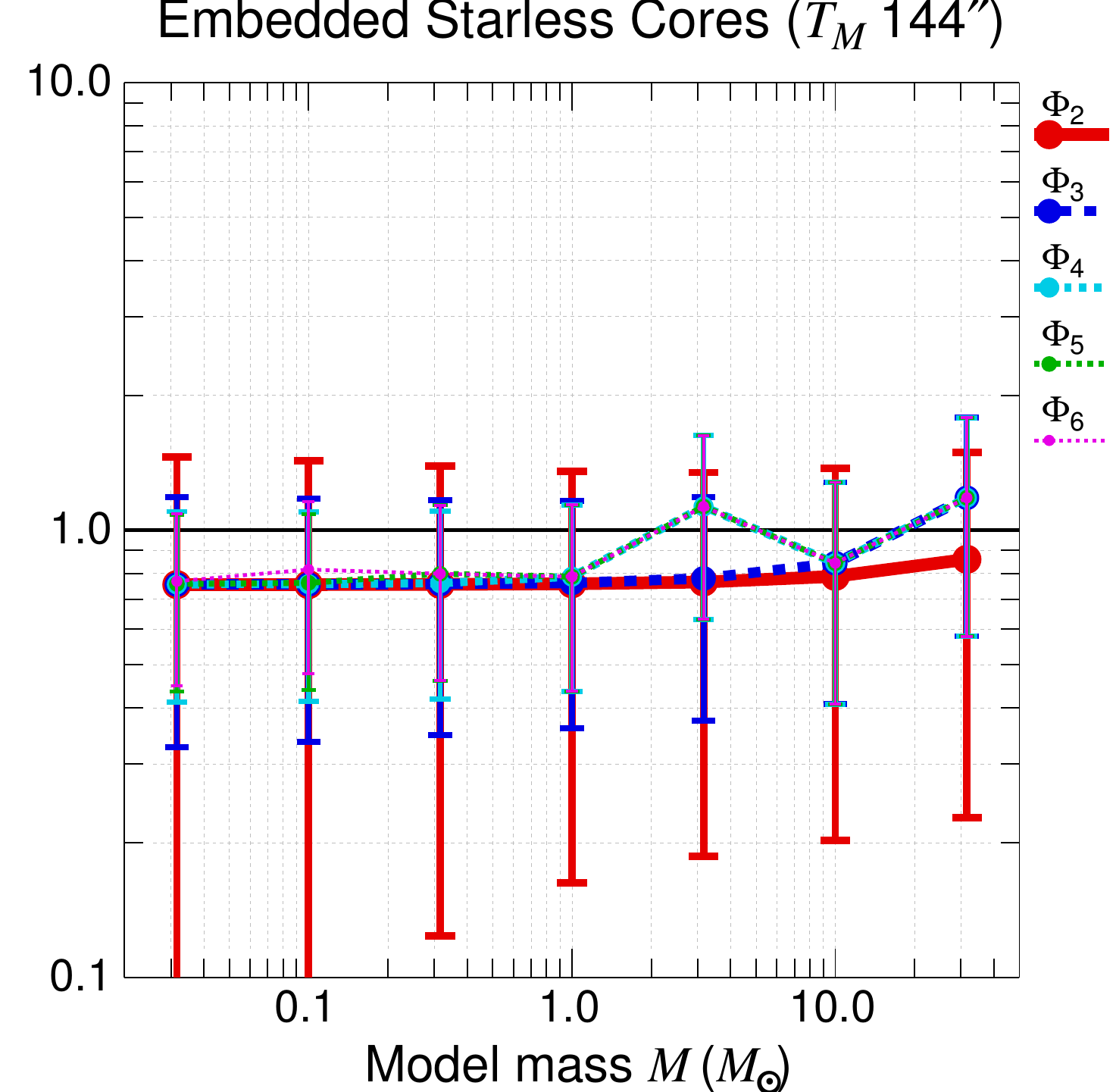}}}
\centerline{\resizebox{0.3327\hsize}{!}{\includegraphics{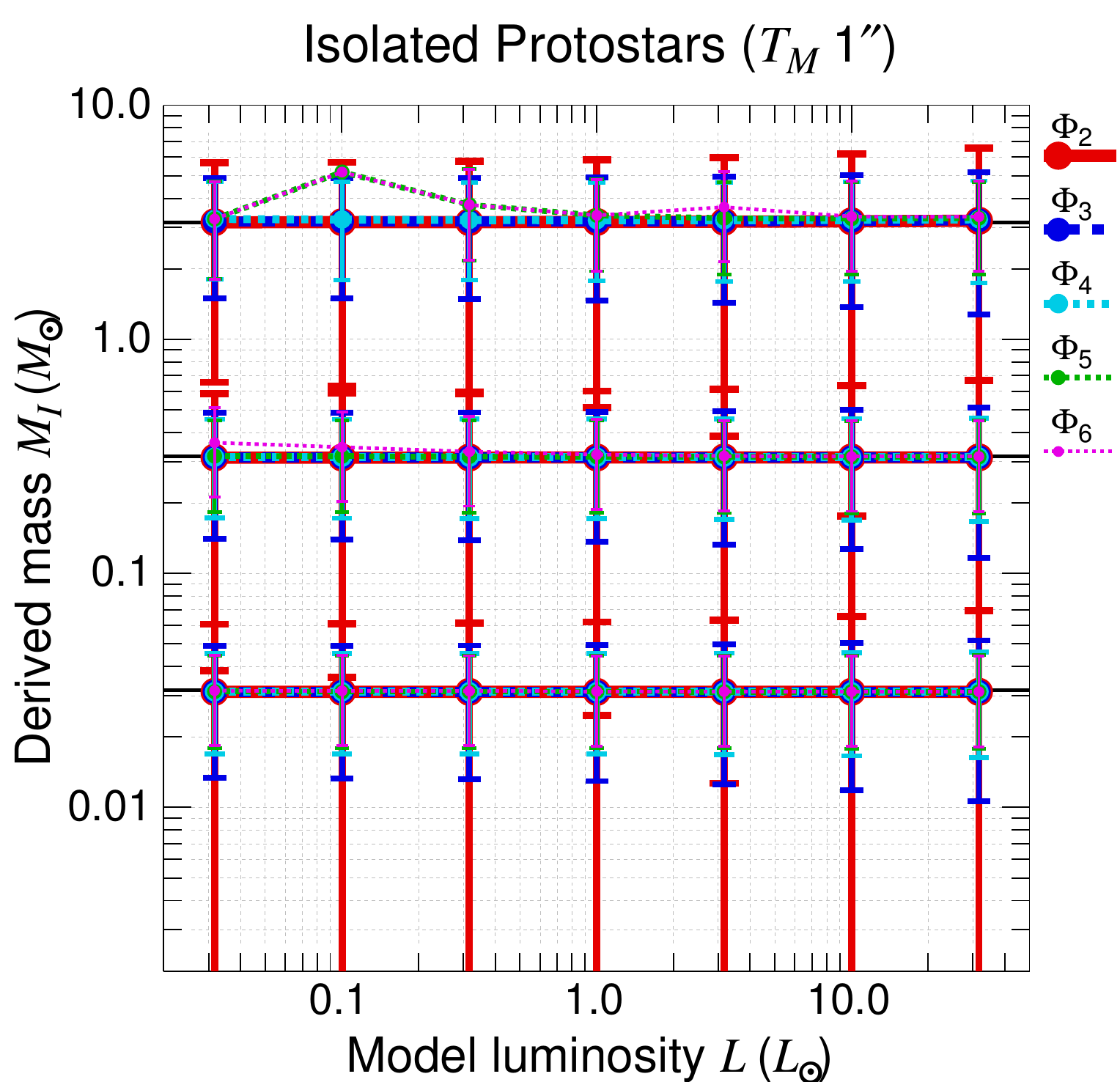}}
            \resizebox{0.3204\hsize}{!}{\includegraphics{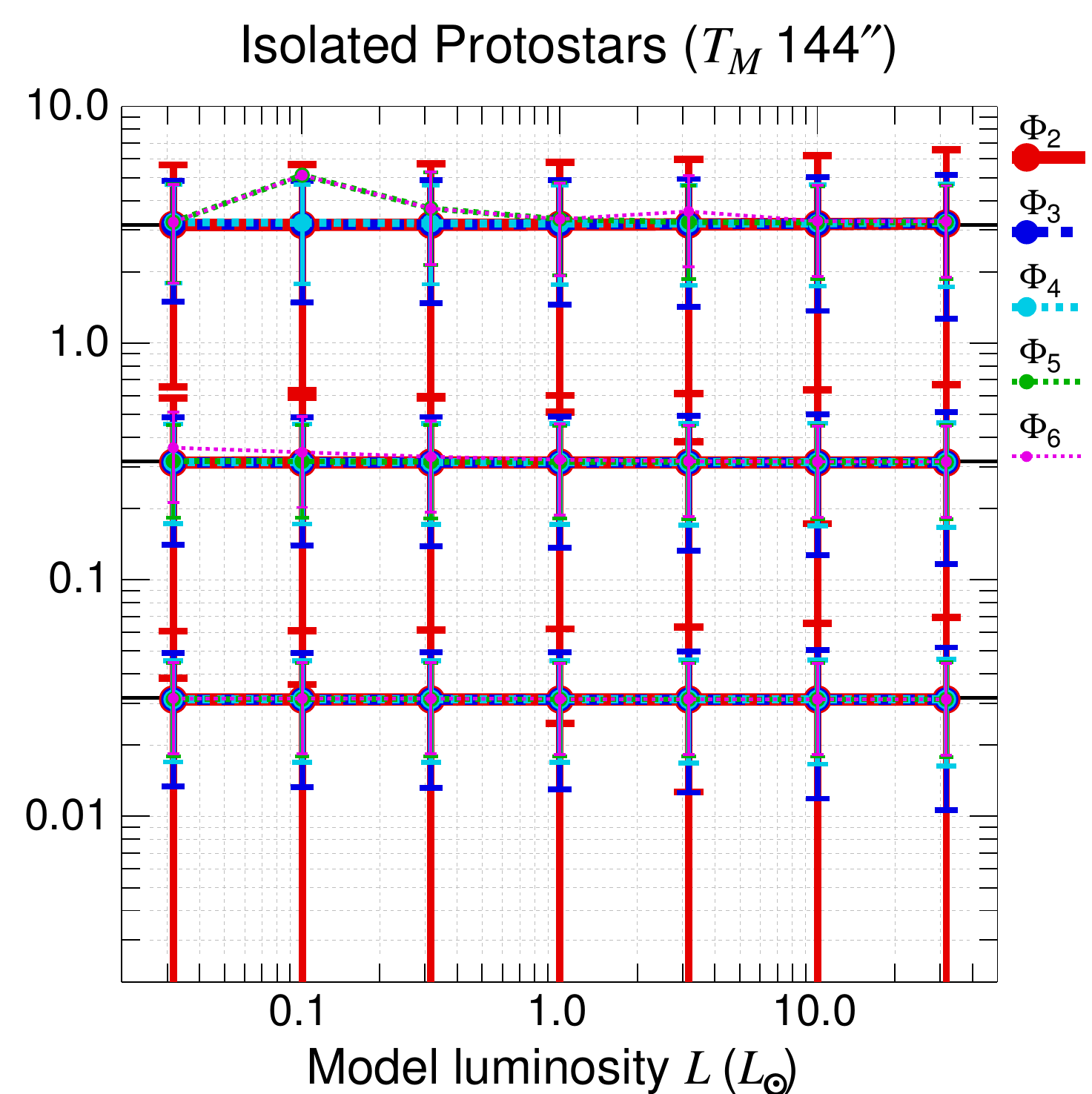}}
            \resizebox{0.3204\hsize}{!}{\includegraphics{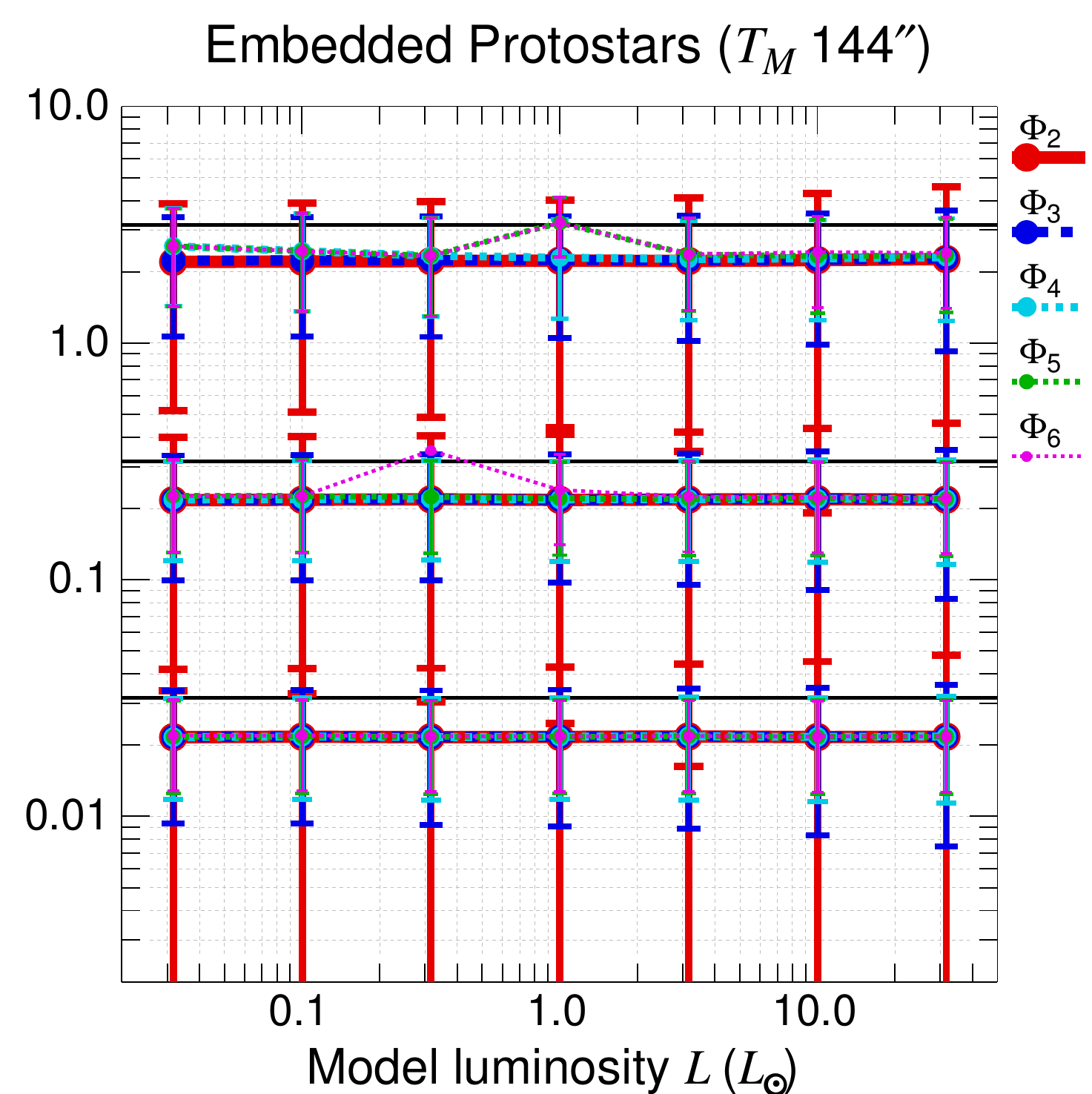}}}
\caption{
Masses $M_{\mathcal{I}}$ derived from fitting images $\mathcal{I}_{\nu}$ of the \emph{isothermal} models of both \emph{isolated}
and \emph{embedded} starless cores and protostellar envelopes for the correct value of $\beta\,{=}\,2$. Two columns of panels
(\emph{left}, \emph{middle}) display the masses derived for resolved and unresolved images of the isolated models, whereas the
third column of panels (\emph{right}) present results for unresolved embedded models. Angular resolutions do not make any 
difference for the isothermal models. See Fig.~\ref{mass.bes.pro.tmav} and Fig.~\ref{coldens.bes} for more details.
} 
\label{coldens.bes.pro.tmav}
\end{figure*}

With the correct value $\beta\,{=}\,2$, derived $M_{F}$ of the isolated starless cores and protostellar envelopes derived with
\textsl{thinbody} agree with the true masses $M$ (Fig.~\ref{mass.bes.pro.tmav}). The same results are obtained for
\textsl{modbody}, with the exception of the minimal subset $\Phi_{3}$ for some models. An inspection of the problematic fits show
that $T_{F}$ is somewhat overestimated because of an additional degree of freedom in \textsl{modbody} and a formal search for a
globally best fit in its parameter space. The fits in question have the globally lowest $\chi^{2}$ value, but they correspond to
small values of $\Omega$ and, therefore, to high optical depth $\tau_{\nu}{\,\sim\,}1$. However, there are also other \emph{very}
good fits with somewhat larger $\chi^{2}$ that do produce accurate $T_{F}\,{=}\,T_{M}$ with $\tau_{\nu}{\,\ll\,}1$, consistent with
the models. The problem seems to be just a simple consequence of the finite accuracy of the numerical models and their fluxes.

With the same $\beta\,{=}\,2$, the derived masses of the embedded models (Fig.~\ref{mass.bes.pro.tmav}) are almost uniformly
underestimated by a factor of approximately $1.3$. The reason for the difference with respect to the isolated models is the
conventional approach to background subtraction. Emission of a transparent cloud embedding a physical object tends to be rim
brightened (Appendix~\ref{AppendixB}, Sect.~\ref{bg.subtraction}). An average intensity in an annulus may overestimate background,
from a few percent to a factor of several, hence may underestimate the background-subtracted intensities $I_{\nu}$, fluxes
$F_{\nu}$, and derived masses (see Sects.~\ref{derived.properties} and \ref{coldens.properties}).

For an inadequate fitting model, skewed by $\beta$ values fixed below or above its correct value, the fits are obviously biased to
over- or underestimate $T_{F}$ and hence to under- or overestimate $M_{F}$, correspondingly (Fig.~\ref{mass.bes.pro.tmav}). The
inaccuracy is within a factor of $2$, independently of whether $\beta$ is a factor of $1.2$ lower or higher. Fitting larger subsets
$\Phi_{n}$ ($n\,{=}\,2\,{\rightarrow}\,6$) for the isothermal models may give somewhat better derived parameters, as they better
constrain $T_{F}$ with an incorrect slope $\beta$.

Mass derivation from images $\mathcal{I}_{\nu}$ of the isothermal models delivers results that are similar to those described above
for both isolated and embedded variants (Fig.~\ref{coldens.bes.pro.tmav}). Dependence on the adopted $\beta$ is the same as
described above, hence only the results with correct $\beta\,{=}\,2$ are presented. With no temperature deviations from $T_{M}$ in
the isothermal models, derived $M_{\mathcal{I}}$ are very accurate for all angular resolutions, in contrast to the results with the
self-consistent $T_{\rm d}(r)$ (Figs.~\ref{coldens.bes} and \ref{coldens.pro}). Derived $M_{\mathcal{I}}$ for the embedded models
are practically identical to $M_{F}$, being underestimated by a factor of $1.3$ owing to the background rim-brightening effect
(Appendix \ref{AppendixB}, Sect.~\ref{bg.subtraction}).

\section{Results for free variable $\beta$}
\label{AppendixE}

\begin{figure*}
\centering
\centerline{\resizebox{0.3327\hsize}{!}{\includegraphics{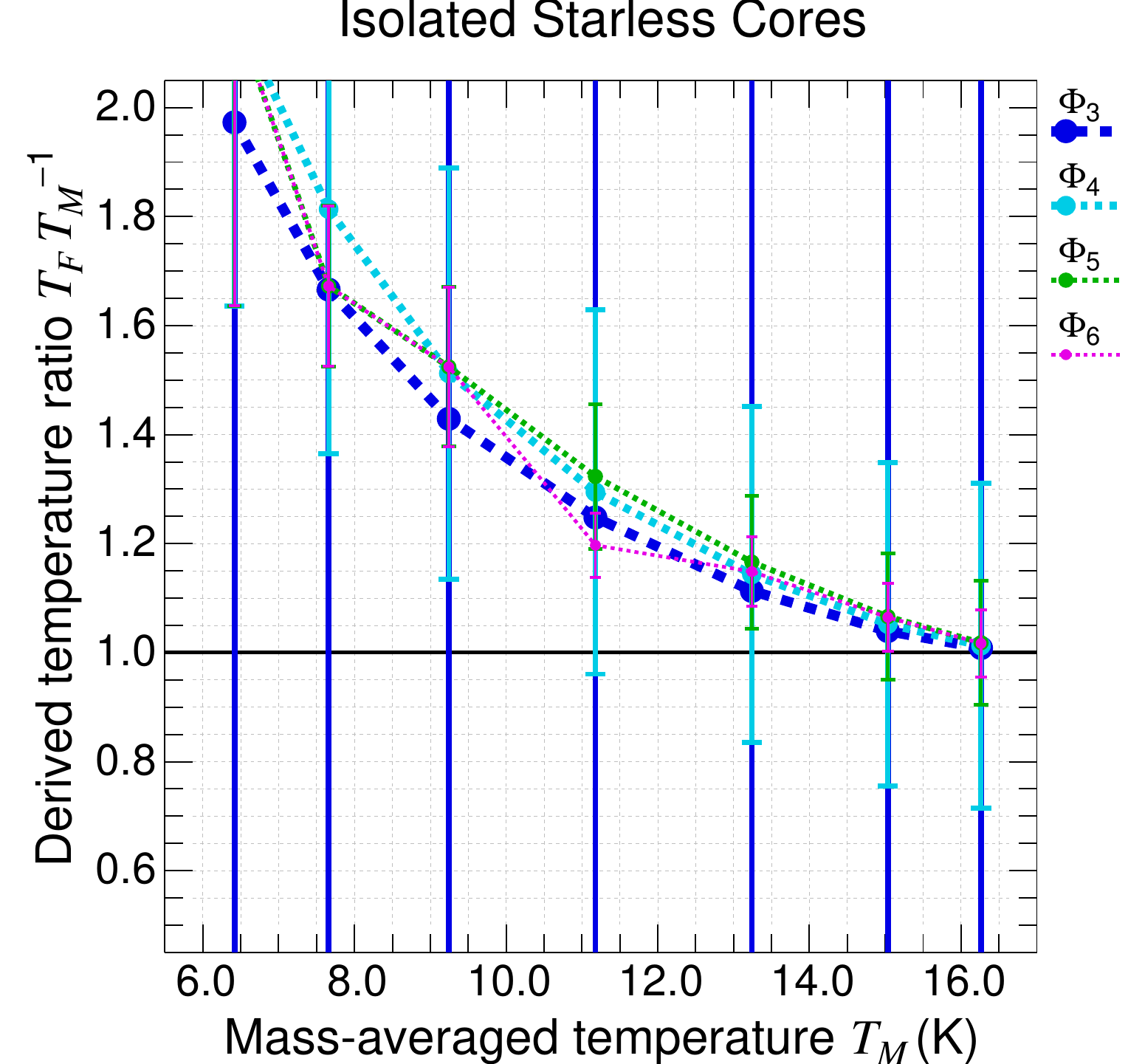}}
            \resizebox{0.3204\hsize}{!}{\includegraphics{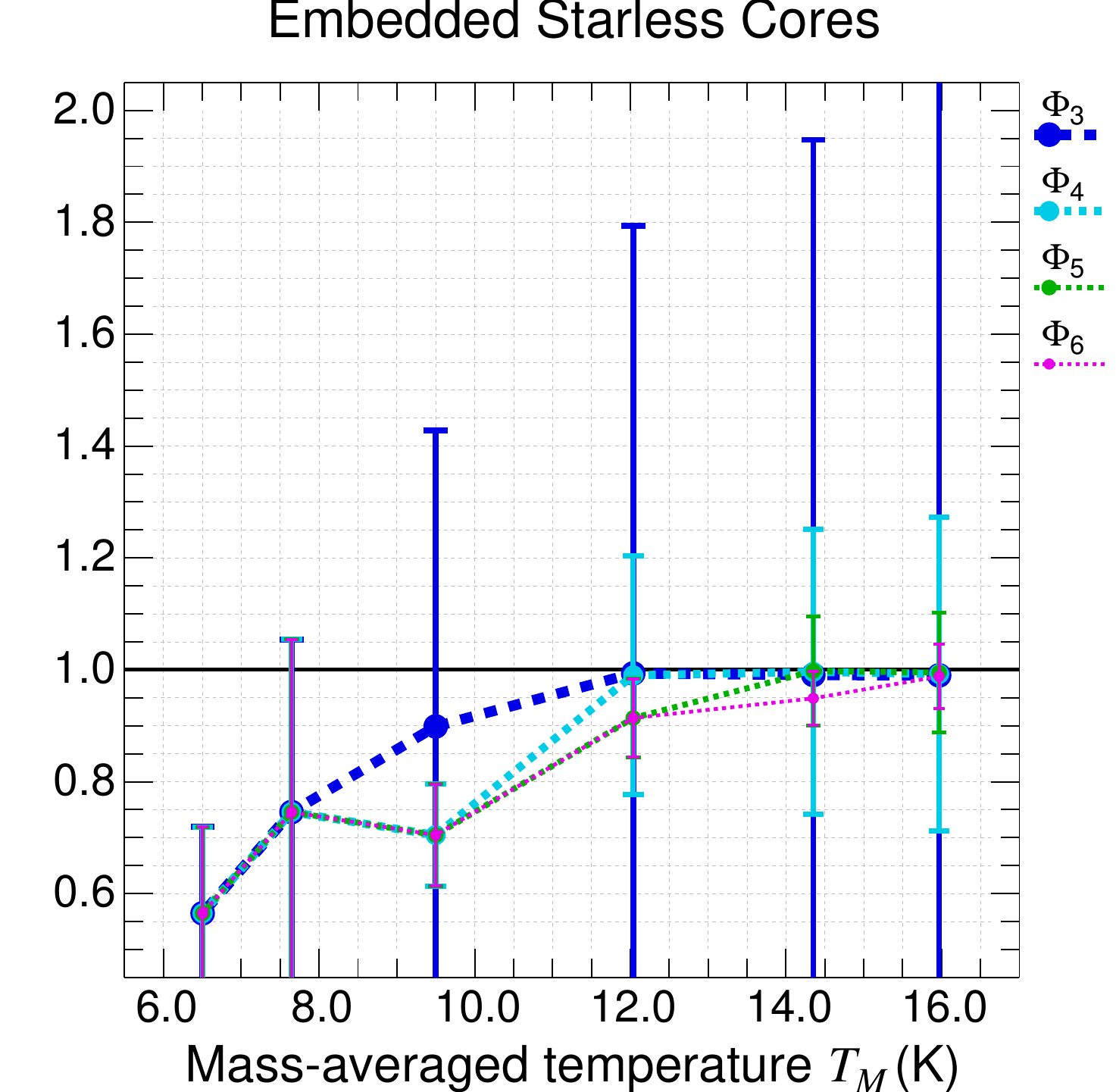}}
            \resizebox{0.3204\hsize}{!}{\includegraphics{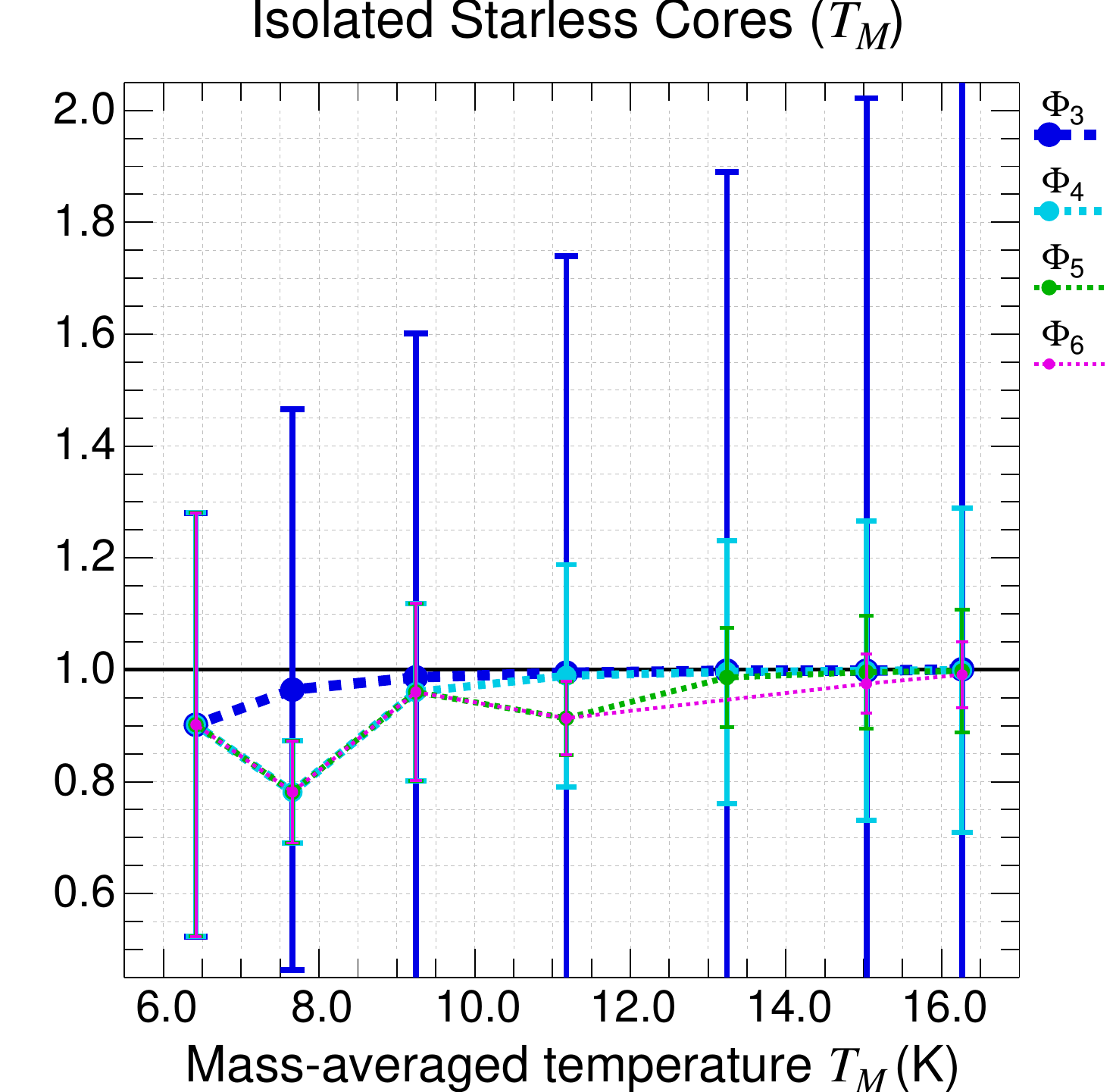}}}
\centerline{\resizebox{0.3327\hsize}{!}{\includegraphics{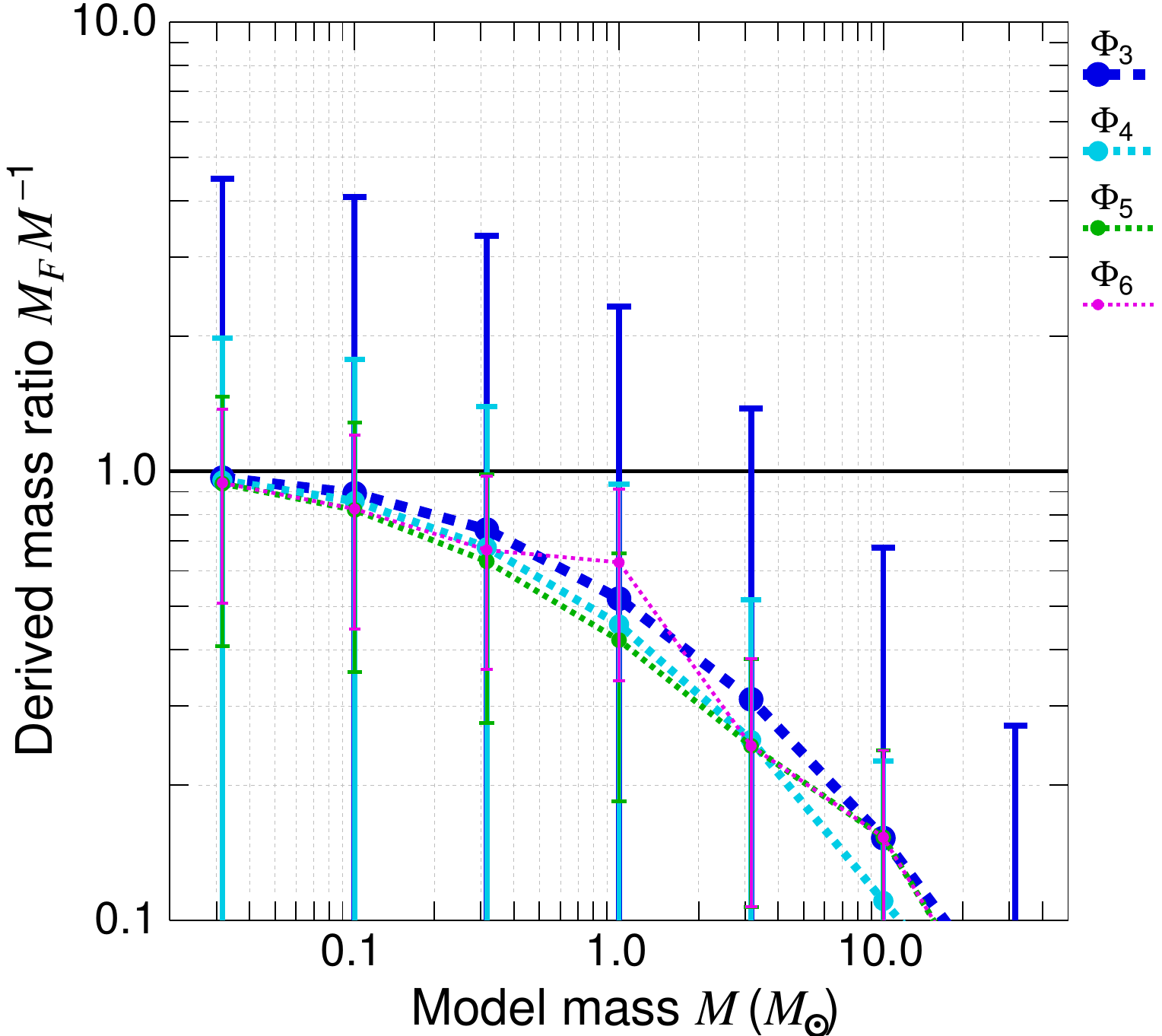}}
            \resizebox{0.3204\hsize}{!}{\includegraphics{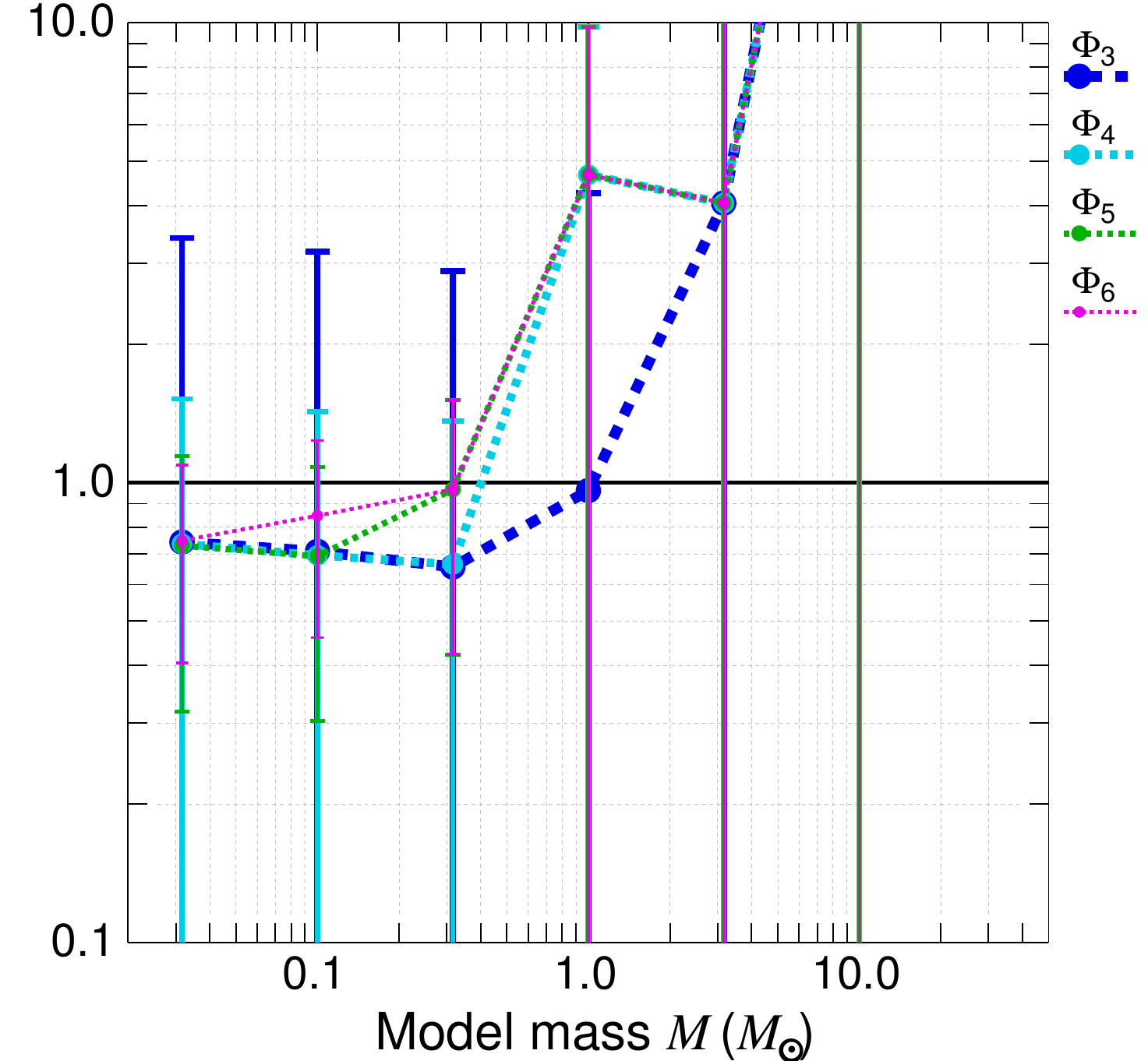}}
            \resizebox{0.3204\hsize}{!}{\includegraphics{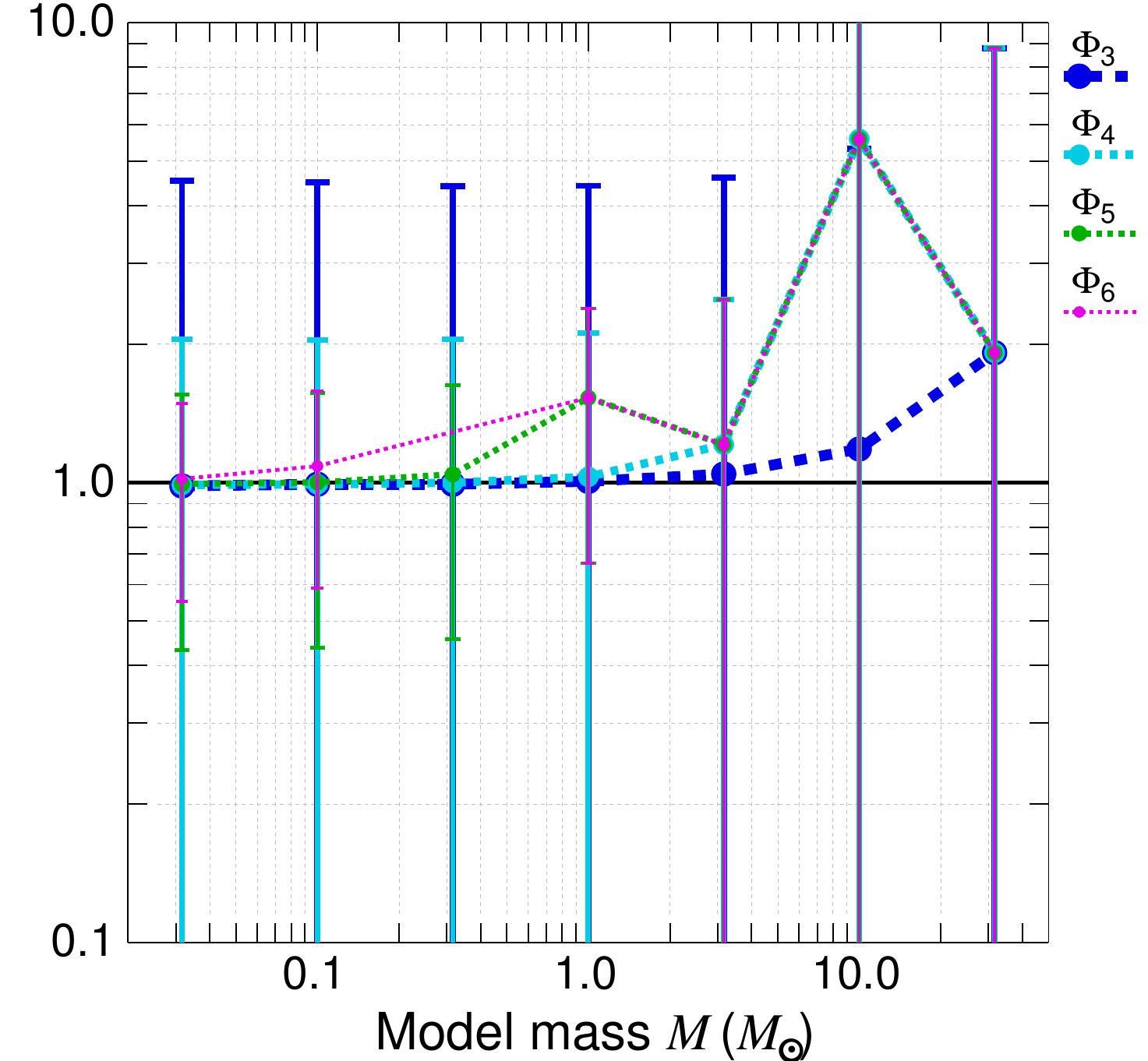}}}
\centerline{\resizebox{0.3327\hsize}{!}{\includegraphics{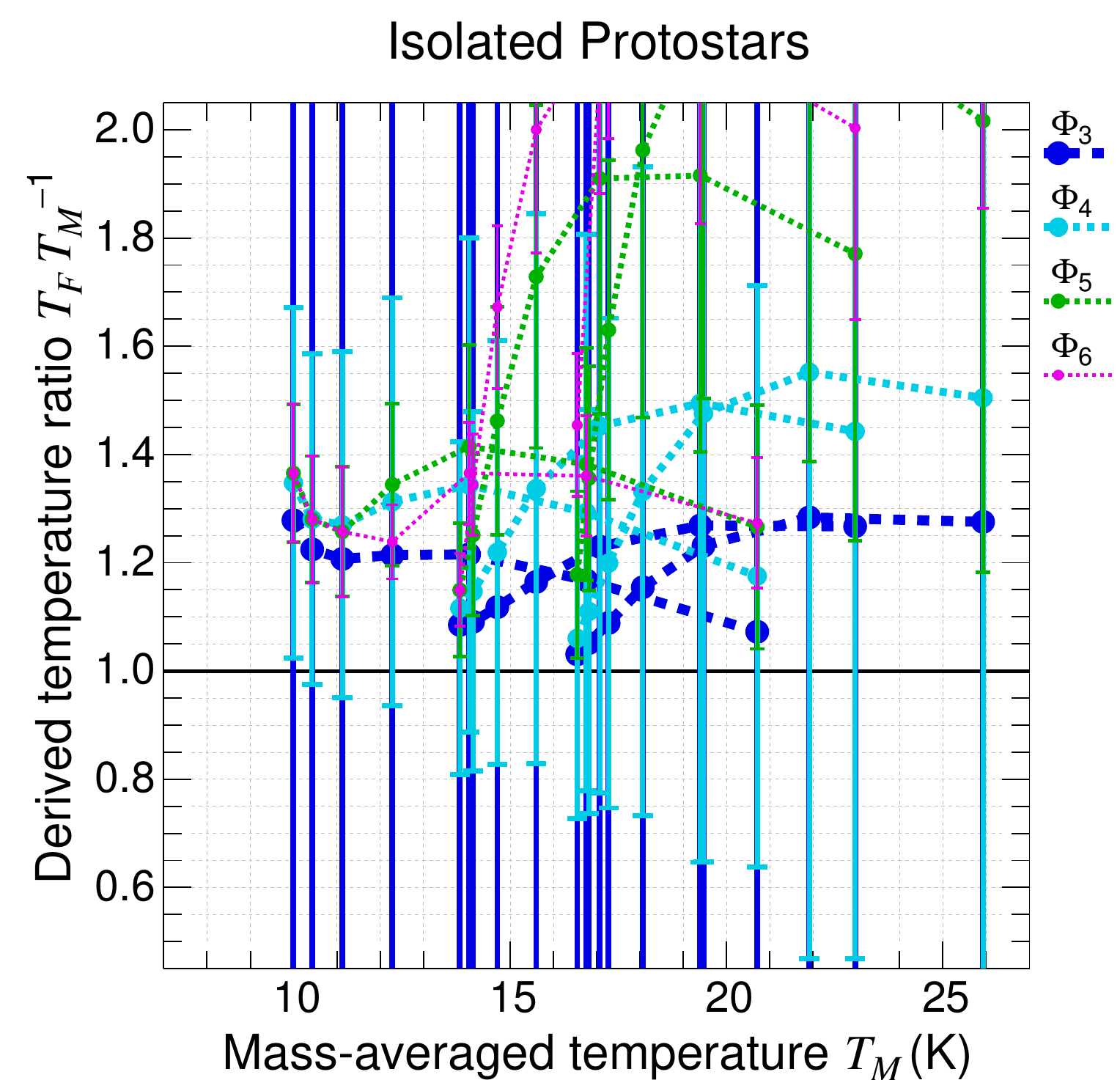}}
            \resizebox{0.3204\hsize}{!}{\includegraphics{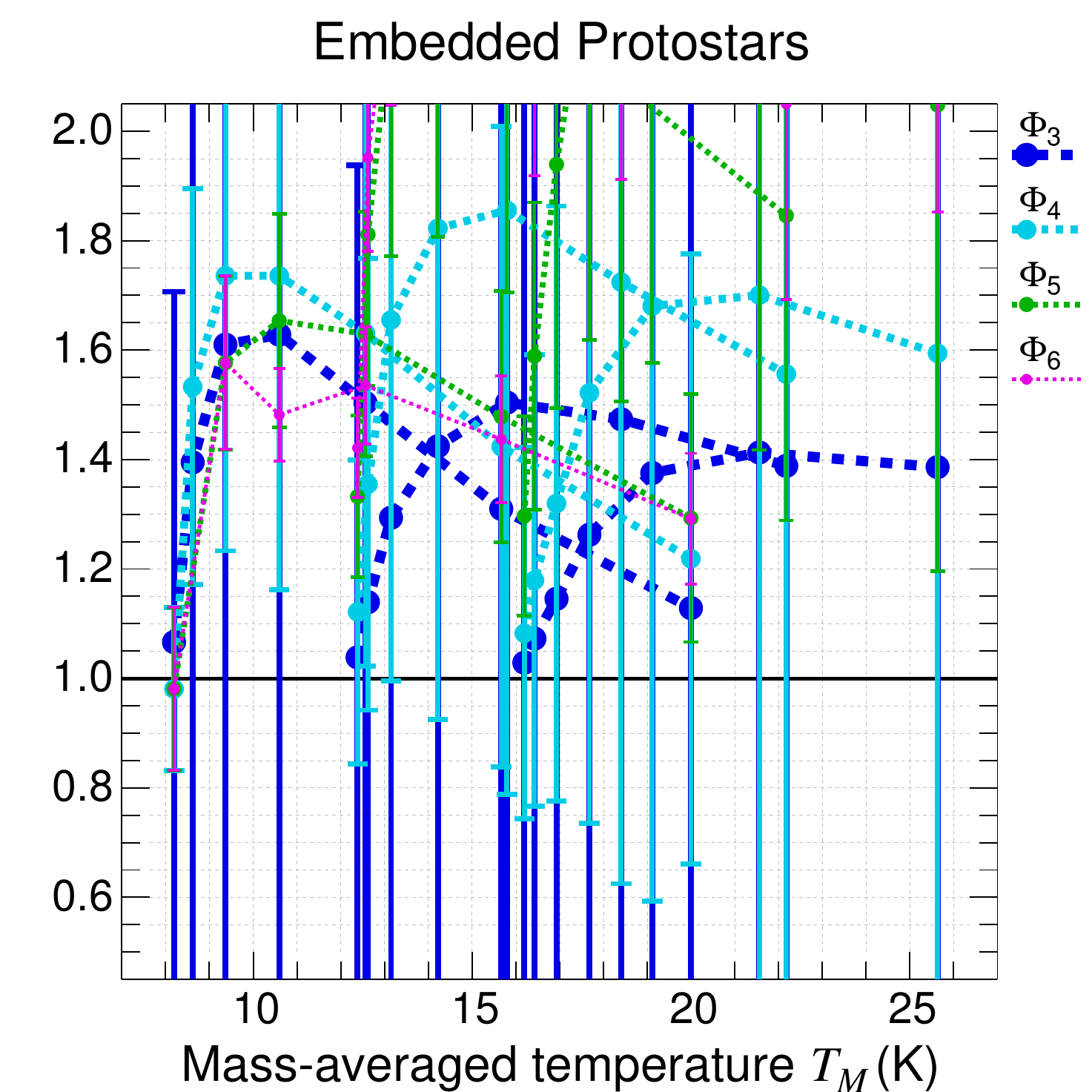}}
            \resizebox{0.3204\hsize}{!}{\includegraphics{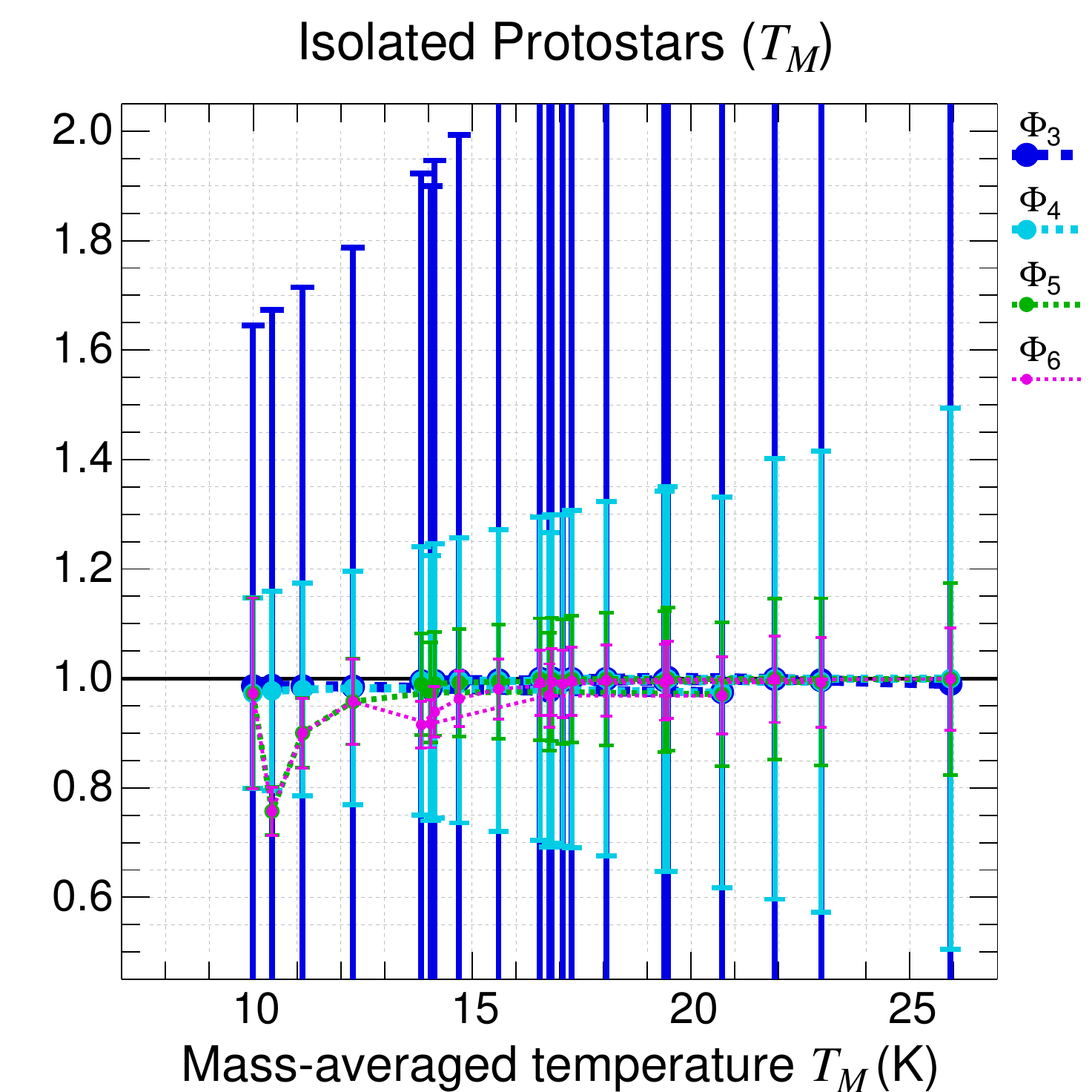}}}
\centerline{\resizebox{0.3327\hsize}{!}{\includegraphics{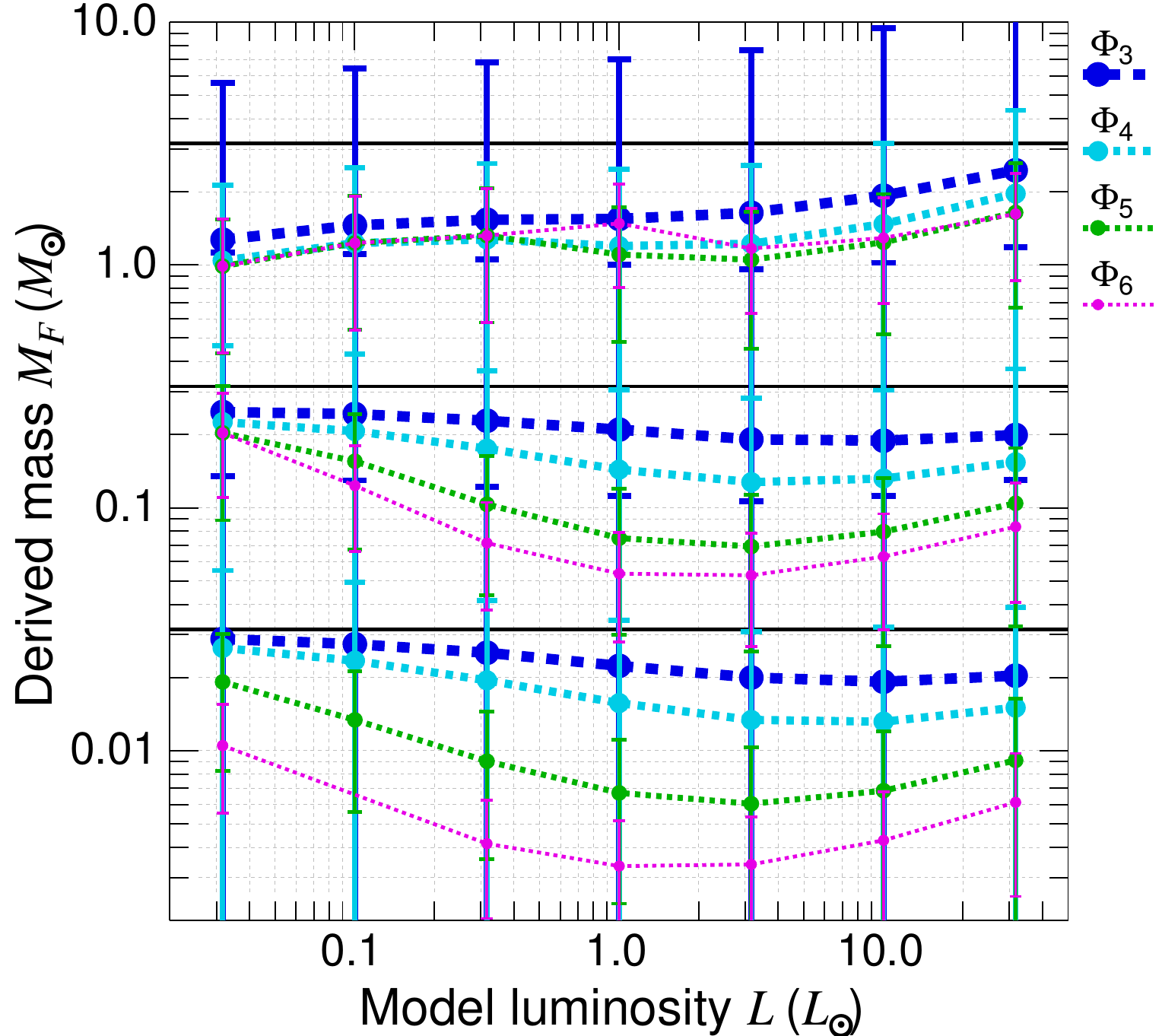}}
            \resizebox{0.3204\hsize}{!}{\includegraphics{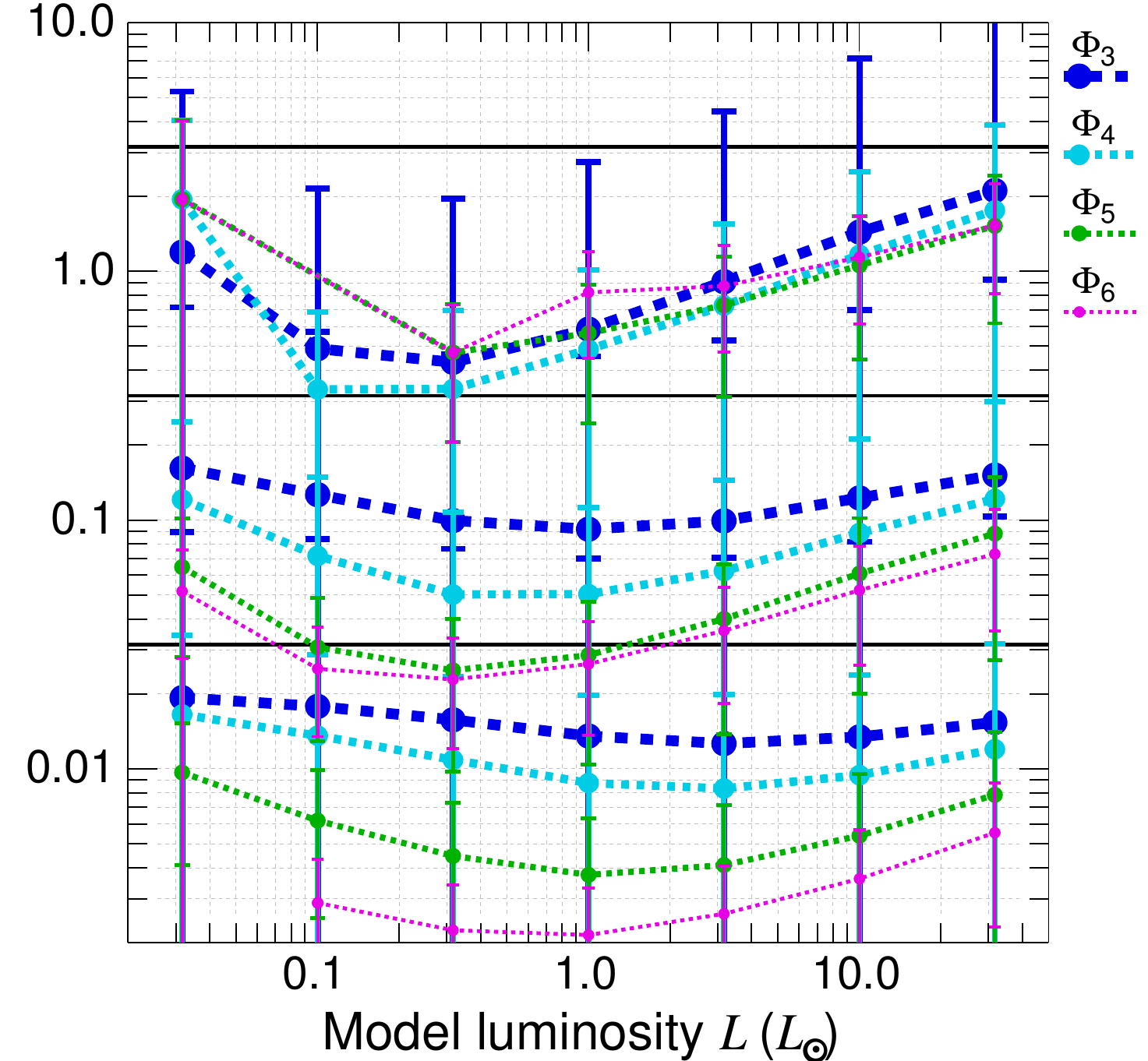}}
            \resizebox{0.3204\hsize}{!}{\includegraphics{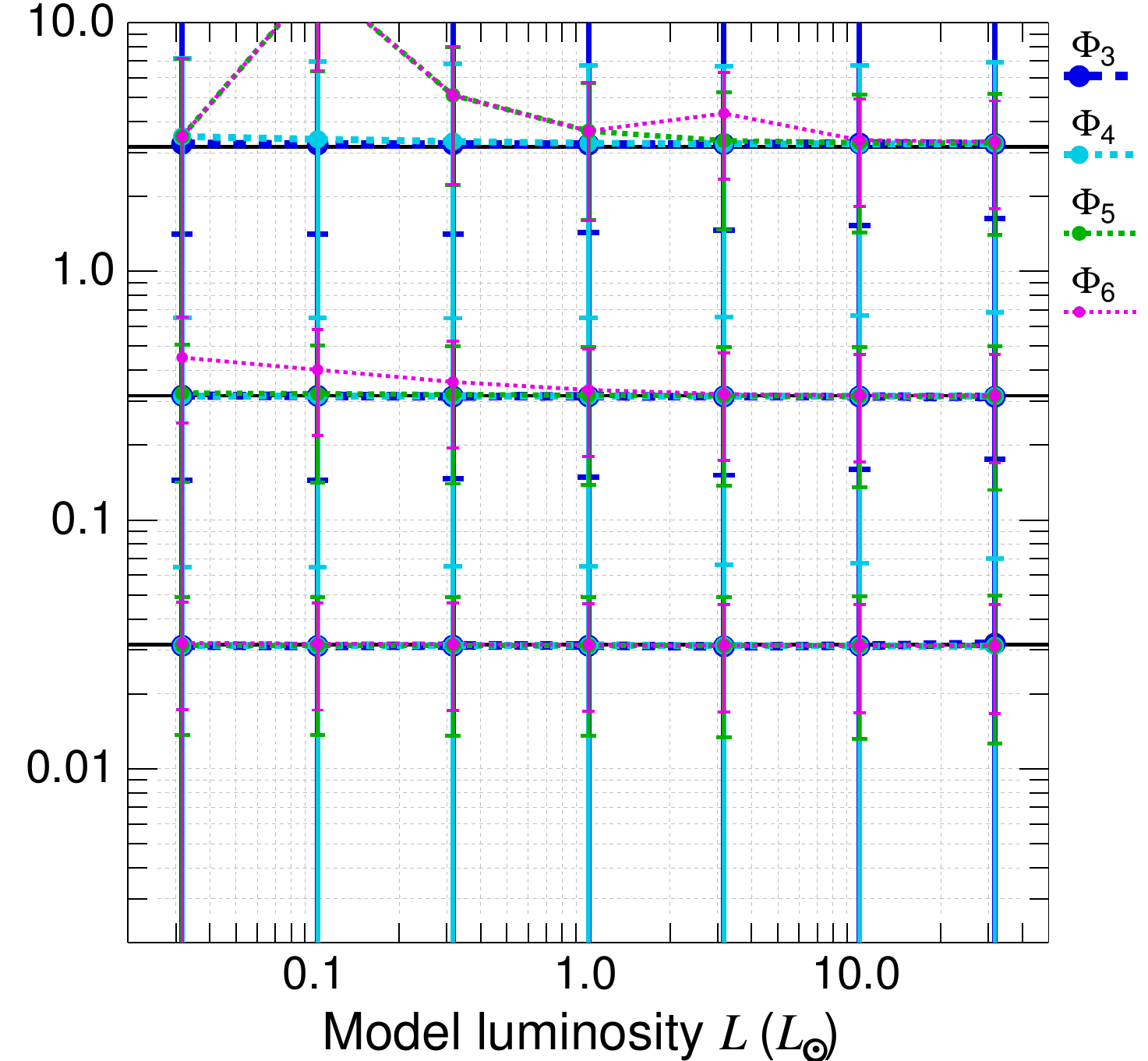}}}
\caption{
Temperatures $T_{F}$ and masses $M_{F}$ derived from fitting $F_{\nu}$ of \emph{isolated}, \emph{embedded}, and \emph{isothermal} 
models (fits with free variable $\beta$) for starless cores (\emph{upper}) and protostellar envelopes (\emph{lower}). See 
Fig.~\ref{temp.mass.bes} for more details.
} 
\label{beta.bes.pro}
\end{figure*}

\begin{figure*}
\centering
\centerline{\resizebox{0.3327\hsize}{!}{\includegraphics{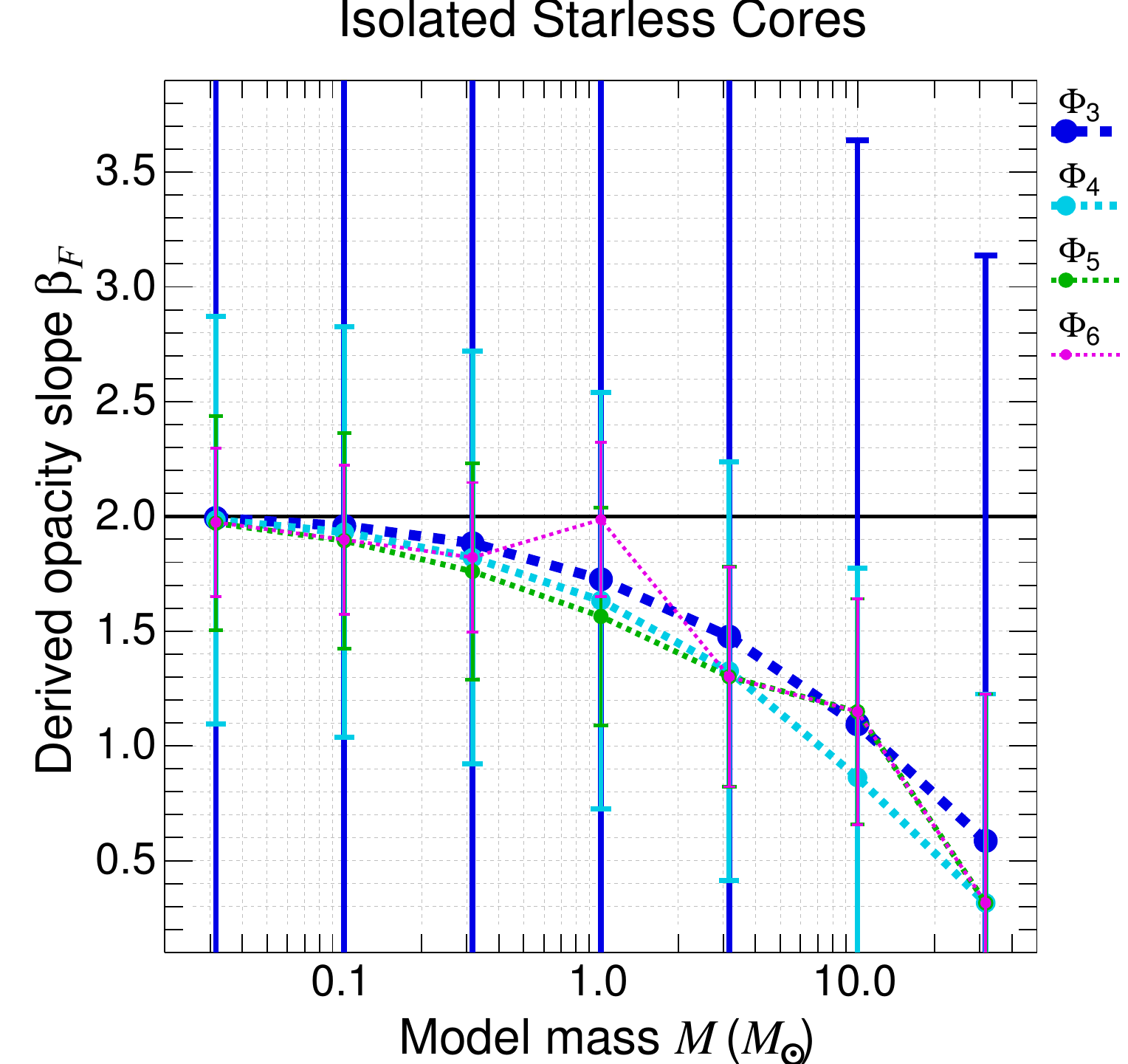}}
            \resizebox{0.3204\hsize}{!}{\includegraphics{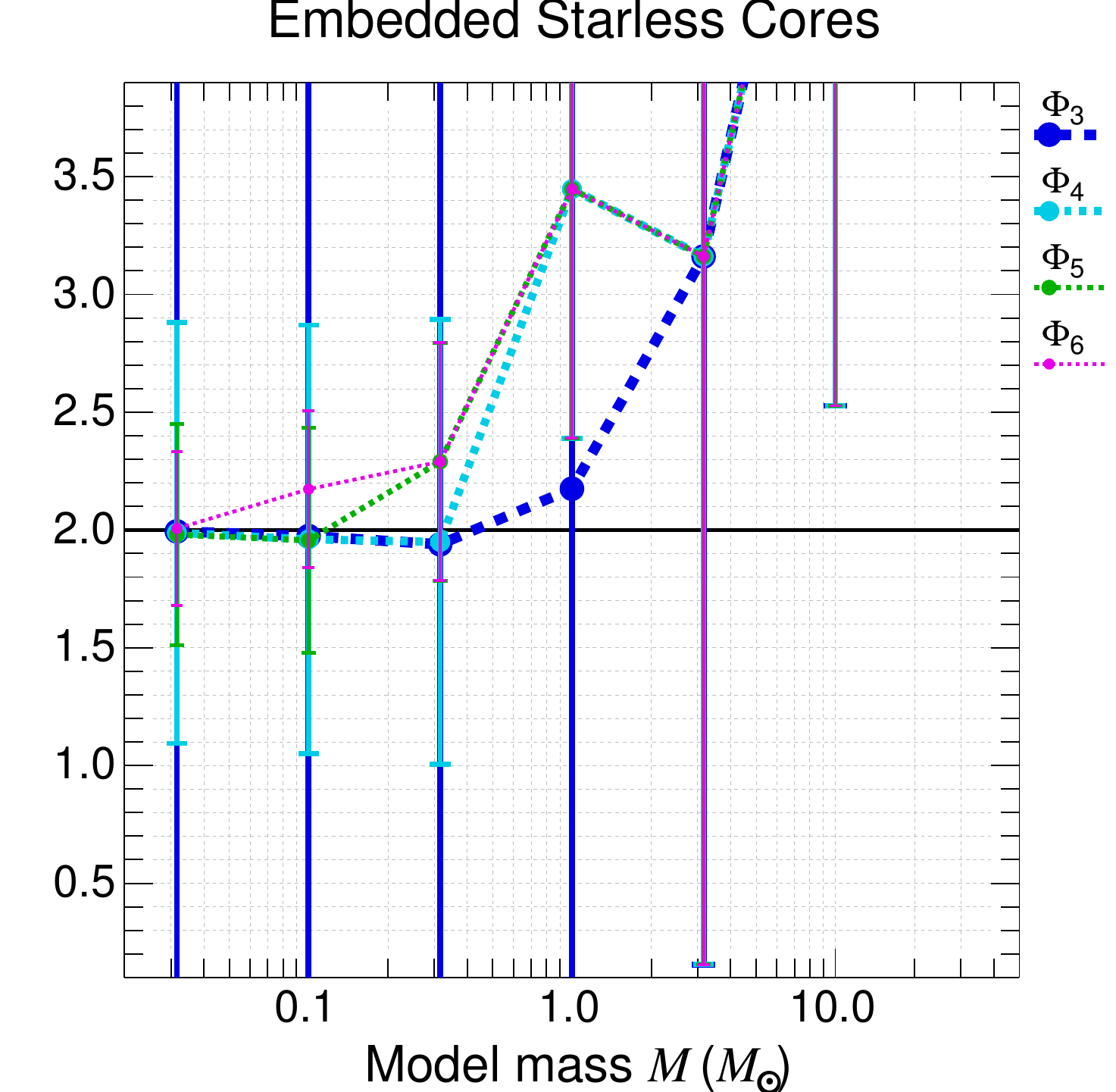}}
            \resizebox{0.3204\hsize}{!}{\includegraphics{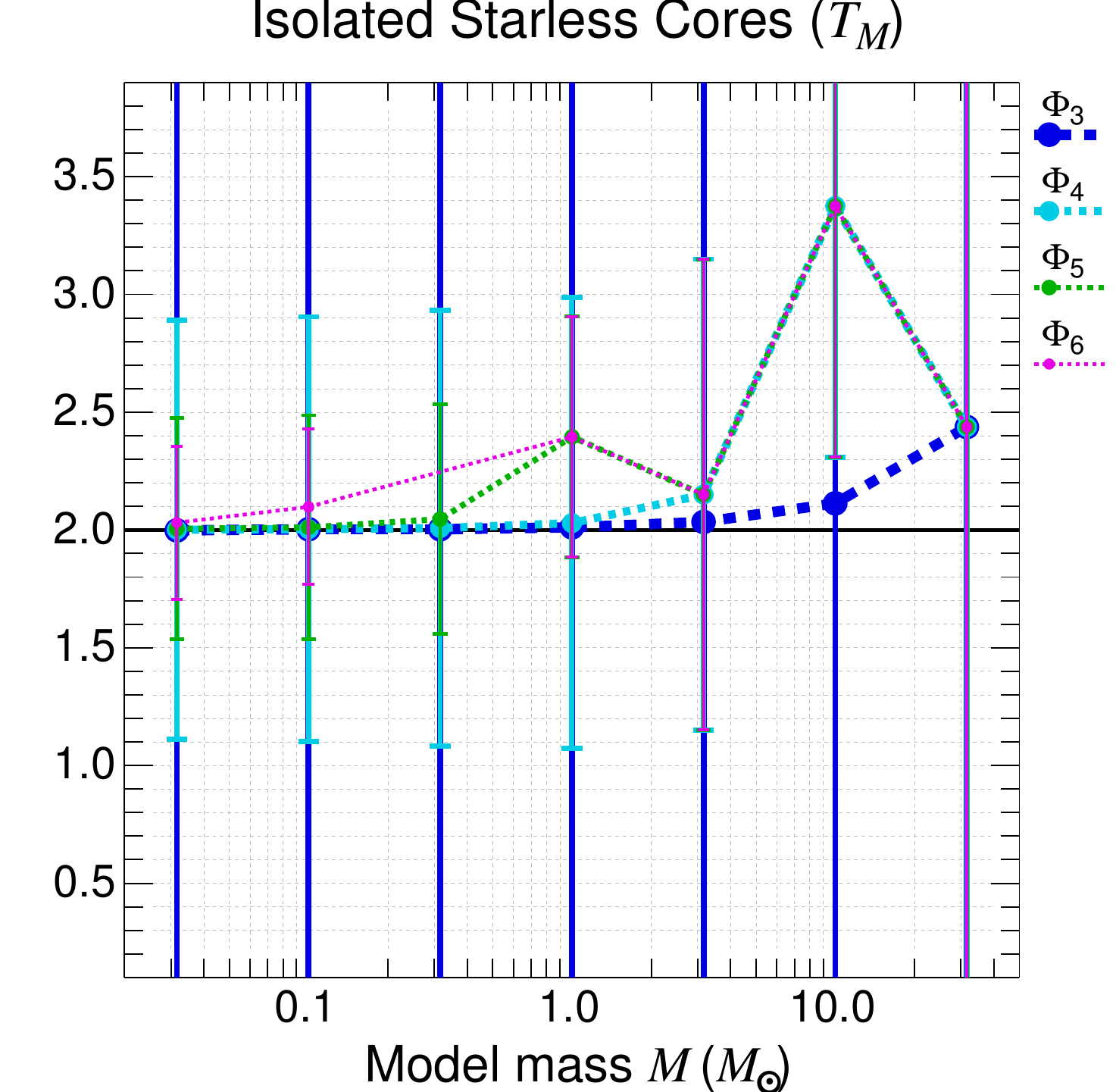}}}
\centerline{\resizebox{0.3327\hsize}{!}{\includegraphics{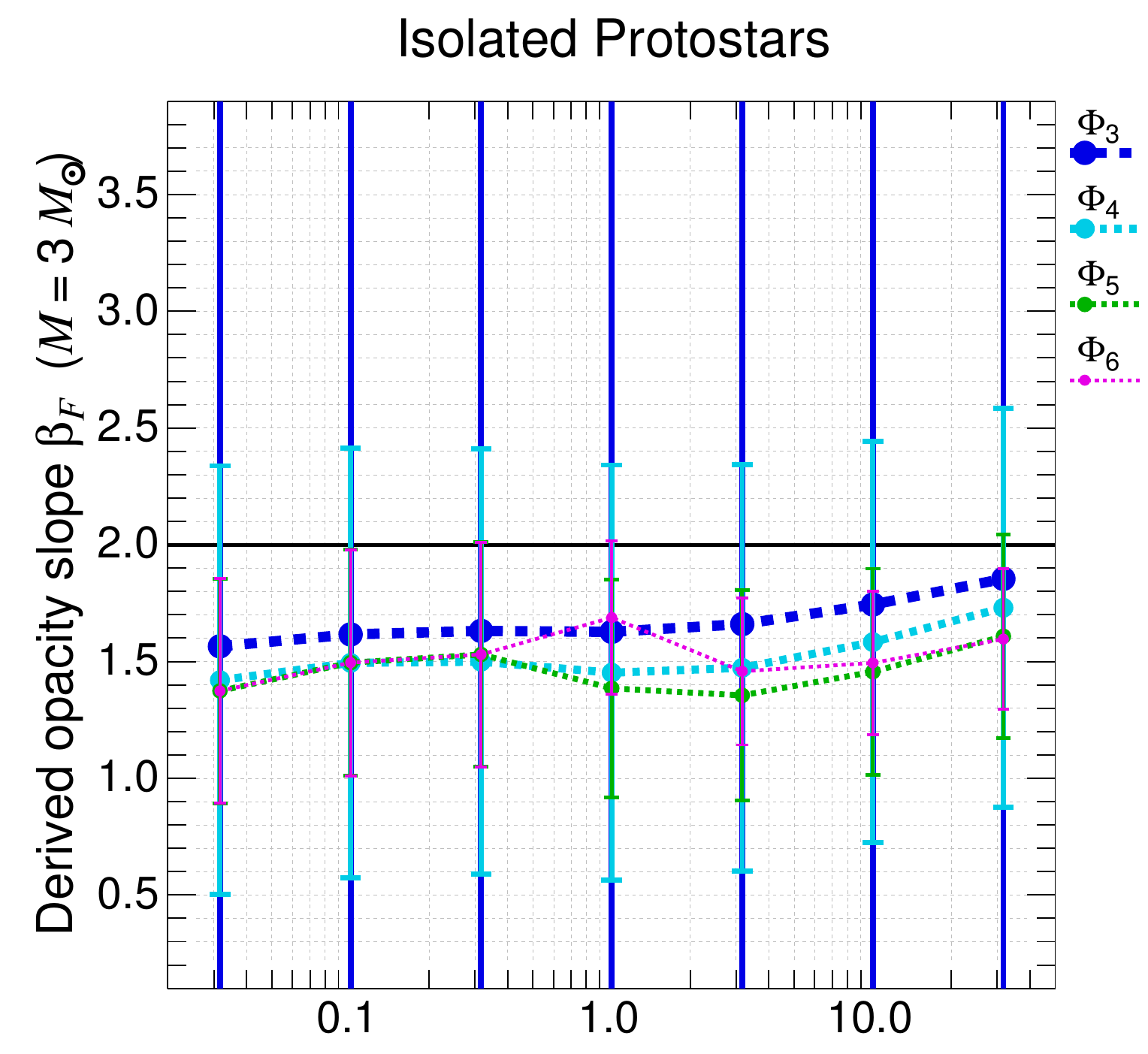}}
            \resizebox{0.3204\hsize}{!}{\includegraphics{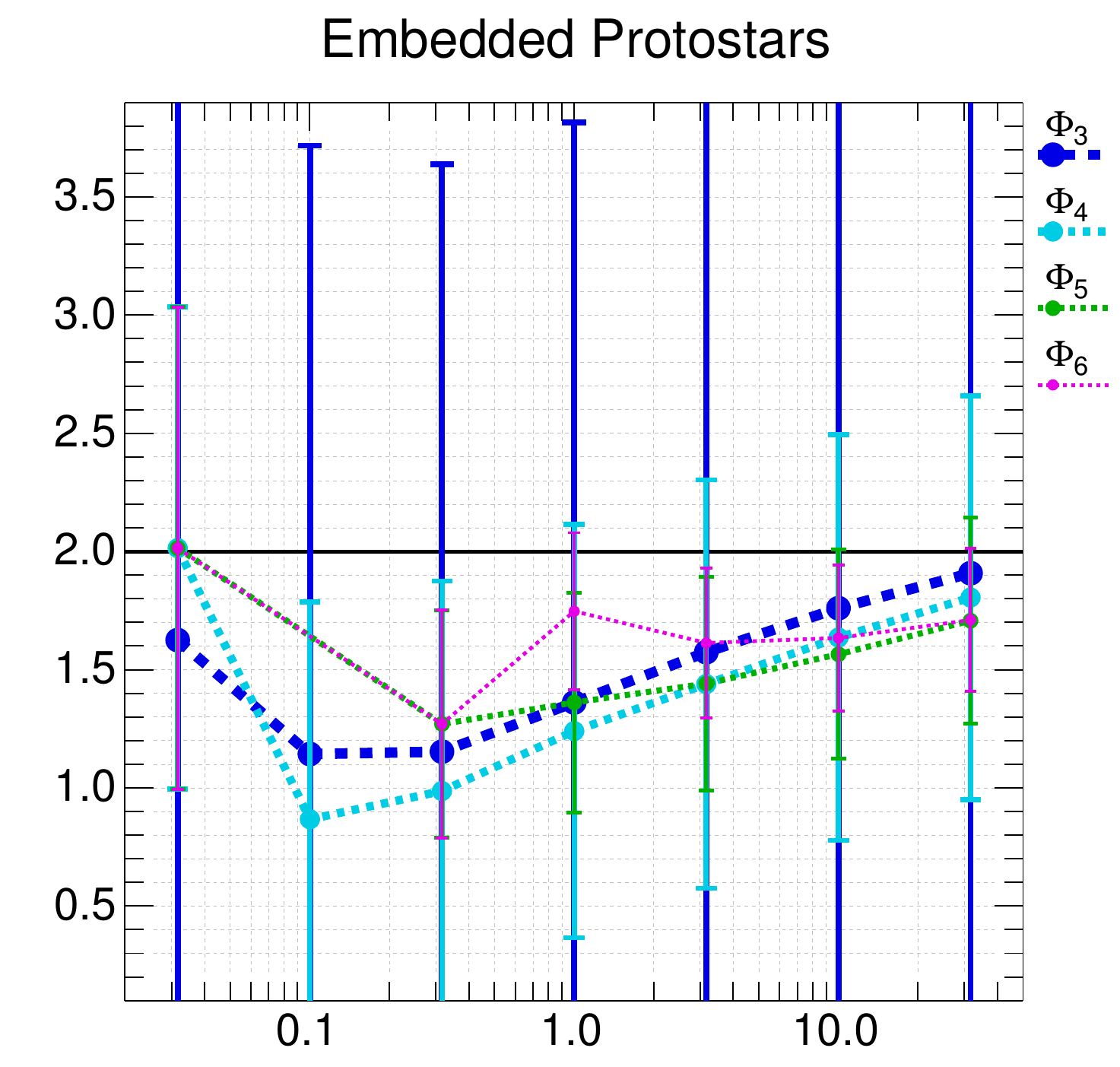}}
            \resizebox{0.3204\hsize}{!}{\includegraphics{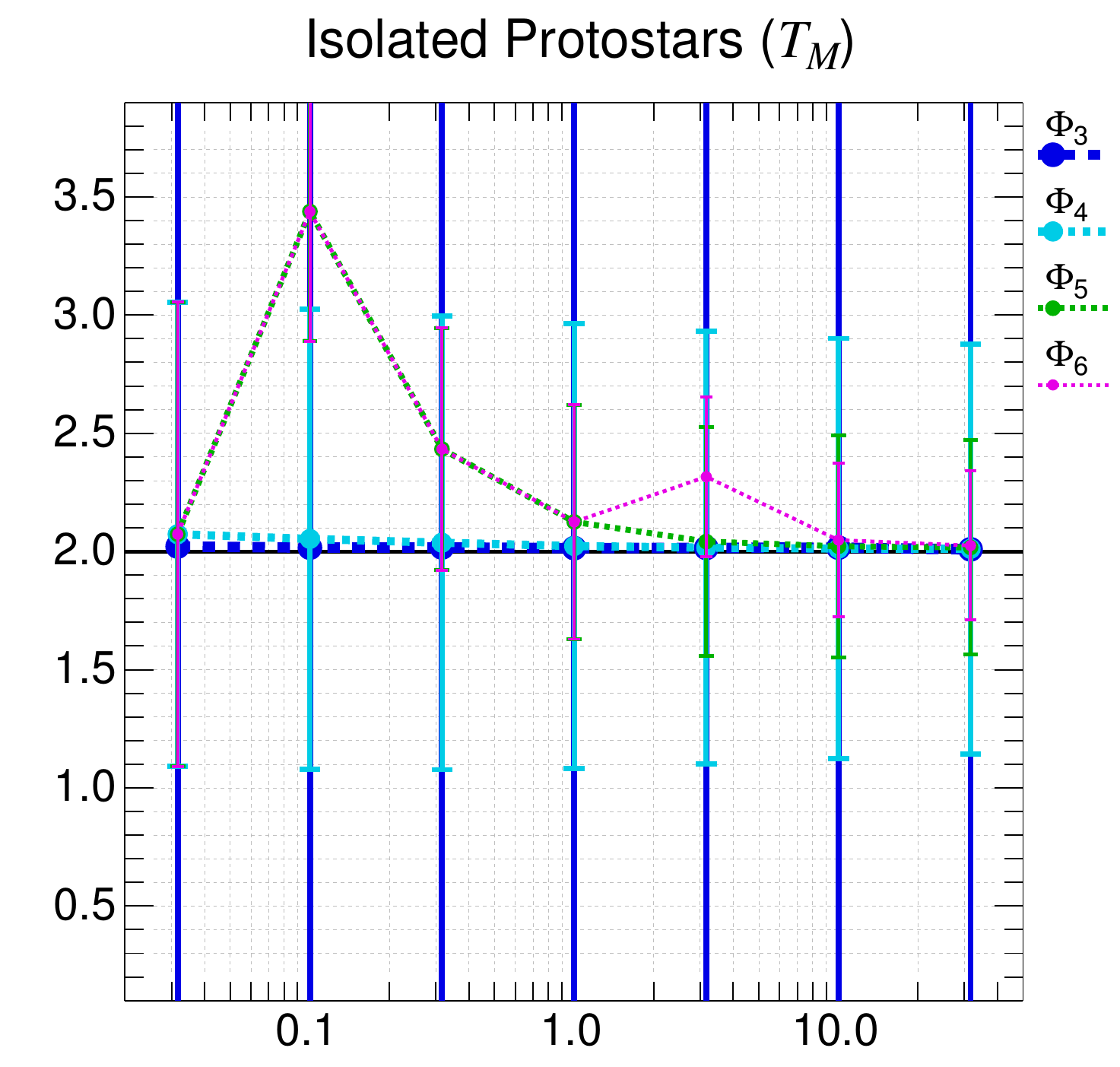}}}
\centerline{\resizebox{0.3327\hsize}{!}{\includegraphics{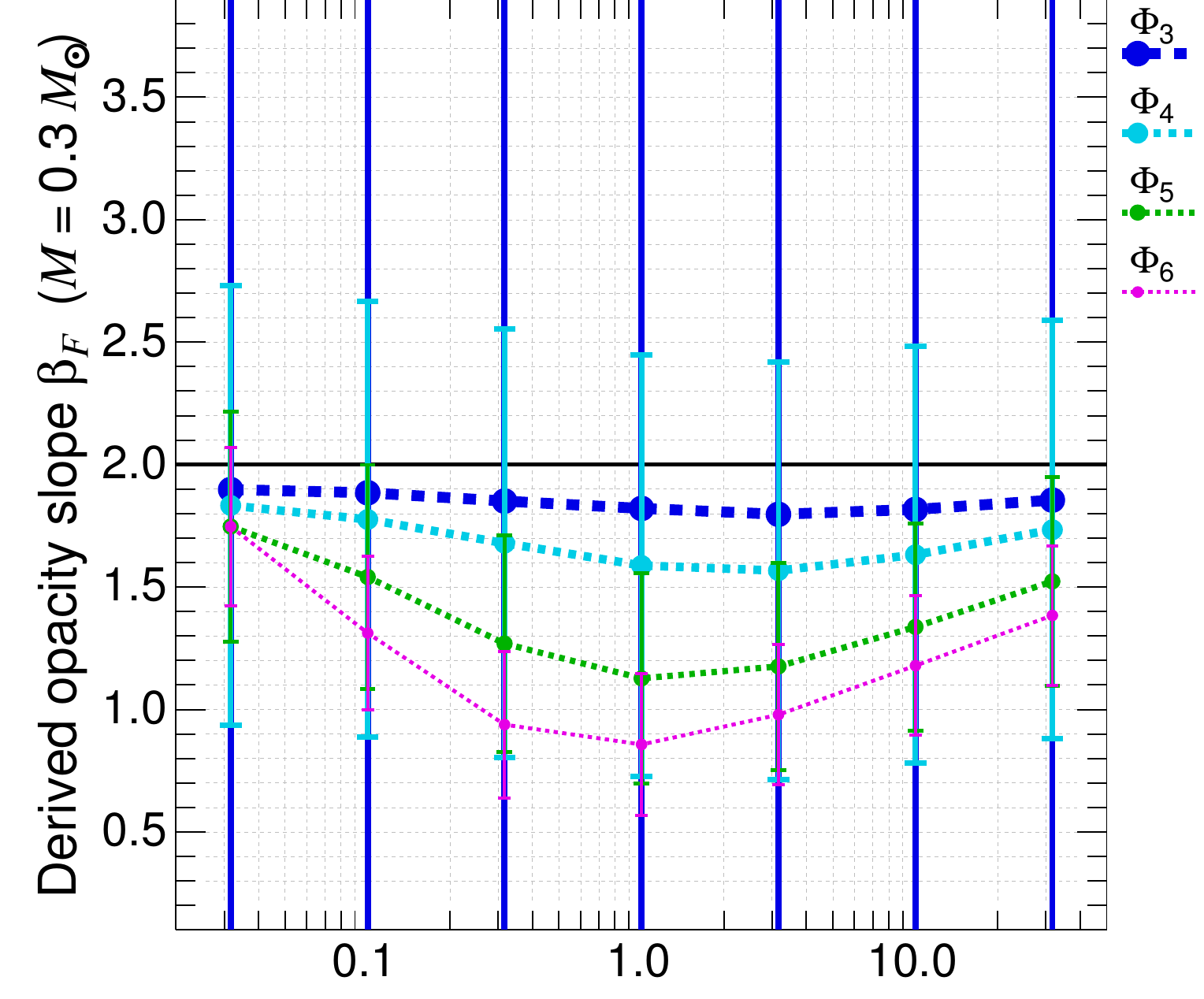}}
            \resizebox{0.3204\hsize}{!}{\includegraphics{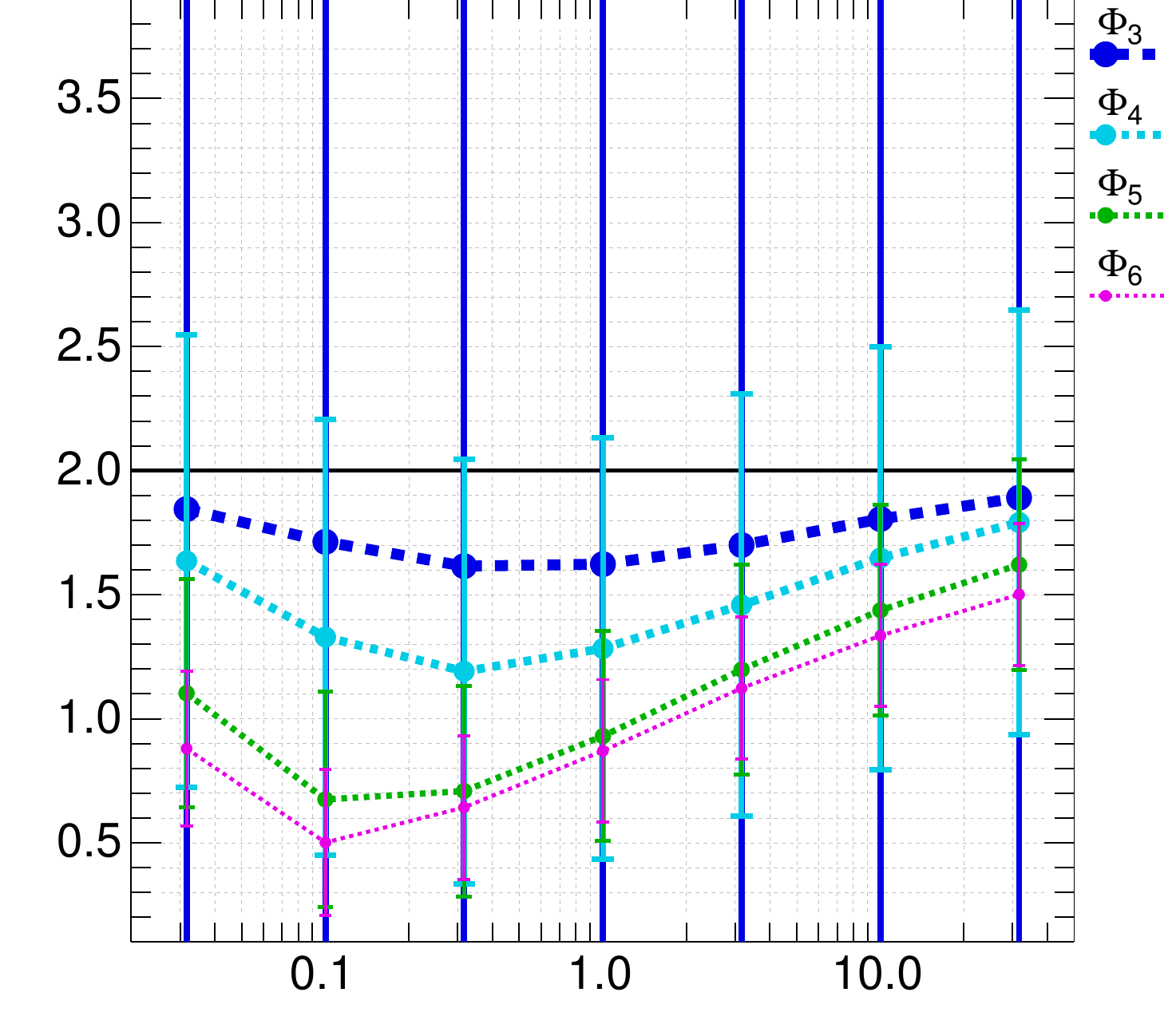}}
            \resizebox{0.3204\hsize}{!}{\includegraphics{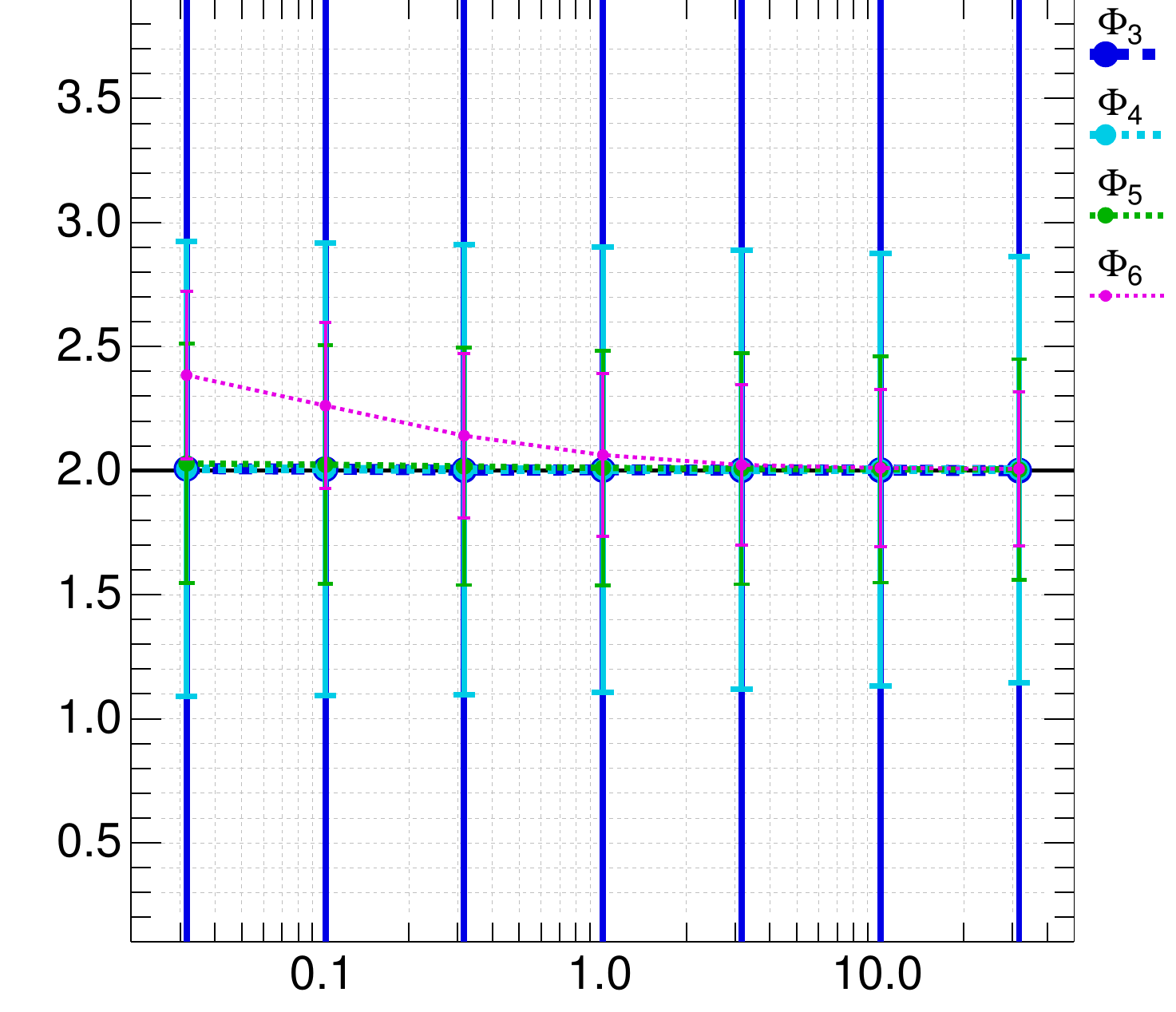}}}
\centerline{\resizebox{0.3327\hsize}{!}{\includegraphics{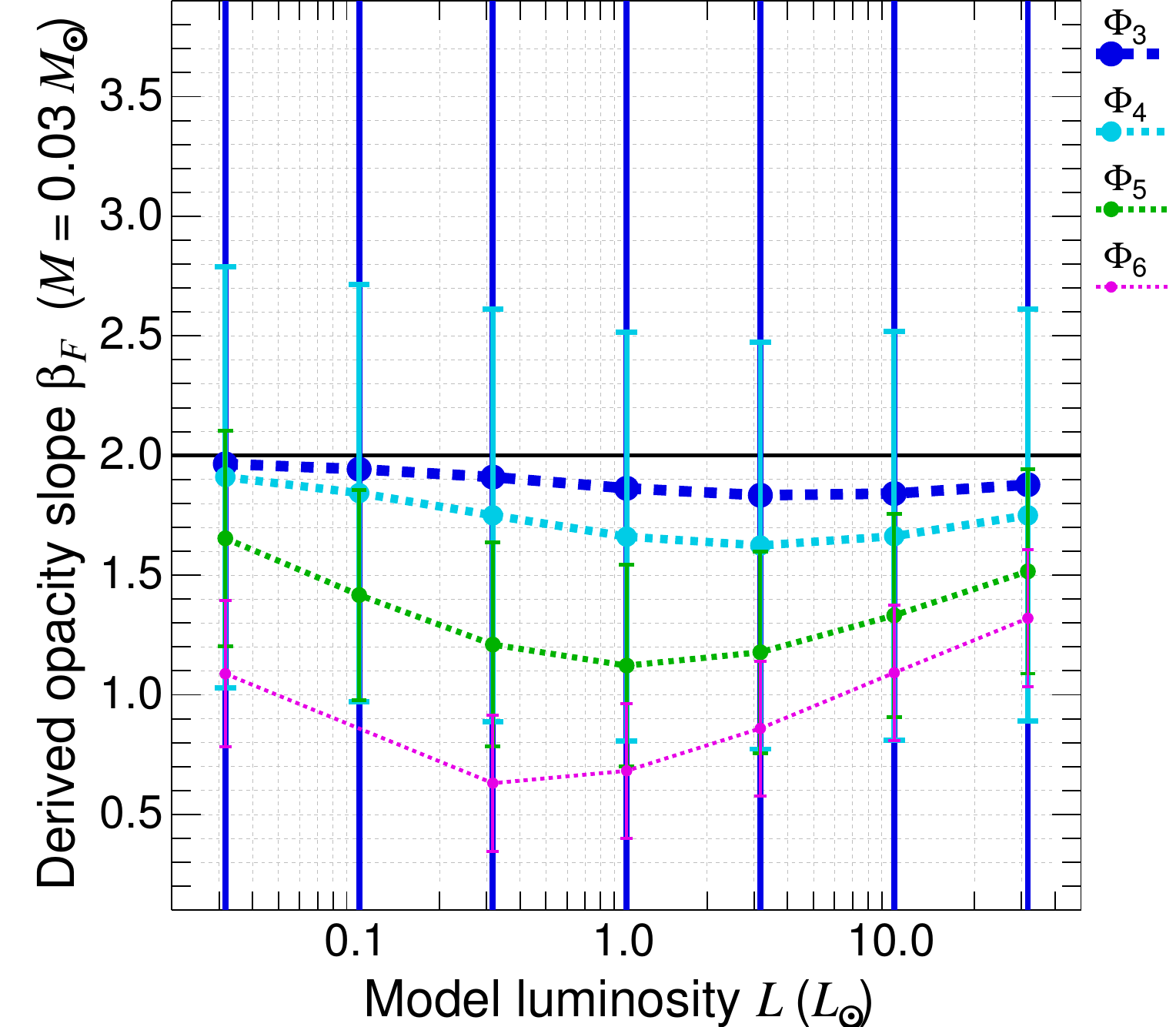}}
            \resizebox{0.3204\hsize}{!}{\includegraphics{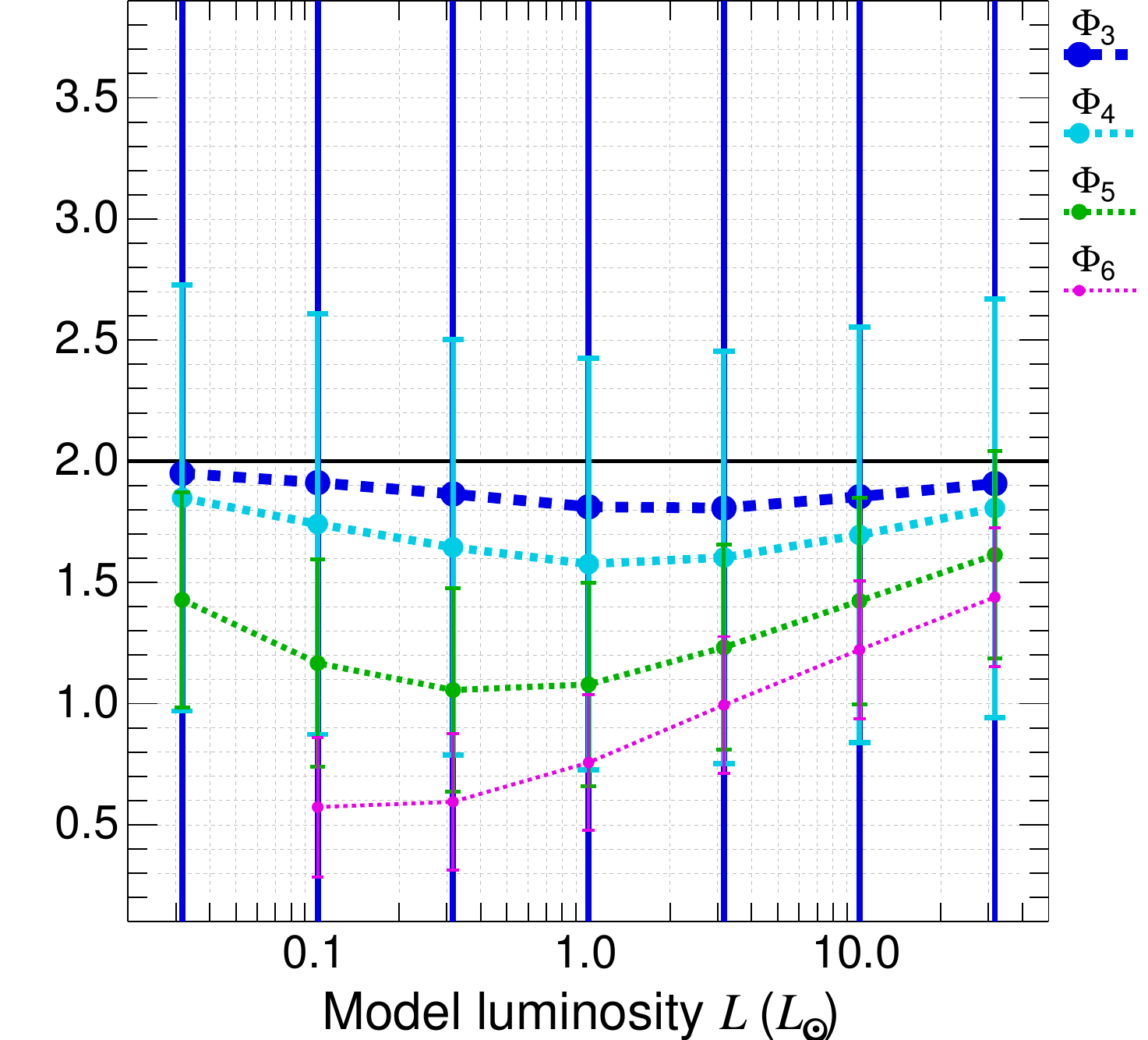}}
            \resizebox{0.3204\hsize}{!}{\includegraphics{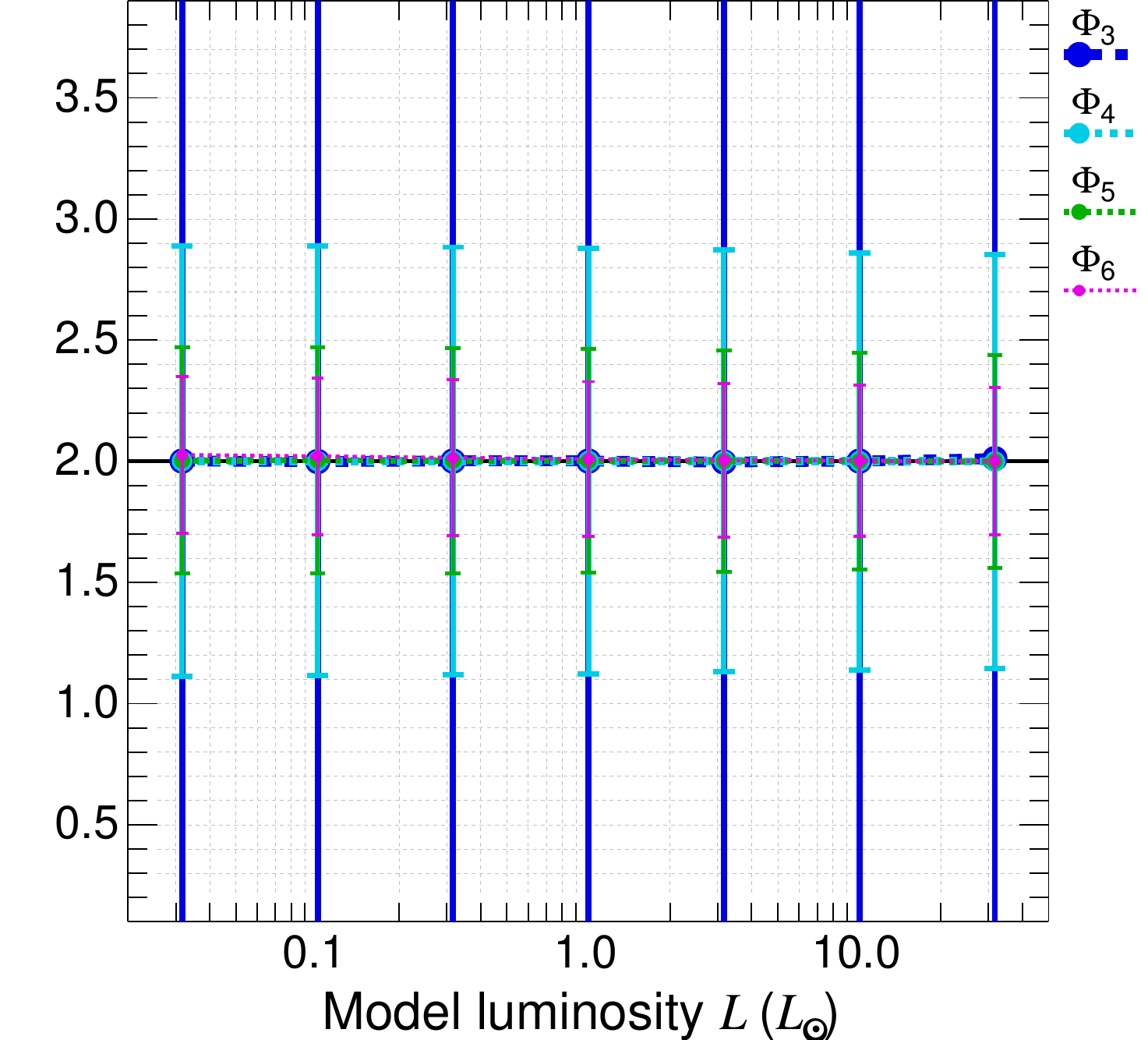}}}
\caption{
Opacity slope $\beta_{F}$ derived from fitting $F_{\nu}$ of \emph{isolated}, \emph{embedded}, and \emph{isothermal} models (fits 
with free variable $\beta$) for starless cores (\emph{upper}) and protostellar envelopes of selected masses ($3$, $0.3$, and 
$0.03\,M_{\sun}$, \emph{lower}). See Fig.~\ref{temp.mass.bes} for more details.
} 
\label{beta.beta}
\end{figure*}

\begin{figure*}
\centering
\centerline{\resizebox{0.3327\hsize}{!}{\includegraphics{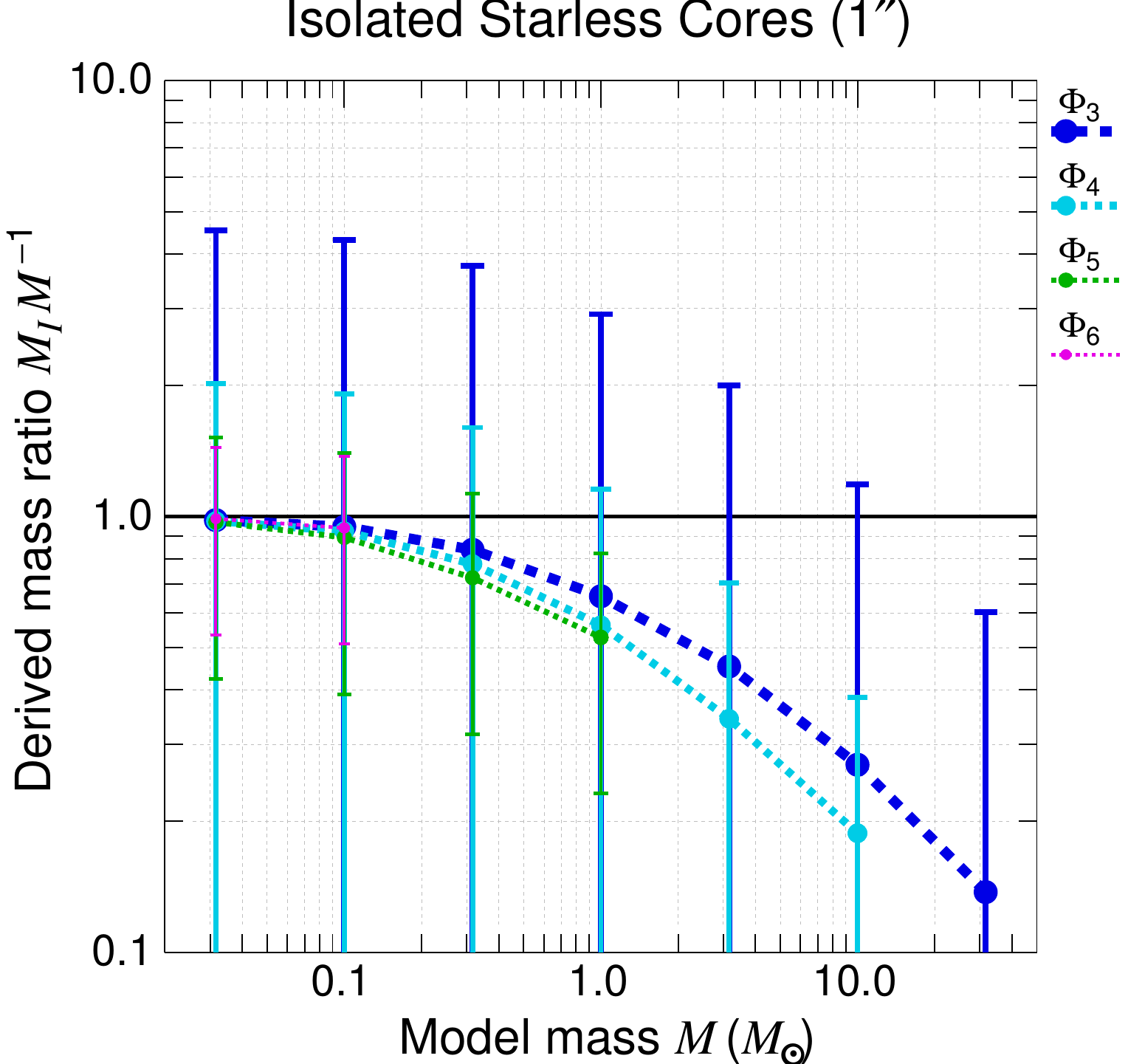}}
            \resizebox{0.3204\hsize}{!}{\includegraphics{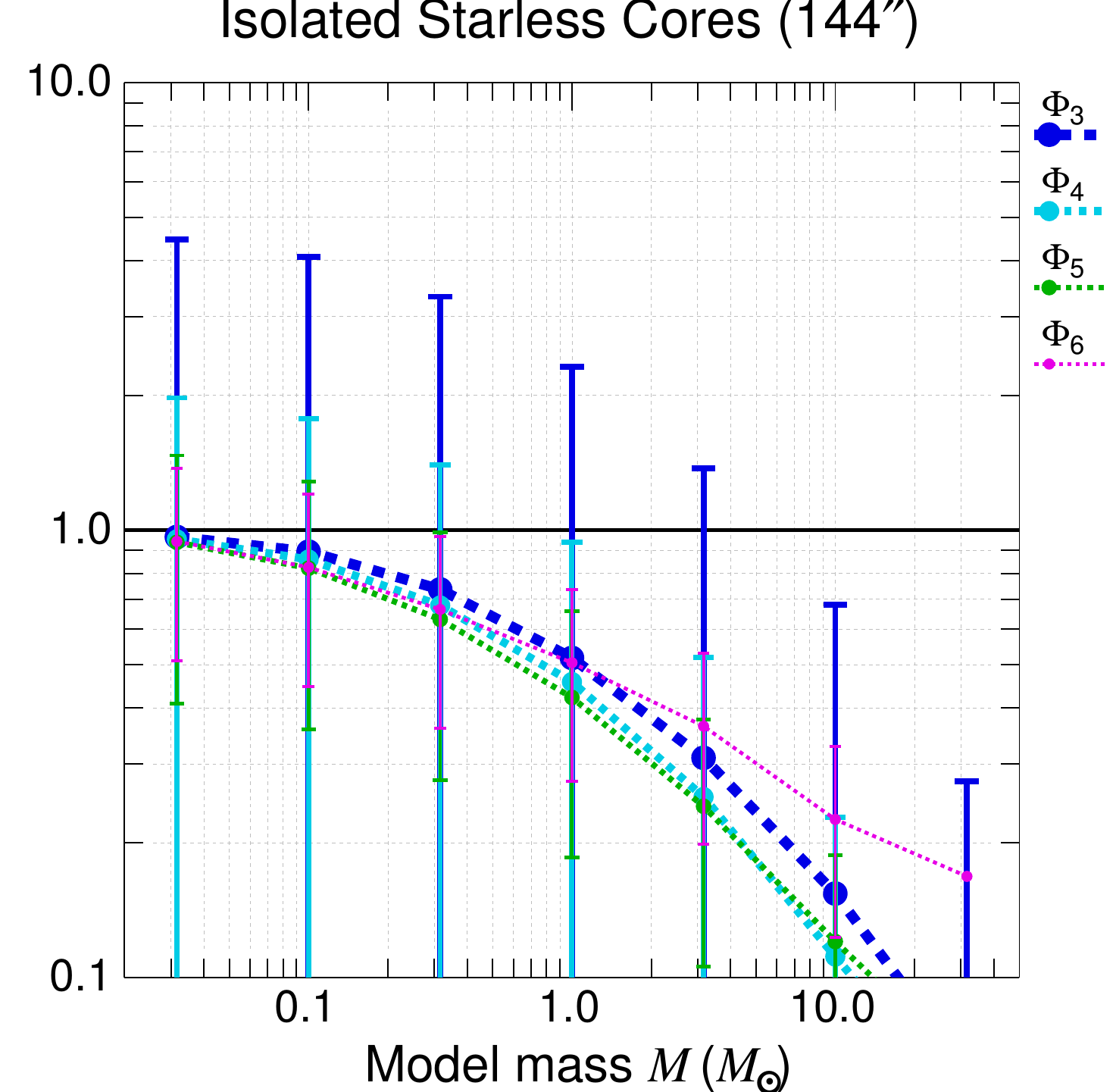}}
            \resizebox{0.3204\hsize}{!}{\includegraphics{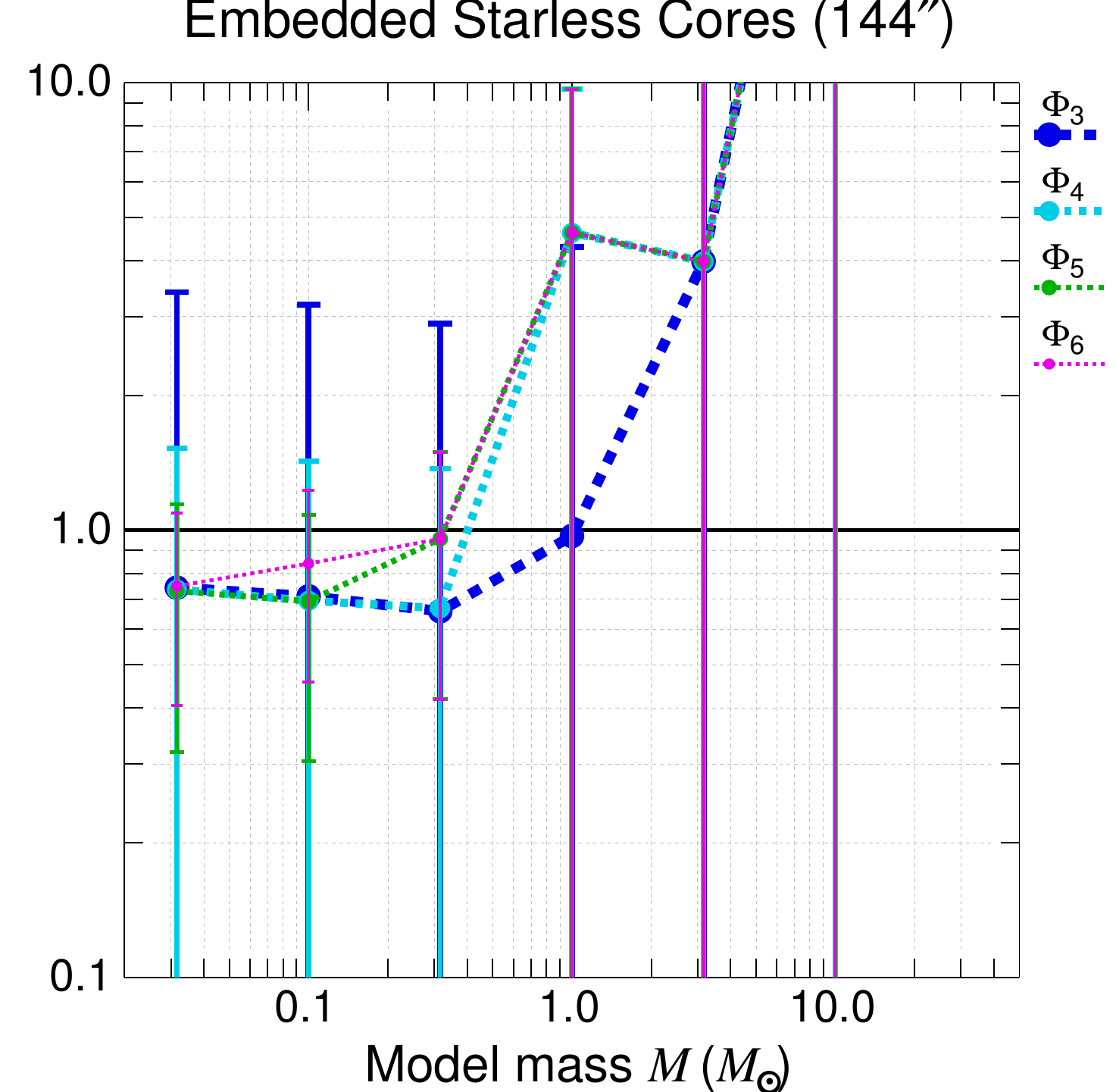}}}
\centerline{\resizebox{0.3327\hsize}{!}{\includegraphics{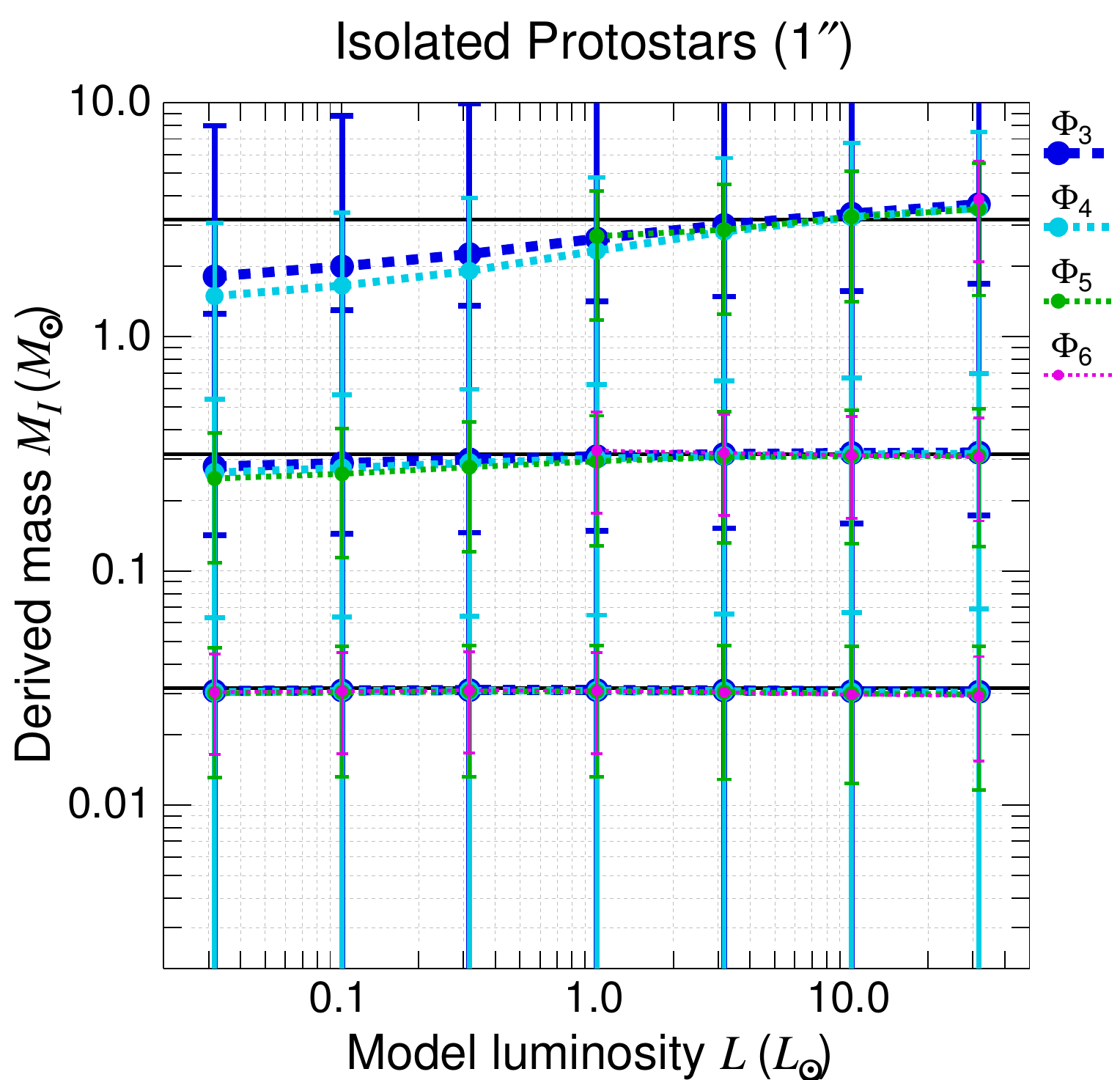}}
            \resizebox{0.3204\hsize}{!}{\includegraphics{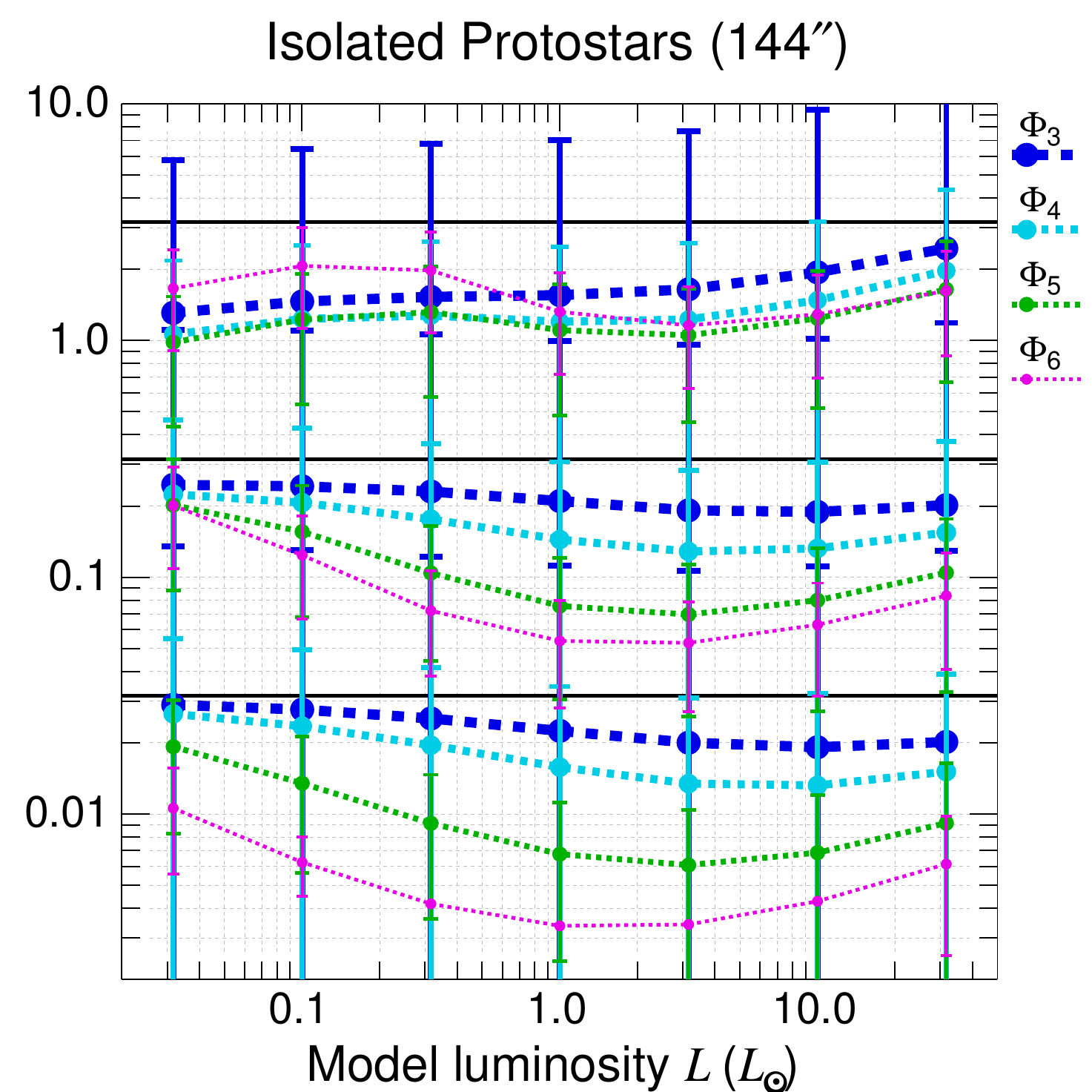}}
            \resizebox{0.3204\hsize}{!}{\includegraphics{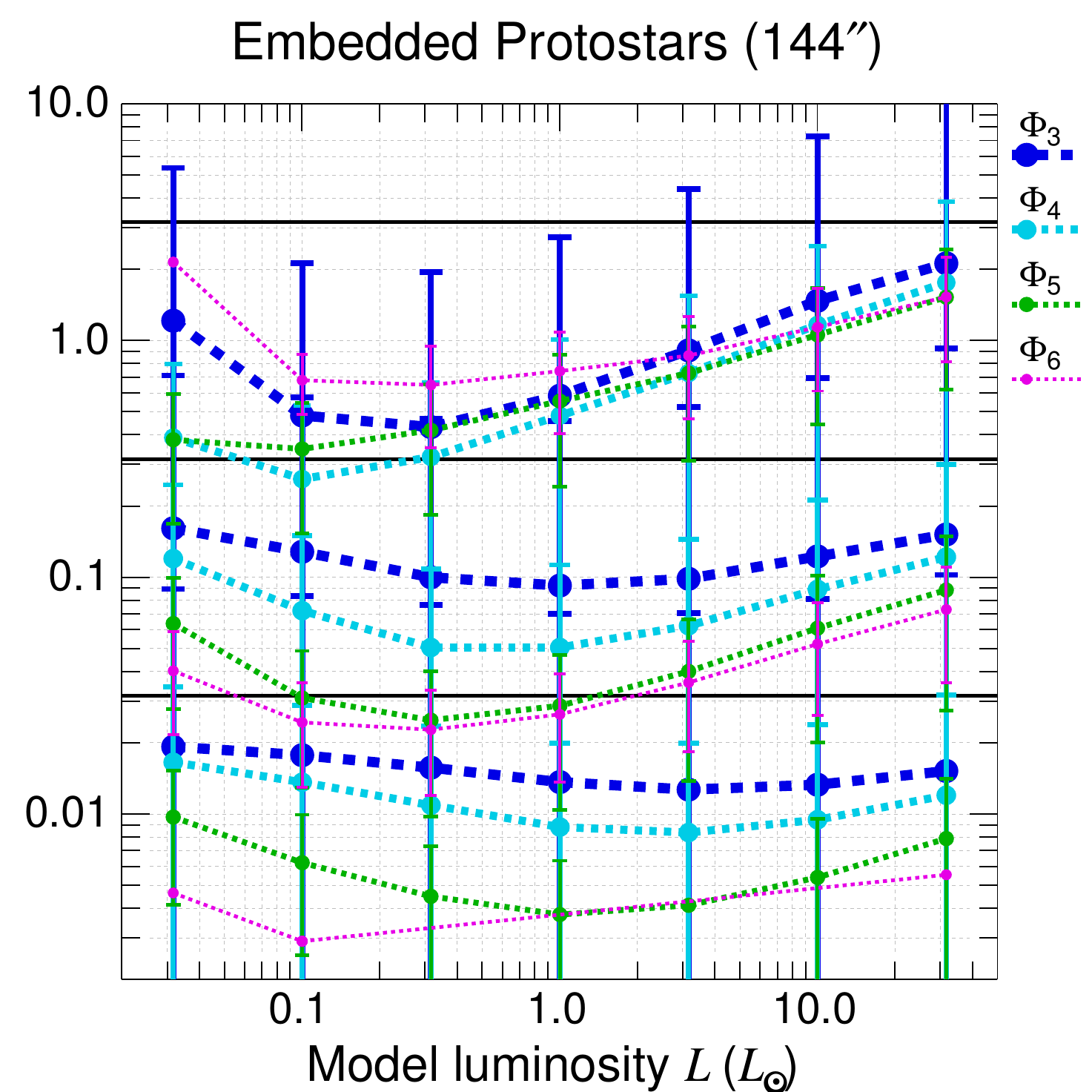}}}
\caption{
Masses $M_{\mathcal{I}}$ derived from fitting images $\mathcal{I}_{\nu}$ of the \emph{isolated} and \emph{embedded} starless cores
and protostellar envelopes (fits with free variable $\beta$). Two columns of panels (\emph{left}, \emph{middle}) display the
masses derived for the resolved and unresolved images of the isolated models, whereas the third column of panels (\emph{right})
presents results for the unresolved embedded models. Derived masses of the latter are underestimated by a factor of approximately
$1.3$ as a result of the conventional procedure of background subtraction (Sect.~\ref{bg.subtraction}). See Fig.~\ref{coldens.bes} 
for more details.
} 
\label{beta.bes.pro.coldens}
\end{figure*}

\begin{figure*}
\centering
\centerline{\resizebox{0.3216\hsize}{!}{\includegraphics{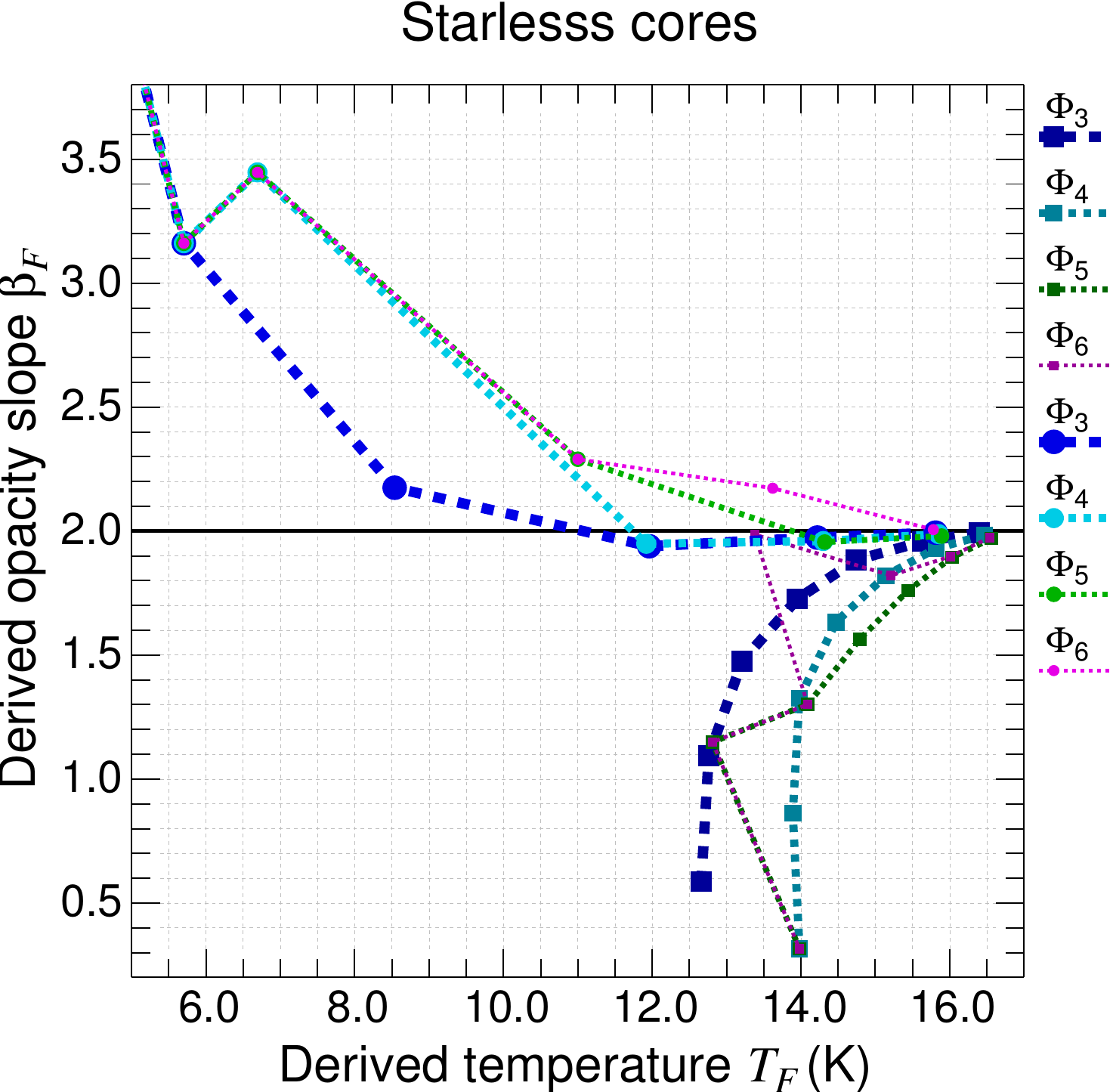}}
            \resizebox{0.3204\hsize}{!}{\includegraphics{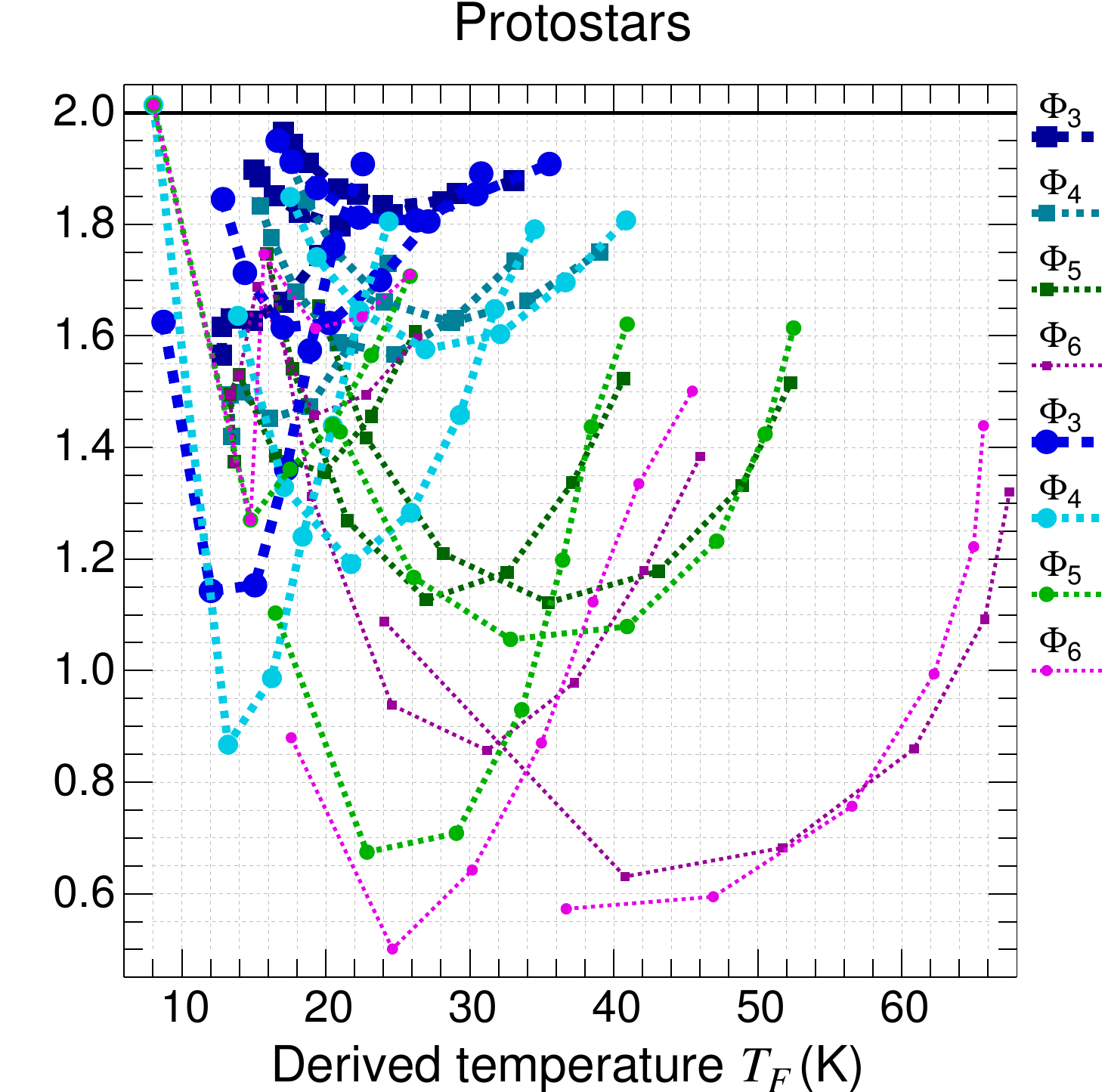}}
            \resizebox{0.3204\hsize}{!}{\includegraphics{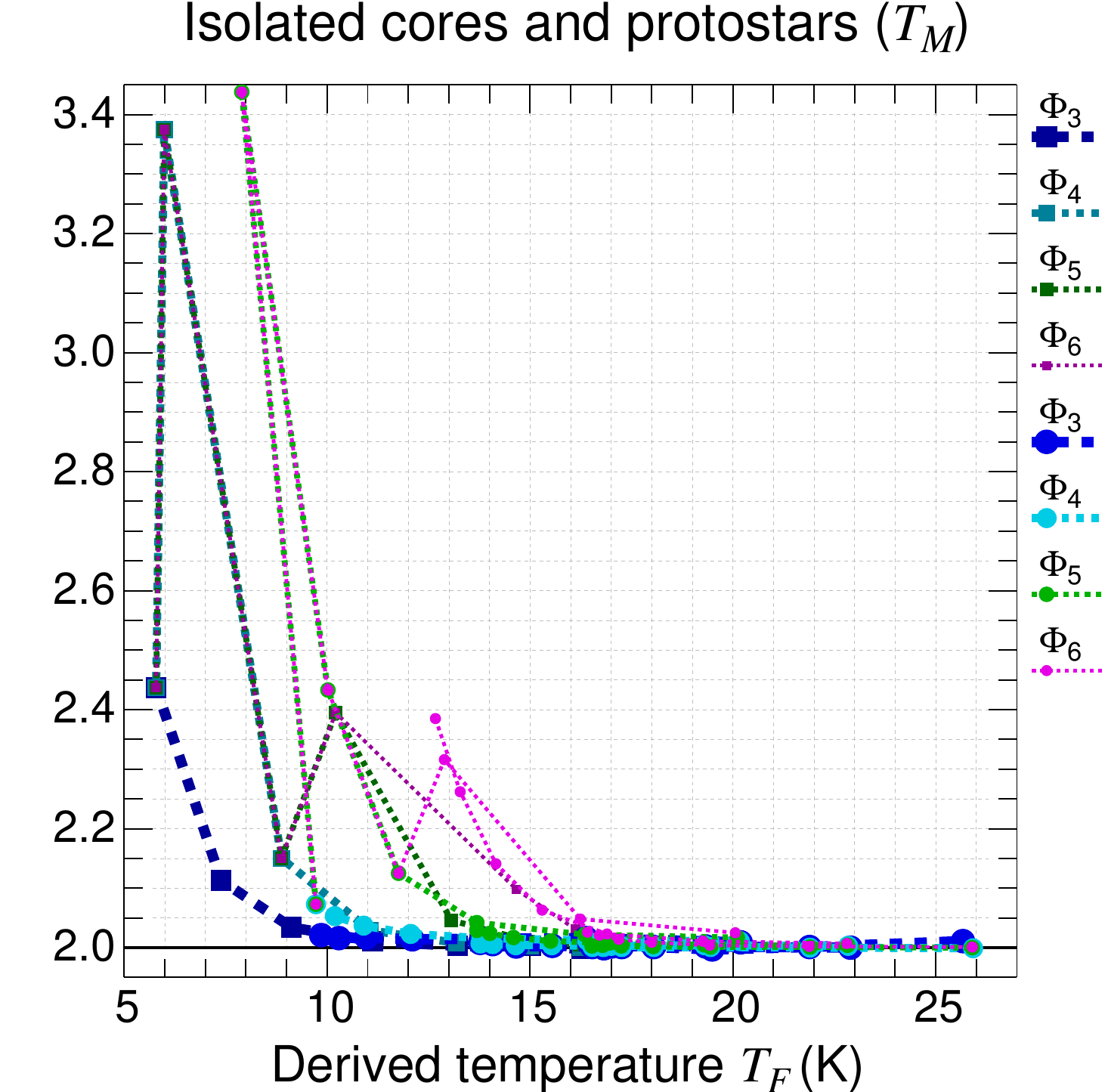}}}
\caption{
Relationships between $T_{F}$ and $\beta_{F}$ derived from fitting fluxes $F_{\nu}$ of starless cores and protostellar envelopes 
(fits with free variable $\beta$). Curves plotted with squares (dark colors) and circles (bright colors) in the left and middle 
panels correspond to the \emph{isolated} and \emph{embedded} models, whereas in the right panel they correspond to the 
\emph{isothermal} versions of the \emph{isolated} starless cores and protostellar envelopes, respectively. Results for the starless 
cores are plotted for all masses $0.03{-}30\,M_{\sun}$. Results for the protostellar envelopes, shown for only selected masses 
($0.03, 0.3$, and $3\,M_{\sun}$, three sets of identical lines), span the entire range of luminosities $0.03{-}30\,L_{\sun}$. Thick 
horizontal lines indicate the true value $\beta\,{=}\,2$. For clarity of the plots, error bars are not shown.
} 
\label{tem.beta.bes.pro}
\end{figure*}

In some applications of the mass derivation methods, the opacity slope $\beta$ has been allowed to vary along with the other
fitting parameters ($T$, $M$, $\Omega$). To quantify effects of the extra degree of freedom, additional fits with variable $\beta$
were performed in this study. Although the parameters for both fitting models were derived and analyzed, only the much less
incorrect \textsl{thinbody} results are presented and discussed here.

Figure~\ref{beta.bes.pro} compares derived $T_{F}$ and $M_{F}$ of the isolated, embedded, and isothermal variants of starless cores
and protostellar envelopes with the true model values ($T_{M}$, $M$). Although the isolated cores and envelopes display behavior
that is qualitatively similar to the $\beta\,{=}\,2$ case (Sect.~\ref{derived.properties}), the biases in $T_{F}$ and $M_{F}$
towards denser (more massive) cores and envelopes, as well as over $L_{\star}$ for the latter, become much stronger. For example,
for the isolated starless cores with $M\,{\ga}\,10\,M_{\sun}$, derived $T_{F}$ are overestimated by a factor of $2$, whereas
$M_{F}$ are underestimated by a factor of $15$. For the embedded protostellar envelopes of $M\,{=}\,3\,M_{\sun}$ with
$L_{\star}\,{\la}\,1\,L_{\sun}$, temperatures are overestimated by a similarly large factor and $M_{F}$ underestimated by a factor
of $8$.

Such errors are caused by the derived $\beta_{F}$ whose values for starless cores are systematically lowered towards higher mass
models, and are underestimated by a factor of $4$ (Fig.~\ref{beta.beta}). In the case of the embedded protostellar envelopes, the
values of $\beta_{F}$ are progressively underestimated towards lower luminosities, up to a factor of $2$ (Fig.~\ref{beta.beta}).
The very large errors in $\beta_{F}$ are, in turn, caused by the temperature excesses over $T_{M}$ (Sect.~\ref{nonuniform.temps}),
which is highlighted in Fig.~\ref{beta.bes.pro} by the accurate results for the isothermal models. As in the fixed $\beta$ case,
errors in derived parameters for the embedded starless cores are smaller than those for the isolated cores, whereas the behavior is
opposite for the embedded protostellar envelopes (cf. Figs.~\ref{temp.mass.bes}, \ref{temp.mass.pro}). However, the biases over the
masses and luminosities in Fig.~\ref{beta.bes.pro} become stronger than in the fixed $\beta$ case, which again is attributed to the
additional biases in the derived $\beta_{F}$ values (Fig.~\ref{beta.beta}).

The method of fitting images $\mathcal{I}_{\nu}$ delivers results that are similar to those described above, for both isolated and
embedded variants (Fig.~\ref{beta.bes.pro.coldens}) of partially resolved and unresolved objects. Derived masses for the
fully resolved objects are much more accurate and they do not depend on the subset of data points $\Phi_{n}$, for the reasons
discussed in Sects.~\ref{coldens.properties} and \ref{the.methods}. With degrading angular resolutions, accuracy of the estimated
parameters deteriorates to the levels obtained from the method of fitting total fluxes $F_{\nu}$ (Fig.~\ref{beta.bes.pro}). Larger
beams heavily blend emission with nonuniform temperatures from different pixels, distorting their spectral distribution towards
shorter wavelengths. In both methods, derived masses become systematically much less accurate (greatly underestimated) when fitting
larger subsets $\Phi_{n}$ ($n\,{=}\,3\,{\rightarrow}\,6$). However, even the smallest subsets $\Phi_{3}$ show very significant
inaccuracies and different biases that depend on the mass and luminosity of an object.

The above results obtained with free fitting parameter $\beta$ can be compared with those from the previous studies focused on the
relationship between derived temperature and opacity slope \citep[cf.][and references
therein]{Shetty_etal2009a,Shetty_etal2009b,JuvelaYsard2012a}. Figure~\ref{tem.beta.bes.pro} shows the intrinsic dependencies
between $\beta_{F}$ and $T_{F}$ for the isolated and embedded starless cores and protostellar envelopes, and for the isothermal
versions of the isolated models. The correlations between $\beta_{F}$ and $T_{F}$ take variety of shapes, from a strongly positive
to a strongly negative correlation, with practically no correlation for the isothermal models. This suggests that they must be
caused by different kinds of deviations of the spectral shapes of $F_{\nu}$ (Fig.~\ref{sed.bes.pro}) produced by the nonuniform
$T_{\rm d}(r)$ (Fig.~\ref{trp.bes.pro}) from $F_{\nu}(T_{M})$ (Fig.~\ref{sed.tmav.bes.pro}). Smaller subsets $\Phi_{n}$
($n\,{=}\,6\,{\rightarrow}\,3$) bring less correlated $\beta_{F}$ and $T_{F}$ than the large subsets do. For the protostellar
envelopes, the correlations are non-monotonic and they may either be strongly negative or positive, depending on the luminosity of
the central energy source.

\end{appendix}


\bibliographystyle{aa}
\bibliography{aamnem99,masses}

\begin{thebibliography}{29}
\expandafter\ifx\csname natexlab\endcsname\relax\def\natexlab#1{#1}\fi

\bibitem[{{Alves} {et~al.}(2001){Alves}, {Lada}, \& {Lada}}]{Alves_etal2001}
{Alves}, J.~F., {Lada}, C.~J., \& {Lada}, E.~A. 2001, \nat, 409, 159

\bibitem[{{Andr{\'e}} {et~al.}(2014){Andr{\'e}}, {Di Francesco},
  {Ward-Thompson}, {Inutsuka}, {Pudritz}, \& {Pineda}}]{Andre_etal2014}
{Andr{\'e}}, P., {Di Francesco}, J., {Ward-Thompson}, D., {et~al.} 2014, in
  Protostars and Planets VI, ed. {H.~Beuther, R.~S.~Klessen, C.~P.~Dullemond,
  \& Th.~Henning}, Space Science Series, 27--51

\bibitem[{{Bernard} {et~al.}(1992){Bernard}, {Boulanger}, {Desert}, \&
  {Puget}}]{Bernard_etal1992}
{Bernard}, J.~P., {Boulanger}, F., {Desert}, F.~X., \& {Puget}, J.~L. 1992,
  \aap, 263, 258

\bibitem[{{Black}(1994)}]{Black1994}
{Black}, J.~H. 1994, in Astronomical Society of the Pacific Conference Series,
  Vol.~58, The First Symposium on the Infrared Cirrus and Diffuse Interstellar
  Clouds, ed. R.~M. {Cutri} \& W.~B. {Latter}, 355

\bibitem[{{Bonnor}(1956)}]{Bonnor1956}
{Bonnor}, W.~B. 1956, \mnras, 116, 351

\bibitem[{{Desert} {et~al.}(1990){Desert}, {Boulanger}, \&
  {Puget}}]{Desert_etal1990}
{Desert}, F.-X., {Boulanger}, F., \& {Puget}, J.~L. 1990, \aap, 237, 215

\bibitem[{{Evans} {et~al.}(2001){Evans}, {Rawlings}, {Shirley}, \&
  {Mundy}}]{Evans_etal2001}
{Evans}, II, N.~J., {Rawlings}, J.~M.~C., {Shirley}, Y.~L., \& {Mundy}, L.~G.
  2001, \apj, 557, 193

\bibitem[{{Hildebrand}(1983)}]{Hildebrand1983}
{Hildebrand}, R.~H. 1983, \qjras, 24, 267

\bibitem[{{Juvela} \& {Ysard}(2012)}]{JuvelaYsard2012a}
{Juvela}, M. \& {Ysard}, N. 2012, \aap, 539, A71

\bibitem[{{Kelly} {et~al.}(2012){Kelly}, {Shetty}, {Stutz}, {Kauffmann},
  {Goodman}, \& {Launhardt}}]{Kelly_etal2012}
{Kelly}, B.~C., {Shetty}, R., {Stutz}, A.~M., {et~al.} 2012, \apj, 752, 55

\bibitem[{{K{\"o}nyves} {et~al.}(2015){K{\"o}nyves}, {Andr{\'e}},
  {Men'shchikov}, {Palmeirim}, {Arzoumanian}, {Schneider}, {Roy}, {Didelon},
  {Maury}, {Shimajiri}, {Di Francesco}, {Bontemps}, {Peretto}, {Benedettini},
  {Bernard}, {Elia}, {Griffin}, {Hill}, {Kirk}, {Ladjelate}, {Marsh}, {Martin},
  {Motte}, {Nguy{\^e}n Luong}, {Pezzuto}, {Roussel}, {Rygl}, {Sadavoy},
  {Schisano}, {Spinoglio}, {Ward-Thompson}, \& {White}}]{Ko"nyves_etal2015}
{K{\"o}nyves}, V., {Andr{\'e}}, P., {Men'shchikov}, A., {et~al.} 2015, \aap,
  584, A91

\bibitem[{{Larson}(1969)}]{Larson1969}
{Larson}, R.~B. 1969, \mnras, 145, 271

\bibitem[{{Malinen} {et~al.}(2011){Malinen}, {Juvela}, {Collins}, {Lunttila},
  \& {Padoan}}]{Malinen_etal2011}
{Malinen}, J., {Juvela}, M., {Collins}, D.~C., {Lunttila}, T., \& {Padoan}, P.
  2011, \aap, 530, A101

\bibitem[{{Men'shchikov}(2013)}]{Men'shchikov2013}
{Men'shchikov}, A. 2013, \aap, 560, A63

\bibitem[{{Men'shchikov} {et~al.}(2010){Men'shchikov}, {Andr{\'e}}, {Didelon},
  {K{\"o}nyves}, {Schneider}, {Motte}, {Bontemps}, {Arzoumanian}, {Attard},
  {Abergel}, {Baluteau}, {Bernard}, {Cambr{\'e}sy}, {Cox}, {di Francesco}, {di
  Giorgio}, {Griffin}, {Hargrave}, {Huang}, {Kirk}, {Li}, {Martin}, {Minier},
  {Miville-Desch{\^e}nes}, {Molinari}, {Olofsson}, {Pezzuto}, {Roussel},
  {Russeil}, {Saraceno}, {Sauvage}, {Sibthorpe}, {Spinoglio}, {Testi},
  {Ward-Thompson}, {White}, {Wilson}, {Woodcraft}, \&
  {Zavagno}}]{Men'shchikov_etal2010}
{Men'shchikov}, A., {Andr{\'e}}, P., {Didelon}, P., {et~al.} 2010, \aap, 518,
  L103+

\bibitem[{{Men'shchikov} {et~al.}(2012){Men'shchikov}, {Andr{\'e}}, {Didelon},
  {Motte}, {Hennemann}, \& {Schneider}}]{Men'shchikov_etal2012}
{Men'shchikov}, A., {Andr{\'e}}, P., {Didelon}, P., {et~al.} 2012, \aap, 542,
  A81

\bibitem[{{Men'shchikov} \& {Henning}(1997)}]{Men'shchikovHenning1997}
{Men'shchikov}, A.~B. \& {Henning}, T. 1997, \aap, 318, 879

\bibitem[{{Men'shchikov} {et~al.}(1999){Men'shchikov}, {Henning}, \&
  {Fischer}}]{Men'shchikov_etal1999}
{Men'shchikov}, A.~B., {Henning}, T., \& {Fischer}, O. 1999, \apj, 519, 257

\bibitem[{{Ossenkopf} \& {Henning}(1994)}]{OssenkopfHenning1994}
{Ossenkopf}, V. \& {Henning}, T. 1994, \aap, 291, 943

\bibitem[{{Parravano} {et~al.}(2003){Parravano}, {Hollenbach}, \&
  {McKee}}]{Parravano_etal2003}
{Parravano}, A., {Hollenbach}, D.~J., \& {McKee}, C.~F. 2003, \apj, 584, 797

\bibitem[{{Pilbratt} {et~al.}(2010){Pilbratt}, {Riedinger}, {Passvogel},
  {Crone}, {Doyle}, {Gageur}, {Heras}, {Jewell}, {Metcalfe}, {Ott}, \&
  {Schmidt}}]{Pilbratt_etal2010}
{Pilbratt}, G.~L., {Riedinger}, J.~R., {Passvogel}, T., {et~al.} 2010, \aap,
  518, L1+

\bibitem[{{Press} {et~al.}(1992){Press}, {Teukolsky}, {Vetterling}, \&
  {Flannery}}]{Press_etal1992}
{Press}, W.~H., {Teukolsky}, S.~A., {Vetterling}, W.~T., \& {Flannery}, B.~P.
  1992, {Nu\-merical recipes in FORTRAN. The art of scientific computing}
  ({Cambridge University Press, 2nd ed.})

\bibitem[{{Roy} {et~al.}(2014){Roy}, {Andr{\'e}}, {Palmeirim}, {Attard},
  {K{\"o}nyves}, {Schneider}, {Peretto}, {Men'shchikov}, {Ward-Thompson},
  {Kirk}, {Griffin}, {Marsh}, {Abergel}, {Arzoumanian}, {Benedettini}, {Hill},
  {Motte}, {Nguyen Luong}, {Pezzuto}, {Rivera-Ingraham}, {Roussel}, {Rygl},
  {Spinoglio}, {Stamatellos}, \& {White}}]{Roy_etal2014}
{Roy}, A., {Andr{\'e}}, P., {Palmeirim}, P., {et~al.} 2014, \aap, 562, A138

\bibitem[{{Shetty} {et~al.}(2009{\natexlab{a}}){Shetty}, {Kauffmann}, {Schnee},
  \& {Goodman}}]{Shetty_etal2009a}
{Shetty}, R., {Kauffmann}, J., {Schnee}, S., \& {Goodman}, A.~A.
  2009{\natexlab{a}}, \apj, 696, 676

\bibitem[{{Shetty} {et~al.}(2009{\natexlab{b}}){Shetty}, {Kauffmann}, {Schnee},
  {Goodman}, \& {Ercolano}}]{Shetty_etal2009b}
{Shetty}, R., {Kauffmann}, J., {Schnee}, S., {Goodman}, A.~A., \& {Ercolano},
  B. 2009{\natexlab{b}}, \apj, 696, 2234

\bibitem[{{Shu}(1977)}]{Shu_1977}
{Shu}, F.~H. 1977, \apj, 214, 488

\bibitem[{{Siebenmorgen} {et~al.}(1992){Siebenmorgen}, {Kruegel}, \&
  {Mathis}}]{Siebenmorgen_etal1992}
{Siebenmorgen}, R., {Kruegel}, E., \& {Mathis}, J.~S. 1992, \aap, 266, 501

\bibitem[{{Stamatellos} {et~al.}(2004){Stamatellos}, {Whitworth}, {Andr{\'e}},
  \& {Ward-Thompson}}]{Stamatellos_etal2004}
{Stamatellos}, D., {Whitworth}, A.~P., {Andr{\'e}}, P., \& {Ward-Thompson}, D.
  2004, \aap, 420, 1009

\bibitem[{{Wolf}(2003)}]{Wolf2003}
{Wolf}, S. 2003, Computer Physics Communications, 150, 99

\end{thebibliography}

\end{document}